%% file: 0.Mainfile.tex
\documentclass[a4paper,12pt]{report}
\usepackage{Stylefile}

\onehalfspace

\title
 {
\vspace*{0.0cm} \Large{\bf High Throughput Software for Powder Diffraction and its Application to
Heterogeneous Catalysis
} \vspace*{3.0cm} \\
\Large{\bf Taha Sochi} \vspace*{4.0cm} \\
\large{\bf  A dissertation submitted to \\
            the Department of Crystallography \\
            Birkbeck College London \\
            in fulfillment of the requirements for the degree of \\
            Doctor of Philosophy}
 }

\date{2010}

\ifpdf
\fi

\makeindex

\begin{document}

\maketitle %


\phantomsection \addcontentsline{toc}{section}{Declaration}
\include{Decl}

\phantomsection \addcontentsline{toc}{section}{Abstract}
\include{Abst}

\phantomsection \addcontentsline{toc}{section}{Acknowledgements}
\include{Ackn}

\phantomsection \addcontentsline{toc}{section}{Contents} %
\tableofcontents

\newpage
\phantomsection \addcontentsline{toc}{section}{List of Figures} %
\listoffigures

\newpage
\phantomsection \addcontentsline{toc}{section}{List of Tables} %
\listoftables

\newpage
\phantomsection \addcontentsline{toc}{section}{Nomenclature}
\include{Nome}


{\setlength{\parskip}{6pt plus 2pt minus 1pt}

\pagestyle{headings} %
\addtolength{\headheight}{+1.6pt}
\lhead[{Chapter \thechapter \thepage}]%
      {{\bfseries\rightmark}}
\rhead[{\bfseries\leftmark}]%
     {{\bfseries\thepage}} 
\headsep = 1.0cm               


\include{Intr}

\include{Soft}

\include{Manu}

\include{Cata}

\include{Nick}
\include{Prob}

\include{Conc}

} 

\phantomsection \addcontentsline{toc}{chapter}{\protect \numberline{} Bibliography} %
\bibliographystyle{unsrt}    


\include{Bibl}
%


\appendix

\include{Samp}

\voffset = 0.0cm
\phantomsection \addcontentsline{toc}{chapter}{\protect \numberline{} Index} %
\printindex


\end{document}

%% file: Decl.tex
%
%
\noindent
{\LARGE \bf \vspace{1.0cm} \\ Declaration} \vspace{0.5cm}\\
\begin{spacing}{2}
\noindent
The work presented in this thesis is the result of my own effort, except where otherwise stated.
All sources are acknowledged by explicit reference.

\noindent Taha Sochi .........................

\end{spacing}

%% file: Abst.tex
%
%
\noindent
{\LARGE \bf \vspace{1.0cm} \\ Abstract} \vspace{0.5cm}\\
\begin{spacing}{1.5}
\noindent
In this thesis we investigate high throughput computational methods for processing large quantities
of data collected from \synrad\ facilities and its application to spectral analysis of powder
diffraction data. We also present the main product of this PhD programme, specifically a computer
software called `\ProgName' developed by the author as part of this project. This software was
created to meet the increasing demand on data processing and analysis capabilities as required by
modern detectors which produce huge quantities of data.

The principal objective of \ProgName\ project was to develop a computer code for visualisation,
batch fitting and bulk analysis of large volumes of \diffraction\ data, mainly those obtained from
\synrad\ sources. Such a tool greatly assists studies on various materials systems and enables far
larger detailed data sets to be rapidly interrogated and analysed. Modern detectors coupled with
the high intensity \Xray\ sources available at \synch\ facilities have led to the situation where
data sets can be collected in ever shorter time scales and in ever larger numbers. Such large
volumes of data sets pose a data processing bottleneck which is set to augment with current and
future instrument development. \ProgName\ has achieved its main objectives and made significant
contributions to scientific research. Some of its functionalities have even surpassed the original
expectations. More importantly, it can be used as a model for more mature attempts in the future.

\ProgName\ is currently in use by a number of researchers in a number of academic and research
institutions, such as University College London and Rutherford-Appleton Laboratory, to process
high-energy \diffraction\ data for various purposes. These include data collected by different
techniques such as \edd\ (EDD), \add\ (ADD) and \CATl\ (CAT). \ProgName\ has already been used in a
number of studies such as Lazzari \etal\ \cite{LazzariJSB2009} and Espinosa-Alonso \etal\
\cite{EspinosaOJBJe2009}. It is also in use by the \HEXITECl\ (\HEXITECs) project \cite{HEXITEC}
and has been commended by the scientists who work on the development of HEXITEC detectors. The
program proved to be efficient and time saving, and hence has met its main objectives in saving
valuable resources and enabling rapid data processing and analysis.

The software was also used by the author to process and analyse data sets collected from
synchrotron radiation facilities. In this regard, the thesis will present novel scientific research
involving the use of \ProgName\ to handle large \diffraction\ data sets in the study of
\alumina-supported metal oxide \catalyst\ bodies. These data were collected using \TEDDIl\ (TEDDI)
and \CATl\ techniques.

\end{spacing}

%% file: Ackn.tex
%
%
\noindent
{\LARGE \bf \vspace{3.0cm} \\ Acknowledgements} \vspace{0.5cm}\\
\begin{spacing}{1.5}
\noindent %
I would like to thank

\begin{itemize}

\item
My supervisors Professor Paul Barnes and Dr Simon Jacques for their kindness, help and advice and
for offering me a scholarship to study crystallography at Birkbeck College London. Dr Simon Jacques
should be accredited for providing some of the ideas of the software developed in the course of
this PhD, namely \ProgName. 

\item
The Engineering and Physical Sciences Research Council (EPSRC) for funding this research.

\item
The Chemistry Department of the University College London and the Crystallography Department of the
Birkbeck College London for in-house support.

\item
The external examiner Professor Mike Glazer from the University of Oxford, and the internal
examiner Dr Dewi Lewis from the University College London for their constructive remarks and
recommendations which improved the quality of this work.

\end{itemize}

\end{spacing}


%

%% file: Nome.tex
%
%
\noindent \vspace{1.0cm} \\
{\LARGE \bf Nomenclature}
\vspace{0.7cm} 

\begin{longtable}{ll}


$\beta$             &       peak broadening due to crystallite size \\
$\SA$               &       \diffraction\ angle \\
$2\SA$              &       \scaang \\
$\kappa$            &       crystallite shape factor \\
$\WL$               &       wavelength \\
$\SD$               &       standard deviation \\
$\VAR$              &       variance \\
$\tau$              &       mean size of crystallites \\
$\GoF$              &       goodness-of-fit index \\

                    &   \vspace{-0.2cm}\\

$\Area$             &       area under peak \\
$c$                 &       speed of light in vacuum (299\,792\,458 m.s$^{-1}$) \\
$\NC$               &       number of constraints \\
$d$                 &       \crystal\ lattice planar spacing \\
$E$                 &       energy \\
$f$                 &       frequency \\
$\FWHM$             &       Full Width at Half Maximum \\
$h$                 &       Planck's constant ($6.6260693\times10^{-34}$ J.s) \\
$\hkl$              &       \Miller \\
$\iu$               &       imaginary unit ($\sqrt{-1}$) \\
$\II$               &       integrated intensity \\
$\Ical$             &       calculated integrated intensity \\
$\Iobs$             &       observed integrated intensity \\
$\funcMin$          &       second norm of residuals \\
$\weiFac$           &       \mixfac\ in \pseVoigt\ function \\
$n$                 &       \diffraction\ order number \\
$\NO$               &       number of observations \\
$\NP$               &       number of parameters \\
$\QF$               &       quality factor \\
$r$                 &       radius \\
$\RB$               &       Bragg's residual \\
$\ER$               &       expected residual \\
$\Rp$               &       profile residual \\
$\SFR$              &       \strfac\ residual \\
$\WPR$              &       weighted profile residual \\
$T$                 &       temperature \\
$\wei$              &       statistical weight \\
$\Pos$              &       position of peak \\
$xyz$               &       spatial coordinates \\
$\y$                &       count rate (intensity) \\
$\ybac$             &       background count rate \\
$\ycal$             &       calculated count rate \\
$\yobs$             &       observed count rate \\

                    &   \vspace{-0.2cm}\\

1D                  &       one-dimensional \\
2D                  &       two-dimensional \\
3D                  &       three-dimensional \\
\ADDs               &       Angle Dispersive Diffraction \\
ASCII               &       American Standard Code for Information Interchange \\
{\AA}               &       angstrom ($10^{-10}$m) \\
$^{\circ}$C         &       degrees Celsius (Centigrade) \\
\CATs               &       \CATl \\
\CCDs               &       \CCDl \\
\CCPs               &       \CCPl \\
DFT                 &       Discrete Fourier Transform \\
\EDDs               &       \EDDl \\
EDF                 &       European Data Format \\
en                  &       ethylenediamine [C$_2$H$_4$(NH$_2$)$_2$] \\
ERD                 &       Energy Resolving Detector \\
\ESRFs              &       \ESRFl\ (Grenoble, France) \\
eV                  &       electron Volt \\
FCC                 &       Face Centred Cubic \\
FFT                 &       Fast Fourier Transform \\
FoM                 &       Function(s) of Merit or Figure(s) of Merit \\
FTP                 &       File Transport Protocol \\
FWHM                &       Full Width at Half Maximum \\
GUI                 &       Graphic User Interface \\
HCP                 &       Hexagonal Close Packed \\
\HEXITECs           &       \HEXITECl\ \\
\ICDDs              &       \ICDDl \\
ICSD                &       Inorganic Crystal Structure Database \\
keV                 &       kilo electron Volt \\
LS                  &       \LSl\ \\
MCA                 &       \MCAl \\
$\mu$m              &       micrometre \\
mm                  &       millimetre \\
nm                  &       nanometre \\
\PDFs               &       \PDFl \\
PDP                 &       Phase Distribution Pattern \\
PEEK                &       Polyether ether ketone \\
\RALs               &       \RALl\ (Didcot, UK) \\
SR                  &       \SynRad\ \\
\SRSs               &       \SRSl\ (Daresbury, UK) \\
\TADDIs             &       \TADDIl \\
\TEDDIs             &       \TEDDIl \\

\end{longtable}

\vspace{0.5cm}

\noindent %
{\bf Note}: Some symbols may rely on the context for unambiguous identification. The physical
constants are obtained from the National Institute of Standards and Technology (NIST) website
\cite{NIST2007}.


%

%% file: Intr.tex
%
%
\chapter{Introduction} \label{Introduction}

In this introduction we present some background materials to outline various aspects related to the
work in this project. These include \pd, radiation sources, data acquisition techniques, and data
analysis and information extraction.

\section{\PDl}

\Pd\ is a powerful and versatile technique for probing the structure and properties of materials.
It is therefore widely used in various fields of science and technology. Its enormous applications
include crystallographic phase analysis, \texture\ and \strain\ examination, and determination of
electronic radial distribution functions. Powder diffraction is used by scientists from various
disciplines as a powerful research technique. The method was originally devised by Debye and
Scherrer in 1916. Important stages in the development of \pd\ are presented in Table
\ref{PDMilestoneTab}.

\begin{table} [!t]
\centering %
\caption{Milestones in the evolution of \pd\ technique.}
\label{PDMilestoneTab} %
\vspace{0.5cm} %

{\normalsize

\begin{tabular}{|l|c|l|}
\hline
Contributor \verb|               | & Year & Progress \\
\hline
R\"{o}ntgen & 1895 & Discovery of \Xray\ radiation \\
Friedrich-Knipping-\Laue\ & 1912 & First \Xray\ \diffraction\ experiment \\
Braggs & 1913 & Formulation of Bragg's law \\
Debye-Scherrer$^*$ & 1916 & First \pd\ experiment \\
Chadwick & 1932 & Discovery of neutron \\
Wollan & 1945 & First neutron \diffraction\ experiment \\
\Rietveld\ & 1967 & Whole pattern refinement \\
SRS & 1980 & First dedicated SR source \\
\hline
\end{tabular}
} %
{\footnotesize *Friedrich, Knipping and \Laue\ may have also conducted powder diffraction
experiments.}
\end{table}

The advent of \whpare\ method by \Rietveld\ in 1967, associated with the computer revolution and
the availability of \synrad, have contributed to the revival and widespread use of this technique
in the last few decades. This opened the door to \abinitio\ structure determination from
polycrystalline samples and resulted in \pd\ becoming the technique of choice, for providing vital
structural insights in various fields, with a massive impact on structural crystallography.

The types of radiation in common use for \pd\ are X-rays and neutrons, where the former is more
popular due to availability, cost, and other practical and technical reasons. Electrons may also be
used for polycrystalline thin film \diffraction\ experiments. The common factor between all these
radiation types is that the radiation wavelength is comparable in magnitude to the interatomic
spacing in \crystal s \cite{BookSchwarzenbach1996}.

Crystalline materials may not necessarily be single crystals but can be made up of a huge number of
tiny ($\mu$m to nm) single crystallites. This type of material is referred to as a polycrystalline
aggregate or powder. In \pd\ the term `powder' may not have the same sense as in common language,
as the powder may be a sample of solid or even liquid substance. The main property for the `powder'
in \pd\ to satisfy is that the sample should be a large collection of very small crystallites,
randomly oriented in all directions. The great majority of the natural and synthetic \crystalline\
materials are polycrystalline aggregates \cite{BookSchwarzenbach1996}.

In most cases, an ideal specimen for \pd\ work consists of a large amount of randomly oriented
crystallites which are small enough for the application in hand and mounted in a manner in which no
preferred crystalline orientation occurs. The crystallites must also be large enough to exhibit
scattering as though from a large single \crystal. When these conditions are satisfied, all
possible orientations are statistically represented and there will be enough crystallites in any
diffracting orientation to yield a complete \difpat. If the crystallites take up some particular
orientation, the specimen is described to be suffering from \preori. In most cases the ideal
crystallite size is between $1-10\mu$m. When the size drops below $1\mu$m the peaks broaden, while
when the particles are larger than $20\mu$m, optimum number of particles are not sampled
\cite{BookPecharskyZ2005}.

There are several reasons for using polycrystalline specimens in \diffraction\ experiments. These
reasons include (a) Non-availability of single \crystal\ in the required size and quality due to
practical or technical difficulties. In general, powders are easier to produce than single crystals
(b) Some materials are microcrystalline powders by nature and cannot be grown as single crystals.
Also, the usable form of the material can be polycrystalline (c) The properties of powder are of
interest in the particular study (d) A single \crystal\ may not survive the extreme non-ambient
conditions such as high temperature and pressure which are usually applied during \insitu\ and
dynamic phase transition studies. Under these circumstances, the powder method can be employed with
no difficulty (e) Single crystals may suffer from effects like extinction and magnetic domain
structures, making a proper interpretation of the \difpat\ unreliable. Many of these systematic
effects either do not arise in the powder method or can be easily circumvented \cite{BookWill2006}.

\Pd\ has many applications across the scientific and technological spectrum. In the following we
present some prominent examples of these applications \cite{BookDinnebierB2008, Moron2000,
MellerHJCJBP2004, BarnesCCHJJCMJOB2000, HallCJJLRAB2000, KockelmannPK2000}.

\begin{itemize}

\item Fingerprint identification: within certain limits, each \crystalline\ substance has a
unique \pd\ pattern which is mainly defined by its chemical composition and \crystal\ structure.
Therefore, the powder \difpat\ of a pure substance can be used as a fingerprint to characterise and
identify polycrystalline substances. This fingerprint can then be used to detect the presence of
material in a pure or mixed form.

\item Qualitative phase analysis: i.e. the identification of phases present in the
material. When the phases in the powder specimen are unknown, comparisons can be made with the
\difpat s of known compounds to find a match. Extensive crystallographic databases, such as Powder
Diffraction File and Cambridge Structural Database, have been developed and maintained for this
purpose. In this regard, \seamat\ computer programs are usually employed to compare experimental
patterns with those stored in these databases.

\item Quantitative phase analysis: i.e. the determination of the abundance of various \crystalline\ phases in
a multi-component mixture by the use of intensity ratios when the phases are known. When a sample
is composed of more than one \crystalline\ phase the diffraction pattern contains a weighted sum of
the \difpat s of the component phases and hence mixtures can be characterised and quantified. The
method is based on the principle of proportionality between the measured \diffraction\ intensities
and the amount of a given \crystalline\ phase in the sample, although the proportionality may be
non-linear in the general case.

\item Phase transformations: one of the most common uses of \pd\ is to monitor phase
transformations in solids in response to pressure, temperature, \stress, electric or magnetic
fields, and so forth.

\item Tomographic imaging: \pd\ can be used to obtain \diffraction\ information from volume
elements within a bulk sample and hence derive 3D images of the interiors of objects in terms of
both density and compositional variations. The availability of intense, energetic and
highly-penetrating X-rays from \synch s assist the recent advancements in tomographic imaging
techniques such as \TEDDIl\ (\TEDDIs).

\item Structure determination: the powder method can also be used for \abinitio\ determination
of \crystal\ structures from \diffraction\ data, although single \crystal\ \diffraction\ is more
suitable for this purpose. However, high quality data from high quality samples are usually
required for successful determination.

\item Probing the state of \crystalline\ materials: \pd\ patterns are commonly used to
determine the physical state of a specimen. This includes \preori\ and \texture\ analysis,
crystallite shape and size distribution, macro and micro \stress\ and \strain, thermal expansion in
\crystal\ structures, degree of \crystalline\ disorder and defects, and so on.

\item \Pd\ is also in use in many other applications such as studying particles in liquid
suspensions or polycrystalline solids, and recognition of \amorphous\ materials in partially
\crystalline\ mixtures.

\end{itemize}

\subsection{Diffraction and Bragg's Law}

`Diffraction' means bending or scattering of waves (e.g. sound, light, radio, electrons, neutrons
and X-rays) due to the occurrence of obstructions or small apertures in their path, followed by an
interference effect which results from the summation of the component waves. In some directions the
diffracted waves combine constructively because they are in phase (i.e. when the path difference is
an integral multiple of a wavelength), while in other directions they interfere destructively
because they are out of phase (i.e. when the path difference is an odd multiple of half
wavelength). This results in a \difpat\ consisting of peaks with various intensities at particular
positions. This pattern can be analysed through the use of physical models supported by
mathematical and computational tools to gain detailed information about the diffracting object.
Diffraction is a characteristic wave phenomenon that occurs on transmission or reflection. It is
well understood and can be explained by a simple wave theory. The effect is observable when the
size of the diffracting object is comparable to the wavelength of the diffracted beam. Unlike
spectroscopy which is based on inelastic scattering, \pd\ is based on elastic scattering processes
\cite{BookSchwarzenbach1996, BookRhodes2000}.

\begin{figure}[!h]
  \centering{}
  \includegraphics
  [width=0.9\textwidth]
  {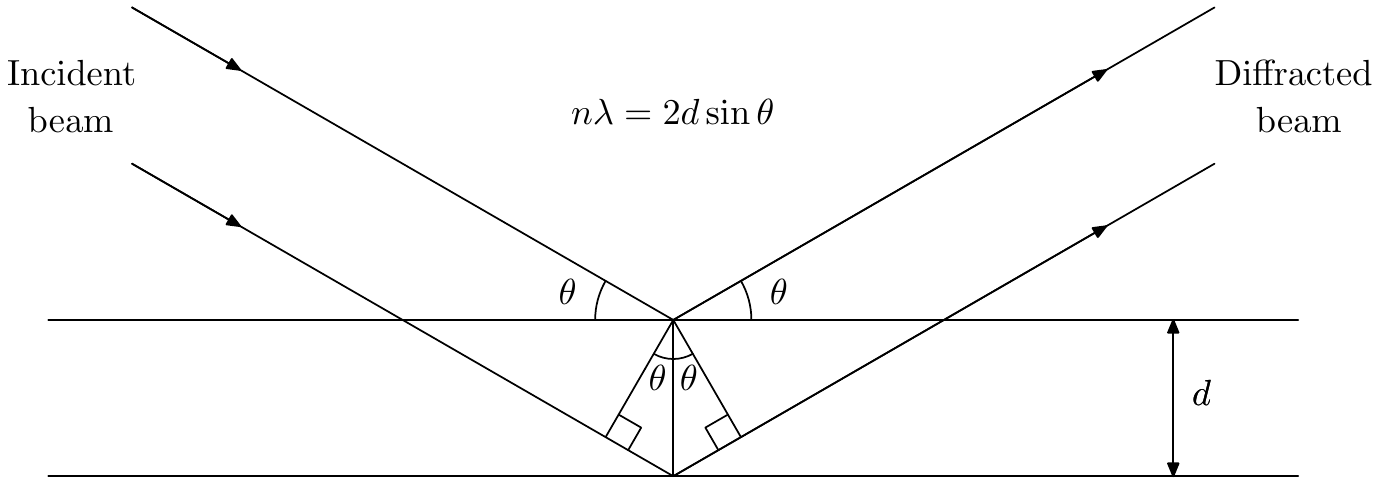}
  \caption{Schematic illustration of Bragg's Law.}
  \label{BraLaw}
\end{figure}

In the context of crystallography and \pd, when a beam of monochromatic radiation of a wavelength
comparable to the interatomic spacing (e.g. X-rays or neutrons) is directed at \crystalline\
material, \diffraction\ will be observed at particular angles with respect to the primary beam. The
relationship between the order of \diffraction, the wavelength of radiation, the interatomic
spacing, and the angle of \diffraction\ is given by the Bragg's law, which can be solved for any
one of these parameters when the other parameters are known. \Bragg's law states that

\begin{equation}\label{Bragg}
    \DO \WL = 2 \PS_{hkl} \ \sin \SA_{hkl}
\end{equation}
where $\DO$ is the order of \diffraction, $\WL$ is the wavelength of radiation, $\PS$ is the
spacing between lattice planes with \Miller\ $hkl$, and $\SA_{hkl}$ is the angle between the
incident beam and the \crystal\ planes characterised by $hkl$. Bragg's law is schematically
demonstrated in Figure \ref{BraLaw}. The \diffraction\ order $\DO$ is a positive integer which in
most cases is unity. The \Bragg\ angle $\SA$ is just half the total angle by which the incident
beam is deflected. In the case of subatomic particle radiation, such as electrons and neutrons, the
wavelength $\WL$ is given by the de Broglie relation

\begin{equation}\label{WLEq}
    \WL = \frac{\PC}{\Mom}
\end{equation}
where $\PC$ is Planck's constant and $\Mom$ is the magnitude of the particles' momentum. Bragg's
law describes the conditions for \diffraction\ in a \crystal\ and depicts the situation for
in-phase scattering by atoms lying in planes that pass through the \crystal\ lattice points. In
qualitative terms, it states that constructive wave interference will occur at certain angles
whenever the path difference of rays, reflected from different planes belonging to a particular
family of lattice planes, is an integral multiple of the wavelength. Despite its simplicity,
Bragg's law is a powerful tool for understanding and analysing \crystal\ \diffraction\ and hence it
forms the basis of crystallography. This law enables the determination of the lattice parameters to
very high accuracy \cite{BookCopley2001}.

On considering the types of radiation (i.e. monochromatic and polychromatic) with the types of
\crystalline\ sample (i.e. single \crystal\ and polycrystalline powder), four types of
\diffraction\ technique can be identified. These are presented in Table \ref{SinPowMonPol}.

\begin{table} [!h]
\centering %
\caption[The four main \diffraction\ techniques.]{The four main \diffraction\ techniques obtained
by combining the two types of radiation with the two types of \crystalline\ sample.}
\label{SinPowMonPol} %
\vspace{0.5cm} %
{\normalsize
\begin{tabular}{|l|l|l|}
\hline
 & Monochromatic Radiation & Polychromatic Radiation \\
\hline
Single Crystal & \Bragg\ & \Laue\ \\
\hline
Powder & Angle Dispersive & Energy Dispersive \\
\hline
\end{tabular}
}
\end{table}

The first mathematical formulation of \diffraction\ by a \crystal\ was developed in 1912 by Max von
\Laue, who described the phenomenon by three simultaneous equations involving a vector dot product
between the three lattice vectors and a vector perpendicular to the reflecting planes. Although the
\Laue\ equations provide a rigorous treatment for the geometry of \diffraction, the concept is not
easy to grasp. In 1913, the Brags (William Henry and his son William Lawrence) developed the
aforementioned formulation which is a much simpler way for depicting \diffraction\ from a \crystal.
They used single crystals in the reflection geometry to analyse the intensity and wavelengths of
X-ray \difpat s generated by different materials. \Laue\ \diffraction\ is based on a stationary
single \crystal\ sample being exposed to a polychromatic beam, so each reflection in the \difpat\
derives from a different wavelength. The \Laue\ technique is useful for determining the orientation
of single crystals and for rapid collection of large amounts of data. Bragg's model treats
\diffraction\ as a reflection of the incident beam from the lattice planes, similar to the
reflection of visible light from a mirror. Although in reality X-rays and neutrons are not
reflected from the lattice imaginary planes, Bragg's model provides a convenient tool for assessing
\diffraction\ phenomena and hence is the more widely used in crystallography. It should be remarked
that the \Bragg\ condition for \diffraction\ is equivalent to the simultaneous solution of the
three \Laue\ equations for a monochromatic radiation \cite{BookSchwarzenbach1996, BookRhodes2000}.

As seen already, \Bragg's analysis treats \diffraction\ as a simple reflection at the lattice
planes. This simplified analysis may be described as kinematical or geometrical theory. The theory
is based on an implicit assumption that the incident radiation is scattered only once with
negligible interaction with the \crystal\ atoms. However, the X-rays penetrate deep inside the
material where additional reflections occur at thousands of consecutive parallel planes, and hence
superposition of the scattered rays occurs. As the incident wave propagates down into the \crystal\
its amplitude diminishes, since a small fraction of the energy is reflected at each atomic plane.
Moreover, the reflected beam can be re-scattered into the direction of the incident beam before it
leaves the \crystal\ by multiple reflection. This means that the atoms of the \crystal\ are not
uniformly illuminated and hence they radiate under different conditions. The theory that takes
consideration of these factors in the analysis is described as dynamical diffraction theory. Though
this more elaborate theory is not needed in most cases, it provides a more rigorous treatment and
should be applied in more complex situations \cite{BookAuthierLT1996}.

The \diffraction\ of incident radiation from a powder sample gives rise to a pattern characterised
by peaks at certain positions. The pattern is usually plotted as diffracted intensity (or count
rate) versus \scaang\ or versus energy of radiation. \Pd\ data may also be plotted as diffracted
intensity versus direct or reciprocal lattice spacing. Several aspects of the pattern (e.g.
position, height and width of peaks) can be analysed to obtain information about the structure of
the material and the state of the sample. The former includes unit cell parameters and atomic
positions, while the latter includes \thervib\ and \microstrain.

\subsection{Powder versus Single Crystal Diffraction}

\begin{figure}[!h]
  \centering{}
  \includegraphics
  [width=0.6\textwidth]
  {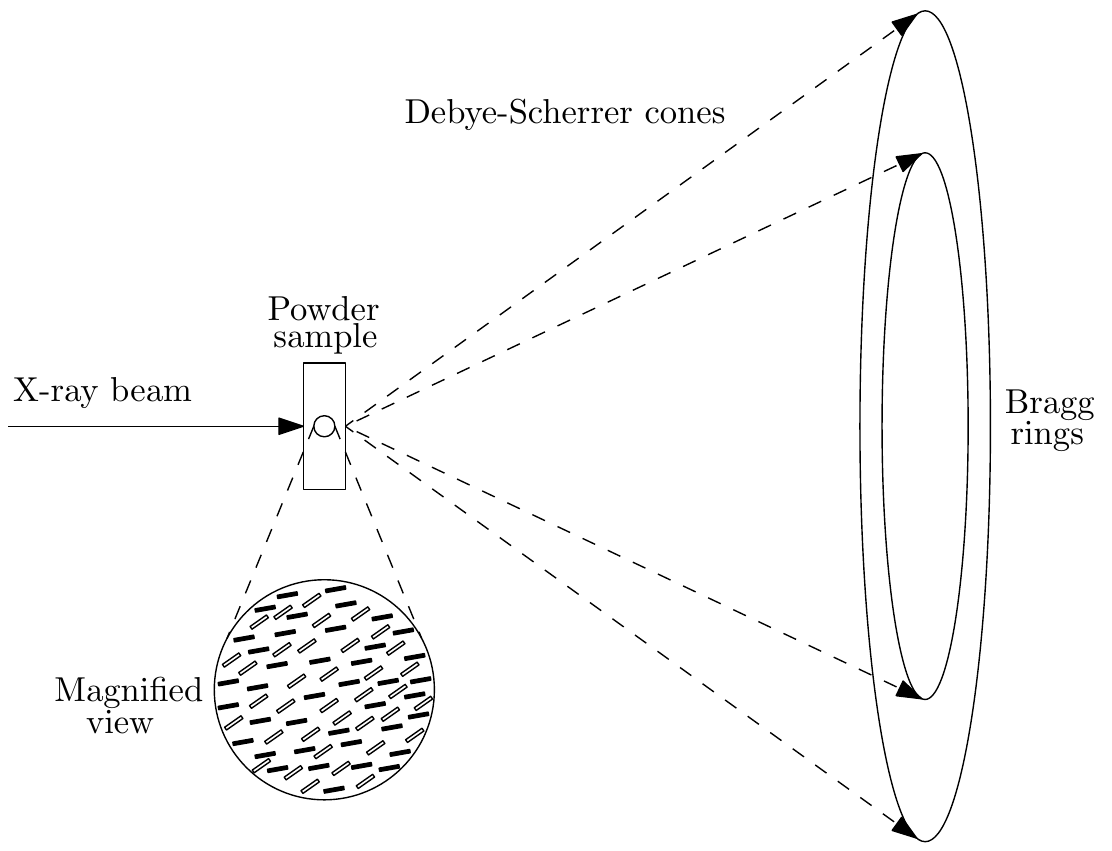}
  \caption{Debye-Scherrer cones and \Bragg\ rings.}
  \label{BraRin}
\end{figure}

The main types of radiation in use in \pd\ experiments are X-rays and neutrons. When the sample is
polycrystalline and not a single \crystal, then normally there will be a great number of
crystallites in all possible orientations. If such a sample is placed in an \Xray\ or neutron beam,
all possible interatomic planes will be seen by the beam with \diffraction\ from each family of
planes occurring only at its characteristic \diffraction\ angle. Thus, on considering all possible
\scaang s, all possible \diffraction\ peaks that can be produced from the differently orientated
crystallites in the powder will be observed. These peaks originate at the crystallites in the
centre and project outward producing cones of diffracted radiation subtending characteristic
scattering angles. Each one of these coaxial non-uniformly spaced cones (called Debye-Scherrer
cones) is associated with a single set of lattice planes. The cones when projected on a flat
surface normal to the incident beam produce a set of concentric circles, called Bragg or Debye or
Debye-Scherrer rings, as seen in Figure \ref{BraRin}. The \difpat s are usually obtained by
rotationally averaging these rings. For each ring, the peak intensity as a function of \scaang\ may
also be computed by integrating around the entire ring \cite{BookCopley2001}.

Although \pd\ may be compared to single \crystal\ \diffraction\ in terms of ups and downs, very
often they are not in competition as single crystals that are suitable for diffraction work may not
be available. One of the characteristic features of \pd, which imposes limitations on its
application, is the collapse of a 3D pattern into one dimension. This leads to both accidental and
systematic peak overlap, and complicates the determination of individual peak intensities. Hence,
while overlaps hardly exist in single \crystal\ \diffraction, it represents a major problem in \pd.
Hence detailed and very precise information usually obtained from single \crystal\ \diffraction\
data is lost when using \pd\ data. Moreover, because \pd\ is usually more affected by background
noise, it produces patterns with worse \signoirat s than single \crystal\ \diffraction. This noise
can be difficult to define accurately and hence data analysis and information extraction become
more problematic. In \pd\ the range of measurable intensities is much smaller than in single
\crystal\ \diffraction. In particular, the weaker intensities at higher angles are measured less
accurately or may not be detectable at all. Both overlapping intensities and the limited \scaang\
range complicate the process of \crystal\ structure determination \cite{BookWill2006}.

Normally, single \crystal\ patterns consist of discrete spots while powder patterns consist of a
smooth continuous curve. The structural information contained in single \crystal\ and \pd\ patterns
is essentially the same. However, due to the above-mentioned complications, solving \crystal\
structures directly from \pd\ data is a more difficult task. Structure solution from powder
patterns becomes particularly difficult for complex compounds and in the presence of impurities.
Therefore, single \crystal\ \diffraction\ is the technique of choice for structure determination,
while the powder technique is used more often for characterising and identifying phases or
performing quantitative phase analysis. \Pd\ is also an ideal tool for investigating a number of
non-structural aspects such as \crystal\ size, micro \stress\ and \strain\ \cite{HarrisT1996}.

An advantage of \pd\ is that it is a non-destructive technique. This is particularly true for
highly penetrating radiation such as neutrons and \synch\ X-rays, as complete bulk objects can be
used without need for preparation or damaging of the specimen by cutting, drilling, or scraping.
This aspect is very important when dealing with high value objects or archaeological artifacts.
Non-destructivity is also important for investigating objects repeatedly over a period of time as
the same sample can be used in a series of experiments. With little or no radiation damage the data
can be compared and analysed to investigate time-dependent properties. Another advantage is that
the technique offers rapid data collection. Since all possible \crystal\ orientations are measured
simultaneously, collection times can be quite short especially when using a strong radiation source
like \synch. This can be essential for \timres\ studies and for samples which are inherently
unstable or deteriorate under radiation bombardment. The use of collective data acquisition
techniques such as \add\ with area sensitive detectors or \edd\ also contribute to the rapidity of
data collection since the collected pattern at each instant is the whole \difpat. For example the
investigation of temperature-dependent changes and reaction kinetics has greatly benefited from
these rapid measurement techniques. Another major advantage of \pd\ is that it is very suitable for
investigating dynamic transformations by measuring changes of the entire \crystal\ structure as a
function of temperature, pressure, time, chemical composition, magnetic field and so forth
\cite{KockelmannPK2000, Barnes1991}.

The disadvantages of using \pd\ include the overlapping of \diffraction\ peaks and this is a major
problem in \pd\ that complicates data interpretation and pattern analysis. The overlap can arise
accidentally from the \diffraction\ geometry and limited experimental resolution or as a
consequence of crystallographic symmetry conditions. As the degree of overlap increases with
increasing \diffraction\ angles and unit cell dimensions (the degree of overlap increases with
$\sin \SA / \WL$), the effective resolution of the data set will be reduced in many cases when even
the best fitting algorithms to extract intensities from a powder pattern will not be able to
determine the separate intensities of completely overlapping peaks. Another disadvantage is
\preori\ which can lead to inaccurate peak intensities and complicate \difpat\ analysis
\cite{BookDinnebierB2008}.

\section{Radiation Sources for \PDl}

The radiation used in crystallography and \pd\ work can be electromagnetic waves, or beams of
subatomic particles. The main requirement for \diffraction\ to occur is that the radiation
wavelength should be comparable in size to the lattice spacings. Several types of radiation are in
use in \diffraction\ experiments. These include X-rays, neutrons, and electrons. However, the main
radiation sources for \pd\ are X-rays and neutrons, as they are more practical to use and more
suitable for obtaining information from powder samples. Although the techniques involved in using
these types of radiation are very different, the resulting \difpat s are analysed using very
similar analytical tools. These radiation sources play complementary roles for \pd. While \Xray\
\diffraction\ mainly provides information about the electronic density distribution, neutron
\diffraction\ is used to obtain information about the mass density distribution and magnetic
ordering. Electron \diffraction\ may be used to provide vital information about surface structure.
Also, the \Xray\ approach is ideal for solving structures, while refinement of some important
details is more accessible with neutrons. The choice of radiation type to use in a particular \pd\
experiment depends on the nature of the required information and the underlying \diffraction\
physics. In most cases, the radiation used in \diffraction\ work for both bulk and thin film
materials is \Xray\ \cite{Brownbook1975}.

Diffraction experiments from powder samples yield a pattern which is a list of reflections
characterised by a number of parameters such as position, shape and intensity. The quality of the
\difpat\ depends on many factors, one of which is the resolution of the collecting instrument. An
important factor for achieving high resolution is the use of intense radiation beams such as those
obtained from \synch s and high flux neutron sources. These sources also offer other advantages
such as high \signoirat\ and excellent collimation. The development of dedicated second and third
generation \synrad\ facilities and nuclear and spallation neutron sources in the last few decades
have triggered profound changes in the design and practice of \diffraction\ experiments in general
and \pd\ in particular \cite{BookRhodes2000}.

X-rays are generated either in the laboratory by \Xray\ tubes or rotating anode devices, or by
\synrad\ (SR) facilities. In the first method, X-rays are generated by bombarding a solid target
with energetic electrons in the range of 5-100 keV, while in the second method electrically charged
particles (normally electrons) are kept revolving in an evacuated \storin. The two main mechanisms
for emission of \Xray\ are acceleration/deceleration of charged particles, and electronic
transitions between atomic energy levels in excited atoms. Both mechanisms rely on the principle of
energy conservation. The production of X-rays from a tube involves both mechanisms (the first for
producing the white continuous spectrum and the second for producing characteristic discrete
lines), while \synrad\ is produced by the first mechanism. In this section we only investigate
\synch s as sources of \Xray s since they are the only source of data that have been used in this
study.

\subsection{\SRSl s} \label{Synchrotron}

\SynRad\ (SR) is the radiation of ultra-relativistic charged particles moving along curved paths
with a macroscopic radius. The physical principle which \synch s rely upon is that accelerated
charges emit electromagnetic radiation. The radiation of \synch s usually covers most parts of the
electromagnetic spectrum from radio waves to hard \Xray s. Synchrotron radiation sources provide
intense beams of X-rays for leading-edge research in a broad range of scientific disciplines.
\Synch\ facilities are very expensive to build, run and maintain. Moreover, they require highly
specialised expertise and very advanced technological infrastructure. Therefore, the number of
\synch s around the world is very limited. The existing \synch s operate as either national or
international facilities. It is estimated that currently (2010) there are about 70 \synch s in 20
countries around the world used by more than 20000 scientists. About 10 of these facilities are
third generation radiation sources. Most of these facilities are based in Europe, Japan and the
United States. Each one is unique in its technical features, available equipment, size, energy
range, operation, and so on \cite{SokolovTKMK1972, Helliwell1998}.

According to the rules of electrodynamics, electrically charged particles accelerated by an
external force emit electromagnetic radiation. For synchrotrons utilising relativistic electrons
the emitted radiation is in the form of a narrow beam tangent to the path of particles in the
direction of travelling, and occupies very wide range of the electromagnetic spectrum. The
radiation output can be calculated from the energy and current of charged particles, bending
radius, angle relative to the orbital plane, distance to the tangent point, and vertical and
horizontal acceptance angles. Synchrotron radiation occurs naturally in many astrophysical systems
throughout the universe. It was first seen in the laboratory in 1947 as a flash of light from a
particle physics accelerator. As this phenomenon results in energy losses from the particle beam,
SR was initially regarded as an undesired parasitic effect. However, it was soon realised that SR
with its exceptional properties is an extremely powerful scientific tool, and this led to the
construction of facilities that are specifically designed and optimised for large scale generation
of \synrad\ \cite{HofmannBook2004}.

The main characteristics of \synrad\ which make it highly valuable tool for research are
\cite{Harding1995, ReySTBLe1995, Barnes1990, BarnesCJJABPHL2001}:

\begin{itemize}

\item Directionality: synchrotron radiation is directed forwards in the direction of (relativistic)
moving charges and is concentrated near the plane of the orbit within a narrow cone with a specific
aperture angle. In simple terms \synrad\ sweeps out a fan of radiation in the horizontal plane with
very small vertical dimension.

\item Very high intensity: the intensity (number of photons per energy interval per unit
time) of SR beams is several orders of magnitude (can be 9 orders) greater than that of
conventional laboratory \Xray\ sources. One consequence of this is that experiments that may take
weeks to complete using laboratory sources can be completed in a few minutes when using a
synchrotron. This time gain has accelerated the research on many frontiers and resulted in huge
advancements in various fields like dynamic transformation studies.

\item Broad energy range: SR has a continuous range of wavelengths across the
electromagnetic spectrum from the radio waves to hard X-rays. This allows for energy tunability to
the wavelength required by the particular experiment. By using monochromators and insertion devices
it is possible to obtain an intense beam of any selected wavelength; alternatively, a polychromatic
radiation spectrum can be used for white radiation experiments.

\item Very low divergence: SR from a bending magnet is highly collimated especially in the vertical
direction, with a divergence of only a fraction of a milliradian. This property facilitates high
resolution measurements required in various investigations.

\item Pulsed time structure: \synrad\ is delivered in pulses with a highly defined time structure. Each
pulse is produced when a bunch of moving charges passes through a bending magnet or an insertion
device. The frequency of these pulses is determined by the spacing and the number of bunches in the
\storin. SR pulses are typically 10-100 picoseconds in length separated by 10-100 nanoseconds. In
some experiments this time structure feature is exploited for time-resolving purposes.

\item Polarisation: SR is highly polarised, that is the electric vector of the
electromagnetic radiation lies in the plane defined by the direction of deflection of the particle
beam. For bending magnets, it is the horizontal orbit plane of the \storin. For insertion devices
the particles can be deflected vertically resulting in vertical polarisation. Apparently, this
property has rarely been exploited in research.

\end{itemize}

\begin{figure}[!h]
  \centering{}
  \includegraphics
  [width=0.6\textwidth]
  {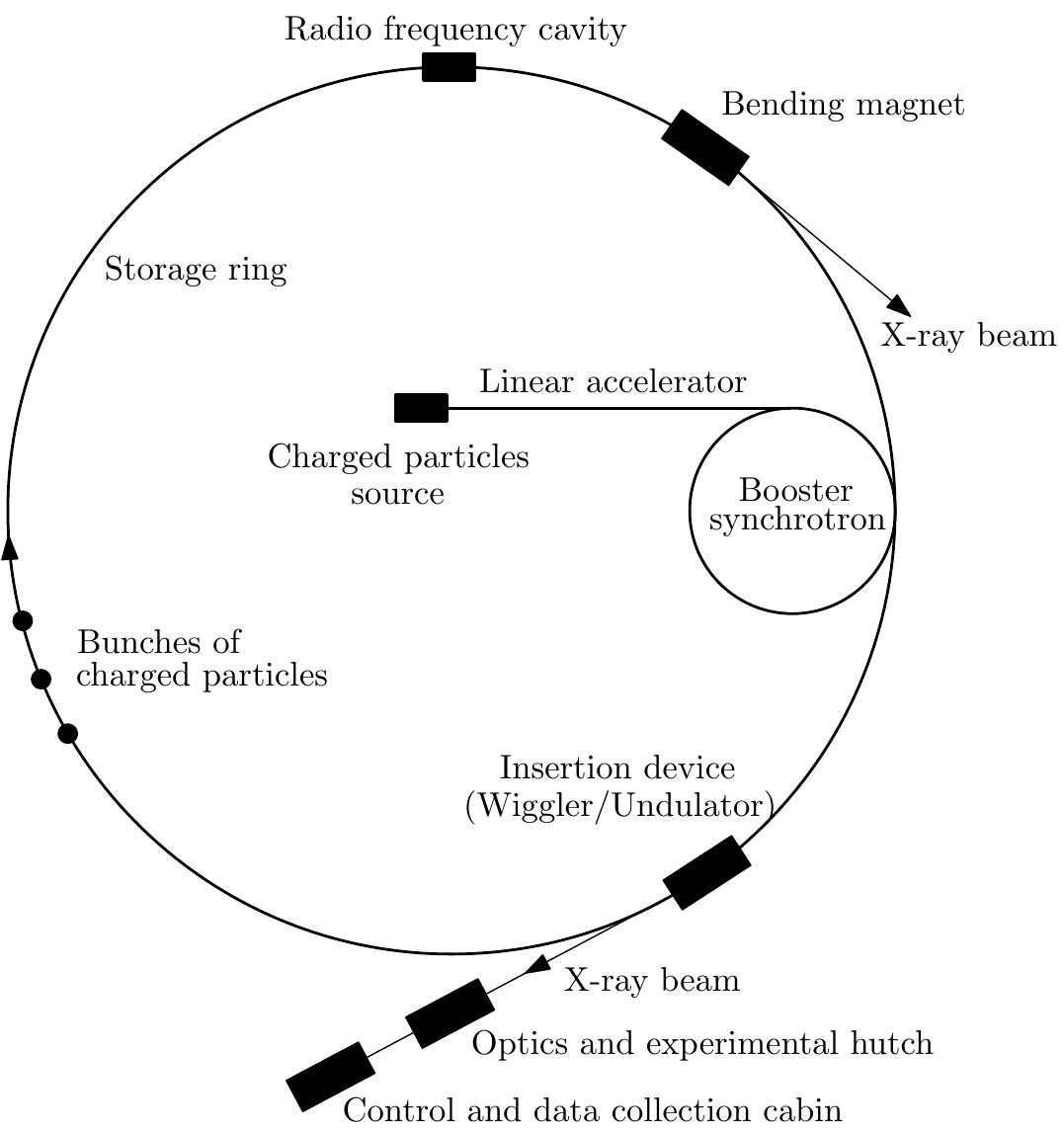}
  \caption{Schematic illustration of the main components of a synchrotron.}
  \label{SynSch}
\end{figure}

The synchrotron is a circular (approximately) particle accelerator that is specifically designed
for the production of electromagnetic radiation. A common feature of \synch s is that they use
microwave electric fields for accelerating the charged particles and magnets for steering them. The
main component of the synchrotron is a \storin\ inside which the circulating charged particles
(e.g. electrons, positrons and protons) at relativistic speeds are maintained in a fixed orbit by a
strong constant magnetic field. Other components include charged particles source, linear
accelerator, booster synchrotron, radio frequency cavities, bending magnets and beamlines, as seen
in Figure \ref{SynSch}. For simplicity, the ring is drawn as a perfect circle in the figure whereas
in reality it consists of straight and curved sections. In general, \synrad\ sources are very large
and highly sophisticated installations. The size of the synchrotron facilities is correlated to the
required radiation energy, that is \synch s designed for generating X-rays tend to be larger than
those designed for generating ultraviolet radiation \cite{HofmannBook2004}.

The radiation from bending magnets and insertion devices is piped off to an experimental station by
a beamline tangential to the path of the charged particles. The radiation beam inside these lines
usually has a very thin cross section; typically a fraction of a millimetre wide. Beamlines are
normally a few tens of metres long. SR travels from the magnet source on the \storin\ along these
highly evacuated beam pipes to an experimental hutch which is heavily shielded to prevent radiation
leak. Beamlines are complex instruments that prepare suitable \Xray\ beams for experiments, and
protect the users against radiation exposure. A number of cabins with highly sophisticated design
and equipment (which are dependent on the type of beamline experimental use) are installed on each
beamline to facilitates harnessing, adapting and exploiting the transported beam. The first is the
optics and experimental hutch which comprises such instruments as slits, collimators, filters,
mirrors and monochromators for controlling and tuning the beam. It also contains the sample, the
sample handling and conditioning equipments (e.g. for alignment and temperature and pressure
control), the computer interface electronics for data acquisition, and the detector system. The
experimental hutch is equipped with radiation shielding, safety interlocks and a radiation
monitoring system. At the end of the beamline, the control and data collection cabin is located
where the station scientist and the users are based with suitable equipment, such as computers and
monitors, to control and scrutinise the experiment and record the measurements. These activities
are usually conducted in shifts around the clock when the ring is operational \cite{DukeBook2000}.

Bending magnets on the \storin\ are used to define the shape of the orbit of charged particles and
generate \synrad. When the particles pass through these magnets they are deflected from their
straight path, and this centripetal acceleration causes emission of \synrad. The \synrad\ produced
by bending magnets is tangential to the trajectory in the form of a horizontal fan as the particles
sweep through the arc of the magnet. The higher the energy of the particles, the narrower the cone
of emission of SR becomes and the emitted spectrum shifts to shorter wavelengths. The energy
spectrum of the emitted radiation, which can be displayed as a universal curve, is proportional to
the fourth power of the particle speed and is inversely proportional to the square of the radius of
the path. Bending magnets are typically electromagnets made of steel. Focusing magnets, placed in
the straight sections of the \storin, are also used to focus the electrons beam and keep them in a
narrow and well-defined path to produce very bright and focused radiation beams
\cite{BookMesserschmidt2007}.

\begin{figure}[!h]
  \centering{}
  \includegraphics
  [width=0.75\textwidth]
  {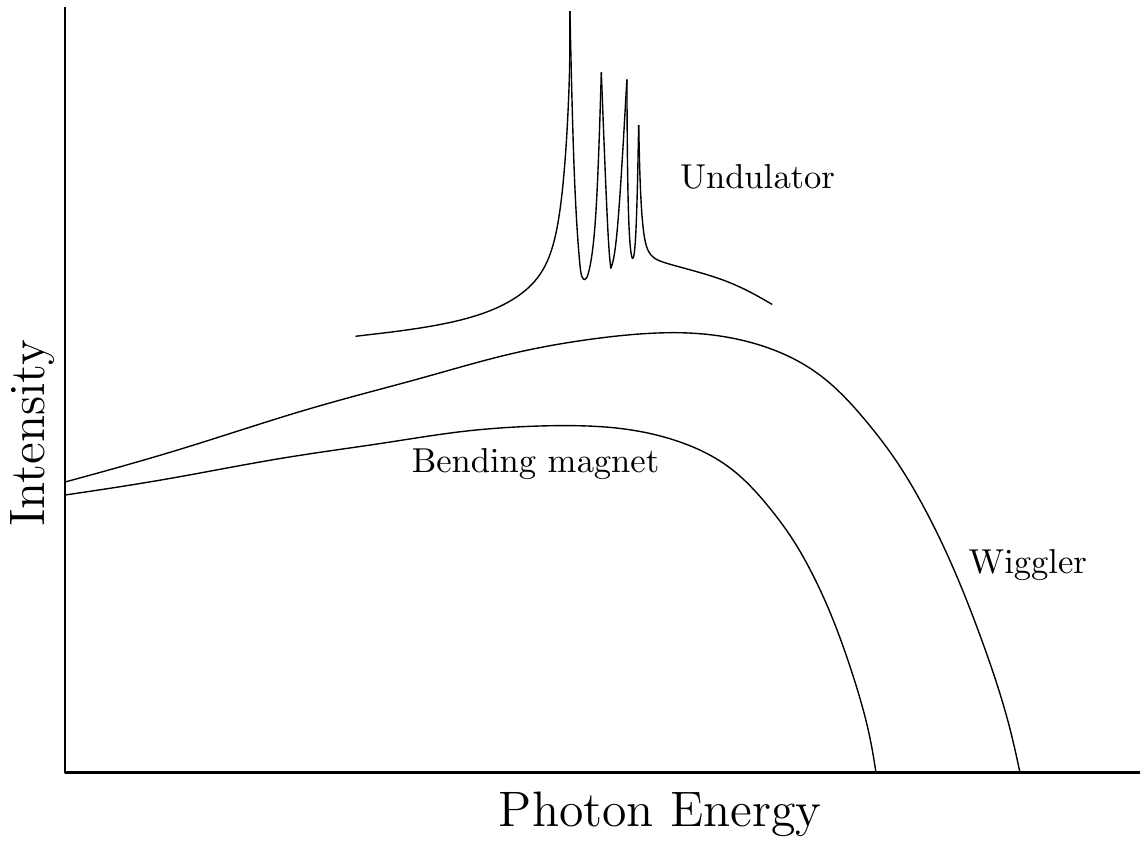}
  \caption{Spectra of bending magnets and insertion devices.}
  \label{BenUndWigSpe}
\end{figure}

Synchrotrons usually include insertion devices as an alternative to bending magnets for generating
\synrad. Insertion devices consist of a string of permanent or superconducting magnets of
alternating polarity designed to deflect the beam of electrons first in one direction and then in
the other. These arrays of magnets produce a magnetic field that is periodically changing in
strength or direction, thus forcing the electron beam to make planar `wiggle' or follow a helical
trajectory. Since the net deflection of the electron beam is zero, these devices are inserted in
the straight sections of the \storin. The principle of radiation production by insertion devices is
the same as that of the bending magnets. However, the radiation from each magnet in the insertion
devices is usually of shorter overall wavelength (wiggler) or concentrated at specific wavelengths
(undulator) due to the various effects of the tight bends within the wiggle/undulation, the number
of magnetic poles and interference effects. The general features of the spectra of bending magnets
and insertion devices are highlighted in Figure \ref{BenUndWigSpe}. Unlike bending magnets, the
properties of insertion devices can be tuned to optimise the radiation beam to meet specific
experimental requirements. By externally manipulating the insertion devices the radiation source
characteristics can be tuned to optimise the radiation delivered to the sample during a particular
experiment. Third generation \synch s heavily rely on insertion devices for their operation. The
two types of insertion devices referred to above have been developed in the last decades; these are
`\wiggler s' and `\undulator s'. Despite the strong similarity between them, the \wiggler\ and
\undulator\ have evolved independently from the beginning \cite{DukeBook2000, Barnes1990}.

The main advantages of using synchrotrons are \cite{AlbinatiW1982, Barnes1990, BassettB1990}:

\begin{itemize}

\item Rapid data collection due to the high intensity of synchrotron sources. This makes SR an ideal
tool in the application of intensity-demanding techniques like tomographic imaging and the study of
time dependent processes. In this regard, samples can be dynamically investigated while
transforming under \stress, \strain, temperature or pressure since the high intensity allows rapid
collection of many data scans as a function of variation under these conditions. Also, high
intensity facilitates the investigation of samples of low elemental concentration, minute samples
and high throughput crystallography.

\item High photon energies of SR allows collecting data to a very high $\QF$-factor $(= 4\pi \sin
\SA/\WL)$. This provides precise determination of positional parameters and temperature factors.

\item Excellent spatial resolution to the micron scale. The highly resolved patterns obtained with \synrad\
can help in resolving more difficult \spagro\ or symmetry problems, and for easier identification
of minority phases present in the sample. This has contributed to the extension of range and
complexity of the materials that can be investigated by X-rays.

\item Superb time resolution which allows dynamic and \timres\ investigations. The pulsed time structure
of \synrad\ can be exploited in this regard.

\item Excellent depth penetration of the highly energetic radiation which allows the
investigation of bulk samples and the application of bulk techniques such as tomographic imaging.

\item High \signoirat\ which, combined with high resolution, provides improved accuracy in
quantitative analysis, structure solution and phase identification.

\item Highly collimated beams with very small divergence which improves angular resolution and data
collection rates. The linear polarisation of SR can also be exploited to remove intensity losses
normally associated with a randomly polarised laboratory \Xray\ source.

\item Tunability of \synrad\ to an absorption edge for anomalous scattering \diffraction\
experiments.

\item Broad wavelength range to choose from to meet the requirements of various types of
experiment. The radiation spectrum is smooth without the superimposed characteristic lines that are
found in the spectra of conventional laboratory sources.

\item Modern synchrotron sources give rise to extremely narrow peak widths in \pd\
patterns, thus reducing the effect of overlapping reflections. Moreover, they allow highly accurate
measurements of peak positions and intensities.

\end{itemize}

In brief, \synrad\ with its unique properties is overwhelmingly superior to the best laboratory
source. On the other hand, \synch s are expensive to construct, operate and use. Moreover, the
access to \synch s is limited as beam time is very scarce and competitive. Hence, to make full use
of available beam time it is essential to prepare the experimental equipment and plan the setup in
advance to minimise time losses, and this may not be easy to do. Another factor is that the use of
\synch s involves inconvenient and time-consuming activities such as travel and moving heavy
equipment. Most synchrotron installations are shared facilities with general purpose tools and
equipment which may not suit the experiment in hand. Furthermore, some synchrotron sources may not
be as stable and reliable as other domestic and non-domestic sources.

\section{Data Collection Techniques}

In this section we present a summary of the data collection techniques that have relevance to this
study.

\subsection{Angle and Energy Dispersive Diffraction} \label{EADD}

\begin{figure}[!h]
  \centering{}
  \includegraphics
  [width=0.9\textwidth]
  {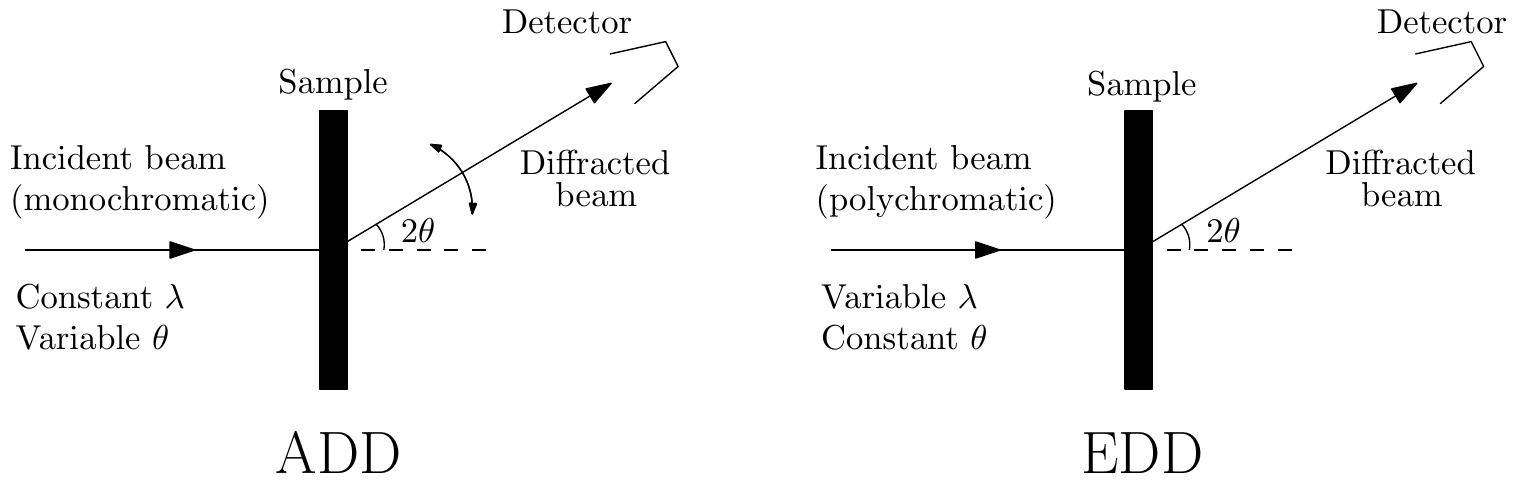}
\caption[Angle dispersive and energy dispersive diffractions.]{Comparison between angle dispersive
and energy dispersive diffractions.} \label{AddEdd}
\end{figure}

Diffraction experiments can be performed either by using white radiation with an
energy-discriminating detector in an energy dispersive mode, or by using a monochromatic radiation
with a position-sensitive detector in an angle dispersive mode. These two modes are presented
schematically in Figure \ref{AddEdd}. \ADDl\ is the more conventional method in powder diffraction
experiments. The \EDDs\ detector sorts the diffracted \Xray\ photons according to their energies
and thus generates \difpat s as a function of energy rather than \scaang. In some experimental
settings the two modes are combined by fitting more than one energy-discriminating detector at
different scattering angles. This setting can allow substantial reduction in data collection time.
In both modes the measured diffraction pattern exhibits peak positions and intensities that
characterise the phases in the sample. Energy dispersive diffraction was first demonstrated in the
late 1960's but has only become prominent since the increased availability of \synch\ \Xray\
sources \cite{BordasGHB1977, BordasR1978, GlazerHB1978, BarnesJCJGRMCC1998}.

As the \scaang, $2\SA$, in the \EDDs\ mode remains fixed, the \Bragg\ equation for the first order
\diffraction\ in angle-dispersive form, i.e.

\begin{equation}\label{Bragg}
    \WL = 2 \PS \ \sin \SA
\end{equation}
is rewritten, using the Planck relation $\Ene = \PC \SoL/\WL$, in its energy equivalent form

\begin{equation}\label{EDDEq}
    Ed \ \sin \SA = \frac{\PC \SoL}{2}
                     = \rm{a \ constant} \ (\cong 6.199 \ keV.{\AA})
\end{equation}
where $\PC$ is Planck's constant, $\SoL$ is the speed of light, $\Ene$ is the energy of the
associated photon, $\WL$ is the wavelength and $\PS$ is the \crystal\ interplanar spacing
\cite{CernikB1995, Barnes1991, ColstonBJJHLDMM2005}.

Energy dispersive \diffraction\ has several advantages over angle scanning \diffraction. One of
these is that \EDDs\ has a fixed geometry and this facilitates the design of industrial and
environmental cells and aids the collection of rapid, \timres\ \diffraction\ data. Another
advantage is that the beam intensity, combined with fixed geometry, leads to fast data collection
rates and allows the collection of high quality kinetic data. The use of a white high-flux beam,
especially from the \synch, is particularly useful for studying reactions under non-ambient
conditions. A third advantage is the presence of fluorescence signals which could provide vital
information about elemental formation and distribution. On the other hand, the \EDDs\ technique
suffers from several shortcomings. One of these is low peak resolution and excessive peak overlap
which make the analysis more difficult and may compromise information extraction. The presence of
scattering bands and fluorescence lines can introduce further deterioration and uncertainty.
Moreover, the energy-discriminating detector, which normally is a semiconductor device, has a
limited count rate and this can impose a limit on the peak intensity and worsen the peak
overlapping problem. Another disadvantage is that in practice the fixed \scaang\ has to be a
compromise and this can be a limiting factor in the $d$-spacing range and overall pattern
resolution. Some of these disadvantages can be eliminated or minimised by using bright radiation
sources (\synch s) and by improving the design of the data collection system
\cite{BarnesJCJGRMCC1998, Moron2000, Clark2002}.

\subsection{\TimRes\ Powder Diffraction}

The general method in \timres\ studies is to initiate the reaction or transformation either
chemically or by varying the physical conditions such as temperature and pressure. This is followed
by collecting a series of \pd\ patterns of equal duration and over a period of time that is
comparable with the process under study. Very large number of \pd\ patterns can be produced during
a \timres\ experiment. This makes \pd\ suitable for following the course of a phase transition or
chemical reaction as it proceeds. The standard methods of phase identification and quantitative
analysis can then be used to give information on the phases present at any particular time and
their relative abundance. It is extremely important in these studies to collect the whole \difpat\
in a short period of time relative to the time scale of the transformation so that the pattern
reflects the state of the system at a certain point along the reaction coordinate \cite{HeBMTK1992,
ClarkB1995, JupeTBCHHH1996}.

Time-resolved studies necessitate the use of \ADDs\ with position sensitive detectors that cover a
large angular range or \EDDs\ with multichannel energy detectors. It is also important to have a
strong radiation source with high flux for rapid data collection so that rapid processes can be
time-resolved. The advent and widespread use of \synch s has therefore revitalised powder
diffraction to study temperature dependent changes, reaction kinetics and so forth. Neutron beams
at reactors and spallation sources are weak in comparison to \synch s though they still have a role
to play in \timres\ studies by virtue of their excellent penetration depth. The development in the
last few decades of high flux radiation sources and rapid data acquisition techniques made it
possible to collect complete \pd\ spectra in a fraction of a second \cite{BetsonBBAJ2004,
ClarkM1990, JohnsonOBCOSBTS2003}.

\subsection{\SpaRes\ Powder Diffraction}

A prominent example of \spares\ techniques is \TEDDIl\ (\TEDDIs) which is based on the \EDDs\ mode
of \Xray\ \diffraction\ as outlined in \S\ \ref{EADD}. This method exploits a well defined
energetic white \Xray\ beam from a \synch\ to gain \diffraction\ information from volume elements
(called lozenges) within a bulk sample. TEDDI can be used to image the interiors of objects in
terms of both density and compositional variations. The volume element sampled is determined by the
geometry of the diffracting lozenge defined by the incident beam, the detector system collimation
and the scattering angle. The sample is moved around this volume element so that \diffraction\
information can be collected at a series of points in a user-defined 1, 2 or 3D grid. In this way,
chemical and structural content of the samples can be derived for each grid point and then
constructed as intensity maps. The use of intense hard white \Xray\ beams (20-125 keV) facilitates
the penetration of bulk objects non-destructively. \TEDDIs\ exploits primarily diffraction, in
preference to spectroscopic, effects to obtain structural/compositional information about the
sample, though detecting fluorescence lines can be added to the imaging capability thereby
supplying specific elemental concentration information. The diffracting region can be made small or
large depending on application, where the ultimate spatial resolution is in the micron range
\cite{HardingNK1990, PileLJRB2006, BarnesJJCCHBHBRSCHWP2001, HallBCCHJJK1998, BetsonBBA2005}.

\subsection{CAT of ADD Type} \label{CatAdd}

In \CATl\ (\CATs) of \ADDs\ type, a pencil beam is employed to collect \diffraction\ signals in
angle dispersive mode from the sample under study, and hence provide information about the
distribution of \crystalline\ phases. The method has been suggested previously in the literature
and has recently been demonstrated by Bleuet \etal\ \cite{BleuetWDSHW2008}. The method has the
advantage over \TEDDIs\ that the quality of \diffraction\ data is superior, since the current
energy-dispersive detector/geometry offers only limited resolution which causes significant peak
broadening. Moreover, the current semiconductor energy-discriminating detectors have limited count
rates and hence the use of Charge-Coupled Devices (CCD) area detectors in the ADD mode can provide
a faster data acquisition mechanism which is vital for monitoring rapid phase transformation
processes.

\CATs\ type \ADDs\ is the basis of some experimental data presented and analysed in this thesis, as
outlined in chapters \ref{Cata} and \ref{Nick}. In these experiments, a pencil beam of
monochromatic synchrotron X-ray is applied on a sample mounted on translational-rotational stage
and a time/temperature slice is collected across each translation-rotation cycle. For each slice
the sample is translated $m$ times across the beam and a complete diffraction pattern is collected
for each translation position. These $m$ translations are then repeated at $n$ angles between 0 and
$\pi$ in steps of $\pi/(n-1)$, and hence $m\times n$ diffraction patterns are collected for each
time/temperature slice. The complete data of a slice represent a sinogram that can be
reconstructed, using a back projection computational algorithm as given in \S\ \ref{MWBP}, to
obtain a (spatial) tomographic image of the slice. A series of slices then give a complete picture
of the dynamic transformation of the phases involved during the whole experiment. As charge-coupled
devices are usually employed in these experiments to collect the diffraction data in angle
dispersive mode, the 2D diffraction images should be transformed to 1D patterns by integrating the
diffraction rings. Curve-fitting can then be used to identify the phases in each stage as the peaks
in these patterns provide distinctive signatures of each phase.

\section{Data Analysis and Information Extraction}

Depending on the required accuracy and the availability of resources and crystallographic
information, several methods are in common use to extract information from raw \pd\ patterns. Some
of these methods require structural data in the form of an initial \crystal\ model while others
allow the extraction of information without presumed structural knowledge. In this section we
present several methods that are widely used to extract structural and non-structural information
from \pd\ patterns. These are \seamat, \curfit, \twstme\ and \whpamo.

\subsection{Search-Match}

Search-match is a recognition technique applied to the diffraction peaks from powder diffraction
patterns. The method is used to compare an experimental pattern with patterns that are stored in
extensive databases of known materials to find a match and hence identify the structure. Two main
components are therefore required to perform computer-based search-match: a database of diffraction
patterns, and a search-match program for that database. A prominent example of a database is the
\PDFl\ (\PDFs). These databases usually store a huge number of standard single-phase patterns. Some
databases have their own dedicated search-match programs. A typical search-match procedure normally
generates a reduced pattern that can be used for phase recognition. Phase identification of
\crystalline\ material is accomplished by comparing the peak positions and relative intensities
from the sample with peak positions and relative intensities from patterns in the database. The
method is powerful and can be used to determine the constituents and proportions of phases in
experimental samples. However, it is of lesser value if the structure of the material is unknown.
In this case any structure must be analysed by an \abinitio\ method. The main merit of search-match
is that it is unbiased by structural information. Moreover, it is simple, fast and requires minimum
effort. These factors made the method very popular since the early days of powder diffraction work.
The technique dates back to the late 1930s from the pioneering work of Hanawalt, Rinn and Frevel
when the method was based on manual searching using indexed cards. However, it has improved
substantially by introducing computer algorithms in conjunction with digitised databases. A
disadvantage of the search-match is that the accuracy of information is very low especially for
weak diffraction peaks, because \seamat\ programs normally use a few strong peaks
\cite{BookPecharskyZ2005, CernikB1995}.

\subsection{Curve-Fitting}

\begin{figure}[!h]
  \centering{}
  \includegraphics
  [width=1\textwidth]
  {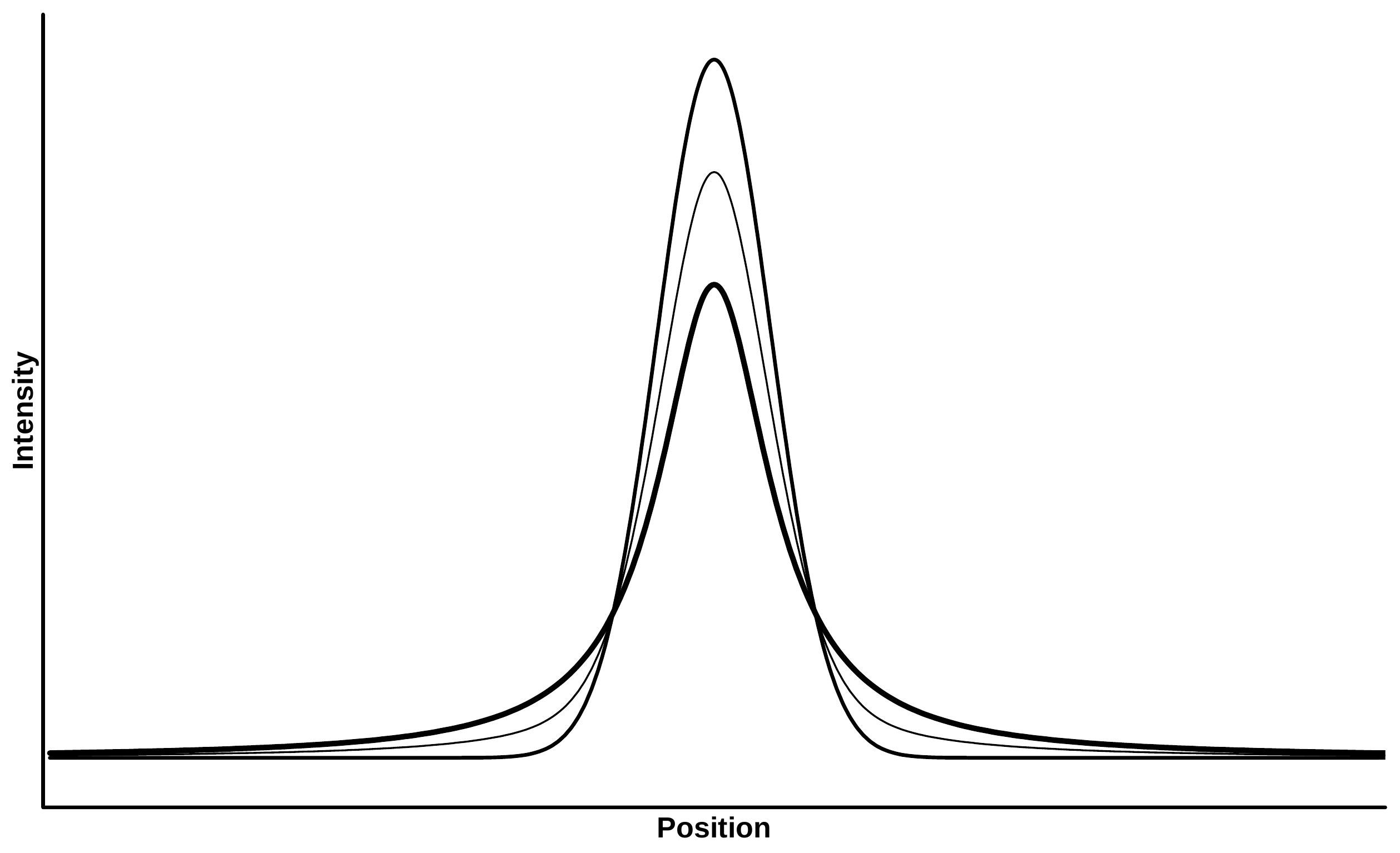}
\caption[Graphs of \Gauss ian, \Lorentz ian and \pseVoigt\ profiles.]{Graphs of \Gauss ian (highest
peak), \Lorentz ian (lowest peak) and \pseVoigt\ profiles having the same parameters with a
\pseVoigt\ \mixfac\ $\MFv = 0.5$.} \label{GauLorVoi}
\end{figure}

The method of \curfit\ (also called peak- or pattern- or profile-fitting) is based on decomposing
the pattern into independent peaks using relatively simple profile models to extract the integrated
intensity and other parameters of the diffraction peaks. In this method no structural or unit cell
parameters are required. Various figures-of-merit are usually used to assess the quality of the
fit. The approach consists of choosing proper functions to describe peak shape, accounting for
background and finally identifying the individual peaks and determining their parameters by a
fitting routine. The calculated profile consists of a sum of the Bragg reflection profiles and a
suitable background function. Three of the most commonly used shape functions to describe
individual peak profiles are the \Gauss ian ($G$), the \Lorentz ian ($L$) and the \pseVoigt\ ($V$);
the latter is a weighted sum of \Gauss ian and \Lorentz ian components. These functions are
presented in Table \ref{lineShapeFuncs} and plotted in Figure \ref{GauLorVoi}. The background
scattering is commonly modelled by an ordinary polynomial of a suitable order, usually up to order
five, or by Chebyshev or \Fourier\ polynomials. The fitting parameters normally include peak
position, full width at half maximum and integrated intensity, defined as the area under the
diffraction peak, of the individual reflections. Constraints may be imposed when the parameters are
highly correlated. The results of fitting may be used as an input to other processes such as
lattice parameters refinement and quantitative phase analysis. They can also be used directly as
signatures to identify phases, for instance in dynamic phase transformation studies
\cite{JansenSW1988, RielloFCC1995, Artioli2001}.

\begin{table} [!h]
\centering %
\caption[Common shape functions used in \ProgName\ to describe peak profiles.]{Some of the common
shape functions used to describe individual peak profiles. $A$ is the area under peak, $\FWaHM$ is
the full width at half maximum, $X$ is the position of the peak and $m$ is a dimensionless \mixfac\
($0 \leq m \leq 1$).}
\label{lineShapeFuncs} %
\vspace{0.5cm} %

\begin{tabular}{ll}
\hline
{\bf Function} \verb|                                | & {\bf Equation} \\
\hline      \vspace{-0.3cm} \\
  \Gauss ian &    $G = \frac{2 \Are}{\FWaHM} \, \, \sqrt[]{\frac{{\rm ln}(2)}{\pi}} \, \, e^{^{\frac{-4 {\rm ln}(2)(x-\PoP)^{2}}{\FWaHM^{2}}}}$ \\ %
            \vspace{-0.3cm} \\
\Lorentz ian &    $L = \frac{2 \Are /(\FWaHM \pi)}{1 + 4(x-\PoP)^{2} / \FWaHM^{2}}$ \\ %
            \vspace{-0.3cm} \\
\PseVoigt &   $V = \MFv L + (1-\MFv) G$ \\ %
            \vspace{-0.4cm} \\
\hline
\end{tabular}

\end{table}

The fit can be performed on the pattern as a whole or on selected regions or selected peaks which
can be fitted together or separately. The procedure is based on minimising the difference between
the observed and calculated profiles using normally a nonlinear \ls\ technique. The technique is
required to minimise the second norm of the residuals given by
\begin{equation}\label{funcMinimise}
    \SNoR = \sum_{i} \SW_{i} \left(  \OCR_{i} - \CCR_{i} \right)^{2}
\end{equation}
where $\OCR_{i}$ and $\CCR_{i}$ are the observed and calculated intensity at each step respectively
and the summation index $i$ runs over all points in the range of the segment to be fitted. The
weights $\SW_{i}$ are taken from the experimental error margins which in most cases are assumed to
be in proportion to the square root of the observed count rate $\OCR_{i}$ following a \Poisson\
counting statistics \cite{SpagnaC1999, BookDinnebierB2008}.

\Curfit\ is the most accurate method for extracting pattern parameters. Moreover, it is
computationally efficient and relatively easy to implement and use. \Curfit\ is also very flexible
since no detailed knowledge about the existing diffraction peaks is required, and hence it is the
best choice when no information about the unit cell and symmetry are available. When the data
quality is reasonable, the fitting can yield very accurate results. These results can then be used
in subsequent processes such as quantitative phase analysis or structure determination, though the
latter can be difficult when the pattern is too complex with severe peak overlapping. However, the
method normally requires considerable time and effort and hence can be very slow and meticulous.
Moreover, it is usually influenced by the results of an automated search routine or biased by the
user judgment about the existence, position and profile of the peaks. The approach can also fail
when the pattern suffers from serious overlapping, as the parameters of peaks occupying the same
position cannot be determined by this type of fitting. In such cases, other fitting techniques such
as whole pattern decomposition should be employed \cite{Clark1995, BookPecharskyZ2005}.

\Curfit\ is the main method used in this study to analyse experimental data collected from a number
of synchrotron facilities in various experimental settings and techniques. \Curfit\ is implemented
in EasyDD as single, multiple, batch and multi-batch processes with high speed and efficiency. This
made possible the analysis of huge data sets by fitting millions of peaks within hundreds of
thousands of diffraction patterns in just a few hours. Examples of such large-scale processing and
analysis of massive data sets are presented in chapters \ref{Cata} and \ref{Nick}. Each peak is
used as a signature for a particular phase that can be used to detect spatial and temporal
distributions. Hence, EasyDD \curfit\ implementation offers a great support to highly important
research areas such as dynamic phase transformation studies.

\subsection{Two-Stage Method} \label{twoStage}

This method for extracting information from powder diffraction patterns requires two-stages. In the
first stage, the diffraction pattern is analysed to separate the peaks and peak clusters of the
pattern into individual reflections to extract their parameters such as position and integrated
intensity. In this stage, \curfit\ or \whpade\ procedures, which do not require structural
information, are usually employed. In the second stage, the individual reflection data are used for
structure determination or for other purposes like \strain\ and \texture\ analysis. This stage
includes the actual crystallographic calculations, which may involve either refinement by using an
optimisation method (usually a \ls\ routine) or procedures like \Fourier\ analysis or Patterson
maps \cite{JansenSW1988}.

The \twstme\ has been used in \Xray\ and neutron diffraction data analysis since early 1960s. The
introduction of the \Rietveld\ \whpare\ with its attractive features diminished the importance of
the \twstme\ and reduced its use. However, it is still in use in some applications that \Rietveld\
refinement cannot substitute. Moreover, those who rejected \Rietveld\ refinement \cite{CooperS1979,
SakataC1979, Cooper1982} regard the two-stage approach as the only legitimate procedure for
information extraction. The method has been constantly modified and extended to incorporate new
developments mainly in peak deconvolution techniques. The method has a number of adaptations and
interpretations which share this common general feature of a two-stage procedure
\cite{BookWill2006, David2004}.

\subsection{Whole Pattern Modelling}

Modern powder diffraction heavily relies on pattern modelling techniques. Whole pattern modelling
is based on fitting the whole diffraction pattern to a model characterised by a number of
parameters and applying a refinement procedure, where `refinement' means adjusting the model
parameters to optimise the fit to the observed data. The best fit is quantified according to some
predefined discrepancy indicator(s). In most cases, \whpamo\ employs a nonlinear \ls\ procedure
which requires sensible estimates of many variables. These normally include peak shape parameters,
background contribution and crystallographic variables such as unit cell dimensions and atomic
coordinates in the unit cell. Various standard indicators (figures-of-merit) are usually used to
measure the quality of fit. Unlike \curfit, \whpamo\ methods require knowledge of the structural
parameters, or at least knowledge of the unit cell parameters, which may be difficult or impossible
to obtain. Moreover, these methods are computationally demanding and usually case-specific and
hence may not be suitable for use in batch processing of large quantities of data
\cite{Rietveld1988, SchwarzenbachAFGHe1989}.

There are two main approaches to the \whpamo : structure modelling and pattern decomposition. In
the first approach a background model and line shape functions are employed to fit the observed
data with presumed crystallographic structural information, while in the second approach no such
information is required. \Rietveld\ refinement procedure \cite{Rietveld1967, Rietveld1969} is the
known example for whole pattern structure modelling, while \Pawley\ \cite{Pawley1980, Pawley1981}
and \Lebail\ \cite{LeBailDF1988} procedures are the prominent examples for \whpade. It is
noteworthy that pattern decomposition techniques were introduced following the advent of the
\Rietveld\ method.

\subsubsection{Whole Pattern Structure Modelling}

As the peak intensities in the diffraction pattern depend on structural factors such as atomic type
and their distribution within the unit cell, measurement of intensities allow quantitative
identification of the phases involved and structure determination. In single crystal \diffraction,
the intensities can be measured, in principle at least, in a straightforward way. The measurement
of intensities in the powder diffraction pattern on the other hand is more complex because of peak
overlap and the involvement of complex sample and instrument factors. Due to these complexities,
structural information from powder diffraction pattern is preferably obtained by a whole pattern
structure refinement technique. The refinement starts by assuming a structural model with variable
parameters which have to be tuned to achieve the best agreement between the observed data and the
calculated values according to some statistical indicators. Background and profile models are also
included alongside the structural model in the fitting procedure to account for non-structural
factors contributing to the experimental pattern. Visual inspection and various figures-of-merit
can be used to assess the quality of the fit \cite{JansenSW1988, AlbinatiW1982, Rietveld1988}.

The \Rietveld\ Method is the most popular of powder diffraction pattern refinement techniques. The
method is based on fitting the structural model directly to the total pattern of \Bragg\
reflections. It extracts the maximum available information from the collected diffraction data. The
\Rietveld\ method is a procedure for structure refinement and not for structure determination.
Therefore, a knowledge of the crystallographic \spagro\ symmetry and unit cell dimensions with
approximate atomic positions is required as the method is not capable of creating a
crystallographic structural model from first principles. In multi-phase samples, crystal structures
of all individual phases must be known. The method also requires high quality experimental
diffraction data and suitable functions to model peak shape and background contribution. In this
regard the type of radiation source, sample quality, experimental settings and instrumental
resolution play a crucial role and impose a limit on the complexity of the problem to be solved.
For example, \Rietveld\ refinement is more likely to succeed when using synchrotron diffraction
data than when using data from domestic X-ray sources. In the \Rietveld\ method a calculated full
profile is generated and a refined list of parameters that best fit the experimental data are
produced. The method relies on a \ls\ routine as a refinement minimisation technique. The solution
should be inspected and assessed by some independent criteria if possible as an apparently
successful \Rietveld\ refinement may not be enough to prove the correctness of a crystal structure
solution \cite{Rietveld1969, Toraya1989, BuchsbaumS2007}.

Although the \Rietveld\ procedure was originally proposed for structure refinement, nowadays it is
widely used for other kinds of analysis as well as structure refinement. The method can provide
information about \crystal\ and magnetic structures of single- and multi-phase samples, and
determine the relative amounts of each phase in quantitative phase analysis from powder samples.
The method can be used to provide a wide range of information about lattice parameters, atomic
positions in the unit cell, fractional occupancy, thermal displacements, average crystallite size,
average \strain, \preori, and so on. It is noteworthy that neutron \pd\ is the technique that
mostly benefitted from the invention of the \Rietveld\ method because of the simplicity of peak
shape produced by the relatively crude resolution of neutron \diffraction\ instruments
\cite{Young1993, LangfordL1996}.

In the literature of powder diffraction refinement there is a number of guidelines and
recommendations that should be followed if good results are to be expected from the \Rietveld\
procedure. These include recommendations about the sample, instruments, radiation source, data
collection method, \Rietveld\ refinement strategy and so on. It should be remarked that the
refinement can lead to a wrong solution even when these guidelines are followed and the refinement
converged with low $R$-values although the likelihood of this occurring is usually small. Apart
from complying with the formalities of the refinement process, such as having good
figures-of-merit, the resultant model should be sensible. Visual inspection and independent checks
from other sources of information, when available, must also be performed \cite{BuchsbaumS2007,
MccuskerDCLS1999}.

\subsubsection{Whole Pattern Decomposition}

Whole pattern decomposition or whole pattern fitting is a widely used method in diffraction pattern
modelling. In many situations the crystal structure is unknown or is not of interest for the
application in hand. In such cases, pattern decomposition can be employed to characterise the
pattern and obtain the required parameters of the individual diffraction peaks. In this method the
whole pattern is deconvoluted into individual \Bragg\ components with no use of a structural model.
A nonlinear \ls\ minimisation technique is usually employed during this process. The parameters
that can be refined by whole pattern decomposition include, integrated intensity, integral breadth,
peak position, full width at half maximum, unit cell parameters, shape factor and line asymmetry
parameter. As a \whpamo\ technique, pattern decomposition can produce accurate individual peak
parameters even when the pattern contains severely-overlapped peaks. The quality of the fit is
assessed using a number of figures-of-merit, as in the case of \Rietveld\ refinement. The
significant advantage of \whpade\ is that it does not require a structural model of the phases.
Furthermore, additional peaks can be included in the refinement to deal with impurities from
unidentified phases \cite{Louer1998, TorayaT1995}.

Whole pattern decomposition has a wide range of applications, for example in lattice parameter
refinement. The method may be used within a two-stage procedure to extract unit cell and profile
information. In this context, the \Bragg\ intensities are obtained in the first stage where the
positions of the individual peaks are constrained by unit cell parameters. These intensities are
then used in the second stage as an input to a refinement process. A shortcoming of pattern
decomposition is that knowledge of the unit cell parameters is required, and hence the method can
be biased towards the user choice. Furthermore, it may not be applicable in situations where such
information is not available. Although both \curfit\ and \whpade\ are profile fitting procedures,
each peak in \curfit\ is normally regarded as independent of the other peaks even when they are in
the same cluster. Moreover, only a limited range of the diffraction pattern is usually considered
in the fitting procedure. On the other hand, in the \whpade\ all peaks of the pattern or a large
number of them are considered simultaneously and fitted as a whole. As pointed out already,
knowledge of unit cell parameters is required for \whpade\ but not for \curfit. The best known
methods in pattern decomposition are the \Pawley\ and \Lebail\ techniques which are based on a \ls\
fitting procedure and are derived from the \Rietveld\ method. Nowadays, various modifications to
the \Pawley\ and \Lebail\ procedures are in use, representing different approaches in extracting
the required parameters from diffraction patterns.

\subsubsection{Statistical Indicators and Counting Statistics}

It is desirable in pattern modelling to have the ability to measure the quality of the fit as a
whole by a single number. In the literature of powder diffraction, a number of standard statistical
parameters have been proposed and used to monitor the convergence process and check the quality of
the fit. They are used as indicators of how the refinement process is progressing and how good the
final result is. These indicators are called functions-of-merit or figures-of-merit (FoM). Although
these parameters are usually associated with the whole pattern structure refinement, they are more
general and are employed in curve-fitting and pattern decomposition as well. The main FoM are the
profile residual $\PR$, the weighted profile residual $\WPR$, the expected residual $\ER$, the
\Bragg\ residual $\BR$, the \strfac\ residual $\SFR$ and the goodness-of-fit index $\GoFI$
\cite{SchwarzenbachAFGHe1989, JansenSW1994}. These are presented in Table \ref{statisticalIndic}.

\begin{table} [!t]
\centering %
\caption{Statistical indicators for pattern fitting and refinement.} \label{statisticalIndic}
\vspace{0.5cm} %

\begin{tabular}{ll}
\hline
{\bf Statistical indicator} \verb|                    | & {\bf Definition} \\
\hline      \vspace{-0.3cm} \\
Profile residual &            $\PR = \frac{\sum_{i} |\OCR_{i} -  \CCR_{i}|}{\sum_{i} \OCR_{i}}$ \\ %
            \vspace{-0.3cm} \\
Weighted profile residual &   $\WPR = \left[ \frac{\sum_{i} \SW_{i}(\OCR_{i} -  \CCR_{i})^{2}}{\sum_{i} \SW_{i}  {\OCR_{i}}^{2}} \right]^{1/2}$ \\ %
            \vspace{-0.3cm} \\
Expected residual &           $\ER = \left[ \frac{\NoO - \NoP + \NoC}{\sum_{i} \SW_{i}{\OCR_{i}}^{2}} \right]^{1/2}$ \\ %
            \vspace{-0.3cm} \\
Bragg residual &              $\BR = \frac{\sum_{k} |\OII_{k} -  \CII_{k}|}{\sum_{k} |\OII_{k}|}$ \\ %
            \vspace{-0.3cm} \\
Structure factor residual &   $\SFR = \frac{\sum_{k} |\sqrt{\OII_{k}} -  \sqrt{\CII_{k}}|}{\sum_{k} |\sqrt{\OII_{k}}|}$ \\ %
            \vspace{-0.3cm} \\
Goodness-of-fit index         & $\GoFI = \left[ \frac{\WPR}{\ER} \right]^{2} =  \frac{\sum_{i} \SW_{i}(\OCR_{i} -  \CCR_{i})^{2}} {(\NoO - \NoP + \NoC)}$ \\ %
            \vspace{-0.3cm} \\
\hline
\end{tabular}

\end{table}

In these relations, $\OCR$ and $\CCR$ are, respectively, the observed and scaled calculated
intensities at step $i$ in the pattern, and $\SW_{i}$ is the corresponding observation weight,
usually assigned the value 1/$\OCR_{i}$ on the basis of counting variance, assuming $\Var_{i}$ =
$\OCR_{i}$ according to \Poisson\ statistical distribution, as given by Equation \ref{weiSigma}.
$\NoO$, $\NoP$ and $\NoC$ are the number of observations, the number of refined parameters in the
calculated model, and the number of applied constraints, respectively. $\OII_{k}$ is the integrated
observed intensity of reflection $k$, and $\CII_{k}$ is the corresponding calculated integrated
intensity. The summation index $i$ runs over all data points measured in the experimental pattern,
while the index $k$ runs over all independent \Bragg\ reflections. As these indicators measure the
agreement between the observed and calculated quantities, they must be closely monitored during the
refinement. When the refinement is progressing in the right direction, they should gradually
decrease and finally settle to a minimum when convergence is reached. If these figures-of-merit
start rising, which is a sign of divergence, the refinement should be stopped and resumed after
imposing suitable constraints on the refined parameters \cite{HillF1990, LangfordL1996, Young1993}.

The profile residual $\PR$ is described as a true quantity because it is based on the discrepancies
between the observed and calculated intensity values. Of these figures-of-merit, the weighted
profile residual $\WPR$ and the goodness-of-fit index $\GoFI$ are statistically the most meaningful
indicators of the overall fit since the numerator contains the residual that is minimised in the
least squares procedure. The goodness-of-fit index, which is inherited from the general statistics
literature and not specific to diffraction pattern refinement, is used and quoted quite often in
the powder refinement literature and considered as one of the most important and widely accepted
figures-of-merit. The expected residual $\ER$ is used in the Rietveld refinement to quantify the
quality of the experimental data. The structure factor residual $\SFR$ is biased towards the
structural model, but it gives an indication of the reliability of the structure. This quantity is
not widely used to monitor the refinement process. The Bragg residual $\BR$ is based on the
intensities deduced from the model and hence is biased in favour of the used model. It is highly
dependent on the procedure in which the observed integrated intensities are estimated, and may be
described as an artificial quantity generated in order to get values similar to the single crystal
and two-stage residual. However, $\BR$ is a quite important figure-of-merit in Rietveld refinement
though it has little or no value in full pattern decomposition because only observed Bragg
intensities are meaningful in both Pawley and Le Bail methods \cite{JansenSW1994, Young1993,
MccuskerDCLS1999, BookWill2006, BookPecharskyZ2005}.

Regarding the counting statistics, in the literature of powder diffraction the count rate is
usually modelled by a Poisson statistical distribution. Consequently, the statistical weights given
to each observation $\SW_{i}$ are obtained from the experimental errors which are regarded to be
proportional to the square root of the observed count rate $\OCR_{i}$, that is
\begin{equation}\label{weiSigma}
    \SW_{i} = \frac{1}{\Var_{i}} = \frac{1}{\OCR_{i}}
\end{equation}
where $\Var$ is the statistical variance. This weighting scheme reflects only the statistical
errors in the observed values with no consideration to the errors in the calculated values. The
latter, which can arise from defects in the structural or profile models and from inadequacies in
the computational technicalities, should be accounted for by other means. Other weighting schemes
for modelling the experimental errors have also been proposed and used. Examples include unit
weight, reciprocal square of the observed intensity, and modulus of the difference between observed
and calculated intensities \cite{AlbinatiW1982, Young1993, SpagnaC1999}.

\subsubsection{Optimisation Methods}

To extract good information from \pd\ patterns, good experimental data and careful modelling should
be associated with a good fitting algorithm. In pattern fitting and refinement, an optimisation
procedure is required to achieve and measure the proximity of the model to the observational data.
The most widely used optimisation procedure is the nonlinear Least Squares (LS) residual method in
its various realisations such as \GauNew\ and \LevMar\ algorithms. Although the \ls\ algorithms are
sufficient, they suffer from some serious problems and are not ideal in all situations. Therefore
other methods have been developed and used in powder \difpat\ modelling. The main rivals to the
\ls\ as an optimisation procedure are the \maxlik\ and \maxent\ methods. These are closely related
as they both rely on the same basic principle \cite{Prince1982, Gilmore1996, GilmoreHB1991}. In the
following, we discuss and compare these optimisation procedures as applied in the \pd\ pattern
refinement.

Least squares fitting is probably the most popular numerical method in science and the main
technique used in powder diffraction as it is an essential element in most pattern refinement
routines. It is widely used to determine the best set of parameters in a model to fit a set of
observational data. Both whole pattern structure refinement (\Rietveld) and \whpade\ (\Pawley\ and
\Lebail), as well as \curfit\ procedures, are based on a nonlinear \ls\ technique to minimise the
difference between the observed and calculated profiles. The refinement of large structures by \ls\
can fail because the required computational resources dramatically increase as the number of \ls\
parameters increase. In these cases, reduction techniques such as block matrix approximation may be
employed where only diagonal blocks of the original \ls\ matrix are used. Theoretically, this
should produce the same solution but with less accurate error estimates. Also \ls\ may suffer from
numerical instabilities resulting in practical difficulties such as unexpectedly large values of
goodness-of-fit statistics in some data sets and unrealistic estimates of standard deviations in
the refined parameters. Although \ls\ can be used to refine any model, the suitability and validity
of the model in the particular case are out of \ls\ scope and hence require independent
justification \cite{BookMesserschmidt2007, SchwarzenbachAFGHe1989}.

The \maxlik\ method was introduced as another possibility for pattern refinement of
crystallographic structures to replace the classical \ls\ and overcome its limitations. The
objective of the \maxlik\ method is to find a raw model that has the best chance to be improved by
applying small steps to achieve full agreement between the observed and calculated models
\cite{RestoriS1995, Bricogne1984, GilmoreBB1990, AntoniadisBF1990}.

The \maxent\ is another optimisation technique that has been used in crystallography and \pd\ to
replace \ls. For example, \maxent\ is employed in some powder diffraction studies to restore the
lost phase angles. It is also used in the pattern refinement as a deconvolution procedure to
determine the relative intensities of the overlapped reflections from first principles. The
technique is found to be efficient with simulated profiles having various noise levels
\cite{DongG1998, GilmoreDB1999}.

%

%% file: Soft.tex
%
%
\chapter{Software for Powder Diffraction} \label{Software}

An overwhelming number of computer programs have been developed over the last few decades for
processing and analysing crystallographic and \pd\ data; most of them are in the public domain. For
instance, Smith and Gorter \cite{SmithG1991} identified over 280 programs developed until 1990 for
the analysis of \pd\ data. These programs can be used for solving almost any kind of
crystallographic and \pd\ problem. The majority of these programs are free of charge for academic
use and can be downloaded from the Internet directly or via links from dedicated websites such as
\CCPs\ \cite{CCP14}. Others can be obtained from the authors on request or require a license. Each
category normally contains a number of programs that vary in quality, licensing, sophistication,
computational efficiency, documentation, user friendliness and so on. The list is substantially
increasing in number and the quality is generally improving all the time, though the rate of growth
varies from one area to another. In each category, some programs are optimised for solving certain
types of problems while others are optimised for different sorts of problems. However, in this
chapter we provide a glimpse into this huge and fast-growing field by presenting some categories of
the available software with very few examples from each. Most of these example programs perform
other functions; however they were included in the category that is more appropriate to the major
type of tasks they perform.

\section{Pattern Simulation}

One way for interpreting the information in a measured powder diffraction pattern is to calculate a
theoretical pattern from a structural model to simulate the experimental data. Digital patterns can
be varied to account for different instrumental and experimental conditions normally present in a
diffraction measurement. Reference data can be adjusted to match observational data for various
purposes such as phase identification, quantitative phase analysis, sample condition tests and so
on. One approach in pattern simulation is to take the integrated intensities, which are calculated
from the crystal structure description, and create a diffraction pattern by calculating profiles
for every peak and combining them to simulate the actual pattern. To generate diffraction patterns,
sufficient information must be supplied by the user regarding the crystal, background, instruments,
peak shape, and so on. The availability of simulated diffraction patterns has facilitated the
indexing and verification of diffraction pattern analysis. However, in some complex experimental
conditions the simulated pattern may not reflect all the factors involved and hence careful
scrutiny is required \cite{SmithG1991}. An example for this category of software is \RIETAN\ (by
Izumi) \cite{Izumi1997} which is a whole pattern refinement program that can also be used for
simulating X-ray and neutron diffraction patterns. Many other \Rietveld\ refinement programs can
also simulate powder diffraction patterns.

\section{Whole Pattern Structure Refinement}

The \Rietveld\ method is the most popular of powder diffraction pattern refinement techniques.
Consequently, there are many programs for doing this sort of structural refinement with a great
diversity in all aspects. Among the most popular of these programs are \FullProf\ (by
Rodr\'{\i}guez-Carvajal), \GSAS\ (by Larson and Von Dreele), \TOPAS\ (by Coelho), \Rietica\ (by
Hunter and Howard).

\FullProf\ is a collection of crystallographic programs mainly developed for \Rietveld\ structure
refinement of neutron and \Xray\ diffraction patterns. \FullProf, which is written in Fortran, can
be used for analysing various experimental data such as magnetic and nuclear scattering, \tofl\ and
constant wavelength neutron experiments, as well as \Xray\ data in energy and angle dispersive
modes. The suite can be run from command line interface and from WinPLOTR or EdPCR graphic
interface \cite{Carvajal2001}. \FullProf\ is one of the most reliable and widely used examples of
software in crystallographic and \pd\ analysis.

The General Structure Analysis System (GSAS) of Larson and Von Dreele is a set of programs for
processing and analysing single crystal and powder diffraction data that are obtained with X-rays
or neutrons. \GSAS\ can handle these types of data simultaneously for a given structural problem.
It can also handle powder diffraction data from a mixture of phases, and refine structural
parameters for each phase. \GSAS\ features a menu-driven editor `EXPEDT' which is used to prepare
all the input for the main calculations. The entire \GSAS\ system is written in FORTRAN. It can run
on Windows, Linux and Macintosh \cite{LarsonV2004}.

\TOPAS\ is a \pd\ analysis software distributed in commercial and academic versions. The program
has a graphic user interface but can also run from command line interface and by text-based input
files, i.e. scripts. Full graphic interface is available only with the commercial version. The
academic version, which is not free, is run from text input files that store a sequence of
commands. \TOPAS\ is praised for its stability, reliability and smooth convergence. The program can
perform pattern decomposition by \Pawley\ and \Lebail\ procedures as well as \Rietveld\ structure
refinement for both \Xray\ and neutron in angle and energy dispersive modes. \TOPAS\ is also
capable of performing powder indexing and structure solution by \simann\ \cite{Coelho2006}. The
text editor jEdit, with its powerful macro language, may be used to drive and control \TOPAS\ in
text-input mode.

\Rietica\ is a \Rietveld\ analysis computer program. The suite consists of the LHPM \Rietveld\
program and Rietica which helps in controlling LHPM as well as creating and updating the structure
input files. \Rietica\ is a Windows (95/98/NT) based program with a simple graphic user interface.
The program can perform neutron and \Xray\ \Rietveld\ and \Lebail\ refinement in constant and
variable wavelength modes, simulation of diffraction data in these modes, data and refinement
parameters plotting, background and region selection input, integrated Fourier plotting, and data
editing. The program has a macro language, similar to Basic, which can be used for complex batch
processing and programming. \Rietica\ can interact with \GSAS\ and \FullProf\ through data
importing \cite{HunterH2000, Hunter2001}.

\section{Whole Pattern Decomposition}

Most \Rietveld\ refinement programs have the capability to do whole pattern decomposition by
\Pawley\ or \Lebail\ methods with the latter being the most common. Examples include \GSAS, \TOPAS,
\RIETAN, and \Rietica. The first two can perform both \Pawley\ and \Lebail\ fittings, while the
others can do \Lebail\ only.

\section{Curve-Fitting}

The method of \curfit\ with no knowledge of unit cell or structural parameters is based on
decomposing the pattern into independent peaks using relatively simple profile models to extract
the integrated intensity and other profile parameters. Usually, a limited range of the pattern is
considered during this process. There are many \curfit\ programs, some are general purpose while
others are specifically developed for \pd. A prominent example of \curfit\ programs is \Fityk\
\cite{Fityk} (by Wojdyr) which is free software with a graphic user interface. It can also be
driven by a command line interface using text file scripts. The program runs under different
platforms including Linux, Windows and Macintosh. \Fityk\ is written in C++ language with Python
bindings. Its main functionality is nonlinear fitting of analytical functions to a set of data
points using a number of commonly used basis functions such as \Gauss, \Lorentz, \Voigt,
polynomials, and \PearsonVII. It offers background subtraction with different nonlinear fitting
methods. \Fityk\ offers three optimisation methods (\LevMar, Genetic Algorithms, and Nelder-Mead
simplex) with error modelling and application of constraints.

A second example of \curfit\ programs is \CMPR\ \cite{CMPR} (by Toby) which is a
platform-independent multipurpose package for analysing, visualising and manipulating \pd\ data. As
well as \curfit, the program can perform manual- and auto-indexing, and offers some handy features
such as graphic display and manipulation of $hkl$ lines. A third example is \Xfit\ (by Cheary and
Coelho) which is used by \pd\ community \cite{CCP14}. \ProgName\ can also perform curve-fitting in
single, multiple, batch and multi-batch modes.

\section{Search-Match}

Search-match programs require a database of diffraction patterns such as Cambridge Structural
Database and Powder Diffraction File. Some of these databases have their own search-match programs.
The databases for \seamat\ routines are generally based on X-ray diffraction patterns. Search-match
programs vary in their underlying search algorithm, speed, reliability, and so on. Some programs
are capable of examining, in a few seconds on a normal computer, tens of thousands of powder
diffraction patterns and proposing a list of candidate patterns to match the unknown pattern. The
results of the search-match depend on the quality of the observational data, the quality of the
database, the search-match algorithm, and the criteria used in the search. Matches are usually
ranked using a figure-of-merit where a large figure-of-merit normally means good match. It is quite
possible that all the suggested patterns are not suitable, and hence the user should accept or
reject the proposed patterns based on clear evidence \cite{BookWill2006}.

Most existing \seamat\ programs do not use the full profile data. Instead, they rely on simplified
patterns in which the full diffraction profile is reduced to a set of the strongest peaks. The main
advantage of this reduced approach is high speed and less computer resources especially when using
large databases. Most \seamat\ programs that are in the public domain are commercial. An example of
the search-match software is \dSNAP\ (by Barr, Gilmore, Dong, Parkin and Wilson) which is a graphic
user interface program for automatic classification and visualisation of the results of database
searches using the \CSDl\ (\CSDs) \cite{BarrGDPW2006}. Another example is Portable Logic Program
(by Toby) which can be used for searching the \ICDDl\ (\ICDDs) database. The program is currently
incorporated within the \CMPR\ suite \cite{CMPR}.

\section{Structure Visualisation}

Visualisation of crystal structure means displaying the arrangement of atoms, ions, and molecules
in the unit cell, usually in 3D space with or without unit cell outlines. The atoms are normally
represented by coloured spheres with their sizes being in proportion to the atomic sizes. The plot
can usually be scaled, translated, rotated, zoomed and manipulated in various ways with possible
labels and explanatory comments. The software may enable the user to measure and display things
like unit cell axes, angles, distances, and torsion angles. The software usually reads the
structure from raw data files with a specific format. Alternatively, the data can be obtained and
mapped directly from within the program following a structure solution or structure refinement
process. These programs usually recognise several data formats and may be able to import and export
between them. Structure visualisation can be a great aid in comprehending and checking the
solution. Examples of structure visualisation programs are \Mercury\ (from Cambridge
Crystallographic Data Centre), \DRAWxtl\ (by Finger, Kroeker and Toby), and \MatStu\ (from
Accelrys).

\Mercury\ \CSDs, which is part of the Cambridge Structural Database System, is a popular program
for visualising crystal structures in 3D using a range of visualisation options and display styles.
The program can read crystallographic structural data in various formats. It is written in C++
language with an object oriented nature. \Mercury\ can display multiple structures simultaneously
and perform \ls\ overlay of pairs of structures. It can also do various transformations such as
rotation and translation \cite{MacraeEMPSe2006, MercuryManual}.

\DRAWxtl\ is an open source computer program for crystal structure drawings. It produces standard
graphical representations such as spheres, ellipsoids, bonds and polyhedra. The drawing output is
produced either in the form of an interactive screen representation, or as Virtual Reality
Modelling Language (VRML) files. The program is supported on a variety of platforms such as
Windows, Mac, Linux and other Unix distributions. It has the ability to plot incommensurately
modulated and composite structures. It can also produce input files for the Persistence of Vision
Raytracer (POV-Ray) rendering program to create high quality images \cite{FingerKT2007}.

\MatStu\ is a commercial multi-module software environment for molecular modelling, simulation and
visualisation. The program is designed for structural and computational researchers in chemistry
and materials science. It provides tools for modelling crystallisation processes and crystal
structure. It can also be used for studying polymer properties, catalysis, and structure-activity
relationships \cite{MaterialsStudio}. It should be remarked that many crystallographic and \pd\
programs are capable of producing visual images of the resolved structures. Several examples can be
found in the category of whole pattern refinement programs such as \TOPAS.

\section{Data Visualisation}

The purpose of data visualisation is to present numerical and abstract data in visual form. Two
main aspects of these visualisation techniques are colour-coding, and graphic plotting in 2D and 3D
space. For example, it is common practice in powder diffraction to use 2D stack plots or 3D surface
plots of a series of diffraction patterns in dynamic phase transformation studies. Another common
form of visualisation in \pd\ is the 2D tomographic colour-coded images of the total intensity of
pixels in tomographic imaging studies. These tomographic images are also used to display the
integrated intensity of a particular peak in phase identification investigations. This sort of
visualisation provides a brief, comparative and informative glimpse of certain parameters of the
individual data sets. This enables the user to perform summary analysis and make critical judgments
about the next step in data collection and analysis. This kind of visualisation is very important
in some situations such as collecting data from \synch\ facilities where beam time is scarce and
should be exploited maximally, or when the available computational resources are limited and do not
allow for detailed analysis.

These days many computer programs have a kind of data visualisation capabilities such as plotting
the measured and refined diffraction patterns. Most whole pattern structure refinement and whole
pattern decomposition programs, such as \GSAS\ and \TOPAS, have this visualisation capacity.
\Matlab\ scripts are also in common use by the scientific community to perform this sort of
visualisation. \ProgName\ includes several visualisation capabilities such as 2D pattern plotting,
3D surface plotting and 2D colour-coded tomographic imaging. The program can also create images in
single and multi-batch modes and save them as image files to the computer permanent storage.

\section{Indexing}

Indexing of powder diffraction patterns means the determination of $hkl$ indices of each
reflection, the dimensions of the unit cell and the crystal symmetry. This is an essential and
limiting step in an \abinitio\ structure determination from powder diffraction data. A list of
indexed intensities obtained from an indexing procedure may be used, for example, as a starting
point for the application of direct methods. There are several indexing methods that vary in their
speed, efficiency and reliability. These include zone indexing, trial, dichotomy, and Monte Carlo
methods. The efficiency of most powder pattern indexing software significantly deteriorates in the
presence of impurity lines \cite{LangfordL1996}. An example of indexing software is \Crysfire\ (by
Shirley \etal) which is a family of programs interconnected by a set of interlocking scripts in the
form of batch files. Its role is to act as an expert system to allow indexing of \pd\ patterns
quickly and smoothly by non-specialists \cite{Shirley2002}. Another example is \Supercel\ (by
Rodr\'{\i}guez-Carvajal) which is for indexing of super-cells and incommensurate structures. A
third example is \Chekcell\ (by Langier and Bochu) which is an indexing program for unit cell and
\spagro\ assignment \cite{CCP14}. As indicated already, \CMPR\ (by Toby) can also perform manual-
and auto-indexing.

\section{Unit Cell Refinement}

On indexing a powder diffraction pattern, the unit cell parameters ($abc, \alpha\beta\gamma$)
become known approximately. These parameters may then be refined by a least squares fitting routine
taking into account constraints due to the crystal system symmetry. Unit cell refinement of
standard materials is also used to check instrument and sample alignment. In principle any whole
pattern refinement program can do a kind of unit cell refinement \cite{Pawley1981}. An example of
dedicated unit cell refinement programs is LAPOD (by Langford) which performs a \ls\ refinement of
cell dimensions from powder data using Cohen's method. Another example is CELREF (by Laugier and
Bochu) which is a unit cell refinement program within the LMGP suite that can calculate lines based
on the \spagro\ and auto-select and auto-match calculated-to-observed peaks \cite{CCP14}.

\section{Structure Solution}

There are several programs that can perform crystal structure determination from powder diffraction
data. These programs use various structure solution techniques such as direct methods, simulated
annealing and Monte Carlo. One example is \EXPO\ (by Altomare \etal) which is the integration of
two programs, one for whole powder pattern decomposition and the other for the solution and
refinement of crystal structures \cite{AltomareBCCCe1999}. Another example is \ESPOIR\ (by Mileur
and Le Bail) which is a reverse \MonCar\ and pseudo \simann\ software for \abinitio\
crystallographic structure determination from a random starting model or from molecule location
\cite{MileurL2000}.

\section{Utility Programs}

These programs do various jobs which are essential for managing, processing and analysing
diffraction data. These jobs include merging files, data format conversion, mass-scale text
replacement, GUI management, background removal, peak finding, graphical interfacing, specimen
displacement, aberration correction, smoothing, peak offset determination, plotting, and so on. The
diversity and usefulness of these utilities cannot be overestimated. Examples of this category are
\DLConverter\ (by Smith and S\'{e}bastien), \PowderF\ (by Dragoe), and \PowderX\ (by Dong).

\DLConverter\ is a program with a simple graphic user interface developed at Daresbury \SRSs\ for
conversion of large amounts of diffraction data produced by modern fast detectors between different
formats. The program has an excellent batch processing capability and hence can be regarded as an
example of batch processing programs \cite{SmithS}. \PowderF\ software is a collection of tools
developed for \Xray\ powder diffraction. Its capabilities include management (e.g. merging) and
conversion of data files between various formats \cite{Powder4Manual}. \PowderX\ is another
software developed for the analysis of \Xray\ powder diffraction patterns. It has a number of
useful processing utilities such as smoothing, background removal and format conversion
\cite{Dong1999}. \ProgName\ can also be included in this category as it can be used, for example,
for data format conversion.

\section{Batch Processing}

There are many programs that do various jobs in batch mode to perform essential duties in managing,
processing and analysing data. These include several \pd\ programs that can be driven in batch mode
and hence can be used for large-scale processing and analysis. An example of these programs is
\PolySNAP\ packages (2 and M) \cite{PolySNAP, BarrDG2004} which are high throughput commercial
software for analysing large data sets. They are designed to match and analyse patterns utilising
their full profiles, and hence allowing for quick and accurate identification of samples. Their
functionalities can be automated for high throughput analysis to allow the interrogation of large
data sets (up to 1500) in a single run. \PolySNAP s provide quantitative phase analysis of mixed
samples using a non-\Rietveld\ approach, and employ several novel statistical methods with a
user-friendly graphic interface. The analysis results are summarised and visualised using a
flexible graphic output.

Among the prominent \pd\ programs that can run in batch processing mode is \FullProf. It is also
reported that \GSAS\ can be driven in batch mode by other purposely-written programs or scripts.
\ProgName\ is another example of batch processing software, as most of its functionalities are
implemented in batch and multi-batch modes, as will be explained in chapter \ref{EasyDD}.

\section{Data Preparation and Pretreatment} \label{DataPrep}

These programs process raw data produced by detectors and acquisitions systems as a first step for
further processing and analysis. Examples of preparation processes include introducing systematic
error corrections, squeezing binary image files to extract ASCII numeric data, and changing the
dimensionality of data by converting 2D diffraction patterns to 1D. An example of this category of
software is \Datasqueeze\ \cite{Datasqueeze} which is a commercial program with graphic user
interface that runs on a number of platforms. It is used for processing and analysing data from 2D
\Xray\ \diffraction\ detectors such as wire, image plate, and \CCDs\ devices. The program can run
in batch mode to do various jobs, and hence it is also an example of batch processing programs. The
batch mode jobs include converting binary image files, such as EDF which contain Debye-Scherrer
rings on 2D image plates, to 1D powder patterns in ASCII numeric format.

Another example of this category is \FITTD\ \cite{FIT2D} which is free software written in Fortran
language for Windows and Unix platforms. It is both a general purpose and specialist 1D and 2D
program for data reduction and analysis. It is used on most beamlines of the \ESRFl\ and by other
research groups around the world. The program supports processing of 1D and 2D data, and recognises
different detector types. It is used to convert data obtained from 2D detectors, such as image
plates, to 1D powder patterns, usually as intensity versus \scaang. \FITTD\ has a graphic user
interface with command line capabilities and macro language which may be used for automating
repetitive processes. Most \FITTD\ functionalities, can be performed in batch mode to process large
data sets in a single run, and hence it is another example of batch processing programs. A third
example of this category is \ProgName\ which has the capability of EDF binary image squeezing. It
also has the capability of manipulating, correcting and performing initial analysis on SRS 16.4
data and binary files from HEXITEC multi-TEDDI detectors.

%

%% file: Manu.tex
%
%
\chapter{\ProgName\ Program} \label{EasyDD}

The idea of high throughput software emerges from the need to process huge quantities of data in a
consistent and repetitive manner using automated procedures. The demand for high throughput
processing capabilities in scientific research has substantially increased in recent years. One
reason is the availability of bright radiation sources, such as synchrotron and neutron facilities,
which facilitate the collection of massive amount of data in short periods of time. Another reason
is the essential improvement in data acquisition technology, and the more radical anticipated
improvement in this technology such as multi-TEDDI detectors which are under development for high
energy \Xray\ applications. The revolution in digital technology has also contributed to the mass
accumulation of data and encouraged this phenomenon. This is due to the growing use of automated
and computerised data collection techniques on one hand, and to the availability of vast electronic
storage with massive processing computational power on the other.

The principal objective of \ProgName\ project was to develop a computer code for batch processing,
visualisation and large-scale analysis of huge volumes of spectral data, mainly those obtained from
synchrotron radiation sources for powder diffraction applications. Such a utility greatly assists
studies on various physical systems and enables far larger and detailed data sets to be rapidly
interrogated and analysed. At the start of \ProgName\ project a number of general objectives were
set; the main ones were

\begin{itemize}

\item The program must be user friendly to minimise the time and effort required to learn and use. A
graphic user interface was therefore adopted in favour of a command line interface although the
latter is more common in scientific computing and much easier to develop.

\item It should be capable of handling several common data file formats including a generic format so that
the program can be used for processing data produced by different detectors and acquisition
systems.

\item The program must lend itself to future development to meet the ever-increasing demand on data
processing and analysis capabilities. Therefore, it should be built on a continuously developing
technology to keep in pace with the advancements in computing and GUI technology.

\item It must be optimised for the commonly available computational resources; most importantly CPU time
and memory. The program therefore tries to set the limits of its data processing capacity to the
limits of the available computational resources.

\end{itemize}

\ProgName\ can be defined as a high throughput software to manage, process, analyse and visualise
scientific data in general and synchrotron \Xray\ data in particular. The name `\ProgName' comes
from the original name `EasyEDD' which was adopted for historical reasons as the software was
developed initially for the users of the energy dispersive diffraction SRS station 16.4. As the
program eventually evolved to be more general and can be used for processing ADD data as well as
EDD data, the name `\ProgName' was adopted to reflect this extension. In fact, currently there is
no restriction on the program being used for general applications not related to synchrotron and
\pd, as the program is capable of processing any data having the correct format. One of these
formats is a generic $xy$ style which can be used for all-purpose data. The program, which consists
of about 30000 lines of code, is written in C++ language and uses a hybrid approach of procedural
and object oriented programming. Its main attribute is the ability to process large quantities of
data files with ease and comfort using limited time and computing resources. Most \ProgName\
algorithms are optimised for speed and hence it can be rightly described as ultra-fast high
throughput software.

\ProgName\ combines Graphic User Interface (GUI) technology, such as dialogs, tooltips, colour
coding, and context menus, with standard scientific computing techniques. Its resources include the
standard C++ library, Qt toolkit \cite{Qt} and its extension QwtPlot3D \cite{QwtPlot3D} for GUI
design, with numerous algorithms, functions and techniques. The ultimate objective of \ProgName\ is
to become a workbench for spectral data analysis. In \ProgName, the main data are stored in a 3D
vector where the basic unit is a `\voxel' object in which all information relevant to an individual
data set are stored. As a user-friendly feature, most components have memory, i.e. the settings are
saved on exiting the program. Another user-friendly feature is that the components have explanatory
tooltips which give brief description of the function (e.g. button tooltips) or reveal information
related to the data represented by that component (e.g. tab tooltips).

Currently, eight input data formats are supported: generic $xy$, MCA of the \DiamonD\ synchrotron,
MCA of the \ESRFs, ERD, \LUCIA, \HEXITECs, \SRSs\ 16.4, and EDF files. The first is a simple $xy$
format where the first column in the file contains the $x$ values (e.g. energy or channel number or
angle) while the second column contains the $y$ values (e.g. intensity or count rate). The second
is the MCA format of the \DiamonD\ Light Source \cite{Diamond}. \label{MCARef} The file contains
the $y$ values (number of counts) only, as a function of an implicit channel number, with possible
redundant header and footer lines. The third is the MCA format of the \ESRFl\ \cite{ESRF}. The file
has only headers to be ignored, and the first data line starts with `@A' with each 16 data entries
occupying a single text line ending with backslash `$\backslash$'. Again, the data in the ESRF MCA
files are number of counts versus an implicit channel number. The fourth is the format of the ERD
multi-pixel 2D detector of the University of Manchester. The fifth is the format of the \LUCIA\
beamline at the \SOLEIL\ synchrotron \cite{LUCIA}. The sixth is the format of the \HEXITECs\
\cite{HEXITEC} multi-pixel 2D detector. The seventh is the format of the data files obtained from
station 16.4 of Daresbury \SRSl\ \cite{SRS} in \EDDs\ mode. The eighth is the format of the EDF
image files of the ESRF. The code can be easily extended to support other data formats.

One of the main functionalities of \ProgName\ is to read files from data sets collectively and map
their information on GUI components. What is required for this operation is to deposit the files in
a directory and invoke the relevant reading function. In the case of the SRS, where the data files
have a highly structured format, the files are read and automatically recognised (i.e. SRS, scalars
and vectors), and therefore non-SRS files in the source directory are identified and ignored. For
the other formats, the file extension is used for type recognition, and hence no foreign files of
the same extension should be mixed. On reading the files, the data are stored in memory and mapped
on a 2D colour-coded tab. Multiple tabs from different data sources and of different data types can
be created at the same time. The tabs can also be removed collectively or individually in any
order. On removing a tab, all the data belonging to that tab are erased from memory and hence lost
permanently. The remaining tabs will be relabelled to reflect the current state.

In the following sections we outline the main components of \ProgName.

\section{Main Window} \label{MWin}

\begin{figure} [!h]
\includegraphics
[scale=0.85] {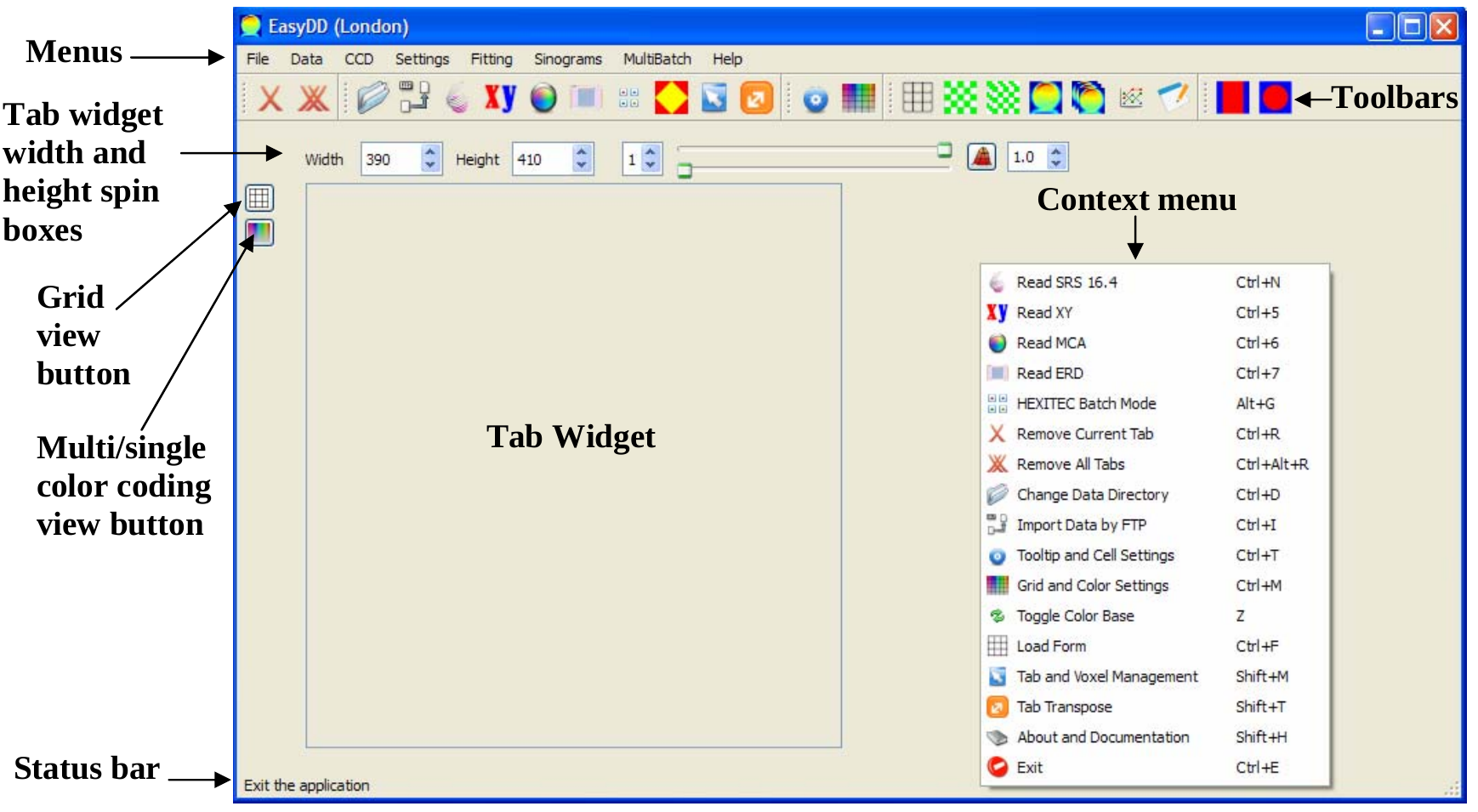}
\caption{\ProgName\ main window.} \label{MainWindow}
\end{figure}

This is a standard GUI dialog with menus, toolbars, a status bar, context menus and so on. The
basic functionality of the main window (seen in Figure \ref{MainWindow}) is to serve as a platform
for accessing and managing the other components.

\subsection{Menus} \label{MWMenus}

The main window has several menus which provide access to the program functions. These menus, some
of which have submenus, contain all the main items of the program. The keyboard shortcut for each
item is shown on the right of the item title, and an icon characterising the item is displayed on
the left of the title.

\subsubsection{File Menu} \label{MWFile}

\begin{figure} [!h]
\centering
\includegraphics
[scale=0.8] {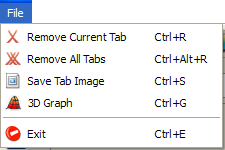}
\caption{The main window File menu.} \label{MWFilMen}
\end{figure}

This menu (Figure \ref{MWFilMen}) includes the following items

\begin{itemize}

\item Remove Current Tab \icon{removeTab}: to remove the current tab and erase data from memory.

\item Remove All Tabs \icon{removeAllTabs}: to remove all tabs and clean the memory.

\item Save Tab Image \icon{saveImage}: to save the image of the current tab in png format.

\item 3D Graph \icon{threeDGraph}: to plot a 3D graph of the current tab.

\item Exit \icon{exit}: for closing the main window and hence exiting the program. All saveable settings
will be saved to the disc, and hence activated on the next launch of the program.

\end{itemize}

\subsubsection{Data Menu} \label{MWData}

\begin{figure} [!h]
\centering
\includegraphics
[scale=0.7] {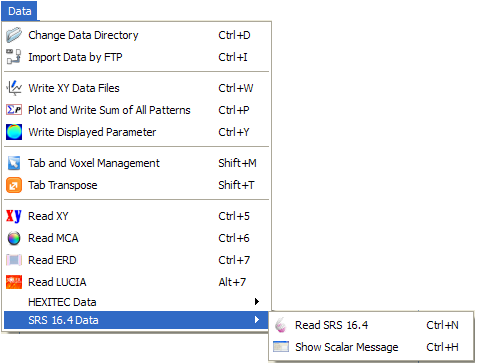}
\caption{The main window Data menu.} \label{MWDatMen}
\end{figure}

\noindent This menu (Figure \ref{MWDatMen}) includes

\begin{itemize}

\item Change Data Directory \icon{changeDir}: to launch `Browse for Folder' dialog (Figure
\ref{browseDia}). From this dialog the user can navigate to the folder of choice where the required
input data files are stored. As a user-friendly feature, the program has default directories for
data reading when the program starts. These default directories are: `DataESRFXY' for $xy$ data
type, `DataMCA' for MCA, `DataHEXITEC' for HEXITEC, `DataSRS164' for SRS 16.4, and `DataEDF' for
EDF. The multi-batch functions in MultiBatch menu also have a default directory called
`MultiBatch'. On changing the data directory by using `Change Data Directory' function, all default
directories are suspended and the newly selected folder will become the default directory for all
functions that require reading input data.

\begin{figure} [!h]
\centering
\includegraphics
[scale=0.8] {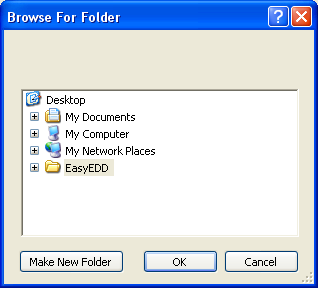}
\caption{`Browse for Folder' dialog.} \label{browseDia}
\end{figure}


\item Import Data by FTP \icon{import}: to launch the FTP dialog (Figure \ref{ftpDialog}). This dialog
(Figure \ref{ftpDialog}) is for importing files by \FTPl\ (\FTPs). The dialog has three line
editors: `Ftp server' to input the address of the remote server, `Username' and `Password' to input
these attributes. The `Connect/Disconnect' push button is for connecting/disconnecting to the
server after entering valid data. On connecting to the server, the folders and files on the server
root directory appear on the list widget beneath the line editors. A folder can be opened and its
contents inspected by double-clicking on its icon in the list widget. On pressing the button
(\includegraphics[width=0.025\textwidth]{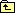}) next to the first line editor, the list
widget view returns to the parent directory. A file can be downloaded from the remote server by
selecting it first followed by pressing the `Download' button. The `Quit' push button is used to
close the FTP dialog. The current functionality of the \FTPs\ utility is primitive and hence
requires major development.

\begin{figure} [!h]
\centering
\includegraphics
[scale=0.8] {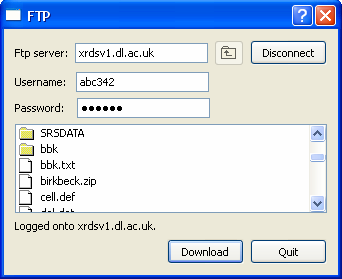}
\caption{FTP dialog.} \label{ftpDialog}
\end{figure}


\item Write XY Data Files \icon{writeXY}: to write the data of the current tab to text files (one file
of *.txt type for each \voxel) in generic $xy$ format. The files of each tab are saved to a folder
named `xyDataTab$N$' where `$N$' stands for the tab number. The naming of files follows the style
`T$t$R$r$C$c$' where $t$, $r$ and $c$ are the indices of tab, row and column respectively. The
indices have constant width (e.g. `01' for `1' as a 2-digit number) to keep the order when
re-reading these files. This routine facilitates saving, converting between data types, and
exporting to other applications for further processing such as visualisation.


\item Plot and Write Sum of All Patterns \icon{plotSum}: to plot the sum of all patterns in the current
tab and write the numeric data to an $xy$ text file named `spectraSumTab$N$' where $N$ is the tab
number. This function is based on the assumption that all common channels of the patterns of the
current tab voxels share the same $x$ coordinates. However, it is general with regard to the number
of data points in each pattern and hence it works even when they are different in size by summing
the existing channels and plotting the result for the maximum number of channels. The curve-fitting
capacity of the plotter used in this function is disabled.


\item Write Displayed Parameter \icon{writeDP}: to write the data of the displayed parameter (e.g. FWHM
and total or integrated intensity) on the current tab to a file named appropriately (e.g.
Tab$N$\_Intensity for intensity) to indicate the written parameter. The value of the parameter of
each voxel in the tab will be written. These values are structured in a 2D matrix depicting the
dimensions of the current tab.


\item Tab and \Voxel\ Management \icon{tabManag}: to perform one of 12 functions for manipulating rows,
columns and cells in the current tab. These functions are: delete row, delete column, set cell to
zero (i.e. zero $x$ and $y$ coordinates), copy cell (source to destination), exchange cells, delete
cell, rotate row anticlockwise, rotate row clockwise, rotate column anticlockwise, rotate column
clockwise, exchange rows, and exchange columns. The copy operation is performed by copying the
source (first cell) to the destination (second cell). An input data file called `TabManagement' is
required for tab and \voxel\ management. To perform each of these functions, a specific keyword is
required to identify the operation. The keyword should be followed by the indices of rows and/or
columns to identify the item(s) that these operations are applied upon. On the first use of this
menu item in the current session, a message (seen in Figure \ref{tabManagMessage}) appears to
display the available functions and their keywords in the required format in the input file with an
example of the required indices. If this message is needed again, a new session, possibly with
another instance of the program, should be launched. The appearance of this message only once in
each session is for the purpose of keeping the user informed with minimum inconvenience. When the
input file contains more than one entry (i.e. a keyword with its parameters) the program will read
the input file line by line and execute each line until it reaches the end of the file. Each line
should contain a single keyword with its parameters as the rest of the line will be ignored and
hence can be used for storing comments or entries that may be needed in the future. It should be
remarked that all these functions are available in the tab context menu (with no need for an input
file) by right-clicking on the current tab and choosing the appropriate option. In this case the
current cell, row or column is identified by the click position. If a second item (cell, row or
column) is required, as in the case of copy and exchange functions, the user will be instructed to
click on the second item. Although the use of a context menu is more convenient, the advantage of
performing tab and voxel management through an input file is to execute a series of these
operations in a single run.

\begin{figure} [!h]
\centering
\includegraphics
[scale=0.8] {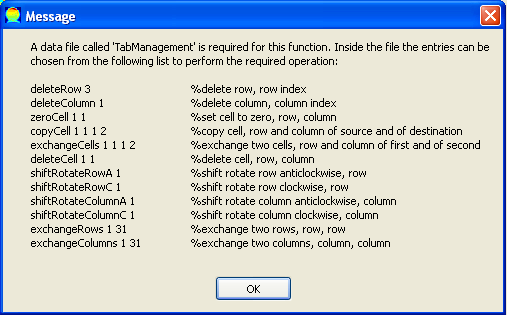}
\caption{Tab management message.} \label{tabManagMessage}
\end{figure}


\item Tab Transpose \icon{tabTrans}: to transpose the current tab, i.e. exchanging rows and
columns.


\item Read XY \icon{readESRFXY}: to read data files that have an $x$ value as a first entry and a $y$
value as a second entry in each row. The rest of the row, which may contain other data such as an
error index, will be ignored. Each file should contain data for a single voxel represented by a
single tab cell. The process starts with the appearance of a dialog (Figure \ref{RowColDialog})
from which the user can choose the number of rows and columns in the tab. This is followed by the
`Change File Order' dialog (Figure \ref{ChaOrdDialog}) from which the user can change the order of
reading files or delete non-required files. The order of files can be changed by first selecting a
file, then moving it up or down in the list using the `Up' and `Down' push buttons. Similarly, a
file can be deleted by selecting it first followed by clicking `Delete' button. It should be
remarked that the reading operation will proceed only when the source directory contains sufficient
number of data files. This is also true for other functions that require reading a certain number
of data files from a source directory.

\begin{figure} [!h]
\centering
\includegraphics
[scale=0.8] {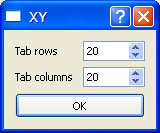}
\caption{XY dialog.} \label{RowColDialog}
\end{figure}

\begin{figure} [!h]
\centering
\includegraphics
[scale=0.8] {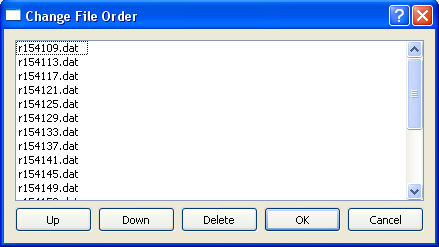}
\caption{Change File Order dialog.} \label{ChaOrdDialog}
\end{figure}


\item Read MCA  \icon{readMCA}: for reading MCA data files. There are two supported MCA formats, as
explained on page \pageref{MCARef}. In both cases, the process starts with the launch of the `MCA'
dialog (Figure \ref{McaFirDialog}) from which the user determines the number of rows and columns in
the tab and the number of header and footer rows to be ignored. This is followed by the `Change
File Order' dialog (Figure \ref{ChaOrdDialog}) as in the case of an XY reading. It is noteworthy
that for the ESRF MCA files the number of footers should be set to zero to avoid ignoring valid
data entries.

\begin{figure} [!h]
\centering
\includegraphics
[scale=0.8] {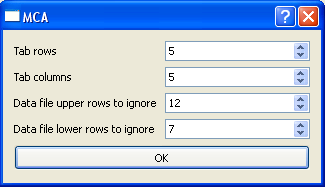}
\caption{MCA dialog.} \label{McaFirDialog}
\end{figure}


\item Read ERD \icon{readERD}: for reading and mapping data files of the ERD detector. Each
row\footnote{The basic description of ERD data is provided by Conny Hansson of the University of
Manchester.} in these files corresponds to an event taking place, i.e. a photon being detected by a
specific pixel. The data in the rows consist of 5 columns. The first column contains the value in
volts of the energy of event. The second column contains the index of the pixel in which the event
took place. The pixels are indexed by a 2D array that depicts the physical layout of the detector
channels. The third column contains a reference voltage used by the detector electronics. The
fourth column identifies the frame in which the event took place. The fifth column contains a
Boolean flag for data validation (`1' for valid and `0' for invalid). In the current implementation
of this function, the third and fourth columns are ignored. Because the energy is not quantised, a
binning process is applied before creating the spectrum. The bin size is determined by the user,
and is fed to the program through the map file which is outlined next.

On invoking this function, a dialog appears to enable the user to choose the ERD data file to be
read and mapped. This operation requires a map file called `ERD.map'. The map file contains the
following data: number of rows, number of columns and voltage bin size. This is followed, on
another text line, by the channels index map which mimics the physical layout of the 2D detector,
as seen in Figure \ref{ERDMap}.

The program ignores all the lines with invalid data, i.e. those whose Boolean flag on column 5 is
0. All the valid lines with the same pixel index (as given on column 2) are then grouped together
and the energies (as given on column 1) for each pixel are processed as spectra. The $x$ values of
these spectra are the bin numbers while the $y$ values are the number of counts, i.e. the number of
events in a specific bin. The pixels are then mapped on a tab with the normal colour coding, tool
tips, 2D and 3D visualisation, curve-fitting capability and so on, like other data formats. When a
channel has less than three events, its cell on the tab is marked with white. The routine is
general regarding the detector dimensions provided that a correct map file is supplied.

\begin{figure} [!h]
\centering
\includegraphics
[width=0.9\textwidth] {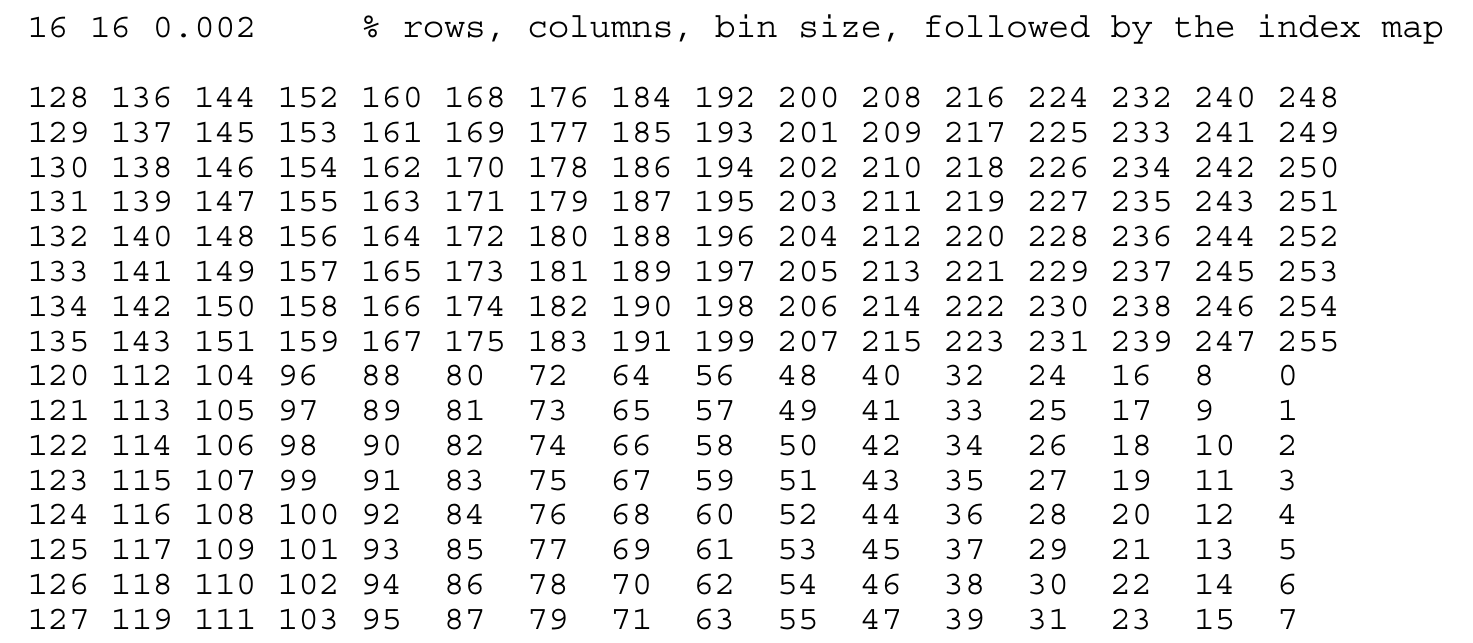}
\caption{A sample of an ERD map file for a $16\times 16$ detector.} \label{ERDMap}
\end{figure}


\begin{figure} [!h]
\centering
\includegraphics
[width=0.9\textwidth] {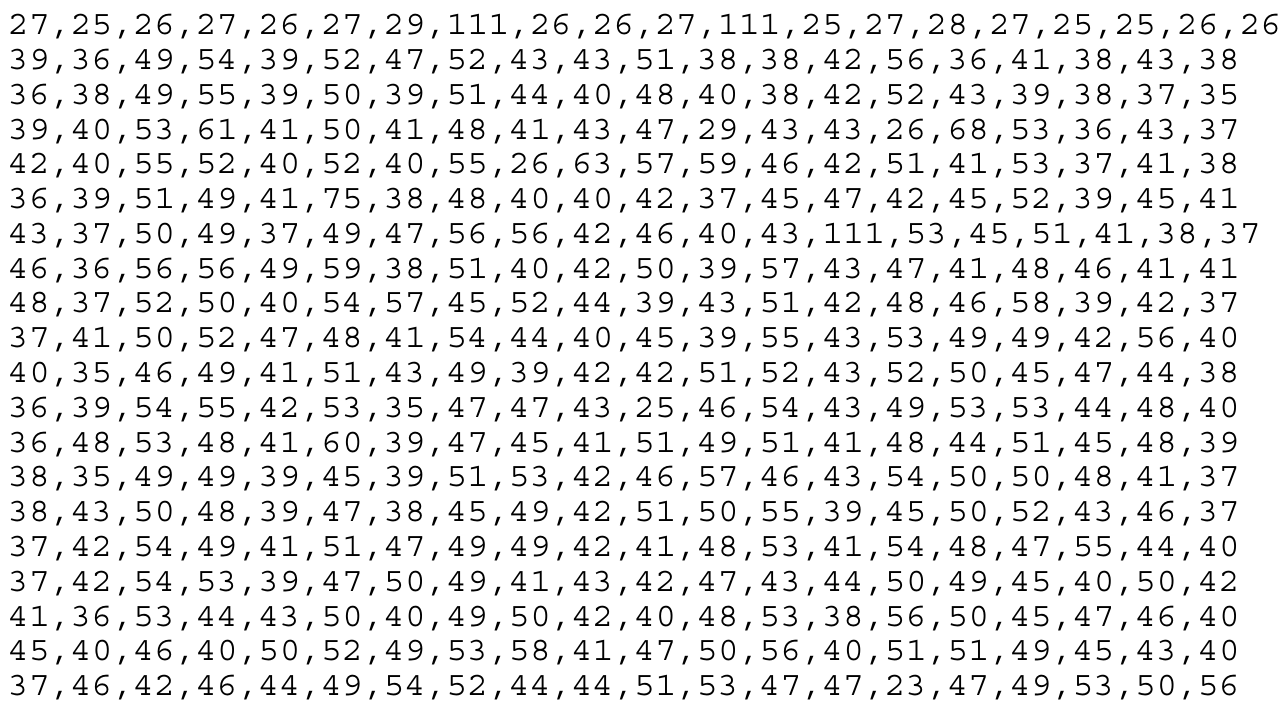}
\caption{A sample of a \HEXITECs\ thresholds file for a $20\times 20$ detector.} \label{thresholds}
\end{figure}

\item
Read LUCIA \icon{readLUCIA}: for reading and mapping data files obtained from \LUCIA\ beamline at
\SOLEIL\ synchrotron \cite{LUCIA}. The file\footnote{The basic description of \LUCIA\ data is
provided by Olivier Lazzari of Birkbeck College.} contains 960 patterns, each with 4000 channels.
The patterns are arranged in columns, where each entry in these columns represents the intensity of
the channel as a function of an implicit channel number. The user has the option to display these
960 patterns in a single row or as 32 rows times 30 columns on the EasyDD tab widget.

\item HEXITEC Data: for reading and mapping data files obtained from the \HEXITECs\ detector of
\RALl. The data\footnote{The basic description of HEXITEC data is provided by Matt Wilson of RAL.}
from the \HEXITECs\ 2D detector is written as a continuous list of 16 bit values to a binary file.
Each file contains a number of frames of data. A single frame contains $N=\mathcal{R}\times
\mathcal{C}$ values where $\mathcal{R}$ and $\mathcal{C}$ are respectively the number of rows and
columns of pixels in the detector. Each pixel in the detector has a different pedestal (offset)
level. Therefore, a list of $N$ pedestal values should be provided in an independent file called
`pedestals.txt'. Also the noise level on each pixel is different. Therefore, another file called
`thresholds.txt' should be provided to define the noise threshold (i.e. upper noise limit) of each
pixel. These thresholds may require scaling by the use of a user-defined constant multiplier. To
discriminate between the different data file formats, the program recognises the \HEXITECs\ raw
data files by the *.dat extension. The pedestal and threshold values should be positioned in their
files in a 2D matrix to mimic the physical layout of the corresponding pixels in the detector. The
neighbouring values in each row should be separated by a comma `,'. An image of a sample threshold
file is shown in Figure \ref{thresholds}. The other data are obtained from a dialog, seen in Figure
\ref{hexitecDialog}. These data include the number of rows and columns, the thresholds multiplier,
the number of bins, the minimum and maximum limits, a Boolean flag to run the end-of-frame
algorithm, a Boolean flag to run charge-sharing algorithm followed by another flag for
keeping/removing charge sharing events, and a Boolean flag to subtract from pedestals or ignore
them. Apart from the *.dat binary files, all the other input data files should be located in the
same directory as the program.

\begin{figure} [!h]
\centering
\includegraphics
[width=0.5\textwidth] {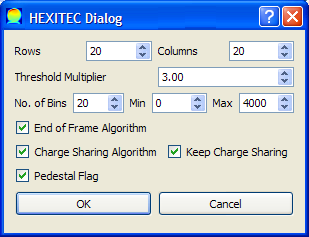}
\caption{\HEXITECs\ dialog.} \label{hexitecDialog}
\end{figure}

Currently, the program can process the \HEXITECs\ detector data in five modes:

\begin{enumerate}

\item HEXITEC Generic Mode \icon{hexiGeneric}: the raw data are read from a single *.dat binary file and
mapped on a tab where the patterns represent uncorrected intensity as a function of frame number.
This mode is normally slow and requires considerable memory space as the number of frames is
usually very high (210000). This mode is useful for carrying initial checks on the raw data. For
this mode, only the *.dat binary file and the number of rows and columns, which are obtained from
the \HEXITECs\ dialog, are required.

\item HEXITEC Single Mode \icon{hexiSingle}: the program reads the raw data from a single file and
corrects the individual entries by subtracting these entries from the corresponding pedestals.
These corrected data are then compared to the corresponding threshold noise and the entries that
fall below the threshold are discarded completely. The data which are below the minimum limit or
above the maximum limit are also removed. The remaining data of each pixel are then binned into a
histogram according to their values. This histogram, which can be plotted on the 2D plotter like
other data formats, represents the number of events as a function of bin number. In addition to the
raw data file, the pedestals and thresholds data files are required for this mode of operation. The
other data (rows, columns, threshold noise multiplier, number of bins, minimum and maximum limits)
are obtained from the \HEXITECs\ dialog. The bins have equal width obtained by dividing the range
(i.e. maximum minus minimum) by the number of bins.

\item HEXITEC Batch Mode \icon{hexiBatch}: the raw data from multiple *.dat binary files are read and
processed sequentially as in the second mode, but the results of these individual files are
collectively displayed on a single tab. The histograms of the voxels in this tab represent the
combined results of the individual files obtained by summing up these files. The program uses a
default directory called `DataHEXITEC' as a data source, unless the user has directed the program
to another directory. All raw data files found in the source directory are read and processed. The
other input files (i.e. `pedestals.txt' and `thresholds.txt') are expected to be in the directory
of the program. Other required parameters are obtained from the \HEXITECs\ dialog. In this mode,
end-of-frame and charge sharing algorithms can be applied. Currently, the HEXITEC batch mode can
run according to one of three options: end-of-frame with charge-sharing, end-of-frame without
charge-sharing, and neither, i.e. charge sharing can be applied only with end-of-frame. It should
be remarked that batch mode can run on a single file as well as multiple files.

{\bf Note about the end-of-frame algorithm}: when a pixel has an event in frame $N$ the user may
want to ignore that pixel value in frame $N+1$. The end-of-frame algorithm identifies these pixels
and applies this correction.

{\bf Note about charge sharing algorithm}: the voltage (signal) generated by a single X-ray photon
in the HEXITEC detector can be shared between up to four neighbouring pixels. The reason is that a
charge generated by an X-ray photon can be detected by more than one pixel. The purpose of the
charge sharing algorithm is to identify and correct such charge sharing events. Charge sharing has
two options: either allocate the event to a single pixel (i.e. the one with the largest voltage) or
remove charge sharing by setting the voltage of all pixels involved in charge sharing events to
zero and hence removing these events. Charge sharing corrections are conducted after removing the
pixel offsets and the end-of-frame corrections, but before the data reduction.

\item
HEXITEC Batch Mode with Energy Calibration \icon{hexi4}: this is similar to the previous item (i.e.
HEXITEC Batch Mode) but with performing linear energy calibration to the data by matching spectral
peaks to their known energy. The energy calibration is used to convert the raw data from the
detector into energy values as they are read into EasyDD. The calibration uses a linear relation
given by:

\begin{equation}\label{}
    E(x,y) = m(x,y) X(x,y) + c(x,y)
\end{equation}
where $E(x,y)$ is the calibrated value (in keV) for the pixel in row $x$ and column $y$ from the
raw data $X(x,y)$ with linear calibration coefficients $m(x,y)$ and $c(x,y)$. The calibration
coefficients are provided in two text files (`m.txt' and `c.txt') each containing an array of
$m\times n$ values (where $m$ and $n$ are the number of rows and columns respectively) similar to
the format of the thresholds and pedestals files. Energy calibration is performed after low energy
threshold cutting and before charge sharing correction/discrimination. A message appears at the
beginning to inform the user about the nature of this function and the sequence of operation. In
the case of end-of-frame with charge-sharing, an output file called `Stat.txt' which contains the
number of valid and discarded events is created. It should be remarked that this function requires
an input file called `binNumber2.txt'. This file (which was temporarily introduced for the purpose
of testing) contains the number of bins which should equal or exceed the number of bins in the
HEXITEC dialog.

\item
HEXITEC Batch Mode with Energy Calibration (New Format) \icon{hexi5}: this function is similar to
the previous one (i.e. HEXITEC Batch Mode with Energy Calibration) but uses the new HEXITEC data
format. In this format, the binary data files of *.dat type contain spectral data for a number of
frames. The sequence `255 255 255' is used to signify a new frame. This is followed by the frame
index which is a 48-bit integer. A new row is signified by the sequence `0 192', which is followed
by the row index (8-bit), the event magnitude (16-bit), and the column index (8-bit), in this
order. This is followed by one of three things (a) `255 255 255' for a new frame (b) `0 192' for an
event on a new row in the same frame (c) Another 16-bit event magnitude for a new event on the same
row in the same frame but on a different column, so the 16-bit event magnitude will be followed by
an 8-bit column index of the new event. On running this function, an output file called `Stat2.txt'
which contains statistical data about each input file is created. These data are the number of
frames (number of `255 255 255' sequence) and the number of discarded events when the row or column
address exceeds the dimensions of the grid. The `binNumber2.txt' input file is also required for
this function.

\end{enumerate}


\item SRS 16.4 Data: to read and process data files obtained from station 16.4 of Daresbury \SRSl. This
item contains two functions:

\begin{enumerate}

\item Read SRS 16.4 \icon{readSRS164}: for reading and mapping SRS 16.4 data files. The process starts with the
appearance of `SRS Scalar' dialog (Figure \ref{SrsScaDialog}) if this option is on, as will be
explained next. This is followed by the appearance of `Change File Order' dialog (Figure
\ref{ChaOrdDialog}) as in the case of XY reading.

\begin{figure} [!h]
\centering
\includegraphics
[scale=0.8] {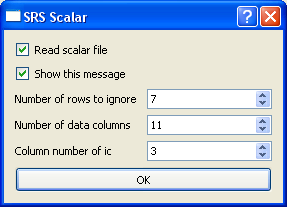}
\caption{SRS scalar dialog.} \label{SrsScaDialog}
\end{figure}

\item Show Scalar Message \icon{showScaMess}: this launches the scalar dialog (Figure
\ref{SrsScaDialog}). The purpose of this dialog is to control the scaling of the count rate by the
\storin\ current `ic'. This dialog contains two check boxes: the first is for choosing between
scaling to `ic' by reading the scalar files or not, while the second is for determining if this
dialog box should appear when invoking `Read SRS 16.4'. The dialog also contains three integer spin
boxes. The first is for the number of header rows in the scalar data files to be ignored, the
second is for the number of numerical data columns in the bottom of the scalar files, and the third
is for the ordinal column number of the `ic' data. The ic scaling correction is based on
normalising all ic's in the tab to the ic of the first \voxel\ (1,1) in the tab.

\end{enumerate}

\end{itemize}

\subsubsection{CCD Menu} \label{MWCCD}

\begin{figure} [!h]
\centering
\includegraphics
[scale=0.8] {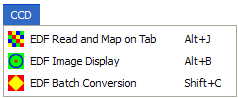}
\caption{The main window CCD menu.} \label{MWCCDMen}
\end{figure}

This menu (Figure \ref{MWCCDMen}) is for reading and processing CCD image files of EDF format.
These are normal binary files with text headers. The binary data represent the CCD pixel intensity
as an implicit function of the pixel position in the CCD array. The header, which occupies the
first 24 lines, contains, among other things, information about the binary data type (e.g. long
integer), the number of rows and columns in the CCD array, and the endian type. This information is
used to read and process the data adequately. Hence, all processes are completely automated as long
as the EDF file header contains the required information and has the right format. The CCD menu
includes the following items

\begin{itemize}

\item EDF Read and Map on Tab \icon{edfRM}: to read and map a number of EDF files. A dialog first appears
to enable the user to define the tab dimensions. A default directory named `DataEDF' will be
inspected by the program to look for EDF files if a different source directory was not selected by
the user. This function can also be used for converting EDF files by mapping them first, followed
by invoking `Write XY Data Files' from Data menu. However, this function uses a basic EDF
conversion technique (e.g. tilt and other corrections are not considered) and hence it is
recommended to use multi-batch numeric conversion from MultiBatch menu.

\item EDF Image Display \icon{edfIC}: the purpose of this function is to display an image of a single EDF
file (Figure \ref{edfImaDisDialog}) and to identify the centre and radius to be used for extracting
the pattern by other processes (i.e. read and map on tab, batch conversion, and multi-batch
conversion). The image plotting technique is simple, that is the square of the intensity is plotted
at the pixel position according to a single colour coding scheme. The purpose of using the square
rather than the intensity itself is to intensify the image features. The centre of the image is
indicated by an `$\times$' mark. By default, the position of the centre is taken at the centre of
the image rectangle and the radius is zero. By click-and-drag action, the centre will be moved to
the position of click while the radius is taken as the distance between the click and the release
positions. A yellow ring is plotted dynamically during the drag action. The 1D pattern will be
extracted only from the part of the 2D pattern represented by the pixels inside the yellow ring.
When the radius is zero the pixels of the entire image are used to generate the 1D pattern. On
clicking on another position within the image, the ring centre is moved to the click position and
the ring image is updated. All following batch processes that involve 1D pattern extraction will be
based on the centre and radius that are identified by this function. The radius and centre data can
be adjusted at any point during the session by invoking this function again and making the
necessary amendments. These data will not be saved on closing the program. There are other
parameters that can be adjusted within the image display dialog (see `EDF Numeric Conversion' in
\S\ \ref{MWMultiBatch}). Also, this dialog is launched automatically on invoking the functions that
require these EDF parameters.

\item EDF Batch Conversion \icon{edfBC}: this function performs batch conversion on a number of EDF
files stored in a specific directory to normal images or to text files containing $xy$ data of 1D
patterns. The image and text files are saved to the directory of EDF files. The program uses the
default directory `DataEDF' as a source for EDF files unless the user made a different choice.
Three image formats are supported: png, jpg and bmp. The image and text files bear the same name as
the source EDF file with proper extension, i.e. png, jpg and bmp for image files and txt for text
files. A scale factor which ranges between 0.05 and 1.0 can be used to adjust the size and hence
the quality of the images. By increasing this factor, the image size increases and the quality
improves, with consequent increase in processing time. Based on a trial, the user should determine
the scale factor as a compromise between quality on one hand and speed and size on the other,
because at certain points the quality improvement may not justify the extra cost in processing time
and file size. It should be remarked that jpg images require a plugin which if it is not available
on the system the jpg production fails. On invoking this function a dialog (Figure \ref{edfSetDia})
appears to enable the user to choose the image type and scale factor and to determine if ASCII text
files should be written. These options are independent and hence they can be chosen in any
combination including all and none. In the last case the function will terminate with no further
action. If numeric conversion is selected the EDF image display dialog (Figure
\ref{edfImaDisDialog}) will appear. In this case the program will remind the user about the image
radius if the default value (i.e. 0) is used. It is noteworthy that the ASCII numeric conversion
uses a basic technique and hence it is recommended to use `EDF Numeric Conversion' from the
MultiBatch menu, which can perform numeric conversion even on a single data set.

\begin{figure} [!h]
\centering
\includegraphics
[scale=0.8] {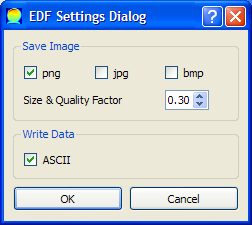}
\caption{EDF settings dialog.} \label{edfSetDia}
\end{figure}

\end{itemize}

\subsubsection{Settings Menu} \label{MWSettings}

\begin{figure} [!h]
\centering
\includegraphics
[scale=0.8] {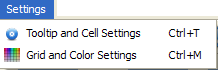}
\caption{The main window Settings menu.} \label{MWSetMen}
\end{figure}

This menu (Figure \ref{MWSetMen}) includes

\begin{figure} [!h]
\centering
\includegraphics
[scale=0.8] {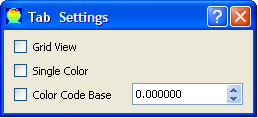}
\caption{Tab settings dialog.} \label{tabSetDialog}
\end{figure}

\begin{itemize}

\item Tooltip and Cell Settings \icon{settooltips}: to launch the Settings dialog (Figure
\ref{SetDialog}). The purpose of this dialog is to control the tooltips and grid size of the tabs.
The dialog has two main sections:

\begin{enumerate}

\item The first is for setting the tab tooltips. This section contains two radio buttons:

\begin{enumerate}

\item Graph Tooltip: to display a graphic tooltip, that is the graph of the spectrum of the cell to which
the mouse is pointing. If this option is selected, the size of the graph can be adjusted from the
`Graph Size' integer spin box and the associated slider.

\item Text Tooltip: to display a text tooltip. If this option is chosen, two groups of tooltips are accessed;
one for the file properties of the cell, and the other for its \voxel\ properties. The latter
include the initial properties like intensity, and the acquired properties such as statistical
indicators and fitting parameters. The properties in both groups can be selected/deselected
individually or as a group. Most of these properties belong to SRS 16.4 data files.

\end{enumerate}

\item The second section is for adjusting the size and aspect ratio of the tab cells. The size is
modified from the `Cell Size Factor' spin box and slider, while the aspect ratio is modified from
the `X Scale Factor' and `Y Scale Factor' spin boxes and sliders. When the width of the current tab
is too narrow, the tab widget handler at the top-left corner can become too small and hence the
other tabs cannot be accessed. In this case, the tab width should be increased by increasing the
cell size factor or the x-scale factor to gain access.

\end{enumerate}

\begin{figure} [!h]
\centering
\includegraphics
[scale=0.65] {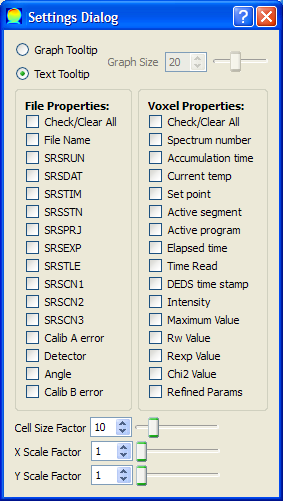}
\caption{Settings dialog.} \label{SetDialog}
\end{figure}

\item Grid and Colour Settings \icon{tabSet}: to launch the Tab Settings dialog (Figure
\ref{tabSetDialog}) which controls the settings of tabs. This dialog contains three check boxes:
`Grid View' to turn the grid on/off for all tabs, `Single Colour' to switch between single- and
multi-colour coding schemes for all tabs, and `Colour Code Base' to select a colour coding scheme
that is based on a certain minimum base value chosen by the user. This base value can be adjusted
from the double spin box next to this check box. When `Colour Code Base' box is unchecked, the
colour coding scheme will be based on the minimum of all values in the tab. It should be remarked
that `Grid View' and `Single/Multi Colour Coding' tool buttons (see \S\ \ref{MWToolsBut}) in the
main window can be used to adjust the settings of the individual tabs independently.

\end{itemize}

\subsubsection{Fitting Menu} \label{MWFitting}

\begin{figure} [!h]
\centering
\includegraphics
[scale=0.8] {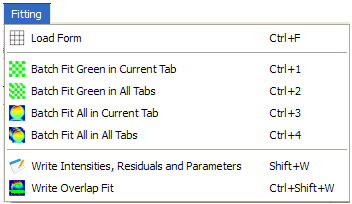}
\caption{The main window Fitting menu.} \label{MWFitMen}
\end{figure}

This menu (Figure \ref{MWFitMen}) facilitates curve-fitting and writing parameters. In EasyDD,
curve-fitting is performed by the use of Levenberg-Marquardt algorithm which is an iterative
nonlinear least-squares optimisation numerical technique. Thanks to its efficiency and good
convergence, the Levenberg-Marquardt algorithm is widely used by scientists and engineers in all
disciplines, and hence it became a standard for nonlinear least squares minimisation problems. The
algorithm can be regarded as a combination of the Gauss-Newton method and the steepest descent
method. It behaves like the former near the solution and like the latter far from it. The need for
the least-squares numerical routines (including Levenberg-Marquardt) arises when a model that
depends on a number of unknown parameters requires optimisation so that it best fits a given set of
data according to a predefined merit function. The best fit model parameters are then obtained by
varying the parameters to minimise the merit function. In the least-squares algorithms, what is
required is to minimise the second norm of the residuals as given by Equation \ref{funcMinimise}.
When the parameters' dependence is nonlinear an iterative procedure is needed where the parameters
obtained from one iteration cycle are fed to the next cycle until a predefined error margin is
reached or a particular number of cycles is exceeded.

In addition to the possibility of fitting all cells in a single tab or multiple tabs, it is
possible to fit selected cells of a tab or tabs. With regard to batch curve-fitting, there are
three ways of selecting/deselecting individual cells for batch curve-fitting:

\begin{enumerate}

\item `Ctrl' + left mouse click to select and deselect a cell.

\item `Shift' + two left mouse clicks to select a rectangular range of cells defined by the two clicked
cells as diagonal corners.

\item `Alt' + mouse hovering to select/deselect the cells that are hovered on.

\end{enumerate}

The selected cells for fitting are marked with green colour. The keyboard `Esc' button can be used
to deselect the green cells.

Fitting menu contains the following items

\begin{itemize}

\item Load Form \icon{table}: to load a spreadsheet form (see \S\ \ref{Form}) that defines and
controls the \curfit\ process.

\item Batch Fit Green in Current Tab \icon{batchRefineOGCT}: to curve-fit all selected (green) cells in
the current tab.

\item Batch Fit Green in All Tabs \icon{batchRefineOGAT}: to curve-fit all selected (green) cells in all
tabs.

\item Batch Fit All in Current Tab \icon{batchRefineACT}: to curve-fit all cells in the current tab.

\item Batch Fit All in All Tabs \icon{batchRefineAAT}: to curve-fit all cells in all tabs.

\item Write Intensities, Residuals and Parameters \icon{writeParams}: to write the intensities,
fitting routine residuals ($\ER,\WPR,\GoF$) and refined parameters of the refined cells in the
current tab. Each one of these will be written to a separate file bearing a proper name (that is
`Intensity', `Rexp', `Rw', `Chi2' and `Par$m$' where $m$ is the parameter index). The files are
saved to a directory called `ResParsTab$N$' where $N$ is the current tab index. The data are
structured in a 2D matrix with the same dimensions as the tab. When the fitting of the tab is
partial, the non-refined cells entries will be marked with `NA'. It is noteworthy that the data
files are formatted as in the case of using `Curve Fitting' in MultiBatch menu (see \S\
\ref{MWMultiBatch}) if the check box in the multi-batch fitting dialog (Figure \ref{MBFDialog}) is
checked.

\item Write Overlap Fit \icon{overlap}: to write the refined parameters in overlapping mode. This means
that the fitting data belongs to various fitting cycles with the latest cycle overwriting the
refinement data only for the cells selected in the last fitting cycle without affecting the data
from the previous cycles. The fitting parameters for each tab are written to a file named
`overlapFit$N$.dat' where $N$ stands for the tab index. The structure of this data file is similar
to the structure of `batch' file (see \S\ \ref{MWCF}). This routine is useful when applying
different fitting basis functions from different forms on various parts of the same tab.

\end{itemize}

\subsubsection{Sinograms Menu} \label{MWSinograms}

\begin{figure} [!h]
\centering
\includegraphics
[scale=0.8] {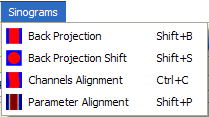}
\caption{The main window Sinograms menu.} \label{MWSinMen}
\end{figure}

This menu (Figure \ref{MWSinMen}) includes

\begin{itemize}

\item Back Projection\footnote{Simon Jacques contributed to the implementation of this algorithm.}
\icon{backProj}: to perform \bacpro\ algorithm on the current tab which represents a \sinogram\ of
rotation versus translation data. Back projection is a reconstruction technique for a 2D image from
a number of 1D projections and their angles of measurement. The purpose of this algorithm, which
relies on the application of the inverse Radon transform and the Fourier Slice Theorem, is to
reconstruct tomographic images in real space from sinograms consisting of a set of intensity values
for rotation versus translation measurements, with possible application of Fourier transform and
filtering. The algorithm takes the sinogram, which is a matrix of intensities, rotates it, and sums
up these values. The sinogram may require padding with zeros to a power of 2 when applying a
Fourier transform. The algorithm works only when the rotational dimension is less than or equal to
the translational dimension. On calling this function a dialog (Figure \ref{backProjDialog}) will
appear. In this dialog there is a `Fourier' section which contains a check box to choose performing
a Fourier transformation or not, and two radio buttons to choose fast (FFT) or discrete (DFT)
transformation\footnote{The DFT algorithm is a contribution of Simon Jacques. FFT and DFT produce
different results.}. This is followed by a `Data Setting' section which includes two radio buttons
to choose the sinogram type, i.e. rotation (rows) versus translation (columns) or vice versa. This
is followed by an `Angles' section which contains two double spin boxes to define the start and the
end angles in degrees. The rotation angles are assumed to be equally spaced between these two
limits with the angle step size being equal to the range (i.e. difference between the limits)
divided by the number of rotations. This is followed by a `Read File' check box, with a
neighbouring line edit to enter the name of the angles file. This option is used when the angles
are not equally spaced. In both cases the number of angles is assumed to be equal to the rotational
dimension. This section is followed by a `Channels' section with three radio buttons to choose one
of three options: `\Bacpro\ for Individual Channels', `Back Projection for Total Intensity' and
`Back Projection for Partial Intensity'. The first option performs a \bacpro\ on all common
channels of all spectra in the tab individually. On performing this operation a vertical line at
the position of the mouse pointer will be added to the 2D plotter (see \S\ \ref{2DPlotter}). On
hovering the pointer over the plotter the tab will be updated to display the back-projected
intensity of the current channel, i.e. the channel at the pointer position. On moving the pointer
off the plotter, the tab will continue displaying the intensity of the last visited channel on the
plotter. The second option will perform a \bacpro\ once on the total intensity (i.e. the sum of
intensities of all channels) of the tab spectra. Similarly, the third option will perform \bacpro\
once on the partial intensity (i.e. the sum of intensities of some channels) in a certain channels
range. The start and the end of the channel range are identified by the two integer spin boxes
below the third radio button. Finally, there are two push buttons: `OK' to proceed and `Cancel' to
abort the operation. The code snippet which implements the \bacpro\ algorithm is given in \S\
\ref{SampleCodeBP} in Appendix \ref{SampleCode}.

\begin{figure} [!h]
\centering
\includegraphics
[scale=0.7] {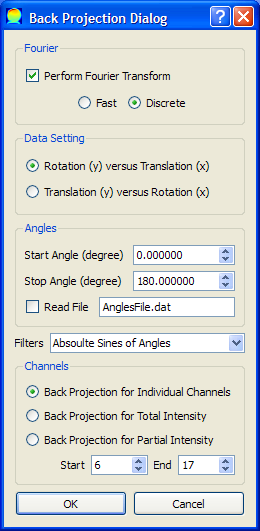}
\caption{\Bacpro\ dialog.} \label{backProjDialog}
\end{figure}

\item Back Projection Shift \icon{backProjShift}: to toggle between the normal sinogram view and \bacpro\
view for the current tab after performing the \bacpro\ in one of the three available modes.

\item Channels Alignment \icon{alignChann}: to perform a sinogram alignment on individual channels of
spectra. The function aligns the edge of the sample in the sinogram to a straight line using an
interpolating technique that relies on a fitting parameter (usually area) of a certain
peak\footnote{This technique is proposed by Simon Jacques.}. The fitting parameter serves as an
alignment reference from which a transformation matrix for the tab rows is obtained. The technique
divides the cells of each row, and assigns linearly-interpolated values to the new divisions. It
then rotates these divisions clockwise and anticlockwise (i.e. displaces them left and right with
move of outliers to the other side) taking in each step the sum of the square differences in the
alignment parameter between the corresponding cells in the current row and the reference row. The
final alignment of the row is that of the least squares. The algorithm requires applying
curve-fitting on all voxels in the tab beforehand. On invoking this function a dialog (Figure
\ref{alignDialog}) appears. This dialog contains a number of spin boxes to identify the algorithm
parameters. These are: inflation factor (number of divisions of each cell), displacement number
(number of divisions to be displaced left and right), parameter index (index of fitting parameter
in the fitting form to be used for alignment), and upper cut-off limit (maximum value for the
intensity to be set to when it exceeds that limit). The inflation factor should be between 1 and 50
inclusive. The displacement number should be between zero and a number of divisions occupying half
the tab width to avoid computational waste. The purpose of the upper cut-off limit is to obtain an
alignment image with sharp features. It should be remarked that this algorithm does not conserve
the intensity of spectra channels as the values of the inserted divisions will be added to the
intensity of the related cells. The code snippet which implements the channels alignment algorithm
is presented in \S\ \ref{SampleCodeCA} in Appendix \ref{SampleCode}.

\begin{figure} [!h]
\centering
\includegraphics
[scale=0.7] {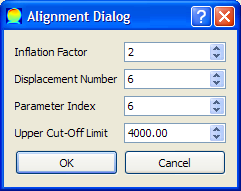}
\caption{Alignment dialog.} \label{alignDialog}
\end{figure}

\item Parameter Alignment \icon{alignParam}: this function is similar to the previous one but
the alignment is performed on the total intensity, and the aligned intensity is displayed on the
tab. This can help in testing the algorithm with a certain set of alignment parameters before
applying channels alignment.

\end{itemize}

\subsubsection{MultiBatch Menu} \label{MWMultiBatch}

\begin{figure} [!h]
\centering
\includegraphics
[scale=0.8] {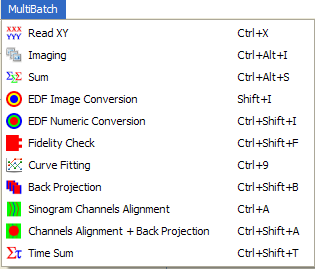}
\caption{The main window MultiBatch menu.} \label{MWMBMen}
\end{figure}

The items in this menu (Figure \ref{MWMBMen}) share a number of common features. First, they work
on a number of data sets (including a single set) each one of which is stored in a particular
directory. Second, they look to a default directory called `MultiBatch' as the source of data where
the data sets are stored unless the user selected another directory by using `Change Data
Directory' from Data menu. Third, they work on an $xy$ file format except `Curve Fitting' which
works on a number of formats including $xy$, and EDF functions which by nature work on EDF files.
The files used by the `Time Sum' function have a single user-defined name, and hence are not
identified by type. Another common feature is that MultiBatch functions store the processed data
either in the source directory or in its subdirectories in an organised fashion using meaningful
names to facilitate managing and post-processing of results. This menu includes the following items

\begin{itemize}

\item Read XY \icon{readXYMB}: to read a number of data sets in $xy$ format and map them on separate
tabs. A maximum of 10 data sets are read in each run if the source directory contains more than 10.

\item Imaging \icon{saveMB}: to produce png images representing the total intensity of spectra, similar
to what the tabs display. The dimensions (rows and columns) of the images are required when
invoking this function. No image will be produced for a data set if the number of data files in its
directory is less than the product of the two dimensions. The image files are saved to the source
directory, each with a name identical to the name of the data directory of that set with a `.png'
extension.

\item Sum \icon{sumMB}: to produce sum spectra by adding the intensity of the corresponding channels.
There is no restriction on the number of files in each data set. The number of channels in the sum
spectrum is the maximum of that of the individual spectra in the data set. This function is based
on the assumption that all common channels of the data set share the same $x$ coordinates. The sum
files are saved to the source directory, each with a name identical to the name of the data
directory of that set with `\_Sum' extension.

\item EDF Image Conversion \icon{edfMB}: to convert EDF binary files to normal images, similar to that
presented in `EDF Batch Conversion' in CCD menu (see \ref{MWCCD}). A sample of these images is
displayed (within a dialog) in Figure \ref{edfImaDisDialog}. The images are saved to a directory
named as the data directory of that data set with an `\_out' extension, and stored inside this data
directory. Each image is named after its EDF source file with an `edf' extension replaced by a
suitable extension (i.e. `png', `jpg' or `bmp'). A scale factor can also be used to adjust the size
and quality of the images as explained previously.

\begin{figure} [!h]
\centering
\includegraphics
[width=0.9\textwidth] {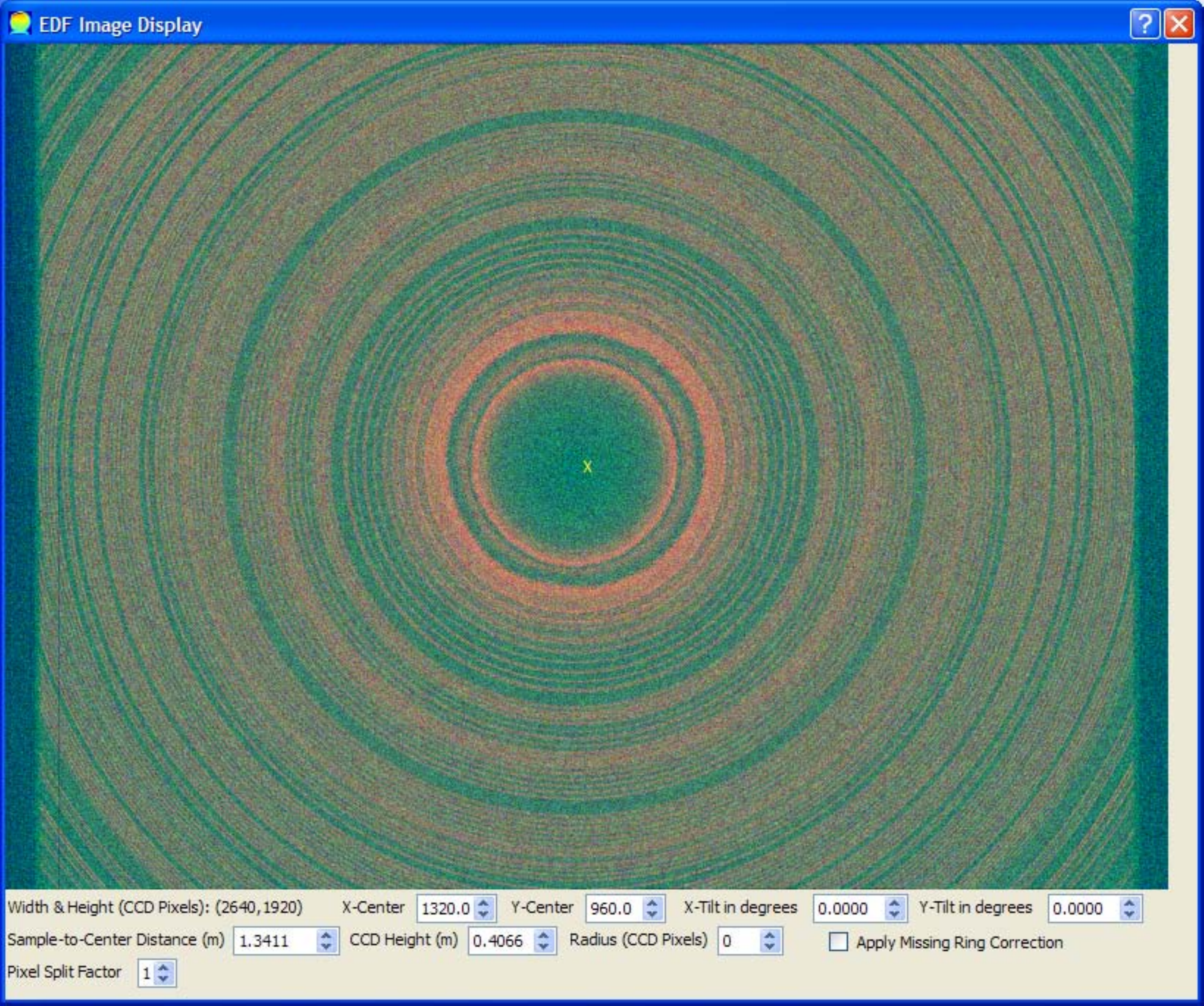}
\caption{EDF image display dialog.} \label{edfImaDisDialog}
\end{figure}

\item EDF Numeric Conversion \icon{edfMB2}: to convert EDF binary files to ASCII numeric data files
containing 1D patterns. The name of each text file is the same as its EDF parent with `edf'
extension replaced with `txt'. These files are saved to the directory of the EDF files unless a
fidelity check is performed (see the next item). The function requires a number of parameters that
determines the nature of conversion. Most of these parameters are defined from the EDF image
display dialog (Figure \ref{edfImaDisDialog}) which appears on selecting a representative EDF when
invoking this function. For computational efficiency, \ProgName\ assumes that all files processed
in a single multi-batch run are obtained from the same batch, and hence have the same dimensions
and the same data type (i.e. short or long integer, and big or little endian). This information
(i.e. dimensions and type) is obtained from the header of the representative EDF. The user will
also be offered the opportunity to run a fidelity check on the text files.

The EDF image display dialog contains a number of widgets:

\begin{enumerate}

\item
Label to display the width and height of the EDF image in pixels.

\item
Two spin boxes to define the $x$ and $y$ coordinates, which correspond to the horizontal and
vertical dimensions respectively, of the centre of the image rings in pixels. By default, these
coordinates are set at the centre of the EDF rectangle at the start of the program. The centre is
indicated on the image by a yellow `$\times$' mark which moves to the new position dynamically when
the coordinates are adjusted.

\item
Two spin boxes to define the $x$ and $y$ tilts in degrees within the range -89$^{\circ}$ to
89$^{\circ}$.

\item
Spin box for the distance in metres between the sample and the centre of the CCD detector.

\item
Spin box for the height (i.e. vertical dimension as displayed on the dialog) in metres of the CCD
detector.

\item
Spin box for the radius (in pixels) of the circle that identifies the part of the EDF image to be
used for numeric conversion, i.e. only the rings inside the circle will be used for extracting the
pattern. The default value for the radius at the start of the program is zero meaning that the
whole image will be used. A yellow ring representing the circle is displayed on the dialog when the
radius is greater than zero. This ring is dynamically resized on adjusting the value of the radius
from the spin box.

\item
Check box for applying the missing-rings correction based on the continuum approximation algorithm
which scales the incomplete rings to the continuum value at that radius.

\item
Spin box to define a split factor for the pixels within the range 1-5. This factor was introduced
because the pixel can be too coarse as a unit of length and storage due to its finite measurable
size resulting in uneven distribution of intensity in the data points of the 1D pattern. The pixels
of the CCD grid can therefore be split by using this factor to improve the resolution and obtain
smoother patterns.

\end{enumerate}

The settings of some of these widgets are saved on exiting the program. It should be remarked that
the EDF image display dialog has tooltips showing the $x$ and $y$ coordinates (in pixels) of the
mouse pointer position. Moreover, the centre of the image can be moved to a new position by mouse
click, and hence the yellow ring is moved and the image centre spin boxes are updated. The radius
can also be adjusted by a mouse click-and-drag action, and hence the radius spin box will be
updated.

Regarding the tilt correction, a formula was derived using a standard Cartesian coordinate system
in a plane perpendicular to the original beam direction that passes through the centre of the image
and hence the origin of the coordinate system. For simplicity, a unit circle centred at the origin
and lying in this plane was used. Here, we outline this derivation making use of the following
symbols:

$\phi$ is the angle of an arbitrary point on the unit circle ($0 \leq \phi < 2\pi$)

$\theta_x$ is the tilt in the $x$ direction

$\theta_y$ is the tilt in the $y$ direction

$\theta$ is the actual tilt at an arbitrary point on the tilted CCD plane. It is obvious that
$\theta$ is a function of $\phi$ and can have negative as well as positive values.

\begin{figure}[!b]
  \centering{}
  \includegraphics
  [width=0.65\textwidth]
  {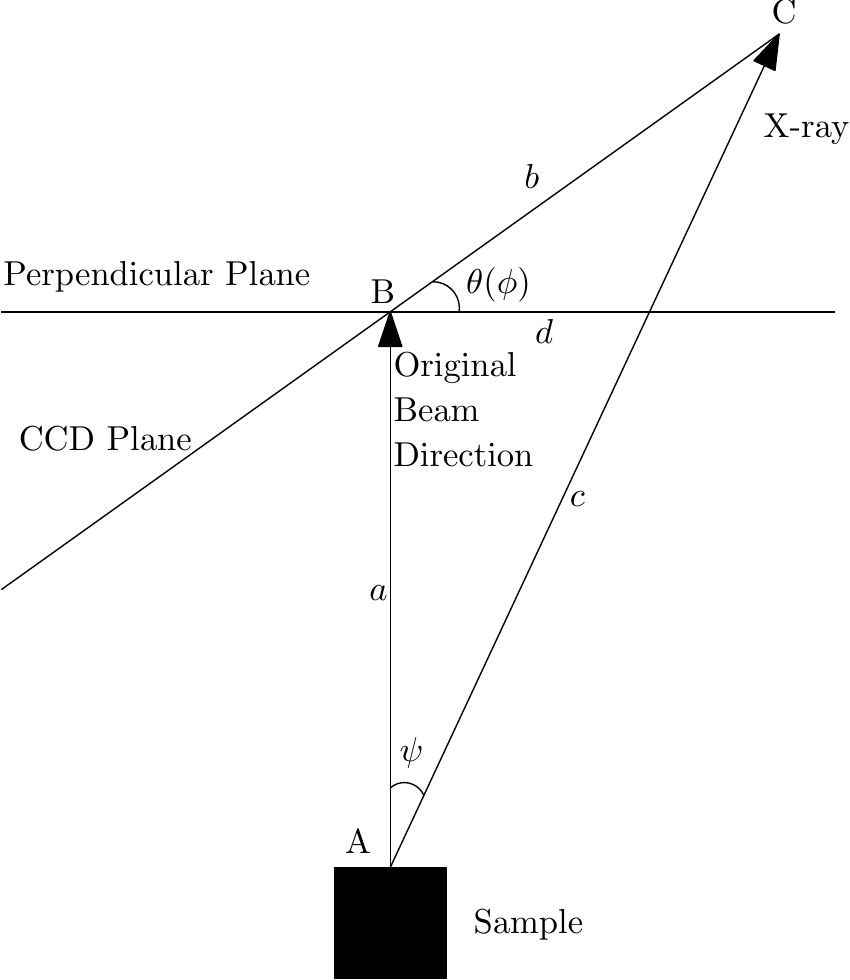}
  \caption{A schematic diagram to demonstrate the derivation of tilt correction.}
  \label{tiltCor}
\end{figure}

From a simple geometric argument, the actual tilt as a function of $\phi$ is given by
\begin{equation}
    \theta(\phi) = \arctan[\cos \phi \times \tan \theta_x + \sin \phi \times \tan \theta_y]
\end{equation}

As can be seen, this formula retrieves the correct (and obvious) tilt in the $x$, $y$, $-x$ and
$-y$ directions.

Now referring to Figure \ref{tiltCor}, which is in the plane identified by the original beam and
the ray directions, and on applying the cosine rule to the triangle ABC we have

\begin{equation}
    c = \sqrt{a^2 + b^2 - 2ab \cos \left(\theta + \frac{\pi}{2} \right)}
\end{equation}

Using the sine rule we obtain the scattering angle
\begin{equation}
    \psi = \arcsin \left[b \times \frac{\sin (\theta+\pi/2)}{c} \right]
\end{equation}

And finally the required quantity, $d$, can be obtained from
\begin{equation}
    d = a \times \tan \psi
\end{equation}

It is obvious that there are other methods for finding $d$. Although this derivation is
demonstrated for $\theta > 0$, as seen in Figure \ref{tiltCor}, it is general and valid for
$-\frac{\pi}{2} < \theta < \frac{\pi}{2}$.

Concerning the missing-rings correction, because the charge-coupled device detector has a
rectangular shape, the rings whose radii exceed a certain limit, by having a radius that exceeds
the distance between the ring centre and one or more of the rectangle sides, will not be complete
and therefore the pattern will have a reduced integrated intensity at high scattering angles. To
compensate for the loss of intensity from these incomplete diffraction rings, a simple and time
saving technique is used. This technique scales the incomplete rings to the continuum value at that
radius by multiplying the total intensity of the incomplete rings by the ratio of the continuum
circumference at that radius to the actual number of pixels in that ring. In multi-batch
operations, this number is computed only once while serially reading the values of the pixels of
the representative EDF file. These scale ratios are stored in a 1D vector and used during the
application of the numeric conversion routine.

\begin{figure} [!t]
\centering
\includegraphics
[width=0.8\textwidth] {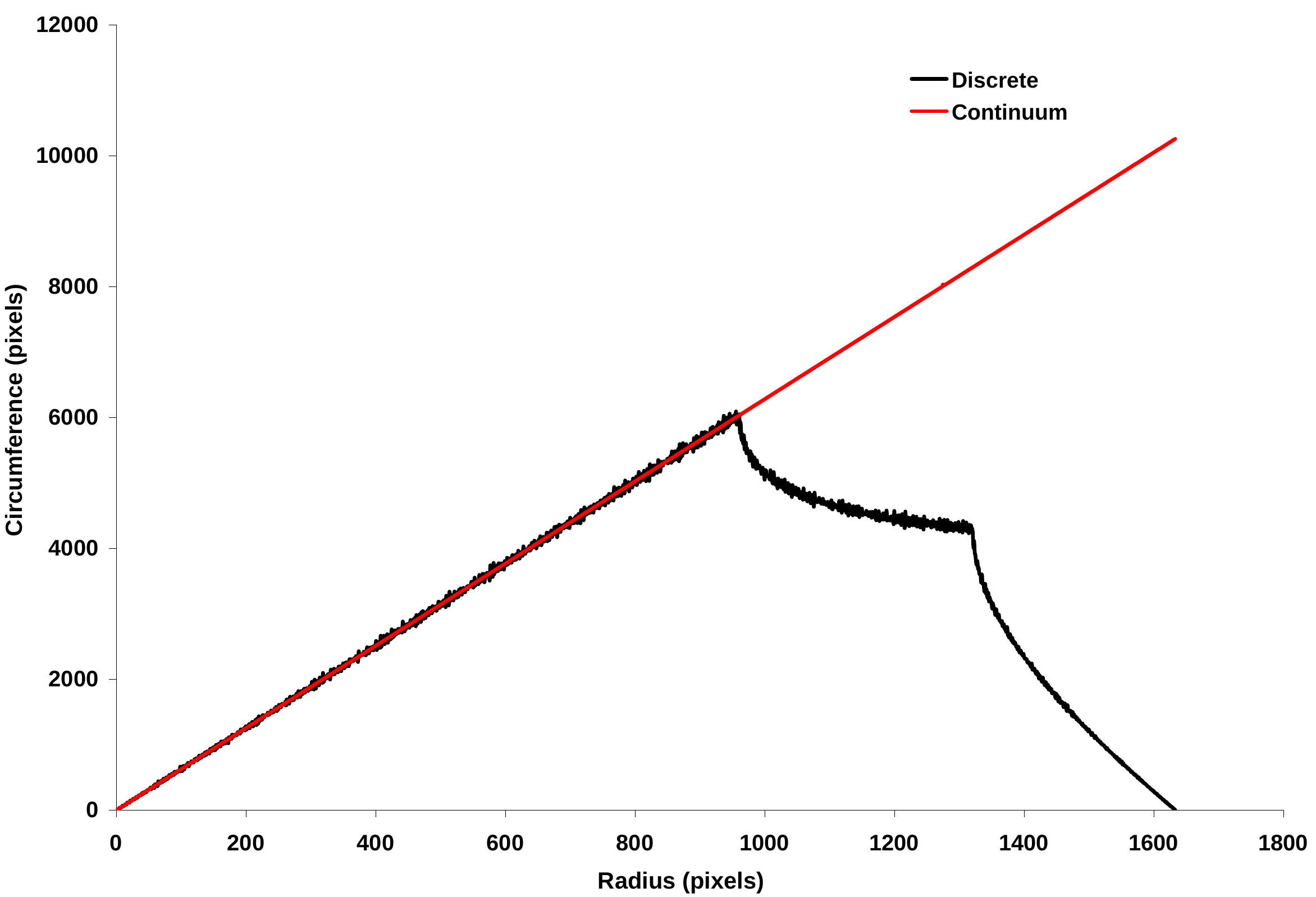}
\caption[The continuum and discrete values for the number of pixels in a ring.]{The continuum
approximation alongside the discrete value for the number of pixels in a ring.} \label{disCon}
\end{figure}

\begin{figure} [!h]
\centering
\includegraphics
[width=0.8\textwidth] {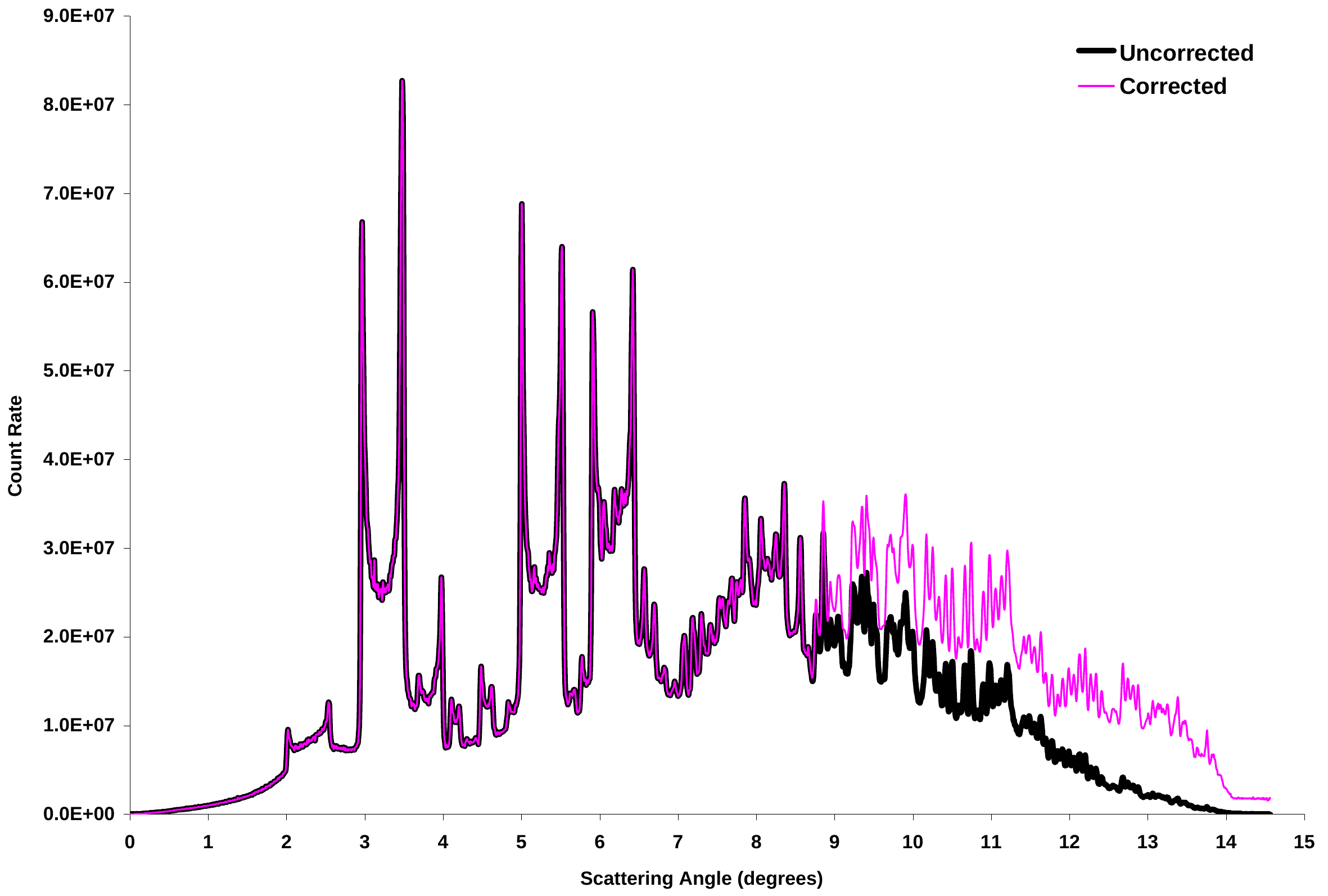}
\caption[A sample of 1D diffraction pattern extracted from a 2D EDF pattern.]{A sample of 1D
diffraction pattern extracted from a 2D EDF pattern. The pink curve is obtained with the
missing-rings correction while the black is obtained without this correction.} \label{disCon2}
\end{figure}

Figure \ref{disCon} displays an example of the actual (discrete) number of pixels of the image
rings as a function of radius in pixels alongside the continuum value of $2\pi r$ represented by
the straight line. As seen, the two curves match very well for the complete rings. This indicates
that the continuum is a very good approximation, except for the rings which are too close to the
image corners. The meaning of the two cusps is obvious as the radius of the rings increases and
exceeds the rectangle sides in one direction (i.e. vertical or horizontal) and then the other.
Figure \ref{disCon2} presents a sample diffraction pattern obtained from an EDF image with and
without the application of the missing-rings correction.

\item Fidelity Check \icon{fidelity}: to check if some expected files, according to a guessed pattern in
the naming of the data set files, are missing. These files will then be created and filled with
zeros (both $x$ and $y$ coordinates) for all channels. The purpose of filling with zero is to mark
these `void' voxels and keep track of them as false voxels created to conserve the rectangular
pattern of the slice and map it properly. The number of entries (channels) in these void files is
equal to the number of points in the ordinary files (three files at the start are inspected to find
the number of entries in these files and take the minimum of them if they are different). On
performing the fidelity check, independently or following EDF numeric conversion, the text files
are moved to a directory named after the directory of the EDF files with `\_txt' extension and
located inside this directory. In the following we outline the fidelity check algorithm in the form
of a pseudo-code:

\begin{itemize}

\item
Find the number of lines in a sample of three files in the data set (or all files if the data set
consists of less than 4 files) and take the minimum (say $l_{min}$).

\item
Obtain the name of the first file in the data set and use it as a reference string.

\item
Create a flag string of 0's with the same size as the reference string. This flag string is used to
test which character position varies in the names of files.

\item
Check if the names of other files have the same size as the name of the first file.

\item
Check if the character at a certain position in the name of any file differs from the reference
string, and set the flag position to 1 in this case.

\item
Loop over all reference flag positions using a seed vector of strings (initially this vector
consists of a single void string):

\begin{itemize}

\item
If the flag position is 1 add a character (where this character varies between the maximum and
minimum found within that position in the names of files) in its position to every entry in the
seed vector and hence the size of the seed vector grows.

\item
If the flag position is 0 add to all strings in the seed vector the character found in the
reference string in that position.

\end{itemize}

\item
Test if any file is missing from the data set by comparing the list of file names to the final seed
vector.

\item
Create the missing files and fill them with $l_{min}$ of zero lines (i.e. lines with $x=0$ and
$y=0$).

\end{itemize}

\item Curve Fitting \icon{batchBatchRefine}: to perform automated \curfit\ on a large scale. The
idea of this function is to deposit a number of fitting forms in the directory of the program, and
store a number of folders containing data sets in the source directory. The program then performs
automated curve-fitting where each pattern in every data set is fitted to each individual form and
the resulting data are saved in a structured directory tree. The intensity, residuals and refined
parameters files (as in `Write Intensities, Residuals and Parameters' in Fitting menu) alongside
`batch' and `OlivierMatlab' files for each data set are saved to a directory named after the
fitting form and located inside the directory of that data set. A report containing information
about the outcome of the fitting process is written to a file called `bbReport' which is saved
inside the source directory.

On starting multi-batch curve-fitting, a dialog (Figure \ref{MBFDialog}) appears, in which there
are three radio buttons to choose the data type, i.e. `SRS 16.4', `XY' and `MCA'. The second choice
requires the number of rows and columns, while the third requires this plus the number of headers
and footers to be ignored. These data are input through the integer spin boxes next to the
corresponding radio button. At the bottom of the dialog there are three push buttons: `Select Data
Directory' to identify the source directory, `OK' to proceed, and `Cancel' to abort the process.
Next to `Cancel' button there is a check box and two integer spin boxes. The check box is used to
select formatting the data in the written files by fixing the width of columns and the number of
decimal places to be displayed. These values are input through the adjacent spin boxes.

\begin{figure} [!h]
\centering
\includegraphics
[width=0.9\textwidth] {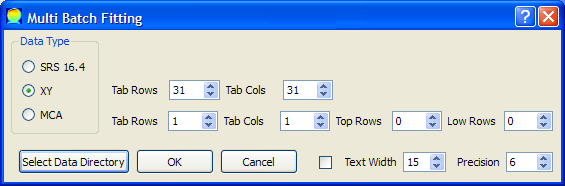}
\caption{MultiBatch curve-fitting dialog.} \label{MBFDialog}
\end{figure}

\item

Back Projection \icon{bpMB}: to perform multi-batch back projection in one of its three modes. The
dimensions of the data sets and the parameters and options of back projection are determined
through dialogs that appear sequentially on invoking this function. Any data set that contains
insufficient number of files will be ignored. For back projecting individual channels, the
resulting data files are saved to a directory having the same name as the parent directory with
`\_BP' extension and located inside this directory. The files are named `BP\_R$r$C$c$.txt' where
$r$ and $c$ are the row and column indices. For back projecting total or partial intensity, the
resulting file of each data set is saved to the main source directory with a name identical to the
name of the directory of that data set with a `\_BP' extension.

\item Sinogram Channels Alignment \icon{alignSinoMB}: to perform channel alignments of sinogram data in
multi-batch mode. Fitting form, dimensions of slice and alignment algorithm parameters are
requested on invoking this function. Data files of `AS\_R$r$C$c$.txt' type will be written to a
directory having the same name as the parent directory with an `\_AS' extension and located inside
this directory. A png image file of the aligned parameter, having the same name as the parent
directory with `\_fittedPeak' extension, is also created and saved to the main source directory.
Moreover, a report file called `saReport' containing brief information is created and saved to the
main source directory.

\item Channels Alignment + Back Projection \icon{alignBPMB}: this is a combination of two functions:
multi-batch channel alignment and multi-batch back projection. This function is provided for
convenience, computational efficiency and saving disc space. The storing and naming of the
resulting data files and directories follow a similar pattern to that of the parent functions.
However, the prefixes and extensions are of `ABP' type. A report file `abpReport' is also created
and saved to the main source directory.

\item Time Sum \icon{timeSum}: to find the sum of a certain parameter obtained from \curfit\ for all
peaks in a number of data sets as a function of the data set number. The data for each peak will be
written to a *.ts type file bearing the name of the peak (usually its position value). This
function assumes that the parameter data of each peak reside in a directory within the directory of
the fitted data set, i.e. similar to the directory tree obtained by running multi-batch \curfit. At
the start of this function, the user is asked to supply the common name of the files (e.g. `Par3')
whose data will be summed. The user will also be offered the option to normalise each sum to the
maximum value within the particular sum file.

\end{itemize}

\subsubsection{Help Menu} \label{MWHelp}

\begin{figure} [!h]
\centering
\includegraphics
[scale=0.8] {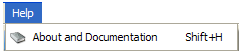}
\caption{The main window Help menu.} \label{MWHelMen}
\end{figure}

\begin{figure} [!h]
\centering
\includegraphics
[scale=0.8] {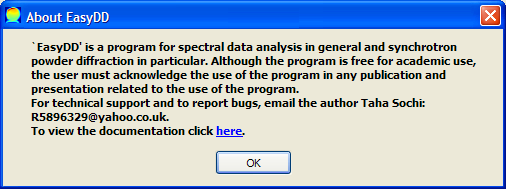}
\caption{`About \ProgName' dialog.} \label{AboutDialog}
\end{figure}

This menu (Figure \ref{MWHelMen}) includes only one item:

\begin{itemize}

\item About  \icon{about}: a dialog (Figure \ref{AboutDialog}) is launched when invoking this function.
This dialog briefly defines \ProgName, offers contact information, and provides a link to access
\ProgName\ documentation which is stored in the `Documentation' directory. The latter should be a
subdirectory of the program directory for normal operation.

\end{itemize}


\subsection{Tools} \label{MWTools}

These include toolbars, tool buttons, spin boxes, sliders, context menus, and a status bar.

\subsubsection{Toolbars} \label{MWToolbars}

Most menus have associated toolbars, as seen in Figure \ref{MainWindow}. These toolbars can be
turned on and off by right-clicking on the toolbar area and choosing from a menu. Each toolbar
contains the main items of the corresponding menu. By default, all toolbars are displayed when
starting the program. The selected setting will not be saved on exiting the program.

\subsubsection{Tool Buttons} \label{MWToolsBut}

The main window contains three tool buttons (Figure \ref{MainWindow}):

\begin{itemize}

\item Grid View Button \icon{grid}: to turn the tab grid on/off for the current tab.

\item Single/Multi Colour Coding Button \icon{rainbow}: to change the colour coding scheme between single-
and multi-colour for the current tab.

\item

3D Graph Button \icon{threeDGraph}: to plot a 3D graph of the current tab (see \S\
\ref{3DPlotter}).

\end{itemize}

\subsubsection{Spin Boxes} \label{MWToolsSpi}

The main window contains four spin boxes (Figure \ref{MainWindow}):

\begin{itemize}

\item Width spin box: to change the width of the tab widget window within the range 0-2000 pixels.

\item Height spin box: to change the height of the tab widget window within the range 0-2000 pixels.

\item Colour code spin box: to select the colour coding range scheme (see \S\ \ref{MWToolsSli}).

\item 3D Graph spin box: to scale the intensity on the $z$-axis of the 3D graph (see \S\
\ref{3DPlotter}).

\end{itemize}

\subsubsection{Sliders} \label{MWToolsSli}

The main window contains two sliders (Figure \ref{MainWindow}). The purpose of these sliders is to
adjust the minimum and maximum of the colour code range. Currently there are two available schemes
which can be selected from the neighbouring integer spin box

\begin{enumerate}

\item `1' (default) to adjust the values of the cells that fall outside the new range so that they have
the value of the adjacent limit.

\item `2' to eliminate these cells from the colour coding scheme and hence mark them with white.

\end{enumerate}

\subsubsection{Context Menu} \label{MWConMen}

The context menu (Figure \ref{MainWindow}) can be accessed by a right-click on the main window
outside the tab widget area. This menu contains a number of items from all menus.

\subsubsection{Status Bar} \label{MWStaBar}

This is a standard bar (Figure \ref{MainWindow}) to display explanatory comments about the item
that is currently hovered upon by the mouse.

\subsubsection{Colour Bar} \label{ColourBar}

The colour bar (Figure \ref{MainWindow2}) appears on reading and mapping data on tabs. The bar
displays the minimum and maximum colour values of the current tab, whether the tab displays the
intensity or the residuals or refined parameters after batch \curfit. These values are instantly
updated on changing the tab selection and on moving the colour sliders to adjust the colour range.
They are also updated when changing the displayed parameter by using the list widgets (see
\ref{MWRW}) following batch \curfit.

\subsection{Tab Widget} \label{MWTW}

\begin{figure} [!h]
\centering
\includegraphics
[scale=0.9] {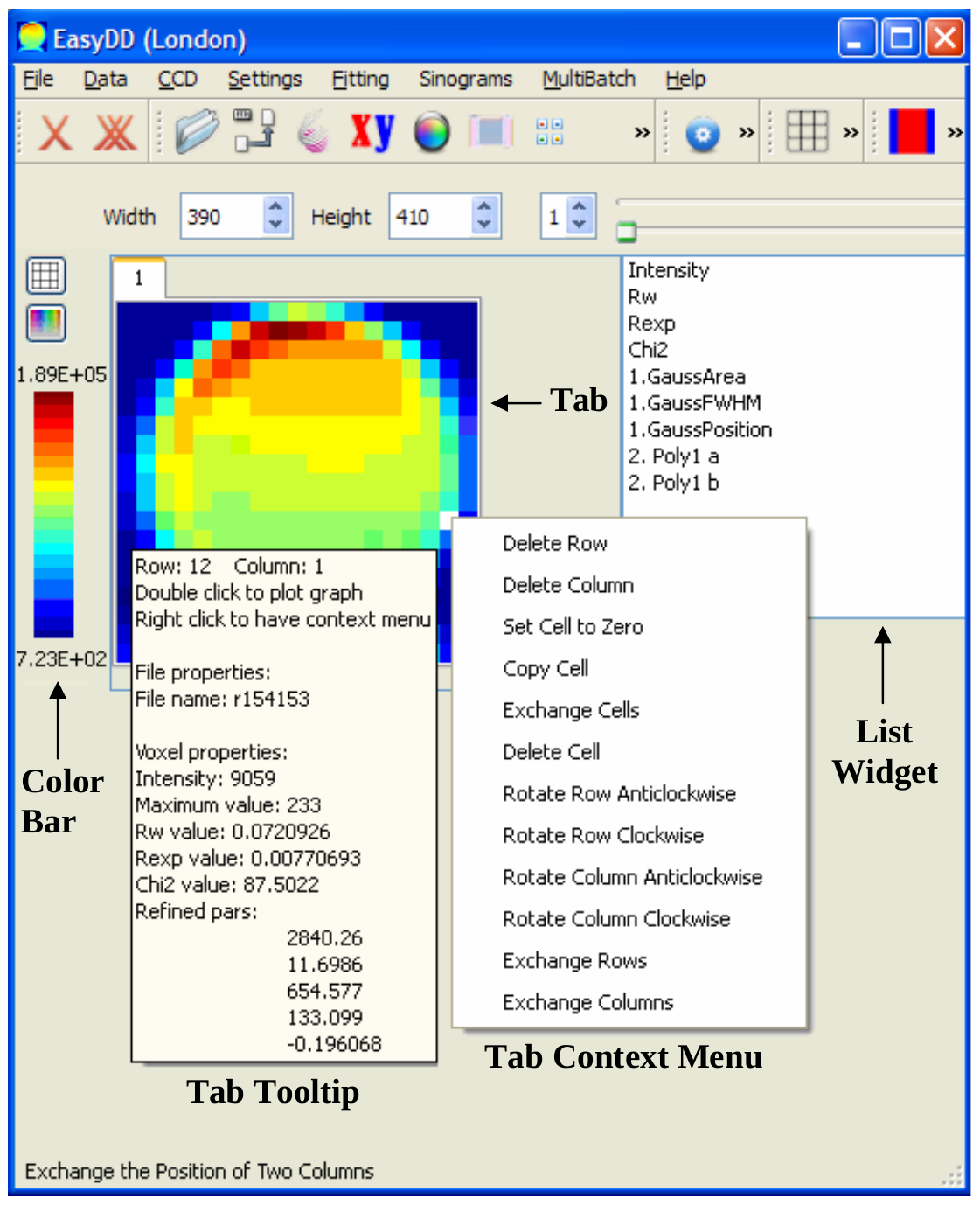}
\caption{\ProgName\ main window following data mapping and \curfit.} \label{MainWindow2}
\end{figure}

This is a widget (Figure \ref{MainWindow}) that can accommodate a number of 2D colour-coded
scalable tabs (Figure \ref{MainWindow2}) for data mapping with graphic and text tooltips to show
all essential file and \voxel\ properties. The tabs also have a context menu with 12 items for
managing rows, columns and cells in the tab. The tab widget space is a scroll area where a
horizontal/vertical scroll bar emerges if the width/height of the current tab exceeds the
width/height of the tab widget. On changing tab selection, the size of the tab automatically
adjusts to fit the current tab.

When a tab cell is double-clicked, the 2D plotter (see \S\ \ref{2DPlotter}) is launched, if it is
not already launched, and the pattern of the clicked cell is plotted. This cell is marked with
white, but the `Esc' key can be used to recover its original colour. The plotter is dynamically
updated when the mouse pointer hovers over the tab so that the plotter displays the pattern of the
hovered cell. By pulling the pointer outside the tab, the plotter displays the pattern of the
double-clicked cell. The plotter is automatically closed when the tab widget is cleared by removing
all tabs.

The tab graphic tooltip is an image of the pattern of the hovered cell while the text tooltips
include its file and voxel properties, as well as the residuals and refined parameters if \curfit\
is performed. The scheme for the tab tooltips is that as long as the plotter is absent the mode of
tooltips (i.e. text or graph) is determined by user choice (see \S\ \ref{MWSettings}). However,
when the plotter is launched the tooltips automatically become text regardless of the user choice
because a graphic tooltip is redundant.

\subsection{List Widgets} \label{MWRW}

List widgets (Figure \ref{MainWindow2}) are created on performing batch \curfit\ on a tab, i.e. one
list widget for each refined tab. These widgets contain items which include intensity, statistical
indicators and fitting parameters. On selecting each one of these items, by mouse or by up or down
arrow keys, the current tab colour coding and colour bar change to display the variation of the
selected item.

\section{2D Plotter} \label{2DPlotter}

\begin{figure} [!h]
\centering
\includegraphics
[width=0.9\textwidth] {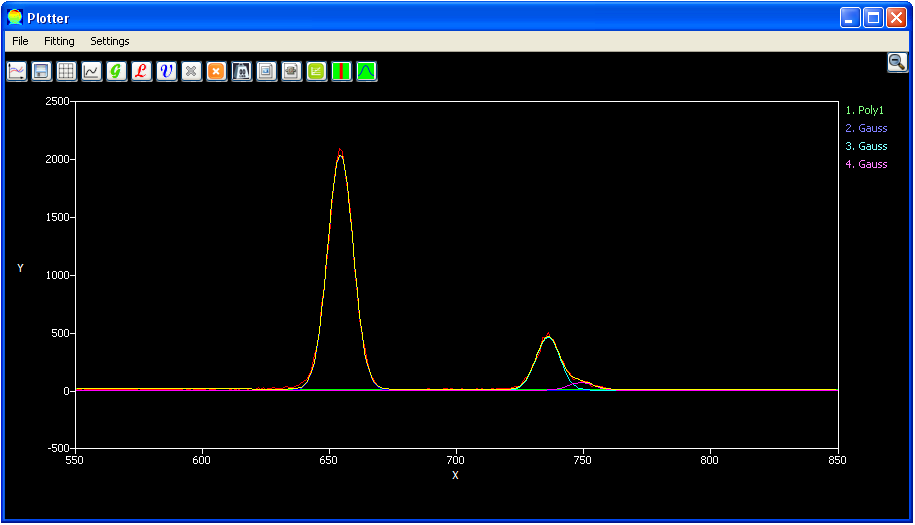}
\caption{The 2D plotter.} \label{plotter}
\end{figure}

This is a plotter (Figure \ref{plotter}) for obtaining a 2D graph of a spectrum (e.g. intensity
versus energy) for any \voxel\ in the current tab. It is also used to create basis functions and
spreadsheet forms for \curfit. The plotter capabilities include:

\begin{itemize}

\item Creating, drawing, modifying and clearing fitting basis functions by simple click or by
click-and-drag actions. The available basis functions are polynomials $\leq$ 6, \Gauss, \Lorentz\
and \pseVoigt. The last three are given in Table \ref{lineShapeFuncs}.

\item Performing \curfit\ on the plotted pattern by a \LevMar\ nonlinear \ls\ algorithm.

\item Saving image of the plotted pattern in 8 different formats, i.e. png, bmp, jpg, jpeg, ppm, xbm, xpm
and pdf.

\end{itemize}

%
%

\subsection{Menus} \label{PloMen}

The plotter has three menus:

\subsubsection{File Menu} \label{PloFilMen}

\begin{figure} [!h]
\centering
\includegraphics
[width=0.5\textwidth] {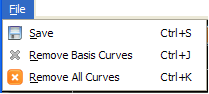}
\caption[Plotter File menu] {The Plotter File menu.} \label{PloFileMenu}
\end{figure}

This menu (Figure \ref{PloFileMenu}) contains three items:

\begin{enumerate}

\item Save image \icon{save}: to save the plotter image in one of 8 formats.

\item Remove basis curves \icon{removeBasisCur}: to remove the basis fitting curves and keep the refined
curve.

\item Remove all curves \icon{removeAllCur}: to remove all curves including the refined curve.

\end{enumerate}


\subsubsection{Fitting Menu} \label{PloFitMen}
This menu (Figure \ref{PloFitMenu}) contains several items and subitems:


\begin{figure} [!h]
\centering
\includegraphics
[width=0.5\textwidth] {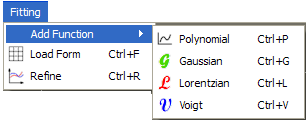}
\caption[Plotter Fitting menu] {The Plotter Fitting menu.} \label{PloFitMenu}
\end{figure}


\begin{enumerate}

\item Add Function: this item includes:

\begin{enumerate}

\item Polynomial \icon{poly}: to plot a polynomial of order $\leq 6$ whose curve passes through a number
of points. The points are selected by clicking on the plotting area. The order of the polynomial
equals the number of points minus one. This function requires activation at the start and
deactivation at the end to plot the polynomial. If the number of clicks is greater than 7 the extra
points will be ignored.

\item \Gauss ian \icon{gauss}: to plot a \Gauss ian curve. This function requires activation at the start
by selecting this item. The curve is plotted by a click-drag-release action using the left mouse
button. The position is determined from the position of click, the area is determined from the
rectangle of the click-release points, and the FWHM is determined from the width of the rectangle.
A dynamic curve is drawn during the drag action.

\item \Lorentz ian \icon{lorentz}: to draw a \Lorentz ian curve, as for \Gauss ian.

\item Voigt \icon{voigt}: to draw a \pseVoigt\ curve, as for \Gauss ian. By
default, the \mixfac\ of the \pseVoigt, $m$, is 0.5.

\end{enumerate}

\item Load Form \icon{table}: to launch a spreadsheet for preparing or loading a form for curve-fitting.

\item Refine \icon{refine}: to start \ls\ curve-fitting by the \LevMar\ algorithm. The
fitting will apply to the part of the pattern within the current $x$ range of the plotter as
defined by the current view on zooming in and out.

\end{enumerate}

\subsubsection{Settings Menu} \label{PloSetMen}

This menu (Figure \ref{PloSetMenu}) contains six items:

\begin{figure} [!h]
\centering
\includegraphics
[width=0.5\textwidth] {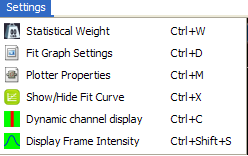}
\caption{The Plotter Settings menu.} \label{PloSetMenu}
\end{figure}

\begin{enumerate}

\item Statistical weight \icon{weight}: to launch a dialog (Figure \ref{StaWeiDialog}) for choosing the
statistical weight scheme in the \ls\ fitting. The dialog contains three radio buttons: a constant
which is in the range 1-10$^{9}$ and can be adjusted from a double spin box, a \Poisson\
distribution with variance being equal to the observed count rate, and a reciprocal \Poisson\ where
the statistical weight equals the observed count rate.

\begin{figure} [!h]
\centering
\includegraphics
[scale=0.8] {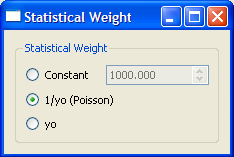}
\caption{Plotter statistical weight dialog.} \label{StaWeiDialog}
\end{figure}

\item Fit Graph Settings \icon{settings}: to launch the plotter settings dialog (Figure
\ref{PloStaDynDialog}) which contains two radio buttons to choose if the fitting curve should be
displayed statically on completing the refinement, or dynamically while the refinement is
progressing. For the second choice a time delay (0-5000 milliseconds) for updating the curve can be
set from the neighbouring integer spin box. This option is useful to monitor how the fitting
converges to the solution.

\begin{figure} [!h]
\centering
\includegraphics
[scale=0.8] {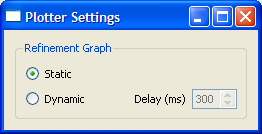}
\caption{Plotter settings dialog.} \label{PloStaDynDialog}
\end{figure}

\item Plotter Properties \icon{plotProp}: to launch the plotter properties dialog (Figure
\ref{PloProDialog}) which contains three main components:

\begin{enumerate}

\item Check box to show/hide plotter tooltips, which are the $x$ and $y$ coordinates of the mouse pointer
position.

\item Another check box to convert the channel number for SRS 16.4 data in the plotter tooltip to
energy using a linear scaling relation. The `a' and `b' coefficients, which are in the range
-10$^{8}$ to 10$^{8}$, stand for slope and intercept respectively and can be adjusted from the
neighbouring double spin boxes.

\item `Change Background Colour' push button to change the plotter background colour. Three colours are
available: black, white, and dark gray. The colour cyclically changes on pressing this button.

\end{enumerate}

\begin{figure} [!h]
\centering
\includegraphics
[scale=0.8] {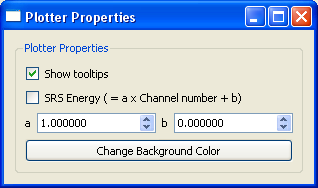}
\caption{Plotter properties dialog.} \label{PloProDialog}
\end{figure}

\item Show/Hide Fit Curve \icon{fitDraw}: to show/hide the fitting curve following a batch \curfit\
operation when the mouse pointer hovers over the cells whose patterns have been fitted. The fitting
curve will vanish when the mouse pointer leaves the tab.

\item Dynamic Channel Display \icon{dynamicPlot}: to display the intensity of the plotter channels
dynamically on the tab when hovering the mouse over the plotting area of the plotter. It should be
remarked that this item is only for non-back-projected tabs. For back-projected tabs the dynamic
display works in a similar way but independent of this function (see \S\ \ref{MWSinograms}).

\item Display Frame Intensity \icon{disSelChan}: to display on tab the total intensity of the channels of
the current frame of plotter when zooming in and out.

The default of the last two functions when launching the program is to be off. Invoking these
functions henceforth toggles their functionality on and off. Currently, these two functions work
independently so they can be both on or off or in a mixed state.

\end{enumerate}

\subsection{Buttons} \label{PloBut}

The plotter contains seventeen tool buttons; fifteen are for accessing the items of the menus and
are included for user convenience. The other two, which are located on the right of the plotter,
are

\begin{enumerate}

\item Zoom out \icon{zoomout}: this button appears on zooming in.

\item Zoom in \icon{zoomin}: this button appears on zooming out.

\end{enumerate}

It is noteworthy that the $xy$ frame of the plotter is fixed on zooming in so that the displayed
pattern on hovering the mouse pointer will be within the zoomed $xy$ range. The dynamic adjustment
to accommodate the whole $xy$ range of the pattern will be resumed on zooming out to the initial
state before zooming in.

\subsection{Labels} \label{PloLab}

\begin{figure} [!h]
\centering
\includegraphics
[scale=0.8] {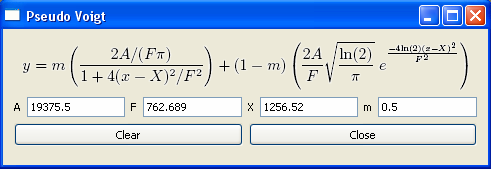}
\caption[\pseVoigt\ dialog] {A \pseVoigt\ dialog.} \label{voigtDialog}
\end{figure}

Associated with each fitting basis curve is a sensitive label (coloured as its curve) on the far
right margin of the plotter. On clicking the label, by left or right mouse buttons, a dialog
corresponding to the basis curve is launched. From this dialog the curve parameters can be adjusted
using the relevant input boxes and hence the plotter is instantly updated. The dialog also contains
an image of the fitting curve equation and two push buttons: one to clear the curve and the other
to close the dialog. An example of these dialogs is presented in Figure \ref{voigtDialog}.

\section{3D Plotter} \label{3DPlotter}

\begin{figure} [!h]
\centering
\includegraphics
[scale=0.4] {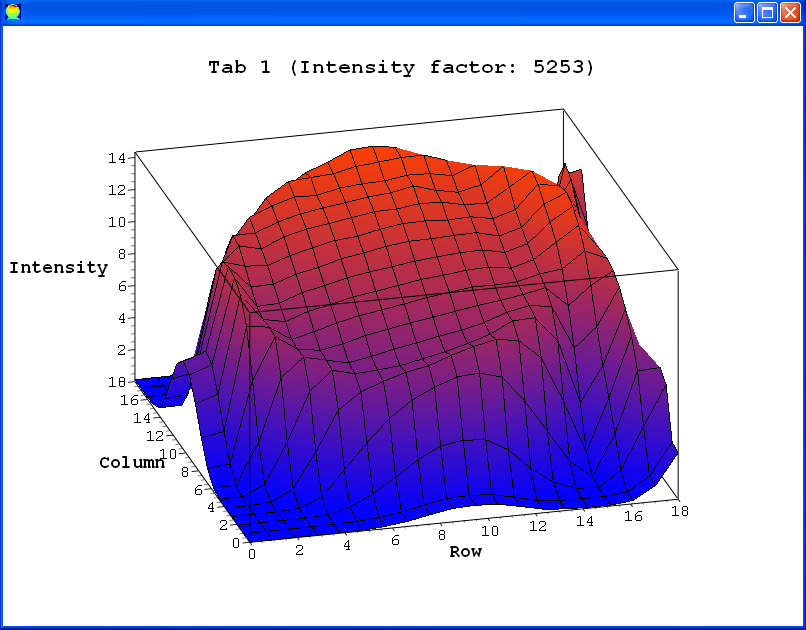}
\caption{A 3D graph of total intensity as a function of \voxel\ position in a tab.} \label{3DGraph}
\end{figure}

This plotter\footnote{The graph plotting capability of the QwtPlot3D library is used in the
construction of the 3D plotter.} (Figure \ref{3DGraph}) is used to create a 3D graph of the current
tab where the $x$ and $y$ axes stand for the number of rows and columns respectively, while the
$z$-axis represents the \voxel\ intensity. The plot can be resized, zoomed in and out, and rotated
in all orientations. To keep proportionality of the plot dimensions, the intensity is automatically
scaled by a scale factor which is displayed beside the tab number. No more than one plotter can be
launched for any tab. The plotter is closed when its tab is removed. Moreover, the label of the 3D
plot is updated when the corresponding tab number changes by removing a tab. The spin box next to
the 3D graph button on the main window is for scaling the $z$-axis. The range of this scale factor
is between 0.1 and 100, with a step size of 0.1. On performing batch curve-fitting in one of its
various modes the residuals and the refined parameters can also be visualised by the 3D plotter
although the scaling may not be adequate.

\section{Form Dialog} \label{Form}

\begin{figure} [!h]
\centering
\includegraphics
[scale=0.6] {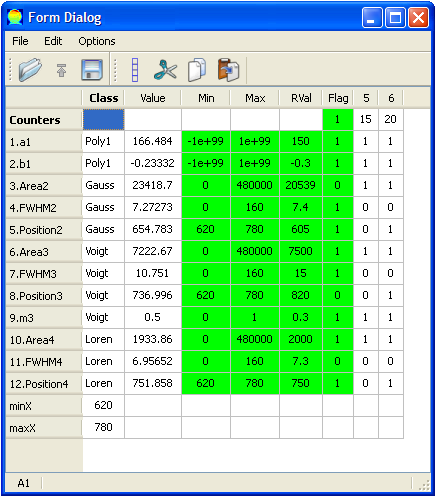}
\caption{Form dialog.} \label{form}
\end{figure}

The main use of spreadsheet forms is in large-scale curve-fitting operations. The idea is to
prepare a form within the plotter and save it to the disc. It is then imported to the main window
for batch fitting a number of cells or tabs in the tab widget, or for multi-batch fitting of a
number of data sets. Importing a previously prepared form to the plotter can also be used for
curve-fitting a single pattern since the basis functions and frame settings, which include zoom
view and $xy$ ranges, are reserved in the form. When the form dialog (Figure \ref{form}) is
launched from the plotter, the dialog interacts with the plotter, i.e. it automatically lists the
existing basis functions in the plotter. Moreover, the plotter is updated on modifying the
parameters of the basis functions in the dialog.

The form dialog consists of a number of menus, toolbars and a status bar similar to those in the
main window. The dialog also has a number of columns:

\begin{enumerate}

\item A gray column for labelling and classifying the parameters of basis functions. Each label starts
with an ordinal index followed by the name or abbreviation of the parameter, followed by the index
of the function to which the parameter belongs.

\item A column, labelled `Class', in the white area for classifying the basis functions, i.e. `\Gauss' for
\Gauss ian, `Loren' for \Lorentz ian, `Voigt' for \pseVoigt, and `Poly$n$' for polynomial of order
$n$. In the end of this column the $x$ range (i.e. minimum and maximum) of the form is displayed.
The single, batch and multi-batch curve-fitting operations are performed only on the pattern data
within this range.

\item Another column, labelled `Value', for the initial values of the parameters. These initial values are
automatically changed to the refined values following a single curve-fitting operation in the
plotter.

\item Four green columns for restrictions. The idea of restrictions is to set the refined parameters to
certain values when they exceed predefined lower or upper limits. The first of these columns `Min'
is for the lower limit. The second column `Max' is for the upper limit. The third column `RVal' is
the restriction value which the parameter should be set to when it exceeds one of the two limits.
The fourth column `Flag' is for the Boolean flags. At the top of this column there is a global flag
for the restrictions to be applied (when it is `1') or not (when it is `0'). The other entries in
the column are the individual flags for each parameter. The restriction on a certain parameter is
applied only when both the global flag and the flag for that particular parameter are true (i.e.
`1'). The default values for these four columns are shown in Table \ref{resDefaults}. The
individual flag of a particular parameter can be set/reset (i.e. toggle between `0' and `1') by
double-clicking the corresponding cell. Double-clicking the cell of the global flag will set/reset
the whole column of individual flags. The value of the global flag can be modified by typing the
value directly inside its cell. This can also be used for modifying the individual flags.

\item Column(s) for counter(s) and Boolean flags for the \ls\ (LS) fitting routine. These columns can be added
by the user to fix the number of LS cycles and determine which parameters should be refined. Hence,
these columns are absent when the form is created. On the top of each column there is a counter of
positive integer type. This counter determines the number of iteration cycles in the LS routine for
that set of Boolean values. Double-clicking the counter cell will set/reset the whole column of
Boolean flags, i.e. they toggle between 0 and 1. The Boolean values can also be changed
individually either by typing directly in the cell, or by double-clicking the cell. A `0' flag
means the corresponding parameter is held constant during the LS routine, while a `1' flag means
the parameter will be refined and adjusted in the LS routine. The default values for the counter
and Boolean flags are 1 and 0, respectively. The number of cycles in the LS routine is fixed when
these columns exist, and this number equals the sum of counters. When these columns are absent, the
number of cycles will vary depending on the pattern, basis functions and initial values of fitting
parameters. In this case, LS cycles continue until convergence is achieved within a certain
marginal error. Moreover, all parameters will be refined in the LS routine.

\end{enumerate}

\begin{table} [!h]
\centering %
\caption[Default values for the form restrictions columns.]{The default values of the form
restrictions parameters for \Gauss, \Lorentz, \pseVoigt\ and polynomial basis functions. The
default value for the global restriction flag is `0'. The `Plotter' in the table refers to the
frame of the plotter, as viewed by zooming in and out, when the form was created.}
\label{resDefaults} %
\vspace{0.5cm} %

{   
\begin{tabular}{|l|c|c|c|c|} 
\hline
 & Min & Max & RVal & Flag\tabularnewline %
\hline
Area & 0 & Plotter area & Initial value & 0\tabularnewline %
FWHM & 0 & Plotter width & Initial value & 0\tabularnewline %
Position & Plotter lower limit & Plotter upper limit & Initial value & 0\tabularnewline %
$m$ & 0 & 1 & Initial value & 0\tabularnewline %
Poly Coefficients & $-10^{99}$ & $10^{99}$ & Initial value & 0\tabularnewline %
\hline
\end{tabular}
}
\end{table}

\subsection{Menus} \label{ForMen}

The form dialog contains three menus: File, Edit and Options.

\subsubsection{File Menu} \label{ForFilMen}

\begin{figure} [!h]
\centering
\includegraphics
[scale=0.8] {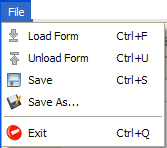}
\caption{The form dialog File menu.} \label{FormFileMenu}
\end{figure}

This menu (Figure \ref{FormFileMenu}) contains five main items:

\begin{enumerate}

\item Load Form \icon{loadForm}: to load a form file to the dialog. The dialog
recognises form files of *.sp type only.

\item Unload Form \icon{unloadForm}: to unload a form from the dialog and erase it from memory.

\item Save \icon{save}: to save a modified form to its *.sp source file.

\item Save As \icon{saveas}: to save a form to a new *.sp file.

\item Exit \icon{exit}: to exit the form dialog and return to the parent widget, i.e. main window or 2D plotter.

\end{enumerate}

This menu may also contain a number of variable items. These are the names of the most recently
loaded *.sp files (up to five) made available for quick load. These items appear, when they exist,
between `Save As' and `Exit'.

\subsubsection{Edit Menu} \label{ForEdiMen}

\begin{figure} [!h]
\centering
\includegraphics
[scale=0.8] {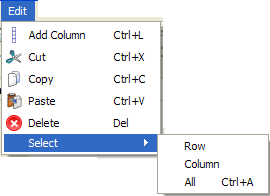}
\caption{The form dialog Edit menu.} \label{FormEditMenu}
\end{figure}

This menu (Figure \ref{FormEditMenu}) contains six items:

\begin{enumerate}

\item Add Column (\includegraphics[width=0.01\textwidth]{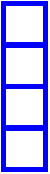}): to add a column for LS counter
and Boolean flags.

\item Cut \icon{cut}: to cut the content of selected cell(s).

\item Copy \icon{copy}: to copy the content of selected cell(s).

\item Paste \icon{paste}: to paste the cut or copied contents.

\item Delete \icon{delete}: to delete the content of selected cell(s).

\item Select: this item contains:

\begin{enumerate}

\item Row: to select the row of the current cell.

\item Column: to select the column of the current cell.

\item All: to select all cells in the form.

\end{enumerate}

\end{enumerate}

\subsubsection{Options Menu} \label{ForOptMen}

\begin{figure} [!h]
\centering
\includegraphics
[scale=0.8] {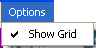}
\caption{The form dialog Options menu.} \label{FormOptMenu}
\end{figure}

This menu (Figure \ref{FormOptMenu}) contains one item:

\begin{enumerate}

\item Show Grid: to turn the grid on and off.

\end{enumerate}

\vspace{0.5cm}

The File and Edit menus have associated toolbars (Figure \ref{form}) which can be turned on and off
by right-clicking the toolbar area and selecting the option. Each toolbar contains the main items
of its menu. By default, all toolbars are displayed when starting the program, and their settings
are not saved on exit. The form dialog has also a context menu which contains the items of frequent
use. The context menu can be accessed by right-click on the form main area.

\section{Modules of \ProgName} \label{Modules}

\ProgName\ contains four main modules; these are

\subsection{Curve-Fitting} \label{MWCF}

One of the main modules of \ProgName\ is curve-fitting by \ls\ minimisation using the \LevMar\
algorithm. Curve-fitting can be performed on a single pattern, or as a single batch process over
multiple patterns, or as a multi-batch process over multiple forms and data sets. Curve-fitting can
be done on single or multiple peaks using a number of profile basis functions with and without
polynomial background modelling. The number of \ls\ cycles can be fixed or vary according to the
convergence criteria. The parameters to be refined can also be selected with possible application
of restrictions. The range of data to fit can be selected graphically or by using a prepared form.
Some relevant statistical indicators for the fitting process, as outlined in Table
\ref{statisticalIndic}, are computed in the curve-fitting routine.

Single \curfit\ is performed from the plotter on the pattern of the cell selected by a
double-click. On running this mode of fitting a file called `aVector' is created. This file
contains a summary of the initial and final values for the fitting parameters and a number of
statistical indicators. As indicated earlier, single curve-fitting applies to the $x$ range in the
current view of the plotter.

Batch \curfit\ is carried out from the main window. Spectral data should be mapped on tab(s) and a
form should be loaded through the main window beforehand. Batch fitting runs only on the parts of
the patterns within the $x$ range of the loaded form. For successful operation, the form $x$ range
should be within the $x$ range of the spectral data. Curve-fitting in this mode can be performed on
a single pattern, a number of randomly selected patterns from one tab or a number of tabs. It can
also be performed on all patterns in a tab or a number of tabs with no need for cell selection. Two
data files are created to summarise the outcome of batch curve-fitting. The first is called `batch'
and contains a summary of the input and output data in general as well as data for each fitted
cell. These data are labelled with `$t$-$r$-$c$' where $t$, $r$ and $c$ are indices for the tab,
row and column of the cell. The second file, called `OlivierMatlab', is a summary intended for use
in post processing. This file contains the refined parameters of the refined cells identified by
`$t$ $r$ $c$' labels.

Multi-batch \curfit\ is carried out from the main window. It is an automated, large-scale fitting
operation on multiple data sets. The use of this mode of operation, to replace repeated manual
application of batch fitting, produces more reliable and organised results and saves a huge amount
of time and effort.

\subsection{Sinogram Treatment} \label{MWBP}

The main item in this module is \bacpro\ which is a tomographic technique for reconstructing a 2D
image in real space from a number of 1D projections and their angles. It relies on the application
of the inverse Radon transform and the Fourier Slice Theorem. The raw data for \bacpro\ are
\sinogram s consisting of a set of intensity values for rotation versus translation measurements.
\Bacpro\ can run in single and multi-batch modes, and can be applied on total or partial intensity
as well as individual channels, with possible application of Fourier transform and filtering.

\subsection{EDF Processing} \label{MWEC}

The purpose of this module is to extract the information from EDF binary files obtained from CCD
detectors. Two forms of extraction are available: conversion to normal 2D images in one of three
formats (png, jpg and bmp), and squeezing to 1D patterns in $xy$ text format.

\subsection{Graphic Presentation} \label{MWGP}

Examples of this module are mapping, visualisation and imaging to produce graphs in 2D and 3D
spaces. These include creating tomographic and surface images in single, batch and multi-batch
modes, as well as $xy$ plots for spectral patterns. Some of these graphic techniques use a direct
display on the computer monitor, while others save the results to the computer storage in the form
of image files.

\section{Context Menus} \label{ConMen}

\ProgName\ contains several context menus which are accessed via right-click:

\begin{itemize}

\item Main window context menu: contains selected items from the main window menus.

\item Main window toolbars context menu: used to turn the main window toolbars on and off.

\item Tab context menu: contains 12 items for manipulating cells, rows and columns of the active tab.

\item Form dialog context menu: contains selected items from the form menus.

\item Form dialog toolbars context menu: used to turn the form toolbars on and off.

\end{itemize}

%

%% file: Cata.tex
%
%
\chapter{Tomographic Studying of Heterogeneous Catalysis} \label{Cata}

In this chapter we outline the experimental part of our investigation which focuses on the study of
tomographic imaging of heterogeneous catalysis.

\section{\Catalysis} \label{Catalysis}

A `\Catalyst' is a chemical agent that affects a chemical reaction by increasing or decreasing its
rate. `\Catalysis' is the involvement of \catalyst s in chemical reactions. \Catalyst s can be in
any one of the three phases (i.e. gas, liquid and solid) although most industrial \catalyst s are
solids or liquids. Important catalytic characteristics that determine their suitability for use in
a given process include activity, selectivity, cost and lifetime. By controlling the rate of
chemical reactions, \catalyst s can play a vital role in the production of important chemicals
under manageable conditions. Another property of \catalyst s which can be more important than
activity is their ability to affect the selectivity of reactions where different products can be
obtained from given reactants by the choice of catalytic system. Reaction routes can be designed
with the help of \catalyst s to maximise useful products and reduce waste and cost
\cite{ChorkendorffN2003}.

The intentional use of catalysts in industrial projects started in the mid 18th century with the
use of lead chambers in the production of sulphuric acid. Since then, the use of \catalyst s in the
chemical industry has become a necessity in most manufacturing processes. Very important industrial
and fine chemicals are produced with the help of \catalyst s. It is estimated that 75-90\% of all
synthetic chemicals are produced using \catalyst s. Examples of chemicals that require \catalyst s
in their production include pharmaceuticals, resins and synthetic fibres. The petrochemical
industry heavily relies on the use of \catalyst s in refining, purification and chemical
transformation processes. \Catalyst s have also very important uses in environmental protection
technology such as emission control. Although \catalyst s have been in use for a very long time,
the theoretical basis for their action had not been established until the last few decades. Indeed,
the use of \catalyst s in many areas is still an art rather than a science \cite{Hagen2006}.

Although \catalyst s do not change the thermodynamics of reaction, they can lower the
transformation activation barrier by introducing a new pathway for a reaction. Despite the
widespread belief that \catalyst s are not affected in the course of reaction, they in fact
participate in a cyclic process of chemical bonding with the reactants. While \catalyst s in
general are not consumed in reactions, they may require replenishment to compensate the losses
which occur by associated processes. They may also undergo chemical changes that reduce their
activity. Different deactivation mechanisms are responsible for the progressive loss of activity
\cite{ChorkendorffN2003}.

According to their production method and structure, catalysts can be divided into three groups:
bulk, shell and impregnated. Industrially used \catalyst s come in various shapes and forms such as
spheres, cylinders, pellets and rings. Sophisticated manufacturing processes which involve physical
and chemical treatment under strictly-controlled conditions are followed in the production of these
catalysts. The purpose is to produce these agents with the required properties as demanded by the
intended process \cite{Richards2006}.

\section{Heterogeneous Catalysts} \label{Heterogeneous}

\Catalyst s can be either homogeneous or heterogeneous. In homogeneous catalysis the reactants and
the \catalyst\ are in the same phase, while in heterogeneous catalysis they have different phases
such as gas and solid. A typical case of \Hetcat s is in the use of a solid support that contains
traces of active catalytic metal such as \molybdenum\ or \nickel\ particles. They facilitate the
reaction by allowing the reactants to diffuse and adsorb onto the surface, coming in contact with
the catalyst. The products then desorb and diffuse away on completing the reaction. The rate of
reaction is determined by the speed of transportation of reactants and products in this
bidirectional process. Highly porous materials with high internal surface area, such as \alumina,
zeolites and \silica, are therefore favoured for use as \catalyst s and catalytic supports. The
surface properties of these agents are very important when selecting a \catalyst\ for a particular
process \cite{Bond2005}.

\Hetcat s have very important scientific, environmental and technological applications. They are
widely used in various organic and inorganic industries such as oil refinement, bio-fuel
production, hydrocarbon cracking, fat hardening, environmental protection and fine chemicals
synthesis. One of the main advantages of using heterogeneous catalysis is the ease of separation
from the reactants. This partly explains the exclusive use of these functional materials in most
industrial catalytic processes over homogeneous catalysis. Another advantage is that they can be
maintained at higher temperatures than the \homcat s. Various properties are considered in the use
of \hetcat s; these include chemical and mechanical stability, porosity, high internal surface
area, ease of handling, separation, safety and cost \cite{Richards2006}.

Metal oxides play a vital role in many industrial heterogeneous catalytic processes. \Molybdenum\
is widely used as a \catalyst\ in many industrial applications. This is partly due to its diverse
oxidation and reactivity states. \Nickel\ is one of the most catalytically active metals and hence
is widely used in various industrial processes. Its applications include desulphurisation of
organic compounds, hydrogenation and hydrogenolysis. \Molybdenum, \nickel\ and \cobalt\ on
$\gamma$-\alumina\ supports are commonly used in industrial hydro-treatment processes
\cite{Serp2009, SmithN1999}.

\Alumina\ is commonly used as a \hetcat\ and as a support for a large number of industrial
\catalyst s; in fact it is the most widely used catalytic support. As an example, alumina is widely
used in environmental protection technology. \Alumina\ is either extracted from natural mineral
deposits or synthesised from aluminium hydroxide and oxide. It is stable at high temperatures with
excellent mechanical and thermal properties. It also has a large internal surface area per unit
volume thanks to its highly porous structure (see \S\ \ref{Example1Sec} for some typical values).
This surface area can be used to support tiny particles of active materials. \Alumina\ supports are
produced by thermally dehydrating Al(OH)$_3$ or AlOOH. \Alumina\ exists in a crystalline $\alpha$
form and other transitional forms such as $\beta$ and $\gamma$. The crystalline form is not usually
used as a catalytic support due to its low surface area \cite{ChorkendorffN2003}.

\Extrudate s are a form of catalyst support material, shaped as cylinders and manufactured by
forcing a mass to flow through a hole using high pressure. \Extrudate s are generally used for
impregnated \catalyst s. Typical sizes of \extrudate s used in the catalysis industry are 10-30 mm
in length with 1-5 mm in diameter. Important issues to be considered when using \extrudate s in
catalysis are porosity, ease of manufacturing, cost, mechanical strength and ability to survive the
reaction conditions \cite{ChorkendorffN2003}.

The optimal distribution of active sites in supported catalysts is dependent on the type and speed
of the required chemical reaction. For slow reactions where the speed of transportation is not a
limiting factor, it is desirable to have a uniform distribution to maximise the distribution of
active sites and hence catalytic exposure. For reactions involving heterogeneous catalysis where
diffusion plays a crucial role in determining the reaction rate, it is beneficial to deposit the
active catalytic sites near the surface of the catalytic support body. Hence an egg-shell
distribution is desirable in this case. Egg-shell distribution is also advantageous in the case of
consecutive reactions with unwanted secondary products to maximise selectivity of primary products.
Heat transport mechanisms also impose limitations on the distribution of active sites on
considering the nature of the catalytic reaction as an endothermic or exothermic. Egg-white or
egg-yolk distributions are preferred when the reactants are contaminated with a catalytic poison
\cite{KomiyamaMH1980, MorbidelliSCV1985}.

Catalyst impregnation is one of the most important techniques for catalyst support preparation. The
technique is widely used thanks to its low cost and simplicity. The impregnation methods are
generally divided into pore volume impregnation and wet impregnation. In the first case the infused
solution volume is comparable to the pore volume and the distribution of the active ingredients is
determined by capillary forces. In the second case the volume of solution considerably exceeds the
pore volume and the distribution is influenced by diffusion and adsorption forces. A required
distribution profile of active sites can be engineered by controlling impregnation parameters such
as pH, temperature, concentration of solution, impregnation time, drying conditions, choice of
precursor, and so on \cite{LeeA1985, FujitaniUE1989, PapageorgiouPGV1996, LekhalGK2001,
LekhalGK2007}.

\vspace{0.5cm}

In the rest of this thesis, we present some examples of using \ProgName\ in processing and
analysing \pd\ data obtained from \synrad\ facilities in the context of heterogeneous catalysis
investigations. In fact large parts of \ProgName\ were specifically developed to facilitate
processing, analysing and visualising of these massive data sets. The examples are ordered
chronologically to reflect the evolution of the use of \Xray\ tomographic techniques for the
investigation of metal oxides anchored to \alumina\ supports. In most cases, the role of the author
of this thesis has been in data collection, data processing and analysis. In the earlier short two
sections, \ref{Example1Sec} and \ref{twoDSpaTem}, the main intellectual ownership of the results
lies with the cited authors and the main contribution of this author was to help the data analysis
and develop specific aspects of the software to do that; in the latter section, \ref{Ex3}, the
author is the main contributor to the given results. In respect of the various results outlined in
this thesis, the role of the author will be highlighted as necessary.

\begin{figure} [!h]
\centering
\includegraphics
[scale=0.35] {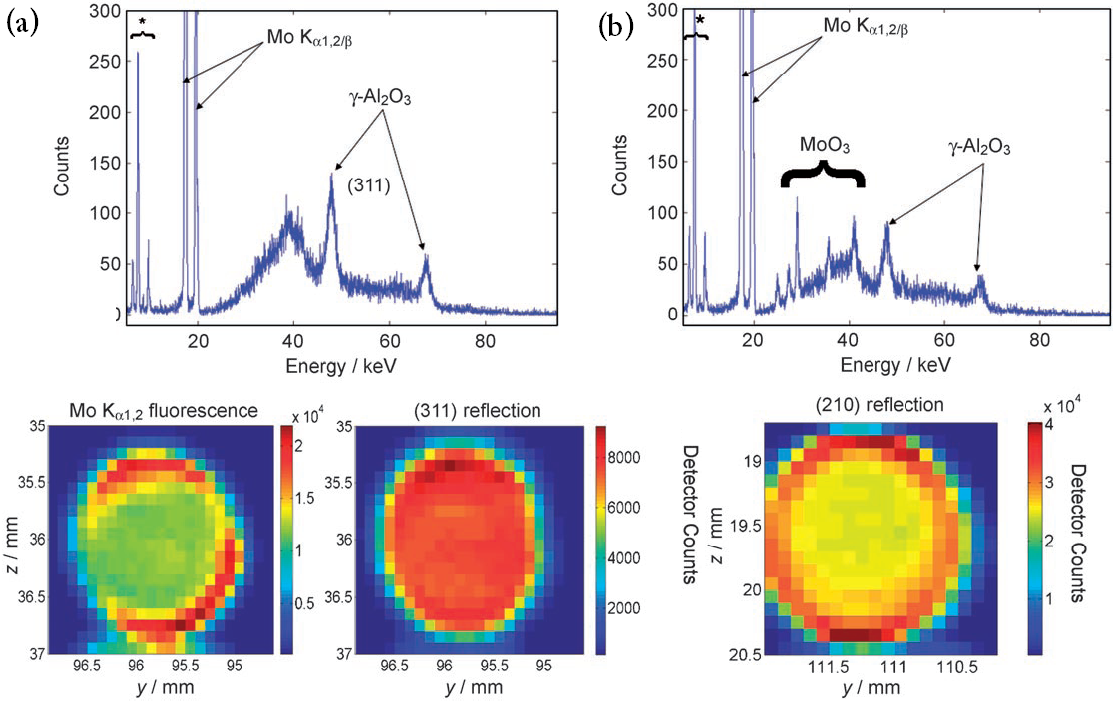}
\caption[Detector signal from a Mo/Al$_2$O$_3$ \catalyst\ body]{Detector signal from a
Mo/Al$_2$O$_3$ \catalyst\ body (a) after drying at 120$^\circ$C and (b) after calcination at
500$^\circ$C (after Beale \etal\ \cite{BealeJBBW2007}).} \label{Ex1}
\end{figure}

\subsection[2D \TEDDIs\ Study of Distributions]{2D \TEDDIs\ Study of Distribution of Precursors and Active \Catalyst s} \label{Example1Sec}

This investigation was carried out by Beale \etal\ \cite{BealeJBBW2007}. The samples used in this
study are cylindrical objects of \alumina\ (Al$_{2}$O$_{3}$) as a base impregnated with active
metallic catalysts. Here, we quote Beale \etal\ for the description of the samples: ``The
cylindrical $\gamma$-Al$_{2}$O$_{3}$ \extrudate s possessed a diameter of 1.5 mm, a length of 10
mm, a pore volume of 0.86 mLg$^{-1}$, and a surface area of 245 m$^{2}$g$^{-1}$. The
Mo/Al$_{2}$O$_{3}$ sample was prepared using pore-volume impregnation with a 1.8M
Mo(NH$_{4}$)$_{6}$Mo$_{7}$O$_{24}$.4H$_{2}$O solution at pH 6.0. Drying was started 1 h after
impregnation by passing hot air over the sample until the temperature reached 120$^{\circ}$C. The
Co\textendash{}Mo/Al$_{2}$O$_{3}$ sample was prepared using a 1.0M
Mo(NH$_{4}$)$_{6}$Mo$_{7}$O$_{24}$.4H$_{2}$O and 0.6M Co(NO$_{3}$)$_{2}$ solution at pH 5.0. This
sample was allowed to age for 15 min before drying. Subsequently, both samples were calcined at
500$^{\circ}$C for 1 h in static air.''

The measurements were carried out at station 16.4 of the Daresbury \SRSl. This station operates
with three coupled MCA detectors and uses a multi-chromatic beam to gather data in EDD mode. The
use of a white beam also allows for the detection of fluorescence signals. In this investigation,
\TEDDIs\ was used to collect information about phase and elemental distributions on supported metal
oxides in \catalyst\ bodies during different stages of \catalyst\ preparation. The distribution of
the active \catalyst\ (i.e. \moltri\ \MoO), as a crystalline phase present only in the
post-calcined sample, was determined from the \diffraction\ signal, while the distribution of the
precursor (amorphous) phase in the pre-calcined sample was determined from the Mo/Co fluorescence
signal. Some of the results which have been obtained from this investigation are presented in
Figure \ref{Ex1}. \ProgName\ was used in the visualisation and analysis of the data collected in
these measurements. The images were reconstructed directly after beam and count rate corrections
with some of these corrections being implemented in \ProgName.

This study highlighted the potential of TEDDI to acquire 3D information about the elemental and
phase distributions of metal oxides embedded in catalytic supports. The stages of catalyst
formation from the precursor phase was partly displayed. It also highlighted the prospects of using
TEDDI with \insitu\ and dynamic phase transformation studies. The results of this study emphasised
the advantage of EDD as a viable data acquisition technique with the benefit of utilising
fluorescence signals which contain extra information about the phases involved in the reactions. It
is noteworthy that the two peaks, seen in Figure \ref{Ex1}, at $E \simeq$ 24.9 and 27.4 keV, which
were not identified previously, can now be identified as the ($hkl$, $d$ \AA) = (004, 2.2759) and
(241, 2.0671) Bragg reflections of aluminium molybdate.

\begin{figure} [!t]
\centering
\includegraphics
[scale=0.45] {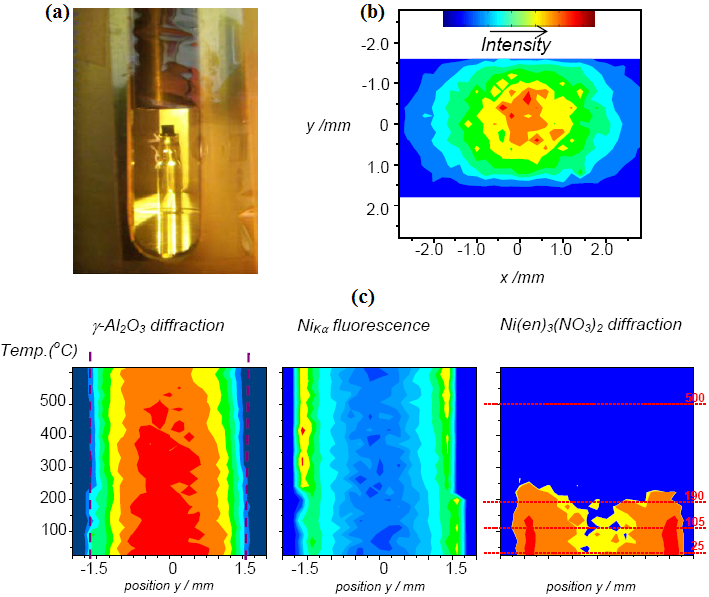}
\caption[A sample of results of 2D spatial/temporal \TEDDIs\ study]%
{(a) 3 mm \extrudate\ mounted in an environmental cell and (b) signal distribution (2D slice)
through a supported \catalyst\ are shown. In (c) the signals from a spatial 1D time/temperature
scan, depicted as a 1D map, are shown in which both diffraction (from alumina and precursor) and
fluorescence (from nickel) signals are collected at each point.} \label{Ex2}
\end{figure}

\subsection[2D Spatial/Temporal \TEDDIs\ Study]{2D Spatial/Temporal \TEDDIs\ of \Alumina\ Supported \Catalyst\ Bodies Under Preparation} \label{twoDSpaTem} %

This investigation was carried out by Espinosa-Alonso and coworkers, and the results are published
in \cite{EspinosaOJBJe2009}. The main objective was to examine the calcination of a heterogeneous
catalyst, and monitor the genesis of active phase formation during the heat treatment. The sample
used in this study is similar to the one used in the previous example. However, \nickel\ on
\alumina\ was used and the initial precursor Ni(en)$_3$(NO$_3$)$_2$ decomposes at about
190$^\circ$C to leave a supported Ni/$\gamma$-Al$_2$O$_3$ \catalyst\ with an egg yolk distribution.
Again \TEDDIs\ was used in this investigation but this time in both spatial and temporal mode. The
author participated in the data collection of this experiment. Moreover, \ProgName\ was used in
processing, visualising and analysing of these data. A sample of the results is graphically
presented in Figure \ref{Ex2}.

This study demonstrated the suitability of \TEDDIs\ for probing temporal dimensions as well as
spatial dimensions. However, the technique is relatively simple compared to the CAT type ADD
experiments (next example). The reason is that each time frame contains only 1D spatial sample
(row) information. Again, fluorescence signals, as well as diffraction signals, were exploited to
gain information about polycrystalline phase and elemental distributions in the sample, and this
feature is typical of the energy dispersive mode. The study revealed that the spatial and temporal
distributions of the precursor and active catalyst are subject to considerable variation. This
indicates the sensitivity and delicacy of the formation process, and its dependency on the fine
details of preparation and thermal treatment procedure.

\subsection[3D Spatial/Temporal CAT Type ADD Studies (Mo)]{3D Spatial/Temporal CAT Type ADD Studies of \Alumina\ Supported \Catalyst\ Bodies} \label{Ex3}%

\begin{figure} [!h]
\centering
\includegraphics
[scale=0.9] {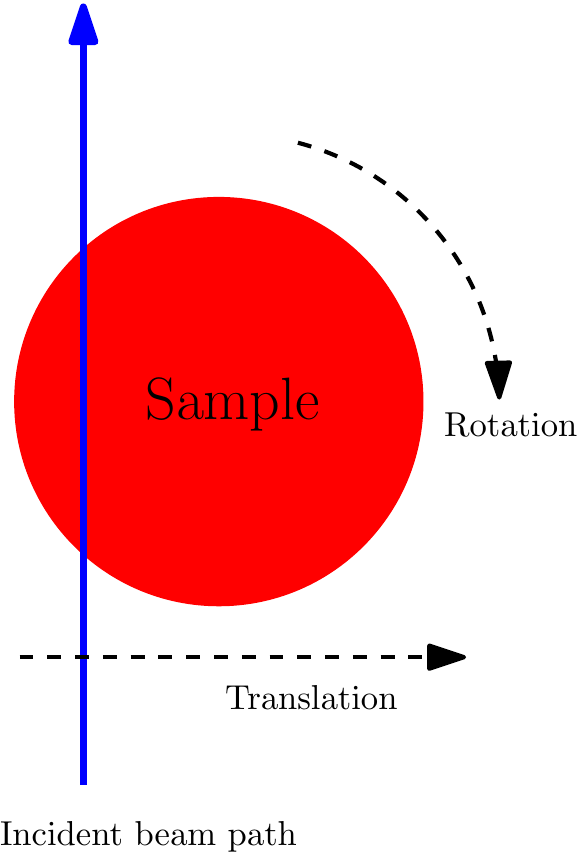}
\caption[CAT layout of the 3D spatial/temporal CAT Type ADD experiment.]%
{CAT configuration used in the 3D spatial/temporal CAT Type ADD experiment.} \label{CAT}
\end{figure}

In the previous study (\S\ \ref{twoDSpaTem}) 2D spatial tomography with a crude temporal aspect was
achieved. In the current study the aim was to have increased temporal resolution, while maintaining
similar spatial resolution, in order to investigate the formation of an active extrudate-supported
\MoO\ catalyst under calcination. In this experiment \CATs\ type \ADDs\ (see \S\ \ref{CatAdd}) was
employed to look at the temporal and thermal behaviour of a slice through an \extrudate\ sample
undergoing calcination to obtain 2D data at various points in time. The temporal aspect to the
study makes it entirely novel and challenging, requiring the data acquisition to be sufficiently
fast to make the process observable. The pertinent information is collected by making individual
\diffraction\ collections from the sample at different positions and different orientations, as
presented in Figure \ref{CAT}.

At the time of data collection, \CATs\ type angle dispersive methods had only very recently been
demonstrated in the literature \cite{BleuetWDSHW2008}. Proportional area detectors, such as CCD and
CMOS-based devices, can be used for recording angle dispersive diffraction patterns, and hence can
support much higher count rates than energy dispersive solid state detectors. Since these modern
devices offer fast reading, they can provide a more thorough insight into the temporal dimension of
the dynamic processes under investigation. The area aspect of such detectors allows the recording
of entire powder rings in angle dispersive mode, which can be integrated to give 1D powder patterns
with good \signoirat. The recorded intensity would be severely reduced if only a strip detector was
employed. Such area detectors, when used in conjunction with very bright sources, allow for very
fast data acquisition even from materials that give fairly poor scattering.

By contrast, the current technology of \EDDs\ detectors does not support high count rates.
Moreover, area type EDD detectors are not realistically available and hence EDD collection is
normally limited to a small solid angle of space. Although the number of photons in a white beam is
considerably higher, the benefits of high intensity cannot be fully exploited due to the limitation
on the count rate of EDD detectors. The CAT type geometry, with the associated algorithms, offer a
convenient method for recording and reconstructing a whole slice through the sample in a short
period of time. Traditionally, CAT is used in medical physics for looking at variations in object
attenuation (e.g. CAT scans in hospitals) and the study of objects by micro-tomography. The view
was that this method could be equally applied in the investigation of supported catalyst bodies to
produce valuable new information on the formation processes.

In this experiment an uncalcined Mo-loaded extrudate, taken from the same batch of the experiment
described in \S\ \ref{Example1Sec}, was used. This system was chosen for a number of reasons

\begin{enumerate}

\item The already existing data from previous experiments showed the expected distribution of the active
phase. This acts as a check on the distribution of the reconstructed images produced from the CAT
type experiment.

\item The amount of formed active catalyst is expected to be relatively high such that one can reasonably
expect to be able to observe active catalyst concentrations within the diffraction patterns.

\item The temporal evolution is unknown, so it was concluded that this experiment would yield new useful
information about this system.

\item The experiment would collectively lead to a proof-of-concept study while also revealing useful information
on a real process.

\end{enumerate}

\begin{figure} [!h]
\centering %
\subfigure []%
{
\begin{minipage}[b]{0.45\textwidth} %
\centering \includegraphics[width=2.8in, trim = 0 0 0 0] {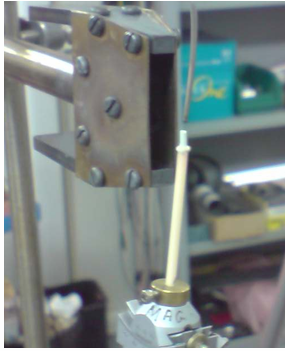}
\end{minipage}}
\hspace{0.5cm} %
\subfigure []%
{
\begin{minipage}[b]{0.45\textwidth} %
\centering \includegraphics[width=2.2in, trim = 0 0 0 0] {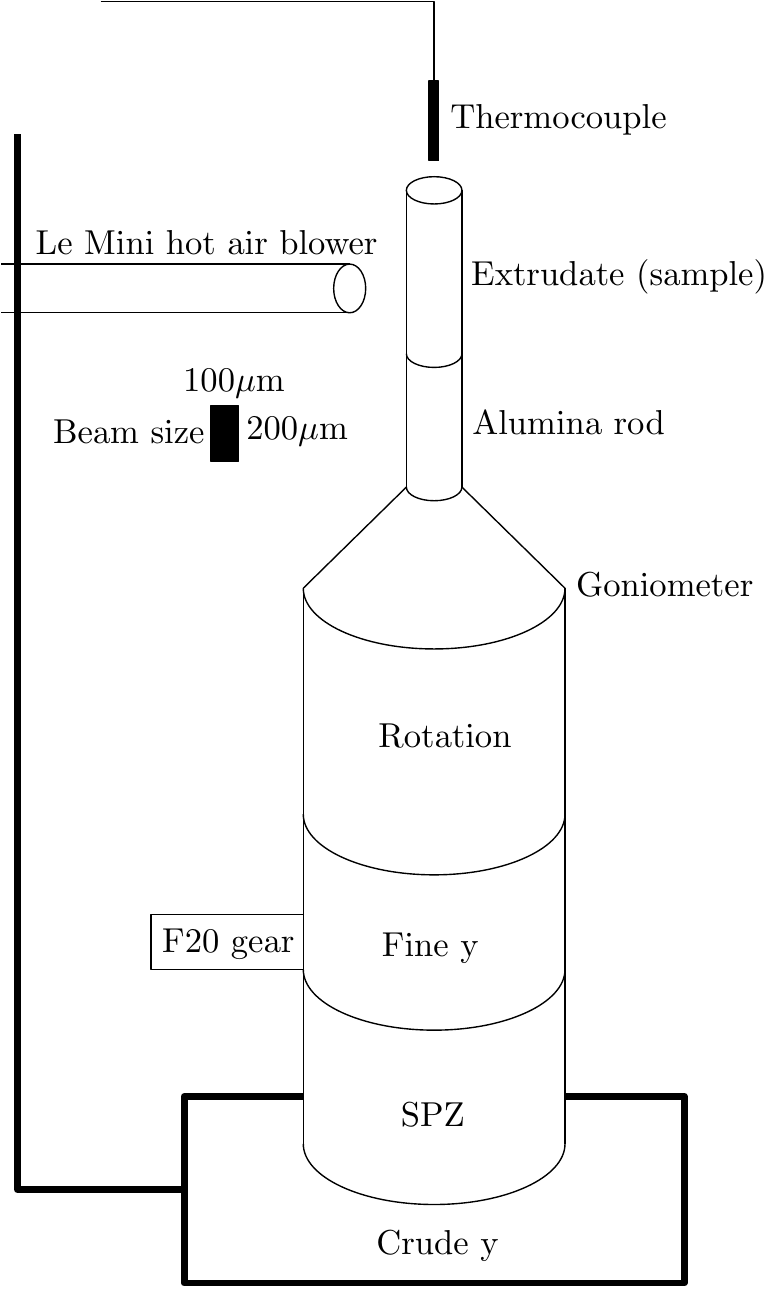}
\end{minipage}} \vspace{-0.3cm}
\caption[Photograph and cartoon of mounted \extrudate\ sample system]%
{(a) Photograph and (b) schematic illustration of the mounted \extrudate\ sample system.
\label{Ex3Photo}}
\end{figure}

The sample is mounted, with its longitudinal axis vertical, on a mounting pole which is then
attached to a \goniometer, as seen photographically in Figure \ref{Ex3Photo} (a) and schematically
in Figure \ref{Ex3Photo} (b). The \goniometer, which serves to align the sample, is mounted on a
rotation stage, which is itself mounted on a translational stage, with the horizontal translation
being perpendicular to the beam direction. The sample is heated by a hot air blower which is
mounted on the translation stage such that during translation the alumina \extrudate\ and the
heater move together, yet the \extrudate\ is free to rotate. During the recording of one slice, and
for the purpose of rapid data acquisition, the sample is continuously translated through -1.5 to
+1.5 mm with respect to the centre of the \extrudate. During this process \diffraction\ is recorded
on a fast area detector in approximately 0.1 mm steps, thereby equating to a total of 31
translation measurements. This is being done at orientations from 0$^\circ$ to 180$^\circ$ in steps
of 6$^\circ$, and hence the rotation measurements are also 31 in number. The total number of
measurements for each slice is therefore 961. Each measurement is contained in a binary EDF file
(see \S\ \ref{MWCCD}). The acquisition time of about 0.5 second per measurement with the associated
motor movement time gives an overall collection time of approximately 10 minutes to record a slice
represented by a single data set. It is important to note that \diffraction\ occurs along the
entire length of the beam-sample interaction and this causes an intrinsic spread in recorded angle.
This effect is worst at the longest path lengths near the centre of the extrudate. The use of high
energy \Xray s mitigates this, since the reflections occur at lower scattering angles.

\begin{figure} [!h]
\centering
\includegraphics [scale=0.5, trim = 0 0 0 0] {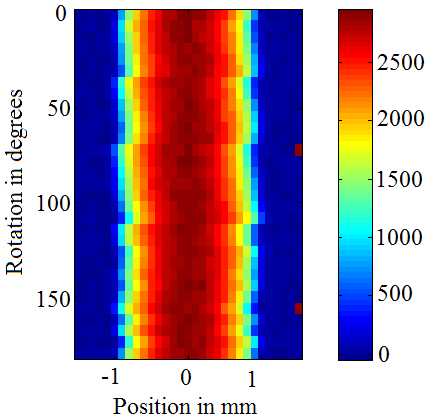}
\caption[A \sinogram\ of integrated intensity of the largest \alumina\ peak.]%
{A \sinogram\ representing the integrated intensity of the largest \alumina\ peak at
$2\SA=5.77^{\circ}$. The peak profile is modelled by a \Gauss ian. \label{Ex3Al1}}
\end{figure}

Figure \ref{Ex3Al1} is a sample \sinogram\ displaying the integrated intensity of the largest
\alumina\ peak, and hence contains information on the distribution of \alumina\ in the extrudate.
It clearly shows the diameter of the \alumina\ cylinder which can be estimated to be about 1.7 mm.
The image also highlights the nature of information conveyed by a sinogram.

\begin{table} [!h]
\centering %
\caption[Time and temperature stamps for the 22 data sets of molybdenum.]%
{List of time and temperature stamps for the 22 data sets of molybdenum experiment. The time is
given in minutes while the temperature is in $^\circ$C.}
\label{stampTable} %
\vspace{0.5cm} %
{\normalsize
\begin{tabular}{|llcc|llcc|}
\hline
     Index &    Data set &       Time & Temperature  &      Index &    Data set &       Time & Temperature  \\
\hline
         1 &    ambient &          0 &         25 &         12 &     ramp11 &        110 &        575 \\
         2 &      ramp1 &         10 &         75 &         13 &     ramp12 &        120 &        625 \\
         3 &      ramp2 &         20 &        125 &         14 &     ramp13 &        130 &        675 \\
         4 &      ramp3 &         30 &        175 &         15 &     final0 &        140 &        700 \\
         5 &      ramp4 &         40 &        225 &         16 &     final1 &        150 &        700 \\
         6 &      ramp5 &         50 &        275 &         17 &     final2 &        160 &        700 \\
         7 &      ramp6 &         60 &        325 &         18 &     final3 &        170 &        700 \\
         8 &      ramp7 &         70 &        375 &         19 &     final4 &        180 &        700 \\
         9 &      ramp8 &         80 &        425 &         20 &     final5 &        190 &        700 \\
        10 &      ramp9 &         90 &        475 &         21 &     final6 &        200 &        700 \\
        11 &     ramp10 &        100 &        525 &         22 &     final7 &        210 &        700 \\
\hline
\end{tabular}
}
\end{table}

\begin{figure} [!h]
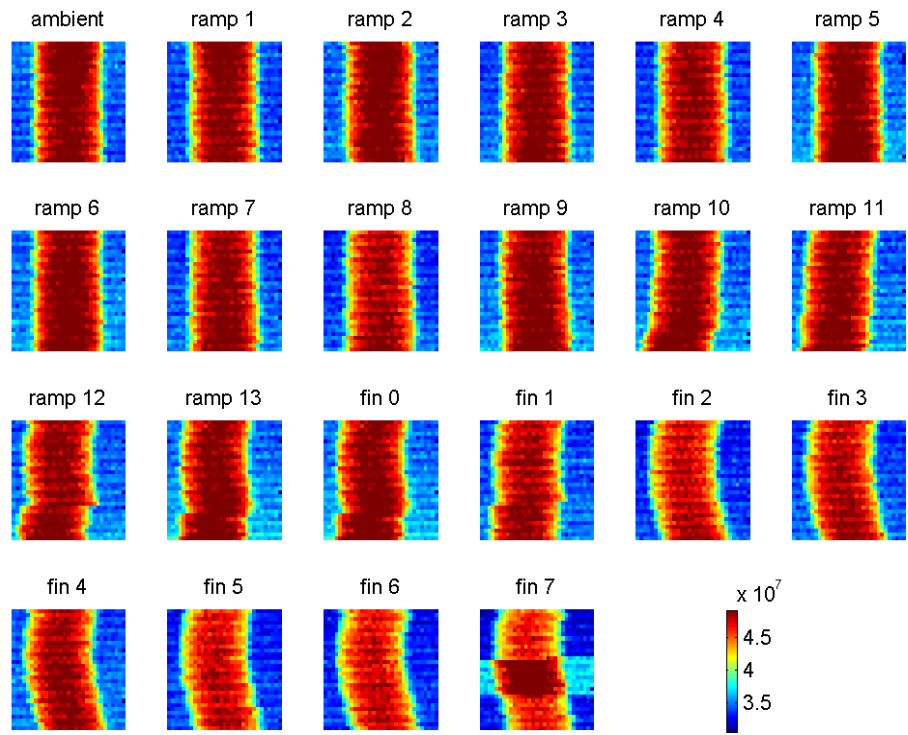

\CIG {Sinograms}
\caption[The 22 data sets of molybdenum displayed as intensity sinograms]{The 22 data sets of
molybdenum experiment displayed as total intensity \sinogram s. \label{Ex3IntensityS}}
\end{figure}

There are 22 data sets: 1 at ambient temperature, 13 at ramp-increasing temperatures, and 8 at a
\stesta\ final temperature. These sets, which are all collected from a single circular slice from
the cylindrical sample, represent the dynamic state of the sample as a function of time and
temperature. The time and temperature stamps of these data sets are presented in Table
\ref{stampTable}. As each set consists of 961 files, the total number of files is 21142. The size
of each EDF file is about 9.5 megabytes, and hence the total size of the EDF files is about 200
gigabytes. In a few cases data files for some \voxel s were missing, and hence void files were
inserted in their positions to maintain the \voxel s positions when mapping and visualising the
slices. These data sets are shown in Figure \ref{Ex3IntensityS}. The images display the total
intensity (sum of counts of individual \voxel s in each data set) of the 22 \sinogram s (which
represent rotation on $y$-axis in degrees versus position on $x$-axis in millimetres).

\vspace{0.2cm}

The experimental setting and data collection were performed by S. Jacques and coworkers, while data
processing, analysis and visualisation were carried out by the author and S. Jacques. \ProgName\
was the main tool used in the analysis of these data. The measurements were performed at the
\ESRFl\ (ESRF) beamline ID15B which is dedicated to applications using very high energy \Xray\
radiation. ID15B houses the \add\ setup using a large area detector \cite{ESRF}. The monochromatic
beam used in this experiment has a wavelength of 0.14110 \AA.

In the following we outline the data processing stage and present results obtained from analysing
these data sets. First, to ensure the consistency of data collection mechanism, the time stamps of
the individual files for each data set were gathered and compared to the device time stamp which is
archived in an independent log file. The comparison revealed that the timing is consistent with no
apparent error. The time stamp also provided a graphical picture of the time-temperature
correlation.

As the original data are in the form of binary EDF images, they have to be converted to 1D patterns
in ASCII numeric format. \Datasqueeze\ (see \S\ \ref{DataPrep}) was used to perform this conversion
because at that time the EDF processing module of \ProgName\ had not been implemented. The
squeezing process (which took about a week to complete on a fast dual-processor pentium
workstation) was repeated three times to ensure consistency and accuracy of the results and
eliminate errors. Each extracted text file contains three columns of data: \scaang\ in degrees,
number of counts computed by integrating around the entire ring, and an error index.

\begin{figure} [!b]
\centering
\includegraphics
[scale=0.75] {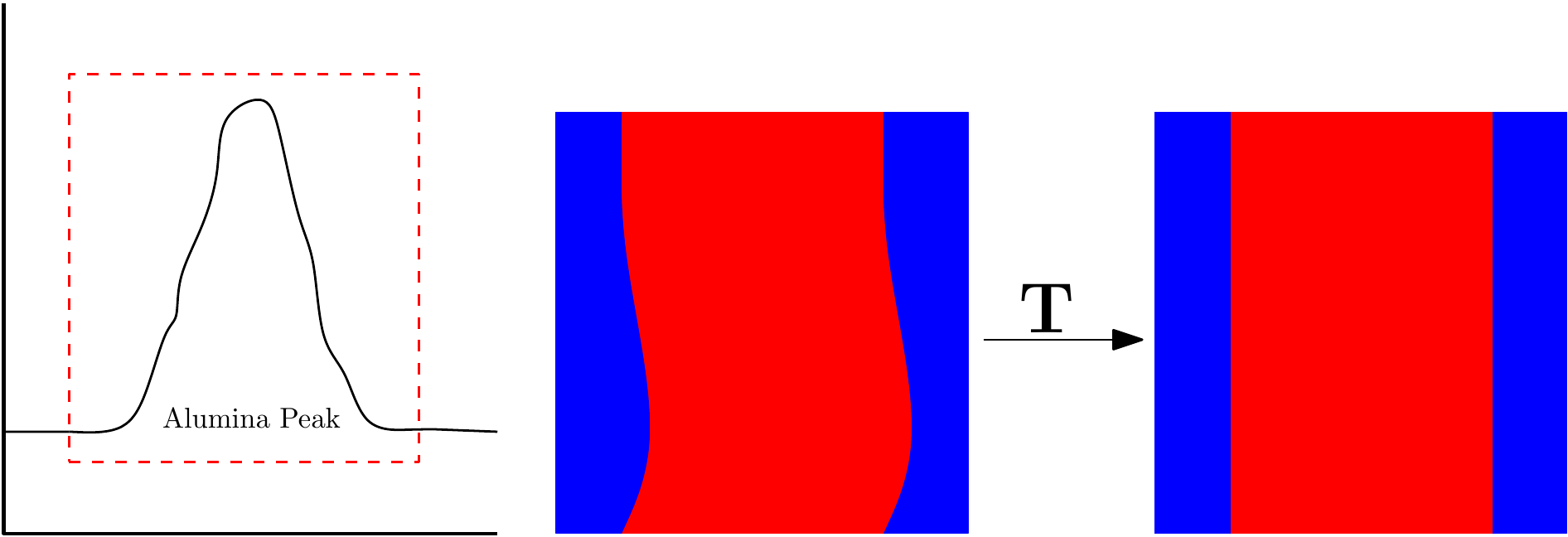}
\caption[Sinogram alignment using a prominent alumina peak.]%
{Sinogram alignment using a prominent alumina peak. The alignment matrix {\bf T} obtained from this
operation is used to align the sinogram channels. In these schematic sinograms, blue represents the
background while red represents the sample voxels.} \label{aluminaCalib}
\end{figure}

\begin{figure} [!h]
\centering %
\subfigure []%
{\begin{minipage}[b]{0.45\textwidth} %
\centering \includegraphics[width=2.5in, trim = 0 0 0 0] {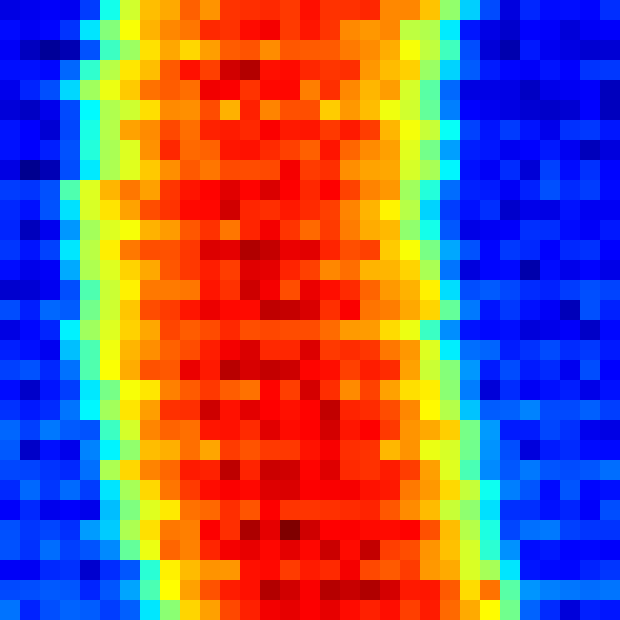}
\end{minipage}}
\hspace{0.5cm} %
\subfigure []%
{\begin{minipage}[b]{0.45\textwidth} %
\centering \includegraphics[width=2.5in, trim = 0 0 0 0] {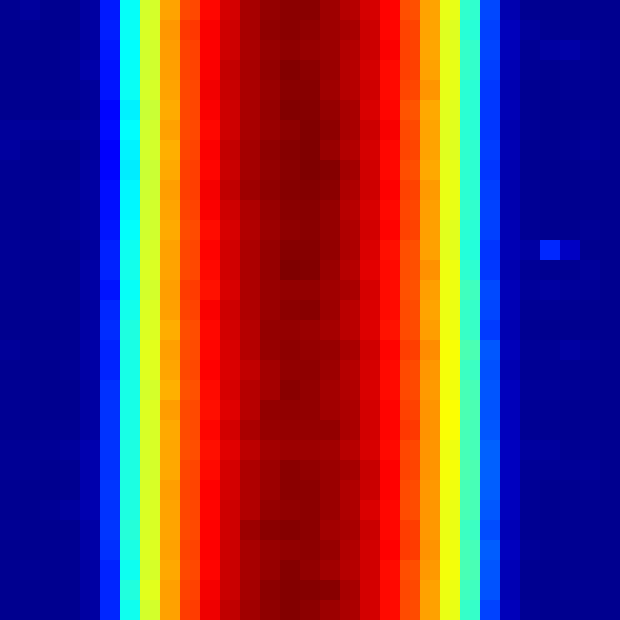}
\end{minipage}} \vspace{-0.3cm}
\caption{An example of a sinogram (a) before and (b) after alignment. \label{sinoAlignment}}
\end{figure}

When constructing the sinograms from the individual slice measurements, it became apparent that
there was a misalignment problem, mild in some data sets and severe in others. The problem is
obviously due to a mechanical error in the translation/rotation system. This resulted in a ragged
edge of the sample voxels in the sinograms, as seen in Figure \ref{Ex3IntensityS}. To correct this
error, a large alumina peak was used to calibrate the positions of the peaks (see Channels
Alignment in \S\ \ref{MWSinograms}). By correctly aligning the sinogram voxels with respect to the
alumina peak, a transformation matrix {\bf T} for each data set was obtained. These matrices were
then applied to all channels of the diffraction patterns of their respective data sets to correct
the sinograms and align the other peaks. This operation is graphically presented in Figure
\ref{aluminaCalib}. Also, a sample sinogram before and after alignment is shown in Figure
\ref{sinoAlignment}.

\begin{figure} [!h]
\centering
\includegraphics
[scale=0.75] {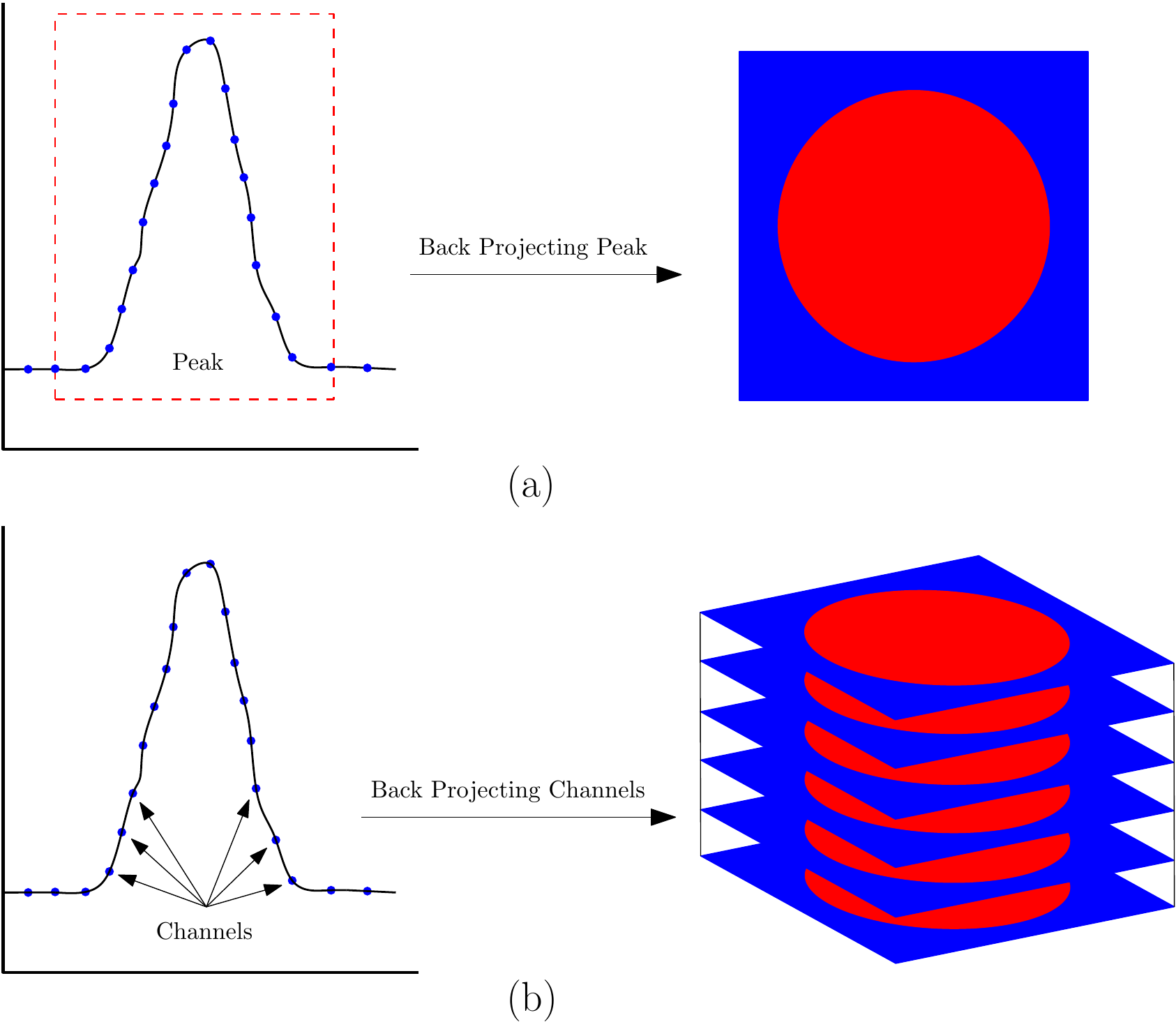}
\caption[Back projection of a peak and channels.]%
{Back projection (a) of a diffraction peak and (b) of a number of channels in a diffraction peak.
In these schematics blue represents the background while red represents the sample voxels.}
\label{backProPeakChan}
\end{figure}

\begin{figure}
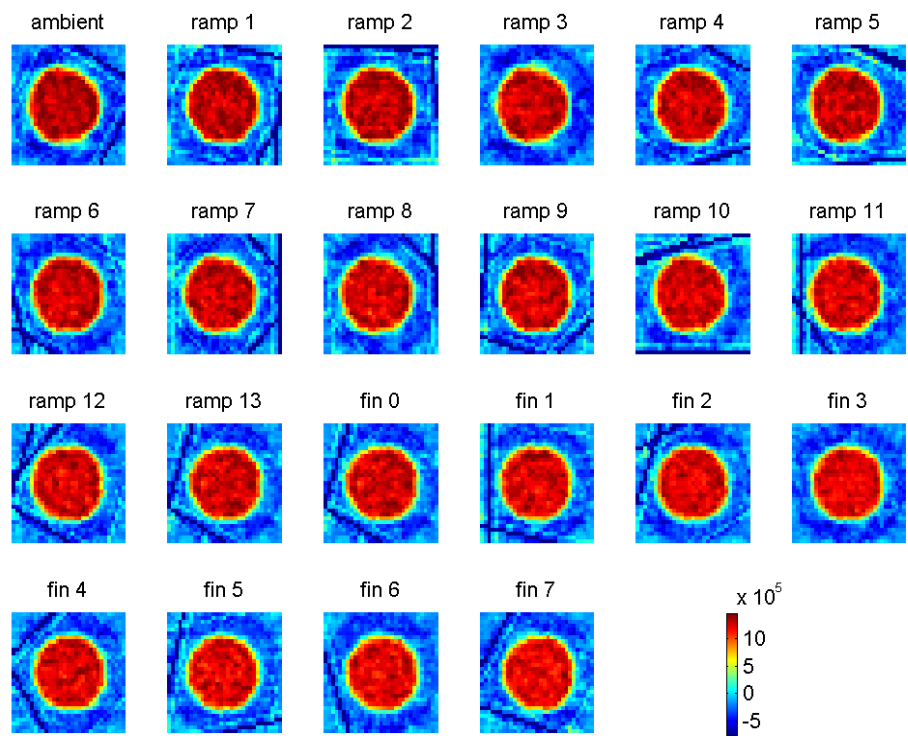

\CIG {Intensity}
\caption[The 22 data sets of molybdenum as intensity of reconstructed images]{The 22 data sets of
molybdenum experiment displayed as total intensity of reconstructed images. \label{Ex3IntensityR}}
\end{figure}

Following this alignment correction procedure, \bacpro\ (see \S\ \ref{MWSinograms} and \S\
\ref{MWBP}) was used to convert the \sinogram s to normal 2D spatial images of slices. In this
regard, there are two ways of applying \bacpro. The first is to apply it to a collective feature
such as the intensity of a given pattern or peak, while the second is to apply it to all the
individual channels. These two approaches are demonstrated schematically in Figure
\ref{backProPeakChan}. \Bacpro\ of individual channels was chosen for our reconstruction procedure.
Figure \ref{Ex3IntensityR} shows the total intensity (i.e. sum of all channels) of the
reconstructed images which represent the local positions of the \voxel s in the 2D slices in real
space.

\begin{figure}
\centering \includegraphics [scale=0.7] {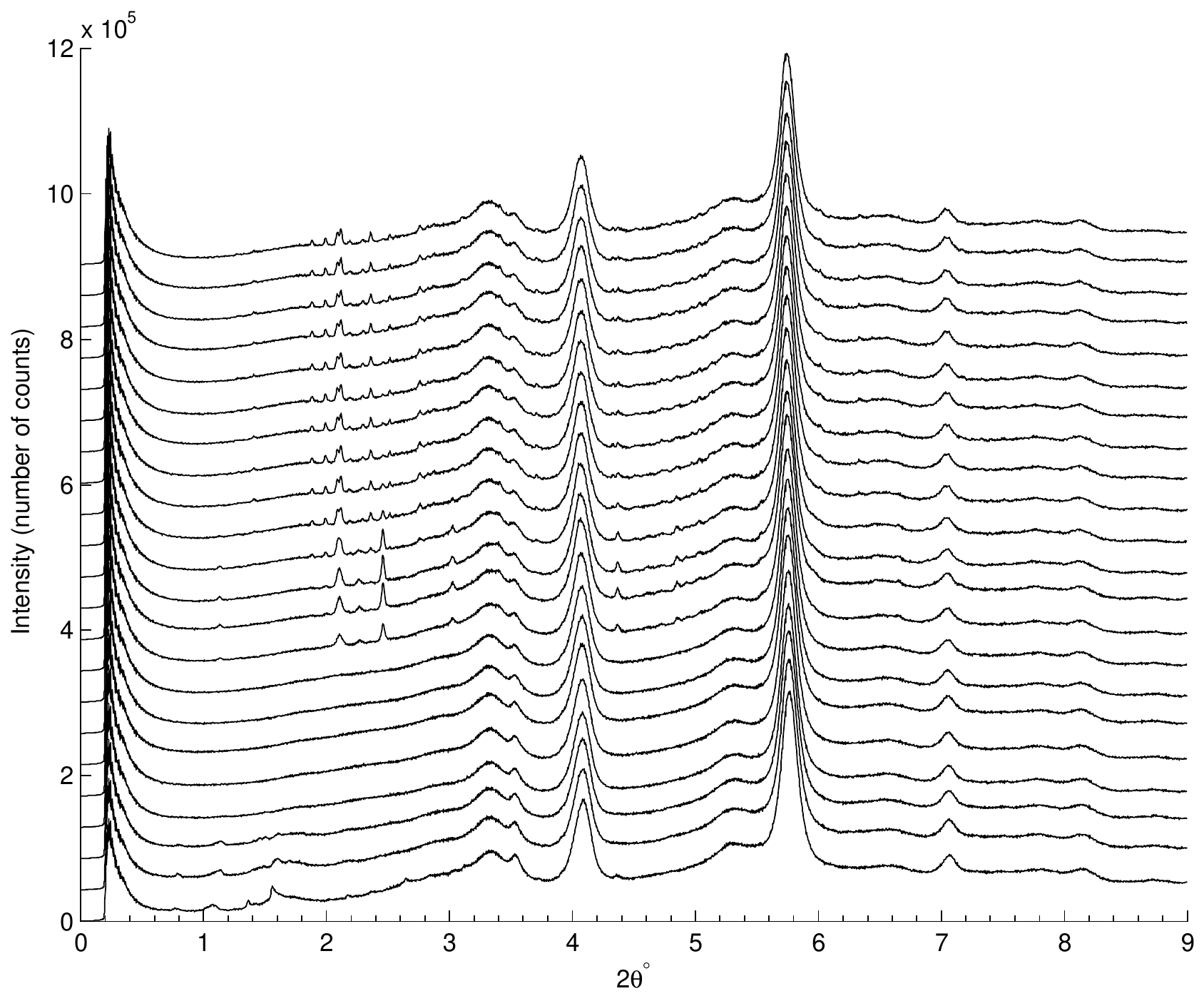}
\caption[Stack plot of the 22 data sets of molybdenum in a 2D graph.]%
{Sums of back-projected diffraction patterns of the 22 data sets of molybdenum experiment presented
as a stack of plots in a 2D graph. \label{MolStack}}
\end{figure}

\begin{figure}
\centering \includegraphics [scale=0.49, trim = 0 0 0 0] {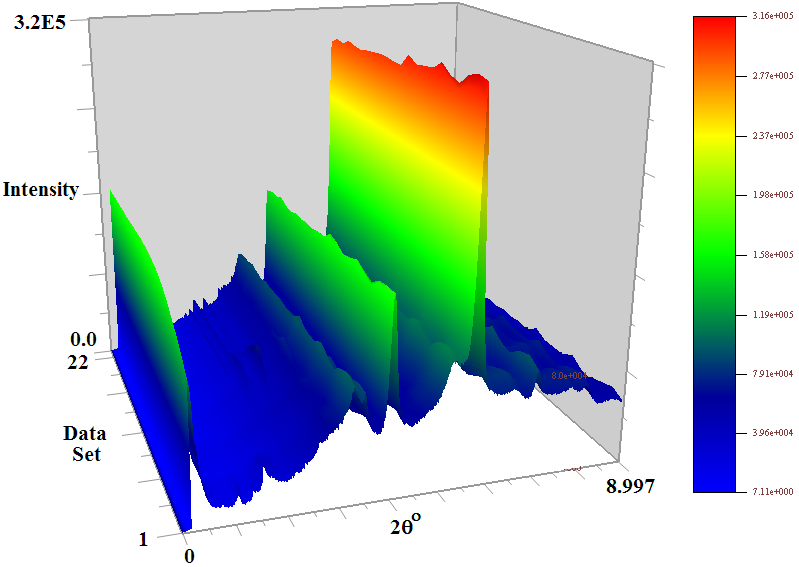}
\caption[Stack plot of the 22 data sets of molybdenum in a 3D graph.]%
{Sums of back-projected diffraction patterns of the 22 data sets of molybdenum experiment presented
as a stack of plots in a 3D graph. \label{sum3DPlot}}
\end{figure}

In Figure \ref{MolStack} the sums of diffraction spectra from these data are displayed. The 22
curves in bottom-up order correspond to the 22 data sets in their temperature and temporal order.
This figure summarises the information contained in each set. The sums are also presented in Figure
\ref{sum3DPlot} as a stack plot in a 3D graph.

Following \bacpro, and guided by the stack plots, \ProgName\ was used in multi-batch processing
mode with restrictions to curve-fit the \diffraction\ peaks. First, a large number of peaks (about
70) were chosen for fitting. These represent the prominent and the suspected peaks from all 22
stages. The purpose of this approach is to capture and curve-fit all important peaks although some
of them are repetitive. As the peak may be more prominent in one stage than in another, and since
peak position may change with temperature, the use of various temperature stages for peak selection
is intended to ensure that all peaks of interest are captured at their ideal state and position. We
also systematically inspected the 2D stack plot to ensure that all the peaks in the sum plots are
included. Most peaks were fitted individually and some were fitted as couples. In all cases, a
simple \Gauss ian profile was used to model the peak shape while a linear polynomial was used to
model the background. It should be remarked that this whole operation has been completed in less
than a day (few hours of manual work to prepare forms and few hours of computer processing using
EasyDD). This stands in sharp contrast to the required time and effort (estimated to be many months
of hard labour) to perform such a task manually using conventional tools.

\begin{figure}
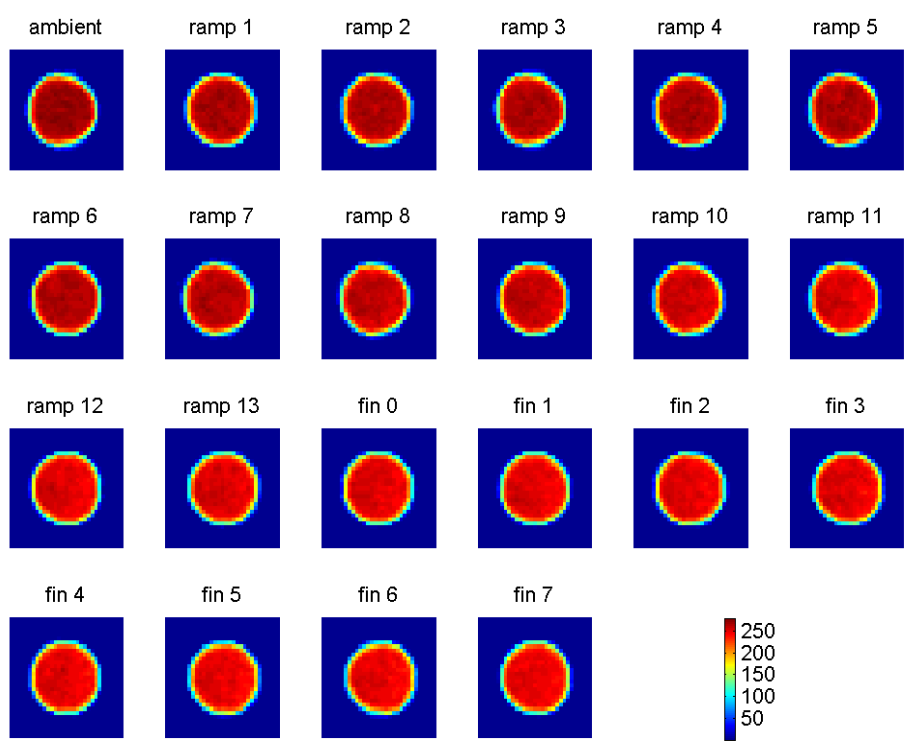

\CIG {5p76706-Par3}
\caption[Gaussian area of the largest \alumina\ peak]%
{Integrated intensity of the largest \alumina\ peak at $2\SA=5.77^{\circ}$ modelled by a \Gauss ian
profile. \label{5p76706-Par3}}
\end{figure}

\begin{figure}
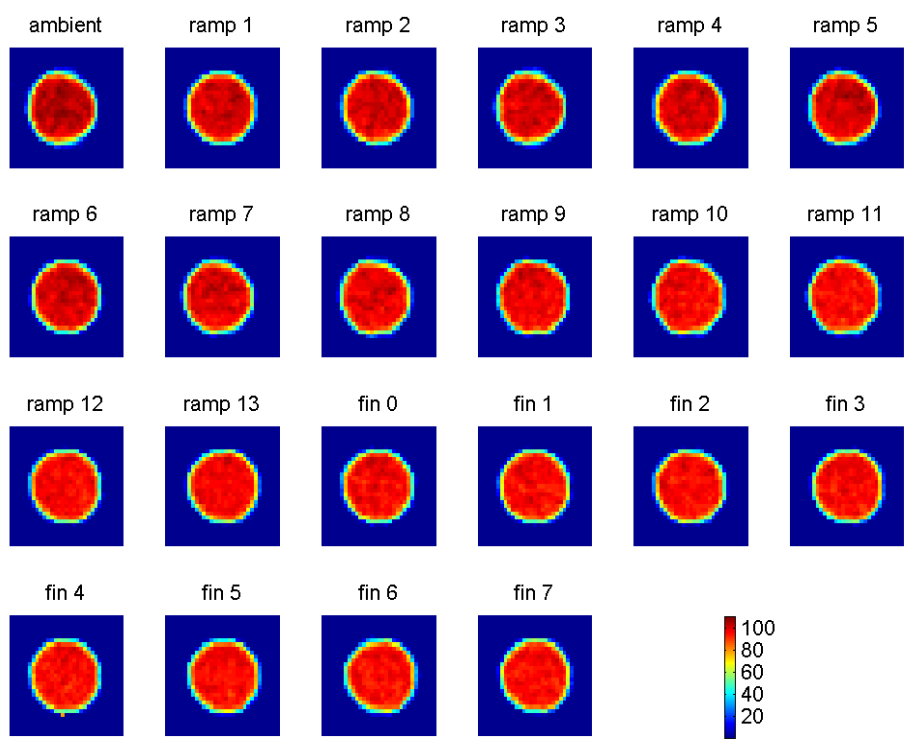

\CIG {4p08027-Par3}
\caption[Gaussian area of the second largest \alumina\ peak]%
{Integrated intensity of the second largest \alumina\ peak at $2\SA=4.05^{\circ}$ modelled by a
\Gauss ian profile. \label{4p08027-Par3}}
\end{figure}

\begin{figure}
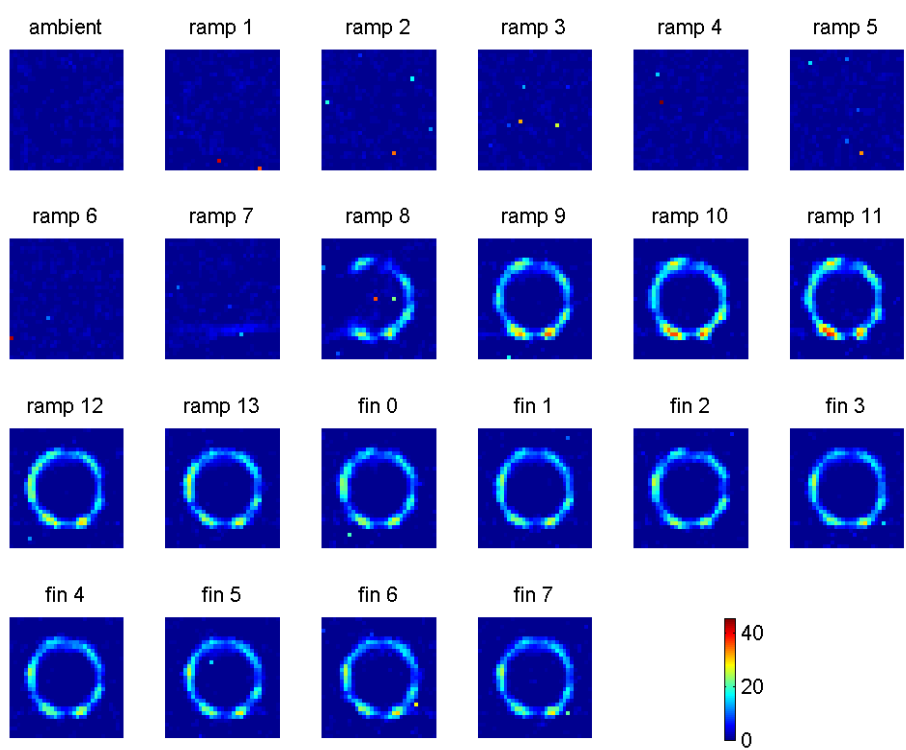

\CIG {2p10395-Par3}
\caption[Gaussian area of a molybdenum compound peak.]%
{Integrated intensity of a molybdenum compound peak which is the \Bragg\ reflection at
$2\SA=2.10^{\circ}$ modelled by a \Gauss ian profile. \label{Ex32p10395-Par3}}
\end{figure}

\begin{figure}
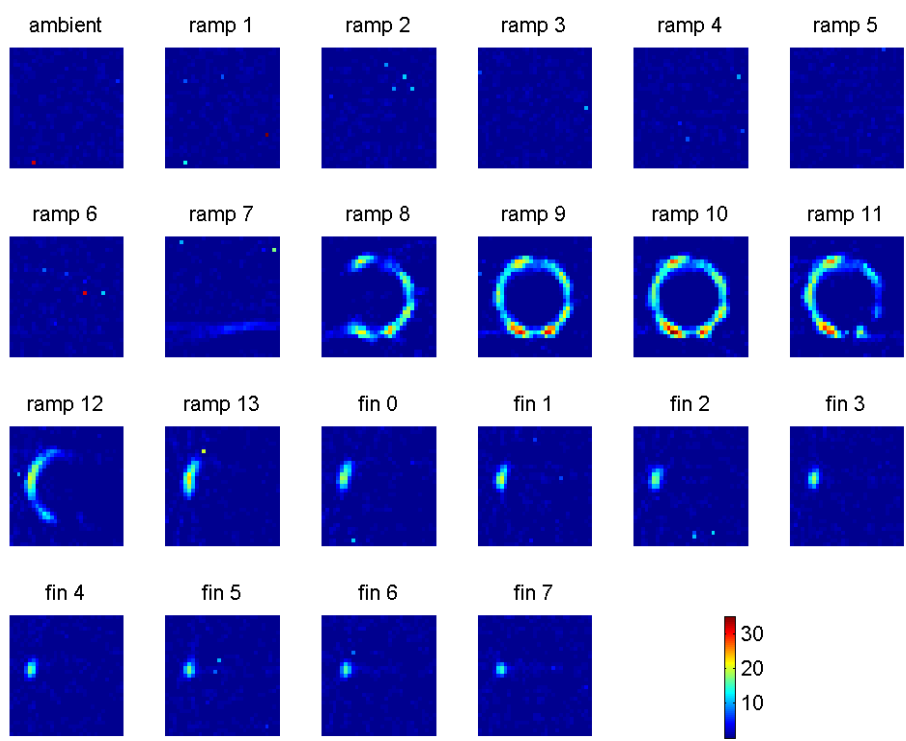

\CIG {2p46046-Par3}
\caption[Gaussian area of another peak of a molybdenum compound.]%
{Integrated intensity of another peak of a molybdenum compound which is the \Bragg\ reflection at
$2\SA=2.50^{\circ}$ modelled by a \Gauss ian profile. \label{2p46046-Par3}}
\end{figure}

\begin{figure}
\centering \includegraphics [scale=0.9, trim = 0 0 0 0] {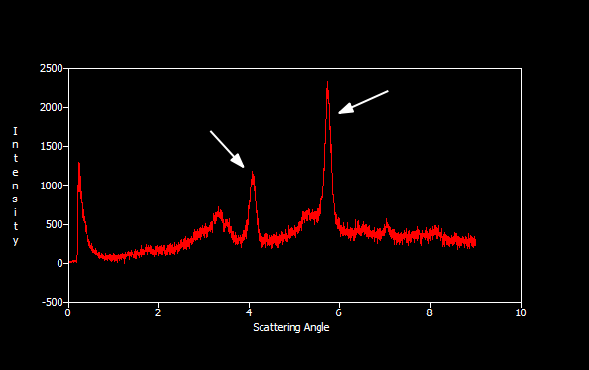}
\caption[The most prominent alumina peaks.]%
{The most prominent \alumina\ peaks in a plot of intensity (in number of counts) versus \scaang\
(in degrees). \label{Ex3Image1}}
\end{figure}

\begin{figure}
\centering \includegraphics [scale=0.9, trim = 0 0 0 0] {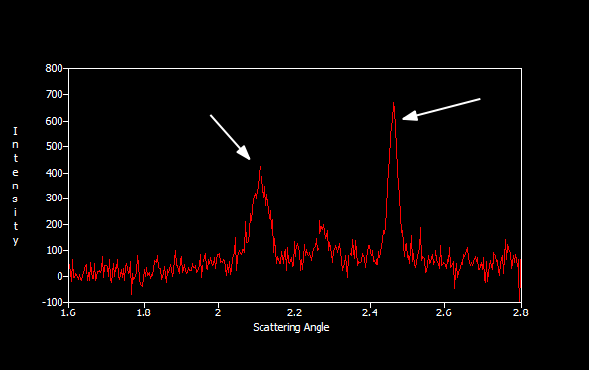}
\caption[The most prominent molybdenum compounds peaks.]%
{The most prominent peaks of molybdenum compounds in a plot of intensity (in number of counts)
versus \scaang\ (in degrees). \label{Ex3Image2}}
\end{figure}

Figures \ref{5p76706-Par3}, \ref{4p08027-Par3}, \ref{Ex32p10395-Par3}, and \ref{2p46046-Par3}
present some of the results obtained from this analysis. Each figure contains 22 colour coded
tomographic images (corresponding to the 22 data sets) of spatial distribution of phases. The
displayed parameter in these colour coded images is the integrated intensity (i.e. \Gauss ian area)
of these peaks. These figures display 4 prominent peaks (2 for \alumina\ and 2 for molybdenum
compounds) in the spectra of these data sets. These 4 peaks are also presented graphically in plots
of intensity versus \scaang\ in Figures \ref{Ex3Image1}, and \ref{Ex3Image2}. In each one of the
tomographic figures the data sets are labelled according to the temperature stage, that is ambient,
ramp and final (abbreviated by `fin') which represents a final stage where the temperature reached
a constant \stesta\ value. Various distinct crystallographic entities with different distributions
can be seen to form during the calcination process. The molybdenum compound images (Figures
\ref{Ex32p10395-Par3} and \ref{2p46046-Par3}) reveal that the phases are distributed unevenly and
only on one side of the sample at some stages. The distribution of molybdenum compound phases is
consistent with an egg-shell configuration.

\begin{figure}
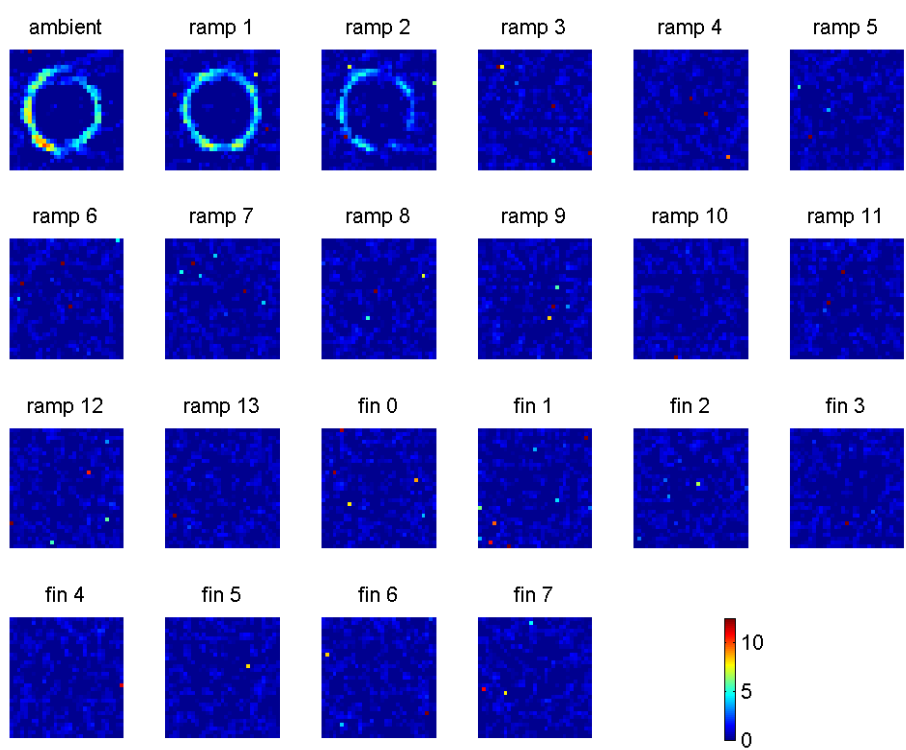

\CIG {0p771782-Par3}
\caption[Disappearance of precursor during calcination.]%
{Disappearance of precursor during calcination as demonstrated by the Gaussian integrated intensity
of the peak at $2\SA=0.771782^{\circ}$. \label{Ex30p771782-Par3}}
\end{figure}

\begin{figure}
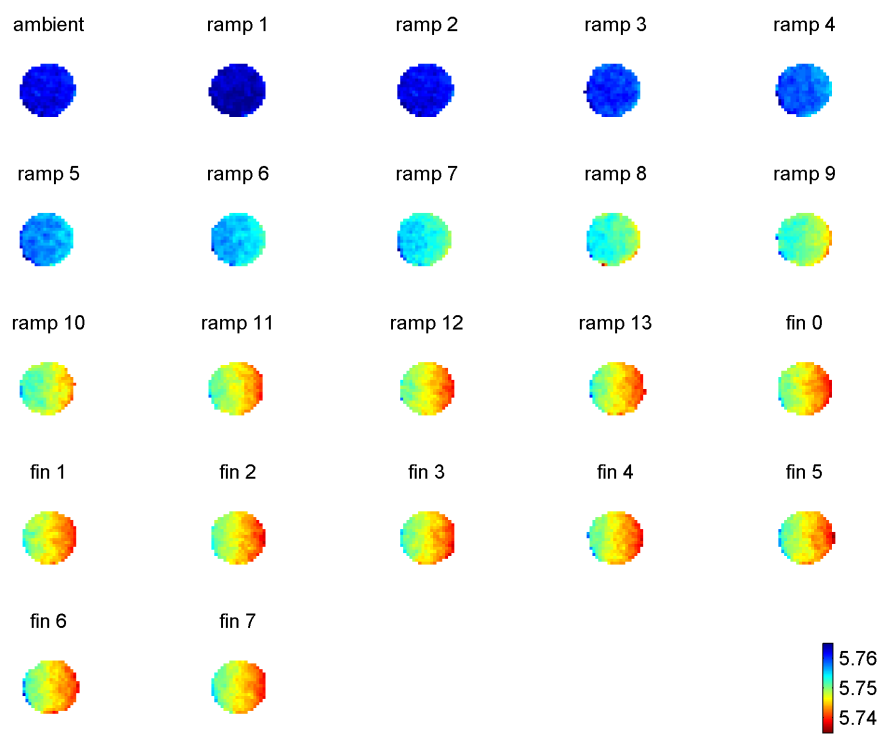

\CIG {AluminaPositionChange}
\caption[Change of position of the largest alumina peak.]%
{Change of position of the largest alumina peak at $2\SA=5.77^{\circ}$ with increasing temperature.
The stability in the last stages indicates a \stesta. \label{AluminaPositionChange}}
\end{figure}

In Figure \ref{Ex30p771782-Par3} the disappearance of a precursor phase during calcination is
shown. Figure \ref{AluminaPositionChange}, which pertains to the largest alumina peak, demonstrates
the observed position shift in the peaks due to temperature change. The parameter displayed in this
figure is the position of the Gaussian fitted peak. This figure shows maps of the peak centre of
the fitted alumina peak as a function of data set and hence temperature. From the colour bar, it
can be seen that the peak centre is moving towards lower scattering angle as the temperature
increases. This obviously corresponds to an increase in the $d$-spacing, consistent with unit cell
expansion that is expected with increasing temperature. In this figure the background is masked and
the colour code range is narrowed to accentuate the position shift feature. The figure also reveals
a thermal gradient across the sample (increasing from left to right) which is due to uneven heating
because of positioning of the heating element on one side of the sample.

\begin{figure} [!h]
\centering
\includegraphics
[scale=2.1] {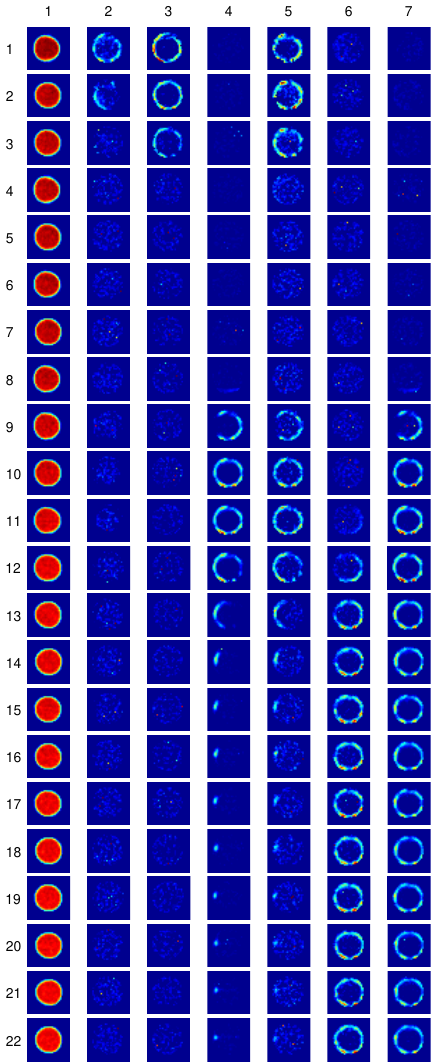}
\caption[Phase distribution patterns of molybdenum data.]%
{Phase distribution patterns of molybdenum data. The rows and columns are labelled with their
indices and the background is masked.} \label{MoPDPsMask}
\end{figure}

To investigate the phases that exist in the molybdenum experiment sample systematically, we
produced Phase Distribution Patterns (PDP), seen as columns in Figure \ref{MoPDPsMask}. The idea of
PDPs is to classify the phases in their spatial distribution in the slices as presented in the 2D
colour-coded tomographic images. The general assumption is that each PDP is a finger print of a
single crystallographic entity, although this assumption may not be valid in general as some phases
may be present in more than one PDP. By inspecting the PDPs and comparing them to the peak
positions in the stack plots, the phases can be identified if standard diffraction pattern data,
such as those obtained from Inorganic Crystal Structure Database (ICSD) \cite{ICSD}, are available.

Several phase distribution patterns are observed in the molybdenum data sets; the main ones are
presented in Figure \ref{MoPDPsMask}. The tomographic images in this figure are obtained from
mapping the integrated intensity of the peaks. Each PDP was observed from a number of peaks with
specific scattering angles. In the following we discuss these phase distribution patterns, as
represented by the columns of Figure \ref{MoPDPsMask} from left-to-right order, trying to link them
to specific crystallographic phases. In all cases, we used standard diffraction patterns obtained
from the ICSD database in phase identification:

\begin{enumerate}

\item
The first PDP (first column from left) obviously comes from alumina. The following peaks of
$\gamma$-alumina \cite{PagliaBRHHe2003} were identified with this PDP: (\sad, $hkl$, $d$ \AA) =
(3.36 , 211, 2.4066), (3.52, 202, 2.2955), (4.05, 220, 1.9983), (5.26, 321, 1.5374), (5.77, 224,
1.4021), (6.49, 116, 1.2464), (7.05, 404, 1.1477), (7.79, 503, 1.0381), (8.10, 440, 0.9991) and
(8.71, 208, 0.9292). One feature of this PDP is the gradual drop in the intensity with time due
apparently to decay in the storage ring current and possibly to transformation of $\gamma$-alumina
to another form. This feature was more evident in some peaks than others due to the relative
intensity of the peaks and possible contamination with neighbouring peaks. A similar PDP to this
one was observed at about \sad\ = 0.23 on a truncated peak due to limits on the measuring device.
This peak, according to the available diffraction patterns, does not belong to $\gamma$-alumina.
However it may originate from another form of alumina. This possibility is supported by the fact
that the spatial distribution of this phase is similar to that of $\gamma$-alumina. Moreover, its
intensity appears to be increasing with time while $\gamma$-alumina peaks are decreasing. This may
suggest possible transformation from one form of alumina to another.

\item
The second and third PDPs belong to the precursor, i.e. ammonium heptamolybdate tetrahydrate
(NH$_{4}$)$_{6}$Mo$_{7}$O$_{24}$.4H$_{2}$O. The following peaks of this compound with space group
P121/C1 \cite{EvansGL1975} were identified with these PDPs: (\sad, $hkl$, $d$ \AA) = (1.07, 100,
7.5466), (1.09, 110, 7.3876), (1.13, 12$\overline{1}$, 7.1472), (1.37, 14$\overline{1}$, 5.8978),
(1.40, 140, 5.7940), (1.53, 15$\overline{1}$, 5.2979), (1.55, 150, 5.2223), (1.59, 061, 5.0769) and
(2.61, 023, 3.0921). This PDP was also observed at \sad\ = 0.79 but the available standard
diffraction patterns of precursor do not include a peak at this position due possibly to
experimental restrictions. It should be remarked that the second PDP could be an anhydrated form of
the precursor which lost some of its water molecules during the preparation and drying process, as
its early disappearance may suggest. However, this cannot be confirmed as diffraction patterns for
these anhydrates are not available.

\item
The fourth PDP comes from \moltri\ \MoO. The following peaks of \MoO-[PBNM] \cite{Wooster1931} were
identified with this PDP: (\sad, $hkl$, $d$ \AA) = (1.16, 020, 6.9700), (2.32, 040, 3.4850), (2.50,
02$\overline{1}$, 3.2404), (3.02, 101, 2.6752), (4.61, 1$\overline{6}$1, 1.7542), (4.91,
11$\overline{2}$, 1.6466), (5.51, 251, 1.4687) and (6.66, 290, 1.2152). This phase has an egg-shell
distribution. As seen, the formation and consumption of \MoO\ on the left and the right sides of
the sample follow different time evolution routes. This is obviously due to the uneven heating
(which is discussed earlier) as the heating element is positioned on one side of the sample.
Therefore, the formation and consumption of \MoO\ at the hotter (right) side is faster than that on
the other side. In fact, a trace of \MoO\ can be seen intact on the left side of the cylinder up to
the end of the experiment. The formation of \MoO\ seems to follow the following route as suggested
by Sharma and Batra \cite{SharmaB1988}:

\hspace{2cm} ((NH$_4$)$_2$O)$_3$7MoO$_3$.2H$_2$O  $\rightarrow$ ((NH$_4$)$_2$O)$_{2.4}$7MoO$_3$ %

\hspace{3.5cm} $\rightarrow$ ((NH$_4$)$_2$O)$_{1.7}$7MoO$_3$ $\rightarrow$ 7MoO$_3$ %

In this case, the loss of two water molecules from the precursor should have happened during the
drying process where the sample reached about 120$^\circ$ or during the early stages of calcination
process.

\item
The fifth PDP is a combination from the precursor and \MoO. It is identified from a single peak at
about \sad\ = 1.14 which is an overlapping peak from the \sad\ = 1.13 from precursor and \sad\ =
1.16 from \MoO.

\item
The sixth PDP belongs to aluminium molybdate, Al$_2$(MoO$_4$)$_3$. The following peaks of
Al$_2$(MoO$_4$)$_3$-[PBCN] \cite{Harrison1995} were identified with this PDP: (\sad, $hkl$, $d$
\AA) = (1.29, 200, 6.2773), (1.42, 111, 5.7075), (1.89, 102, 4.2792), (2.00, 021, 4.0418), (2.19,
202, 3.6850), (2.31, 311, 3.5039), (2.37, 212, 3.4113), (2.52, 022, 3.2040), (2.76, 130, 2.9244),
(2.83, 222, 2.8537), (2.87, 411, 2.8187), (3.11, 231, 2.5991), (3.72, 114, 2.1734), (4.99, 252,
1.6205), (5.23, 135, 1.5456), (6.01, 136, 1.3468) and (6.33, 561, 1.2772). Most of these are the
strongest peaks in their neighbourhood. On comparing the fourth and sixth PDPs, where the
disappearance of \MoO\ is followed by the appearance of Al$_2$(MoO$_4$)$_3$, it can be concluded
that the formation of Al$_2$(MoO$_4$)$_3$ occurs through the following chemical reaction

\hspace{4cm}  \AltOt\ + 3\MoO\ $\rightarrow$ Al$_2$(MoO$_4$)$_3$

This appearance-disappearance correlation is supported by the fact that in the position of the
trace left of the \MoO\ on the left side of the ring there is a gap on the ring of the
Al$_2$(MoO$_4$)$_3$. Moreover, the intensity of the \MoO\ and Al$_2$(MoO$_4$)$_3$ rings in various
parts are strongly correlated. The egg-shell distribution of Al$_2$(MoO$_4$)$_3$ is obviously due
to the egg-shell distribution of \MoO.

\item
The seventh PDP is a combination of the fourth and sixth PDPs. It is identified from peaks at the
following approximate positions: \sad\ = 2.09, 2.13, 3.43, 4.38, 4.73, 4.83 and 5.04. All these
positions have strong peaks from both \MoO\ and Al$_2$(MoO$_4$)$_3$ in their immediate
neighbourhood.

\end{enumerate}

Several other crystallographic structures, which include molybdenum and a number of its oxides in
various space groups, were inspected but could not be traced. It should be remarked that some peaks
of the standard diffraction patterns were not observed in our patterns of the corresponding phases.
The absence of these peaks can be explained by low intensity of those peaks, or masking by other
prominent peaks from the same phase, or another phase, or preferred orientation effects which are
known for example to occur with MoO$_3$. The coexistence of several phases in the experimental
sample complicates the overall diffraction pattern and makes the analysis and identification more
difficult.

In the following we present general conclusions that can be drawn from the analysis of these data
sets:

\begin{itemize}

\item
The formation of various material phases can be tracked during the ramp stages, as can be seen for
example in Figures \ref{MolStack} and \ref{Ex32p10395-Par3}. The entire synthesis of active
catalysis takes place in an egg-shell distribution region.

\item
The initial precursor, (NH$_{4}$)$_{6}$Mo$_{7}$O$_{24}$.4H$_{2}$O, exists only in the first three
frames at low temperatures between $T \simeq 25^\circ$C to $T \simeq 125^\circ$C.

\item
As can be seen in several figures (e.g. \ref{Ex32p10395-Par3} and \ref{AluminaPositionChange}), the
situation is hardly changing during the \stesta\ frames, which is a sign of consistency.

\item
In this experiment the transformation of \MoO\ to Al$_2$(MoO$_4$)$_3$ was observed.

\item
MoO$_3$ starts to form around frame nine at $T \simeq 425^\circ$C, while Al$_2$(MoO$_4$)$_3$ starts
to form around frame twelve at $T \simeq 575^\circ$C.

\item
As indicated already, the sample is not evenly heated as one side can be seen cooler than the other
(see Figure \ref{AluminaPositionChange}) due to the location of the heating element (hot air
blower). Also, from Figure \ref{Ex3IntensityS} it is evident that there is an alignment problem in
some data sets. These problems should be addressed in the forthcoming experiments.

\end{itemize}

%

%% file: Nick.tex
%
%
\chapter{Preparation of Nickel Nanoparticle Catalysts} \label{Nick}

In this chapter we present three data collections of nickel compounds which we have processed and
analysed in a manner similar to the molybdenum data. The samples and data collection methods of
these data are similar to those described in the molybdenum section (see \S\ \ref{Ex3}). The nickel
data are part of a larger data collection which consists of about 254 thousand EDF image files in
179 data sets (sinograms) with a total size of about 2.45 terabytes. The data were collected at the
ESRF beamline ID15B using the \CATs\ type \ADDs\ technique. The wavelength of the monochromatic
beam used in these measurements was $\lambda=0.14272$ \AA. The experimental setting and data
collection were performed by S. Jacques and coworkers, while data processing, analysis and
visualisation were carried out by the author. EasyDD was used in multi-batch mode to squeeze the
data and convert the images to 1D spectral patterns in ASCII numeric format. It was also used to
align, back project, and curve-fit the peaks in the nickel data. A Gaussian profile with linear
background was used for curve-fitting the back-projected patterns. Most peaks were fitted as
singlets while the remainder were fitted as doublets. Each collected data set consists of a
sinogram representing a slice in the cylindrical extrudate sample with 33 rotational and 43
translational steps, while each back-projected data set consists of 1849 (43$\times$43) patterns.
The data sets in each data collection represent temporal stages in dynamic phase transformation
under thermal or chemical treatment.

The nickel nitrate extrudate sample is similar to the molybdenum sample but nickel ethylenediamine
dinitrate, \Nieddn, was used as a precursor. The sample was subjected to a calcination process
under heat treatment, followed by reduction under a flow of nitrogen to avoid contact with oxygen.
These data therefore consist of two collections: calcination collection which contains 43 data
sets, and reduction collection which contains 12 data sets. Similarly, the nickel chloride sample
consists of nickel chloride ethylenediamine tetrahydrate, NiCl$_2$(en)(H$_2$O)$_4$, as a precursor
on a $\gamma$-alumina extrudate cylinder. This collection contains 23 data sets. The heat treatment
of the nickel chloride sample consists of ramp increasing temperature from 25$^\circ$C at the first
frame to 500$^\circ$C at the 20th frame (i.e. 25$^\circ$C increase per frame) followed by steady
state temperature of 500$^\circ$C at the last three frames.

In the following sections we discuss the results of the three nickel data collections focusing on
the stack plots and PDP patterns as the main guide for the analysis and phase identification. In
all cases, standard diffraction patterns obtained from the ICSD database are used in phase
identification unless it is stated otherwise. It should be remarked that the given scattering
angles of peaks in phase identification are approximate values as the standard diffraction patterns
of a particular phase differ from one reference to another.

\section{Nickel Nitrate Calcination} \label{NiNitCal}

\begin{figure} [!h]
\centering
\includegraphics
[scale=0.85] {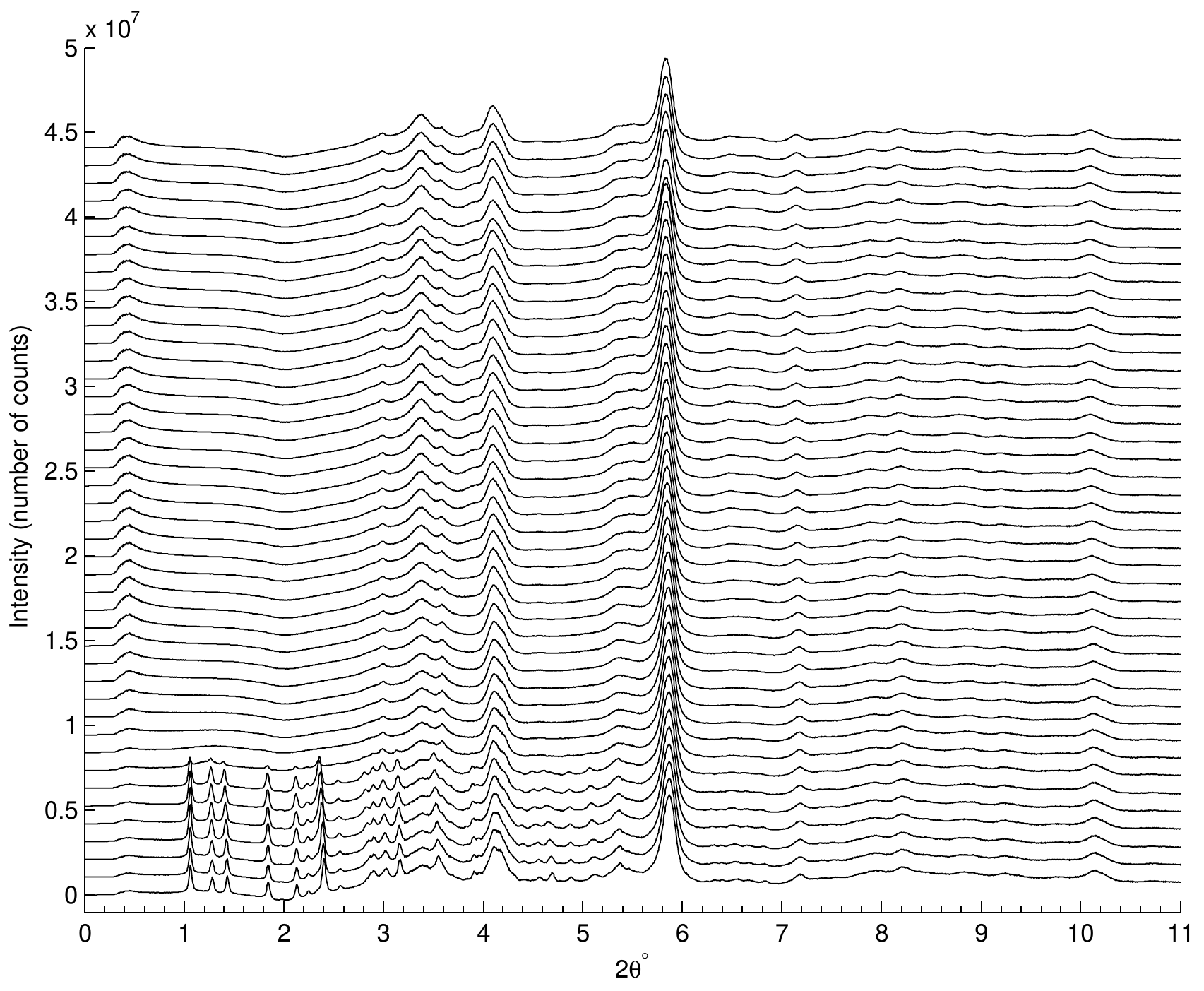}
\caption[Stack plot of the 43 data sets of nickel nitrate calcination.]%
{Sums of back-projected diffraction patterns of the 43 data sets of nickel nitrate calcination
presented in a stack plot.} \label{NiNitCal2ndStackBP1}
\end{figure}

\begin{figure} [!h]
\centering
\includegraphics
[scale=2.1] {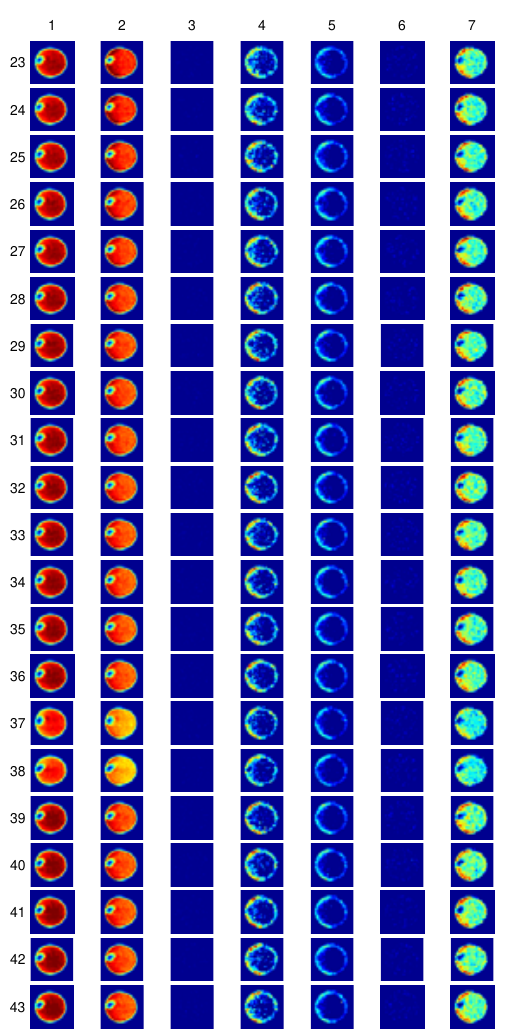}
\caption[Phase distribution patterns of nickel nitrate calcination data.]%
{Phase distribution patterns of the nickel nitrate calcination data. The rows and columns are
labelled with their indices and the background is masked.} \label{NiNitCalPDPsMask2}
\end{figure}

The stack plot of the back-projected data for the nickel nitrate calcination experiment is shown in
Figure \ref{NiNitCal2ndStackBP1}, while the PDP diagram is given in Figure \ref{NiNitCalPDPsMask2}.
Because of its size, the last figure is split into two segments on two pages (pages
\pageref{NiNitCalPDPsMask1} and \pageref{NiNitCalPDPsMask2}). A number of phase distribution
patterns are observed in these data; the main ones can be seen in Figure \ref{NiNitCalPDPsMask2}. A
prominent feature of this figure is the rotation of the sample during the early part of the
calcination process as indicated by the hole position near the cylinder perimeter. This is
obviously due to an unstable mechanical setting which allowed the sample to spin during the scan.
In the following we discuss the main PDPs and try to link them to actual phases:

\begin{enumerate}

\item
The first PDP (first column from left) belongs to alumina (\AltOt). The following peaks of
$\gamma$-alumina \cite{PagliaBRHHe2003} were identified with this PDP: (\sad, $hkl$, $d$ \AA) =
(2.92, 112, 2.8042), (3.40, 211, 2.4066), (3.56, 202, 2.2955), (4.09, 220, 1.9983), (4.16, 004,
1.9677), (5.32, 321, 1.5374), (5.79, 400, 1.4130), (5.83, 224, 1.4021), (6.47, 420, 1.2638), (6.73,
413, 1.2150), (7.13, 404, 1.1477), (7.92, 415, 1.0338), (8.19, 440, 0.9991), (8.76, 336, 0.9347),
(8.87, 611, 0.9228), (9.19, 444, 0.8909), (10.06, 624, 0.8137) and (10.14, 408, 0.8074). It should
be remarked that the fading intensity seen in frames (rows) 37 and 38 is most likely to be from a
sudden and temporary drop in the beam intensity or from an error in the detector, and hence should
not be ascribed to a dynamic transformation in the sample. This view is supported by the fact that
this sudden drop in intensity is observed with the other PDPs of the other phases.

\item
The second PDP displays a spatial distribution similar to that of $\gamma$-alumina but with an
obvious increase in intensity. A peak at about \sad\ = 0.42 with this PDP was observed but could
not be identified as arising from $\gamma$-alumina. This peak is highly asymmetric and apparently
consists of two overlapping peaks. Most probably, this peak originates from another form of
alumina. The observed intensity decrease of $\gamma$-alumina's most prominent peak, at about \sad\
= 5.83, with the observed increase in this unknown peak may suggest that $\gamma$-alumina is
transforming to another form of alumina. It is noteworthy that the gradual drop in the beam
intensity was tested against the gradual decrease in the storage ring current, by comparing the
total intensity of the first and last data sets of all data collections (i.e. molybdenum, nickel
nitrate calcination, nickel nitrate reduction, and nickel chloride). Although this drop was
confirmed in all cases, it cannot alone explain the tangible drop in the $\gamma$-alumina
intensity, since such a big drop in intensity is not observed with the other phases. It should be
remarked that a large number of standard patterns of alumina were inspected, with various space
groups, as found on the ICSD database, but no structure could be found with a peak at such a low
scattering angle. However, one exception is that of Repelin and Husson \cite{RepelinH1990} which
has two peaks at \sad\ = 0.35 and 0.69. This may be a possibility for the phase that is responsible
for this peak which seems to comprise two overlapping peaks.

\item
The third PDP is observed at \sad\ = 1.05, 1.25, 1.41, 1.82, 1.96, 2.12, 2.24, 2.38, 2.55, 2.82,
2.85, 2.88, 3.00, 3.15, 3.52, 3.88, 3.93, 4.37, 4.53, 4.65, 4.84 and 5.07. All these peaks are
identified as originating from the nickel ethylenediamine dinitrate, \Nieddn, precursor according
to the diffraction pattern of this compound which was obtained from Krishnan \etal\
\cite{KrishnanJS2009}, after re-scaling the \scaang s to match the wavelength; no data about
$hkl$'s and $d$-spacings are available for this pattern. This PDP was also observed at higher
scattering angles but these angles are beyond the range of the diffraction pattern of Krishnan
\etal. This PDP suggests that the precursor is consumed at the beginning of the calcination
process. It should be remarked that there is an obvious shift of the precursor peak positions
towards lower \scaang s as the temperature increases. This is consistent with an increase in the
$d$-spacings as the temperature rises. The position shift feature is also observed in other phases
such as alumina. Another remark is that the relative intensity of the diffraction peaks suggests
that the precursor is damaged by radiation as demonstrated by Figure 4-b of Krishnan \etal.

\item
The fourth PDP is observed at about \sad\ = 5.52 and 6.47. Apparently, this phase is bunsenite,
NiO-[FM3-M], \cite{Taylor1984} as there are two strong peaks of this compound at these positions
with ($hkl$, $d$ \AA) = (220, 1.4829) and (311, 1.2647) respectively. Interestingly, some alumina
peaks which are close to expected peaks of NiO appear as if they are contaminated with this phase.
An example of this is shown in the last column of Figure \ref{NiNitCalPDPsMask2} where the
$\gamma$-alumina peak at \sad\ = 8.76 seems to be contaminated with the NiO peak at \sad\ = 8.73
($hkl$ = 420, $d$ \AA\ = 0.9379) which is a strong peak. Also, the alumina peak at \sad\ = 3.40
shows strong sign of contamination with the strong peak of NiO at \sad\ = 3.38 ($hkl$ = 111, $d$
\AA\ = 2.4216). Also the next PDP contains a peak from NiO. Apparently, the formation of bunsenite
is due to thermal decomposition of the precursor at high temperature. The missing link between the
two phases (i.e. frames 9-16 where no transitional phase is observed at this spatial distribution)
suggests that an amorphous form of the precursor or a transitional compound has participated in
this process. The more apparent egg-shell distribution of the NiO may suggest an oxidation process
where a molecular oxygen from air is involved in this reaction.

\item
The fifth PDP is observed at about \sad\ = 3.90. This PDP results from a combination of the last
two PDPs due to overlapping peaks from the two phases. It therefore consists of the (\sad, $hkl$,
$d$ \AA) = (3.90, 200, 2.0972) which is the strongest peak of NiO, and the precursor peak at \sad\
= 3.88 or 3.93.

\item
The sixth PDP is observed at about \sad\ = 2.86, 4.49 and 4.62. It could not be identified,
although it may be a transitional form of the precursor.

\end{enumerate}

It is noteworthy that no pure nickel metal can be traced, in either of the two forms for which
there exists data, i.e. Ni-[FM3-M] and Ni[P63/MMC]; similarly other potential nickel compounds were
considered but not found.

\clearpage \hspace{1.23cm} \includegraphics [scale=2.1] {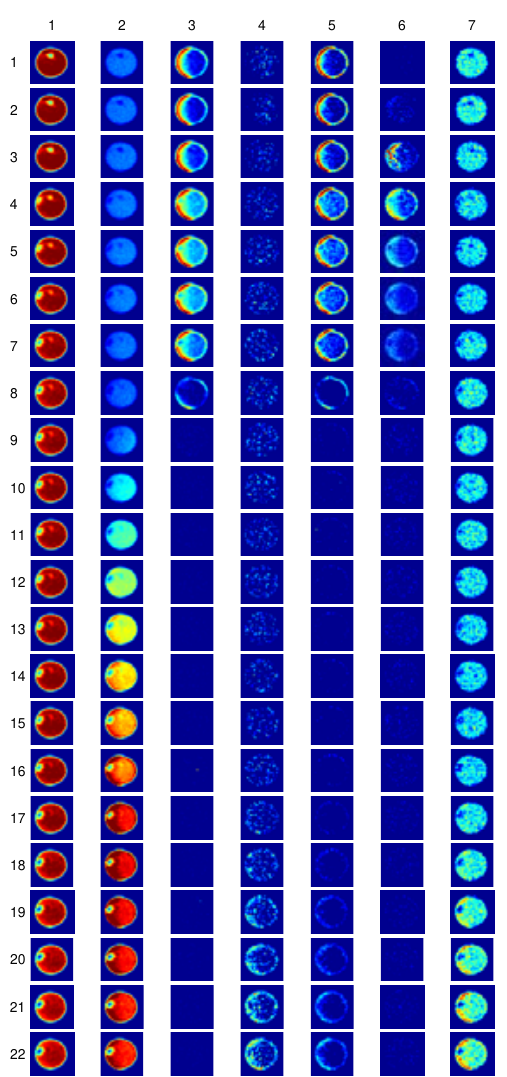}
\label{NiNitCalPDPsMask1}
\clearpage

\clearpage
\section{Nickel Nitrate Reduction} \label{NiNitRed}

\begin{figure} [!b]
\centering
\includegraphics
[scale=0.85] {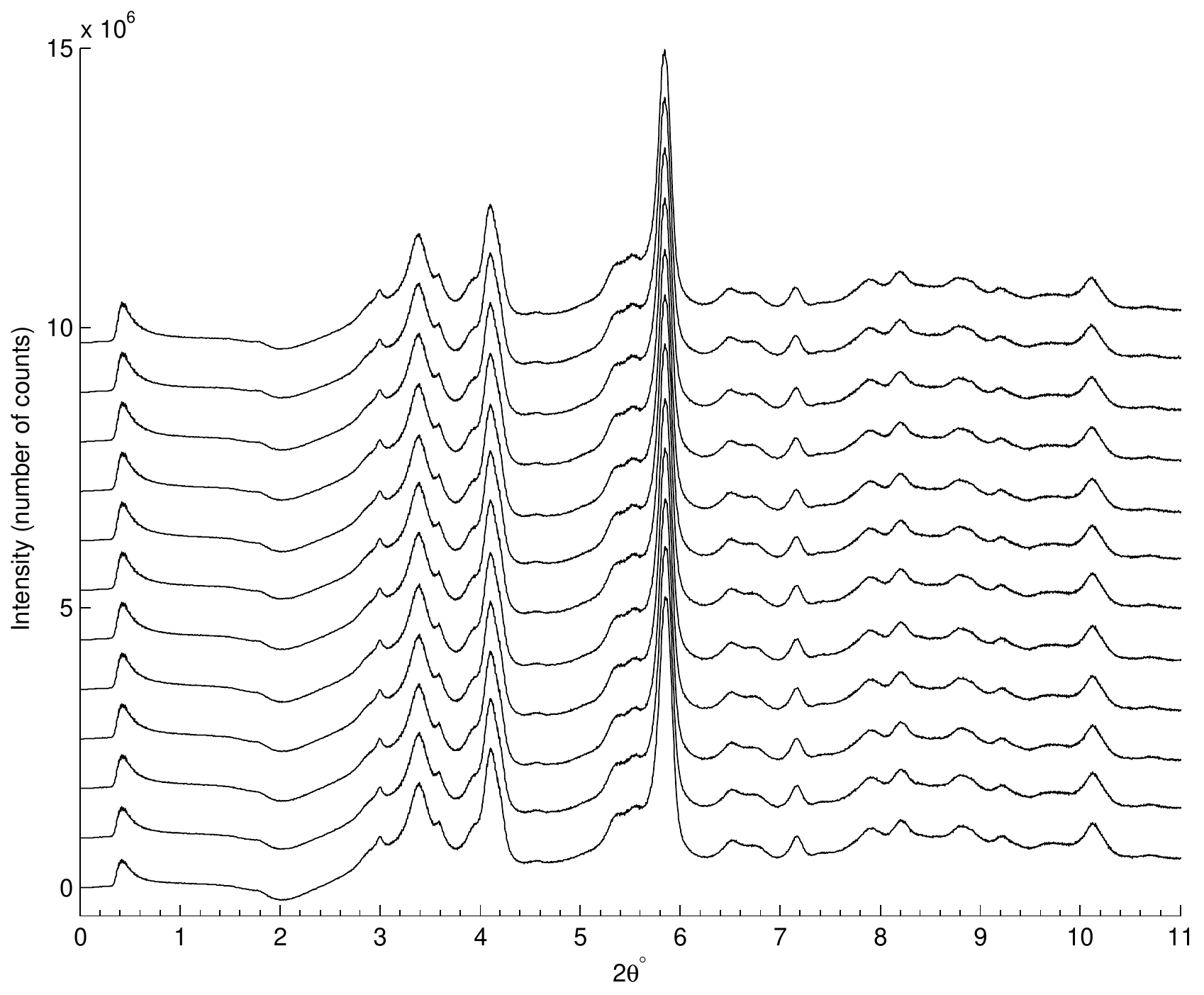}
\caption[Stack plot of the 12 data sets of nickel nitrate reduction.]%
{Sums of back-projected diffraction patterns for the 12 data sets of nickel nitrate reduction,
presented as a stack plot.} \label{NiNitRed2ndStackBP1}
\end{figure}

\begin{figure} [!h]
\centering
\includegraphics
[scale=2.0] {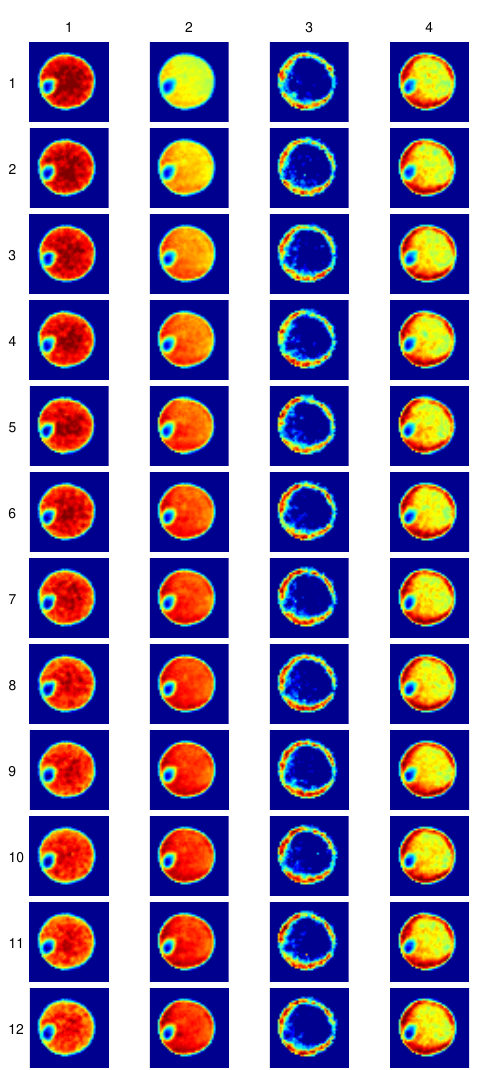}
\caption[Phase distribution patterns of nickel nitrate reduction data.]%
{Phase distribution patterns of nickel nitrate reduction data. The rows and columns are labelled
with their indices and the background is masked.} \label{NiNitRedPDPsMask}
\end{figure}

The stack plot of back-projected data for the nickel nitrate reduction experiment is shown in
Figure \ref{NiNitRed2ndStackBP1}. A number of phase distribution patterns are observed in this
collection; the main ones are displayed in Figure \ref{NiNitRedPDPsMask}. The prominent feature is
that in most cases hardly any changes occur during the reduction process. This includes the
positions of the peaks which are almost constant since the sample is held at constant steady state
temperature. In the following we discuss these PDPs and try to link them to crystallographic
phases:

\begin{enumerate}

\item
The first PDP (first column from left) originates from alumina. The following peaks of
$\gamma$-alumina \cite{PagliaBRHHe2003} were identified with this PDP: (\sad, $hkl$, $d$ \AA) =
(1.78, 101, 4.5910), (2.89, 200, 2.8260), (2.92, 112, 2.8042), (3.40, 211, 2.4066), (3.44, 103,
2.3798), (3.56, 202, 2.2955), (4.09, 220, 1.9983), (4.16, 004, 1.9677), (4.49, 213, 1.8203), (5.03,
312, 1.6274), (5.32, 321, 1.5374), (5.79, 400, 1.4130), (5.83, 224, 1.4021), (6.47, 420, 1.2638),
(6.73, 413, 1.2150), (7.13, 404, 1.1477), (7.42, 107, 1.1028), (7.92, 415, 1.0338), (8.19, 440,
0.9991), (8.76, 336, 0.9347), (8.87, 611, 0.9228), (9.19, 444, 0.8909), (9.64, 604, 0.8497),
(10.06, 624, 0.8137), (10.14, 408, 0.8074) and (10.72, 329, 0.7637).

\item
The second PDP, which is observed at about \sad\ = 0.42, is similar to the second PDP of the
calcination data, and apparently consists of two peaks. As its intensity is increasing while some
of alumina peaks appear to be decreasing, as seen in the first PDP, it is likely to be another form
of alumina transforming from $\gamma$-alumina.

\item
This egg-shell phase distribution pattern is observed at \sad\ = 3.90, 5.52 and 6.47. Similar to
the calcination data, it seems to be originating from bunsenite NiO-[FM3-M] \cite{Taylor1984} with
the following values ($hkl$, $d$ \AA) = (200, 2.0972), (220, 1.4829) and (311, 1.2647)
respectively. Also some alumina peaks seem to be contaminated with peaks expected from NiO at those
positions. An example is presented in the last column where the $\gamma$-alumina peak at \sad\ =
3.40 seems to be contaminated with the second strongest peak of NiO at \sad\ = 3.38 ($hkl$ = 111,
$d$ \AA\ = 2.4216). There is also a strong sign of contamination of alumina peaks at \sad\ = 6.73
and 8.76 with the NiO peaks at \sad\ = 6.76 and 8.73, respectively. The last peaks have ($hkl$, $d$
\AA) = (222, 1.2108) and (420, 0.9379) respectively.

\end{enumerate}

\clearpage
\section{Nickel Chloride} \label{NiChloride}

\begin{figure} [!b]
\centering
\includegraphics
[scale=0.8] {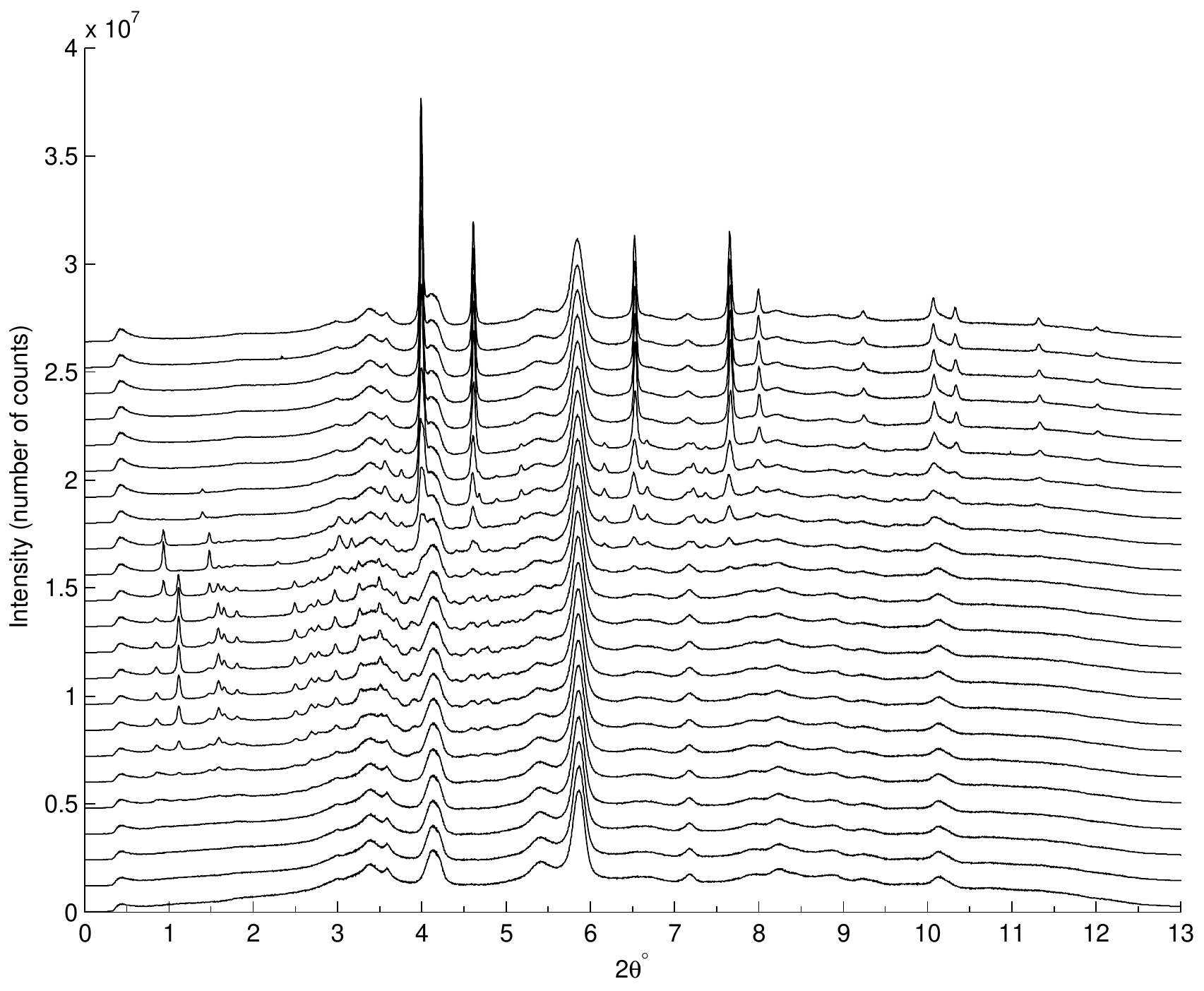}
\caption[Stack plot of the 23 data sets of nickel chloride.]%
{Sums of back-projected diffraction patterns of the 23 data sets of nickel chloride presented in a
stack plot.} \label{NiCl2ndStackBP1}
\end{figure}

\begin{figure} [!h]
\centering
\includegraphics
[scale=1.2] {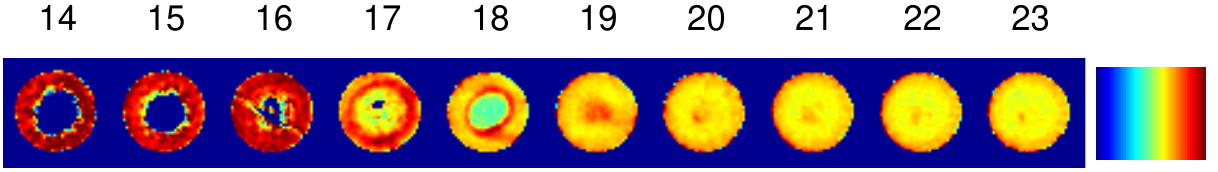}
\caption[Crystallite size distribution of FCC nickel.]%
{Crystallite size distribution of FCC nickel by mapping the FWHM of the (220) reflection at $2\SA =
6.51^\circ$. The colour bar ranges between 0$^\circ$ (blue) and 0.065$^\circ$ (red). The blue maps
the regions where no FCC nickel peaks exist and hence the FCC phase is absent, while the red
represents the regions of smallest crystal size and the yellow represents the regions of largest
crystal size. The numbers shown on the top are the indices of the data sets.} \label{fwhm}
\end{figure}

\begin{figure} [!h]
\centering
\includegraphics
[scale=2.1] {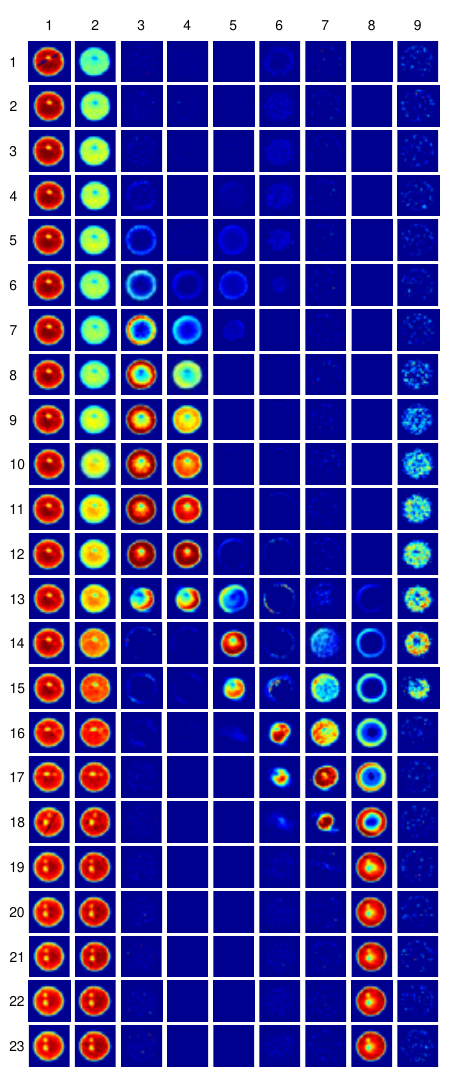}
\caption[Phase distribution patterns of nickel chloride data.]%
{Phase distribution patterns of nickel chloride data. The rows and columns are labelled with their
indices and the background is masked.} \label{NiCl2ndMask1}
\end{figure}

The stack plot of the back-projected data for the nickel chloride experiment is shown in Figure
\ref{NiCl2ndStackBP1}. A number of very interesting phase distribution patterns are observed in the
data with the main candidates displayed in Figure \ref{NiCl2ndMask1}. In the following we discuss
the main PDPs and attempt to link them to actual phases, where the identifications are made jointly
by Simon Jacques and the author of this thesis:

\begin{enumerate}

\item
The first PDP (first column from left) belongs to alumina. The following peaks of $\gamma$-alumina
\cite{PagliaBRHHe2003} were identified with this PDP: (\sad, $hkl$, $d$ \AA) = (1.78, 101, 4.5910),
(2.92, 112, 2.8042), (3.40, 211, 2.4066), (3.56, 202, 2.2955), (4.09, 220, 1.9983), (5.39, 105,
1.5165), (5.83, 224, 1.4021), (7.13, 404, 1.1477), (7.92, 415, 1.0338), (8.19, 440, 0.9991), (8.87,
611, 0.9228), (9.19, 444, 0.8909) and (10.14, 408, 0.8074). Other alumina peaks overlap with peaks
from other phases and hence are masked. Some of these masked peaks have a shadowy trace on the
image of the masking peaks.

\item
The second PDP is observed at about \sad\ = 0.42. As already stated, this seems to be another form
of alumina transforming from $\gamma$-alumina since the increase in its intensity is associated
with a decrease of the $\gamma$-alumina intensity (see Figure \ref{NiCl2ndMask1}). This possibility
is supported by the fact that the spatial distribution of this phase is identical to that of
$\gamma$-alumina.

\item
The third PDP is originating from the nickel chloride ethylenediamine tetrahydrate,
NiCl$_2$(en)(H$_2$O)$_4$, precursor. Interestingly, the polycrystalline form of precursor is absent
in the first few frames as it forms during the heat treatment. This PDP is observed at about \sad\
= 0.85, 1.58, 1.80, 2.49, 2.67, 2.77 and 4.75.

\item
The fourth PDP is observed at about \sad\ = 1.11, 1.58, 3.25, 3.49, 3.69 and 3.89. It seems to be
related to the previous phase. There is also a sign of contamination with this phase at about \sad\
= 3.58, but it is difficult to verify as the position is surrounded by other peaks.

\item
The fifth PDP is Ni$_2$Cl$_2$(en). This PDP is observed at about \sad\ = 0.93, 1.48, 2.29, 2.89,
3.02 and 3.16. The shadow seen at frames 4-7 is an artifact resulting from a neighbouring peak
which is not related to this phase.

\item
The sixth PDP is NiCl$_2$. The following three peaks of NiCl$_2$-[R3-MH] \cite{FerrariBB1963} are
identified with this PDP: (\sad, $hkl$, $d$ \AA) = (1.41, 003, 5.8000), (4.70, 110, 1.7415) and
(4.90, 113, 1.6679). Other peaks are either too faint to observe or they are masked by the presence
of larger peaks from other phases in their position.

\item
The seventh PDP is hexagonal close packed (HCP) nickel. The following peaks of Ni-[P63/MMC]
\cite{HemengerW1965} are identified with this PDP: (\sad, $hkl$, $d$ \AA) = (3.79, 002, 2.1605),
(5.23, 102, 1.5652), (6.24, 110, 1.3110), (6.73, 103, 1.2163), (7.21, 200, 1.1354), (7.45, 201,
1.0981), (9.54, 210, 0.8583) and (9.73, 211, 0.8418). The strongest peak of Ni-[P63/MMC] at \sad\ =
4.07 is masked by the strong $\gamma$-alumina peak at \sad\ = 4.09.

\item
The eighth PDP is face centred cubic (FCC) nickel. The following peaks of Ni-[FM3-M]
\cite{ZhangZHPL2009} are identified with this PDP: (\sad, $hkl$, $d$ \AA) = (3.99, 111, 2.0521),
(4.60, 200, 1.7771), (6.51, 220, 1.2566), (7.64, 311, 1.0717), (7.98, 222, 1.0260), (9.21, 400,
0.8886), (10.04, 331, 0.8154), (10.30, 420, 0.7948), (11.29, 422, 0.7255) and (11.98, 511, 0.6840).
In Figure \ref{fwhm} the full width at half maximum (FWHM) of the (\sad, $hkl$, $d$ \AA) = (6.51,
220, 1.2566) peak of this phase is tomographically mapped for the last ten frames (14-23) where the
phase is formed. According to the Scherrer equation the mean size of crystallites, $\tau$, is given
by
\begin{equation}\label{Scherrer}
    \tau = \frac{\kappa \WL}{\beta \cos \SA}
\end{equation}
where $\kappa$ is a shape factor which ranges between about $0.87-1.0$ depending on the shape of
crystallites, $\WL$ is the wavelength of the radiation, $\beta$ is the peak broadening at half
maximum due to the size of crystallites, and $\SA$ is the diffraction angle. As the actual
broadening involves instrumental as well as size factors, the total full width at half maximum,
$\FWHM_{t}$, is given by
\begin{equation}\label{FWHMeq}
\FWHM_{t}^2 = \FWHM_{i}^2 + \beta^2
\end{equation}
where $\FWHM_{i}$ is the instrumental broadening. In the nickel chloride experiment, $\WL =
0.14272$ \AA, $\FWHM_{i} \simeq 0.021^{\circ} \simeq 0.00037$ radian, $\SA \simeq 3.26^{\circ}$ for
the (220) peak of FCC nickel, and $\FWHM_{t}$ ranges between about $0.025^{\circ}-0.209^{\circ}$
($0.00044-0.00365$ radian). Inserting these values in Scherrer equation with $\kappa = 0.90$, which
is a typical value, we obtain a crystallite size range of about $3.5-54$ nanometre. Figure
\ref{fwhm} reveals that the FCC nickel starts forming at the perimeter of the cylinder with a
subsequent invasion to the inside region. The figure also reveals that the size of the crystallites
at the interior regions eventually grows to become larger than those near the surface of the
extrudate cylinder. It should be remarked that the larger crystallite size found in the interior is
due to the prior presence of NiCl$_2$ which promoted crystal growth. Another remark is that the
disappearance of HCP-nickel with the growth of FCC-nickel (as can be observed by comparing PDP 7 to
PDP 8) suggests that HCP-nickel is transforming to FCC-nickel which is the more stable form.

\item
The ninth PDP, which is observed at about \sad\ = 4.39, is suspected to belong to another phase
which could not be identified. A trace of this PDP was also observed at about \sad\ = 2.96, 4.94
and 5.06. These positions are surrounded by other peaks from other phases and hence cannot be
unambiguously identified.

\end{enumerate}

It is noteworthy that some images show clear sign of sample rotation at frame 18.

\vspace{1cm}

%% file: Prob.tex
\chapter[Problems and Solutions]{Some Problems and Solutions associated with Diffraction-Based Imaging} \label{Problems}

In this chapter a number of problems related to the use of diffraction in X-ray imaging are
discussed and some solutions are offered. The main focus will be on two data sets collected by
Olivier Lazzari and Simon Jacques using CAT-type \TEDDIs\ and \TADDIs\ techniques. Also, a third
data set collected by Conny Hanson, Matt Wilson and Simon Jacques [private communication], using a
simple radiographic technique, will be briefly discussed.

\section[Large Object Problem and Solution]{Large Object Problem and Multi TEDDI Solution} \label{LargeObject}

\begin{figure} [!h]
\centering
\includegraphics
[scale=0.6] {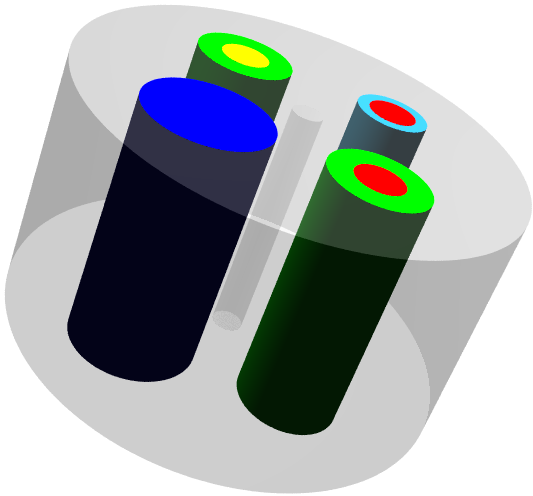}%
\caption[A 3D image of the test object.]{A 3D image of the test object. The colours refer: blue to
aluminium, green to PEEK, red to bone ash, yellow to iron oxide and sky blue to glass.}
\label{Sample3D}
\end{figure}

A key problem associated with the type of imaging employed in this thesis is that significant peak
broadening and splitting are observed when studying objects above a certain size. These effects
arise due to the different paths that X-rays take from sample to detector. Figure
\ref{SizeAngleEffect} illustrates how peak broadening and splitting can arise in a recorded
diffraction pattern. In this section we present two data sets collected by two different
diffraction modes: angle dispersive (ADD) and energy dispersive (EDD) where the sample used is a
large object. Both experiments use the same sample which is a specially designed test object
consisting of known materials. The purpose of these experiments is to investigate sample size
effects and assess the data acquisition and data analysis methods since the ingredients and
morphology of the object are well known. The test object, depicted schematically in Figure
\ref{Sample3D}, consists of a cylinder of wax ($r=18$ mm) with a hollow axis ($r=1.25$ mm). The
cylinder accommodates in each one of its quartets another cylinder with its axis parallel to the
cylindrical wax axis: an aluminium cylinder ($r=5$ mm), an iron oxide cylinder ($r=2$ mm) inside a
polyether ether ketone (PEEK) polymer container of external radius $r=4$ mm, a bone ash cylinder
($r=2$ mm) inside a glass container of external radius $r=3$ mm, and another bone ash cylinder
($r=2$ mm) inside a PEEK container of external radius $r=4$ mm. The details of this test object can
be found in Lazzari \etal\ \cite{LazzariJSB2009}. The purpose of using such a heterogeneous
multi-phase object is to have regions of different X-ray absorption and varying degrees of
crystallinity. This imposes a more rigorous test of the data acquisition and reconstruction
methods, as well as being a more faithful simulation of real life objects.

\clearpage
\subsection{ADD Data} \label{ADDData}

\begin{figure} [!h]
\centering
\includegraphics
[scale=0.2] {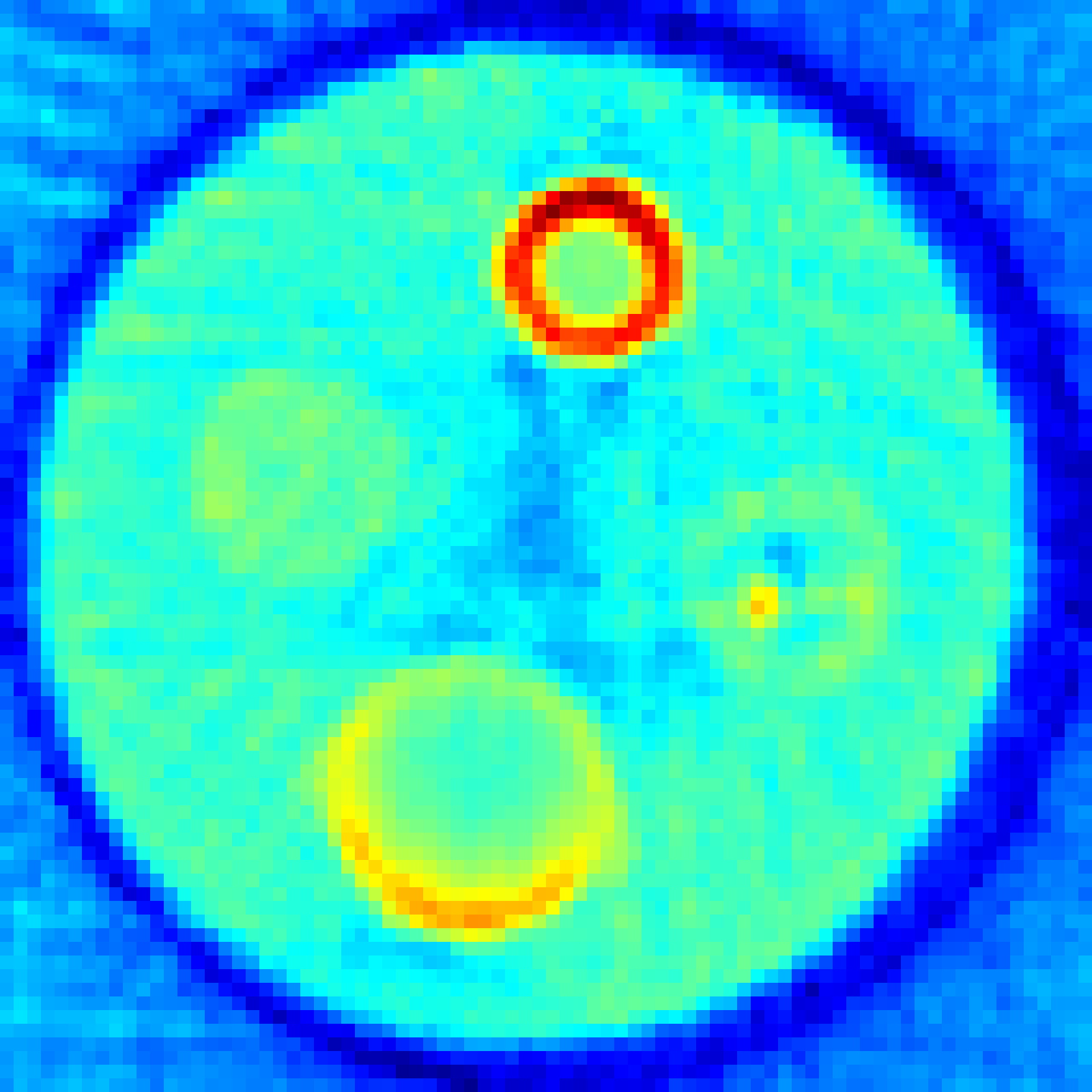}%
\caption[A tomographic image of the back-projected ADD data.]{A tomographic image ($80\times80$)
obtained using the total intensity of the back-projected ADD data.} \label{ADDBP}
\end{figure}

\begin{figure} [!h]
\centering
\includegraphics
[scale=0.7] {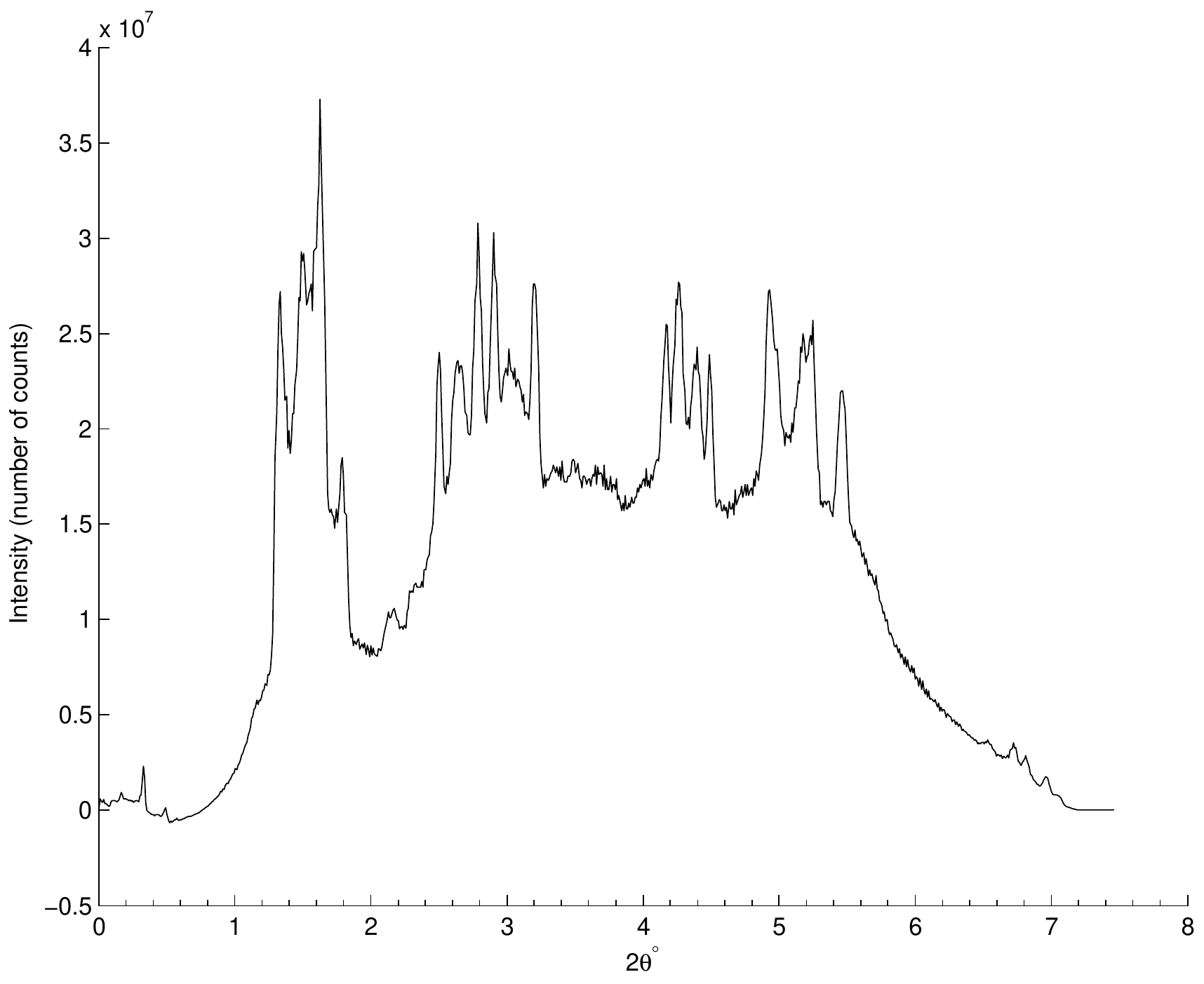}%
\caption{Sum of all diffraction patterns of the back-projected ADD data.} \label{ADDSum}
\end{figure}

\begin{figure} [!h]
\centering
\includegraphics
[scale=0.4] {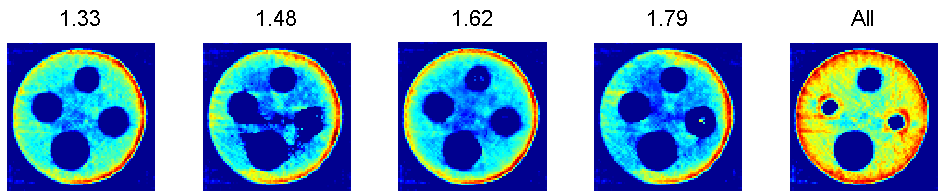}%
\caption[Tomographic images of the wax cylinder from ADD data.]{Tomographic images from ADD data of
the wax cylinder as represented by the Gaussian area of the peaks. The numbers on the top of the
first 4 images are the approximate positions of the peaks in terms of \sad, while the last image
represents the total of the first 4 peaks as modelled by a single Gaussian peak.} \label{TomoWax}
\end{figure}

\begin{figure} [!h]
\centering
\includegraphics
[scale=0.44] {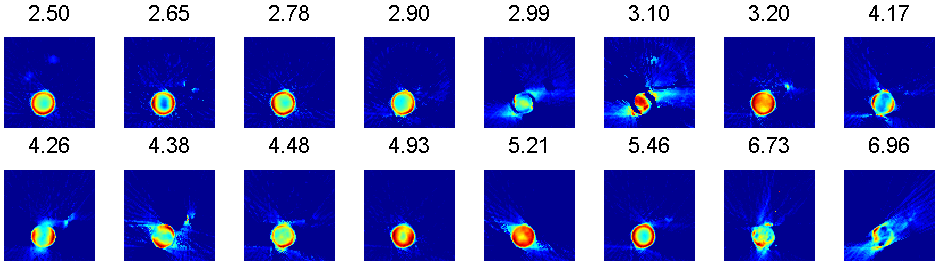}%
\caption[Tomographic images of the aluminium cylinder from ADD data.]{Tomographic images from ADD
data of the aluminium cylinder as represented by the Gaussian area of the peaks. The numbers on the
top of each image are the approximate positions of the peaks in terms of \sad.} \label{TomoAl}
\end{figure}

\begin{figure} [!h]
\centering
\includegraphics
[scale=0.35] {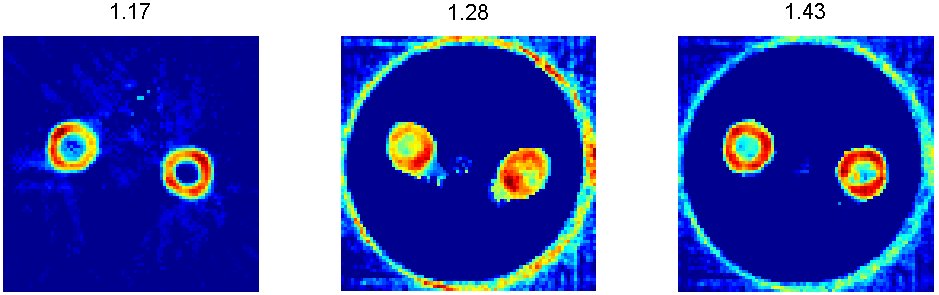}%
\caption[Tomographic images of the PEEK cylinders from ADD data.]{Tomographic images from ADD data
of the PEEK cylinders as represented by the Gaussian area of the peaks. The numbers on the top of
the images are the approximate positions of the peaks in terms of \sad. In frames 2 and 3, traces
of other components are masked for clarity.} \label{TomoPeek}
\end{figure}

\begin{figure} [!h]
\centering
\includegraphics
[scale=0.4] {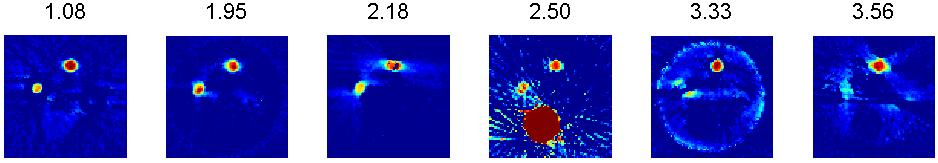}%
\caption[Tomographic images of the bone ash cylinders from ADD data.]{Tomographic images from ADD
data of the bone ash cylinders as represented by the Gaussian area of the peaks. The numbers on the
top of the images are the approximate positions of the peaks in terms of \sad. The fourth frame
includes an aluminium trace.} \label{TomoBone}
\end{figure}

\begin{figure} [!h]
\centering
\includegraphics
[scale=0.4] {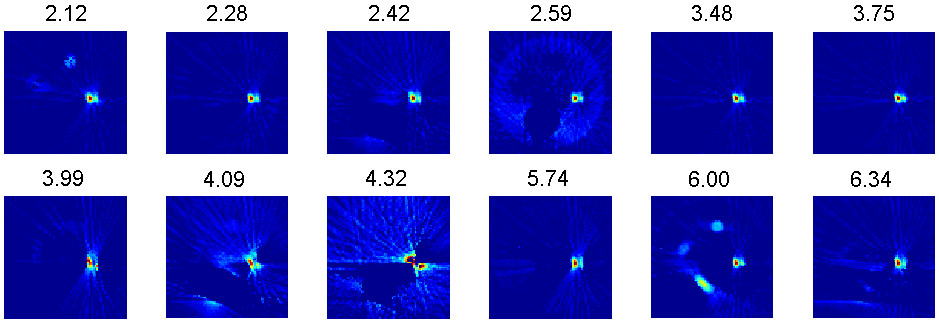}%
\caption[Tomographic images of the iron oxide cylinder from ADD data.]{Tomographic images from ADD
data of the iron oxide cylinder as represented by the Gaussian area of the peaks. The numbers on
the top of the images are the approximate positions of the peaks in terms of \sad.}
\label{TomoIron}
\end{figure}

\begin{figure} [!h]
\centering
\includegraphics
[scale=0.45] {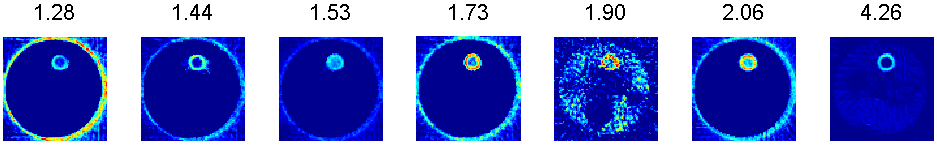}%
\caption[Tomographic images of the glass cylinder from ADD data.]{Tomographic images from ADD data
of the glass cylinder as represented by the Gaussian area of the peaks. The numbers on the top of
the images are the approximate positions of the peaks in terms of \sad. Traces of other components
are masked in most frames.} \label{TomoGlass}
\end{figure}

The ADD data were collected at station ID15A of ESRF which uses a monochromatic beam in the energy
range $30-500$ keV for angle dispersive investigations. The station can also use a white beam in
that energy range for EDD studies \cite{ESRF}. The wavelength of the monochromatic beam used in
this experiment is $\WL=0.17712$ \AA. A circular slice of the object (perpendicular to the
cylindrical axis) was chosen for a CAT-type scan as presented in \S\ \ref{Ex3} and demonstrated in
Figure \ref{CAT}. The measurements consist of 81 rotations around the object axis between
0-180$^\circ$ in regular intervals times 80 translations in steps of 0.5 mm. The total collected
data set therefore consists of 6480 ($81\times80$) patterns. The collection time of each pattern
was about 0.5 second, giving a total collection time for the whole data set of about one hour. This
raw data set, which is a sinogram of rotation versus translation, was back-projected (excluding the
last rotational step) using the EasyDD program. An image of this sinogram, as represented by the
total intensity of the individual patterns, is given in Figure \ref{ADDSino}. The resultant data,
which consist of 6400 ($80\times80$) patterns, represent a tomographic image of the slice in real
space. An image of the back-projected data represented by the total intensity of the individual
patterns is shown in Figure \ref{ADDBP}. A sum of all back-projected diffraction patterns can also
be seen in Figure \ref{ADDSum}. More details about this experiment can be found in Lazzari \etal\
\cite{LazzariJB2010}. Using EasyDD in multi-batch mode, the back-projected peaks of the six phases
are curve-fitted using a Gaussian profile with a linear background. The main results are presented
as tomographic images in Figures \ref{TomoWax}-\ref{TomoGlass}. In the following points we discuss
some issues related to these results:

\begin{figure} [!h]
\centering
\includegraphics
[scale=0.2] {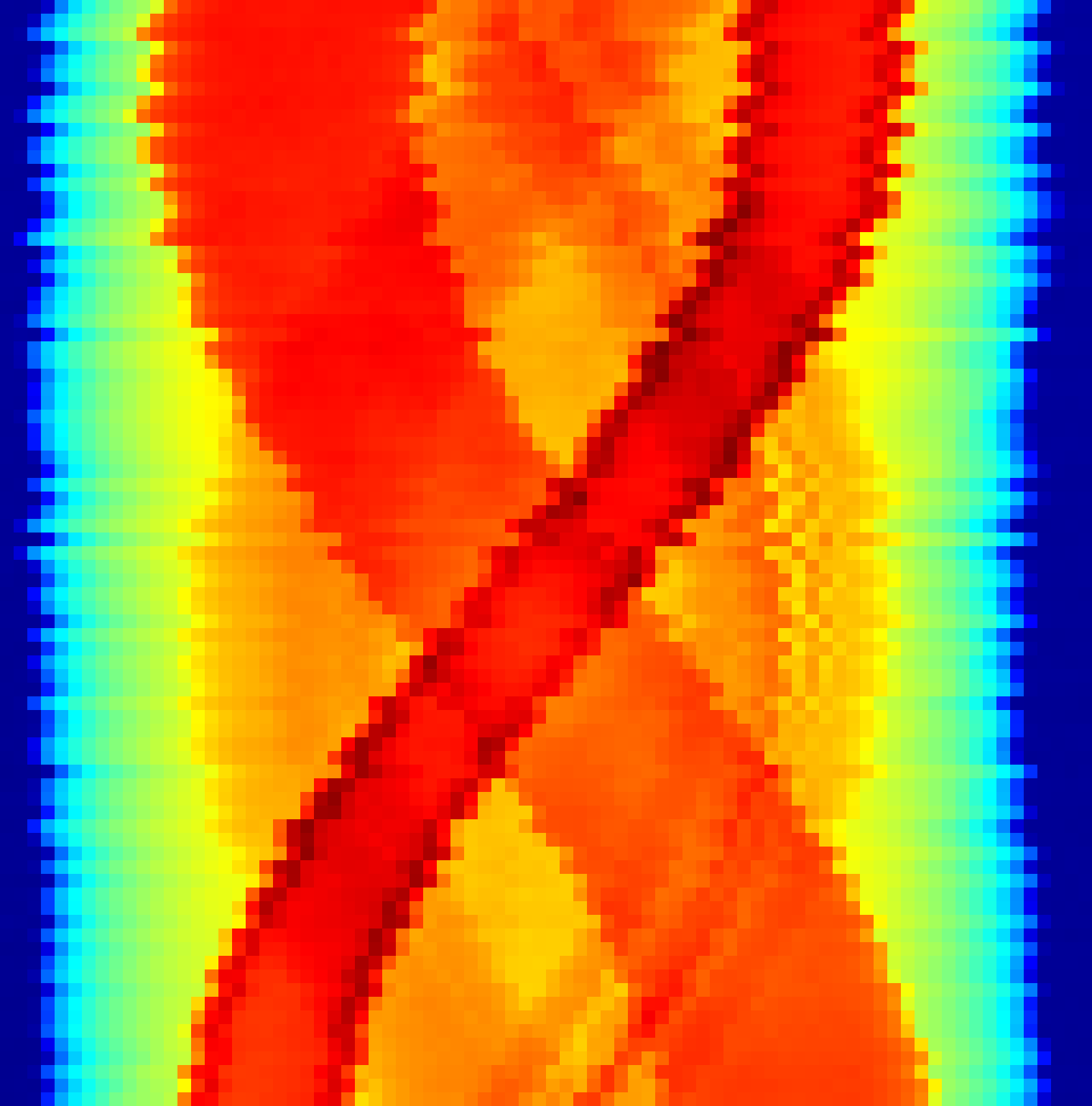}%
\caption[A sinogram of the ADD data.]{A sinogram consisting of 81 rotation rows $\times$ 80
translation columns for the ADD data. The displayed parameter is the total intensity of the
individual patterns.} \label{ADDSino}
\end{figure}

\begin{itemize}

\item
Wax: (Figure \ref{TomoWax}) No standard diffraction pattern was available to compare with the
experimental data and identify the Bragg's reflections. Therefore, the identification of this phase
is based on the spatial distribution in the tomographic images. It should be remarked that the
distortion in the second frame of Figure \ref{TomoWax} is due to peak overlap and restrictions in
curve-fitting.

\item
Aluminium: (Figure \ref{TomoAl}) Peaks at the following approximate scattering angles \sad\ = 2.50,
2.65, 2.78, 2.90, 2.99, 3.10, 3.20, 4.17, 4.26, 4.38, 4.48, 4.93, 5.21, 5.46, 6.73 and 6.96 are
identified with this phase according to the spatial distribution in the tomographic images.
Considering the uncertainties in the standard diffraction patterns and experimental data the peak
at \sad\ = 4.38 can be identified with the following peak of aluminium [FM3-M] \cite{TougaitN2004}:
(\sad, $hkl$, $d$ \AA) = (4.33 , 111, 2.3417) which is the strongest peak.

\item
PEEK: (Figure \ref{TomoPeek}) No standard diffraction pattern was available to compare with the
experimental data and identify the Bragg reflections. Therefore, the identification of this phase
is based on the spatial distribution in the tomographic images.

\item
Bone ash: (Figure \ref{TomoBone}) Peaks at the following approximate scattering angles, \sad\ =
1.08, 1.95, 2.18, 2.50, 3.33 and 3.56, are identified with this phase according to the spatial
distribution in the tomographic images. Considering the uncertainties in the standard diffraction
patterns and experimental data the peaks at \sad\ = 1.95, 2.50, 3.33 and 3.56 can be identified
with the following peaks of calcium phosphate Ca$_3$(PO$_4$)$_2$-[R3CH] (which is the main
component of bone ash) respectively \cite{YashimaSKH2003}: (\sad, $hkl$, $d$ \AA) = (1.95, 110,
5.2176), (2.49, 024, 4.0685), (3.37, 300, 3.0124) and (3.53, 217, 2.8779).

\item
Iron oxide (hematite Fe$_2$O$_3$): (Figure \ref{TomoIron}) Peaks at the following approximate
scattering angles \sad\ = 2.12, 2.28, 2.42, 2.59, 3.48, 3.75, 3.99, 4.09, 4.32, 5.75, 6.00 and 6.34
are identified with this phase according to the spatial distribution in the tomographic images.
Considering the uncertainties in the standard diffraction patterns and experimental data the peaks
at \sad\ = 3.75, 3.99, 6.00 and 6.34 can be identified with the following peaks of hematite [R3-CH]
\cite{PailheWGD2008} respectively: (\sad, $hkl$, $d$ \AA) = (3.76, 104, 2.6998), (4.03, 110,
2.5173), (5.99, 116, 1.6948) and (6.34, 122, 1.6026).

\item
Glass: An amorphous scattering signal was observed at several $2\SA$ positions with no obvious
peak. Examples of this signal (which associates with signals from other phases) are given in Figure
\ref{TomoGlass}.

\end{itemize}

Because a large test object was used in this experiment, peak broadening and splitting were
observed in the raw data and hence in the reconstructed data. As the EDD technique employs a fixed
and usually small scattering angle which could reduce the effect of sample size on the signal
spread (refer to Figure \ref{SizeAngleEffect}), it was decided to perform a similar experiment
using EDD to investigate and assess the potential improvement of data quality offered by this
technique. The EDD experiment, with a fairly detailed analysis, is presented in the next section,
while a comparison between ADD and EDD experiments is given in section \ref{AddEDDCompare}.

\clearpage
\subsection{EDD Data} \label{EDDData}

The EDD data were collected at station 16.4 of Daresbury SRS before its closure in 2008. This
station used a white beam for energy dispersive powder diffraction and was designed for the study
of chemical reactions and phase transitions of materials held at high temperatures and pressures
\cite{SRS}. In this experiment a white beam of diameter 0.5 mm in the energy range 1-110 keV was
used. The collection angle of the fixed EDD detector was $2\SA$ = 4.5$^\circ$. The reason for
choosing this small angle is to collect the diffraction patterns at higher energies to reduce the
absorption of diffracted signals along its path inside the sample. This will also mitigate the
effect of signal spread. A drawback of this shift to high energies is a distortion of the sampling
volume shape (lozenge) which requires correction at the data processing stage.

\begin{figure} [!h]
\centering
\includegraphics
[scale=0.2] {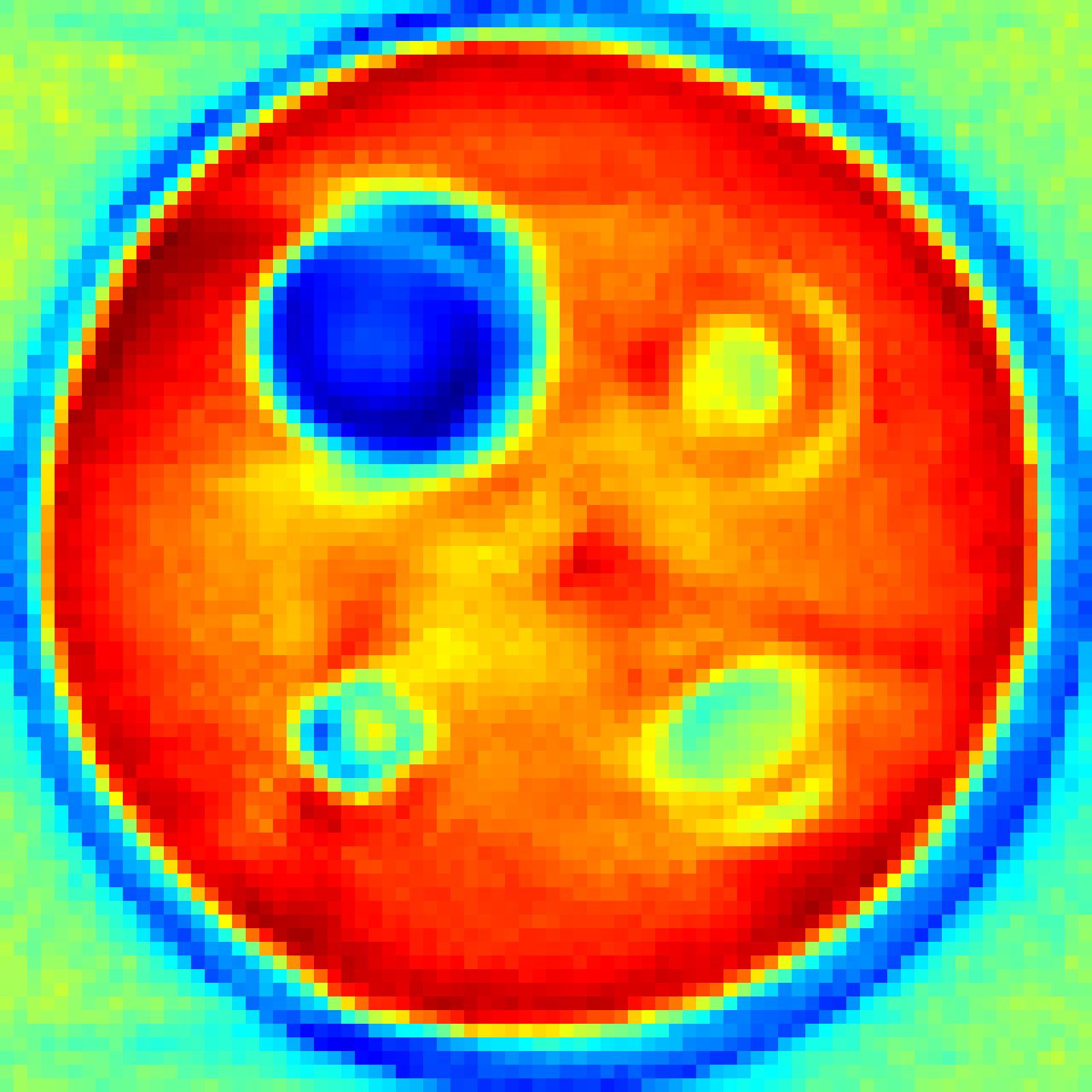}%
\caption [A tomographic image of the back-projected EDD data.]{A tomographic
image ($80\times80$) of the total intensity of the back-projected EDD data.} \label{EDDBP}
\end{figure}

\begin{figure} [!h]
\centering
\includegraphics
[scale=0.7] {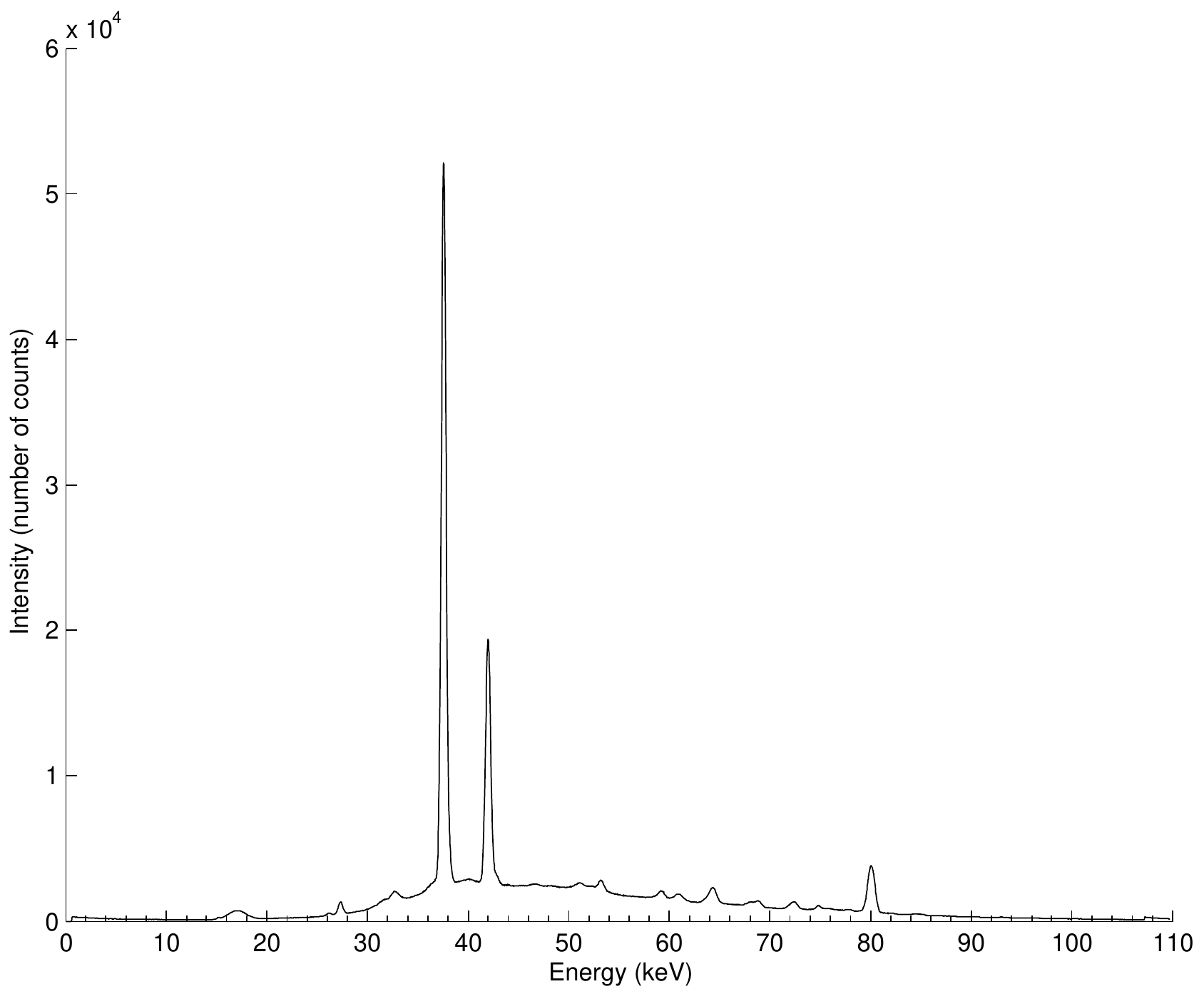}%
\caption{Sum of the diffraction patterns of the back-projected EDD data.} \label{EDDSum}
\end{figure}

\begin{figure} [!h]
\centering
\includegraphics
[scale=0.43] {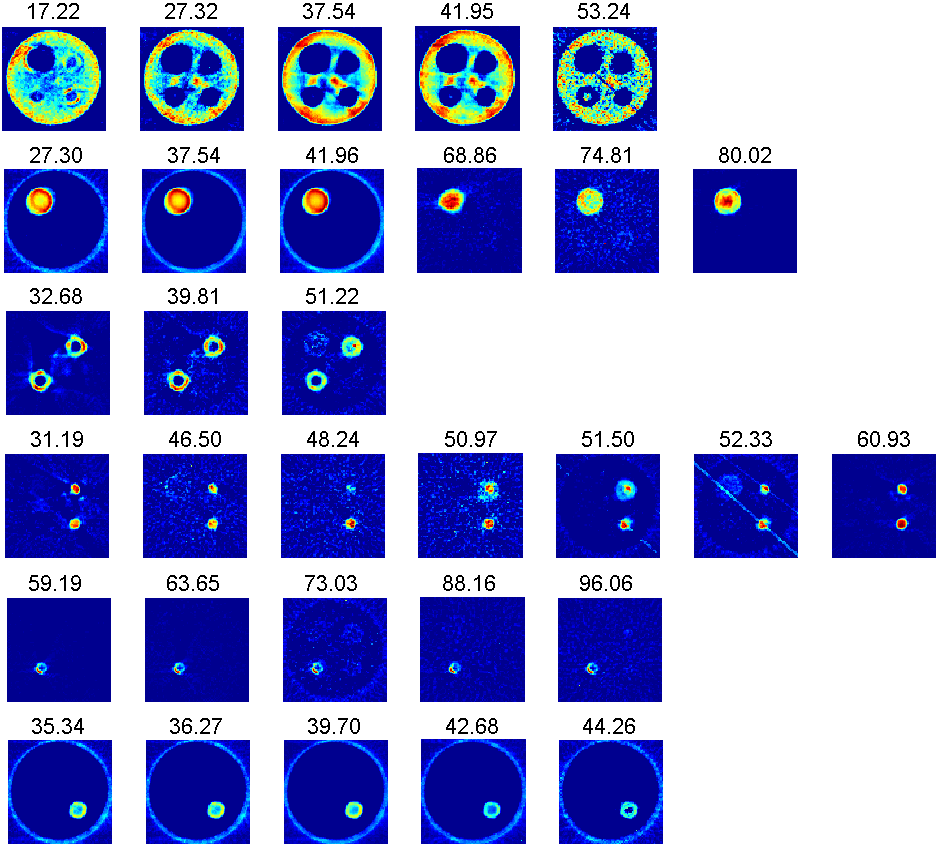}%
\caption[Tomographic images of the test object components from EDD data]{Tomographic images of the
six components of the test object from the EDD data as represented by the Gaussian peak area. The
rows in top-down order stand for: wax, aluminium, PEEK, bone ash, iron oxide and glass. The numbers
on the top of each frame are the energy position of the peak in keV. In some frames, traces of
other components are masked for clarity.} \label{TomoEddAll}
\end{figure}

A circular slice of the object (perpendicular to the cylindrical axis) was chosen for a CAT-type
scan as presented in \S\ \ref{Ex3} and demonstrated in Figure \ref{CAT}. The measurements consist
of 79 rotations around the object axis between 0-180$^\circ$ in regular intervals times 80
translations in steps of 0.5 mm. The total size of the collected data therefore consists of 6320
($79\times80$) patterns. The collection time of each pattern was about 10 seconds, totaling a
collection time for the whole data set to about 18 hours. This raw data set, which is a sinogram of
rotation versus translation, was back-projected using the EasyDD program. An image of this
sinogram, as represented by the total intensity of the individual patterns, is presented in Figure
\ref{EDDSino}. The resultant data, which consist of 6400 ($80\times80$) patterns, represent a
tomographic image of the slice in real space. An image of the back-projected data, as represented
by the total intensity of the individual patterns, is given in Figure \ref{EDDBP}. A sum of the
back-projected diffraction patterns can also be seen in Figure \ref{EDDSum}. It is noteworthy that
the orientation of the test object in the EDD experiment is different to that in the ADD case, and
therefore the position of the elements in the related images are different. More details about this
experiment can be found in Lazzari \etal\ \cite{LazzariJSB2009, LazzariJB2010}. Using EasyDD in
multi-batch mode, the back-projected peaks of the six phases are modelled by a Gaussian profile
with a linear background. The main results are presented as tomographic images in Figure
\ref{TomoEddAll}. In the following points we discuss some issues related to these results:


\begin{figure} [!h]
\centering
\includegraphics
[scale=0.2] {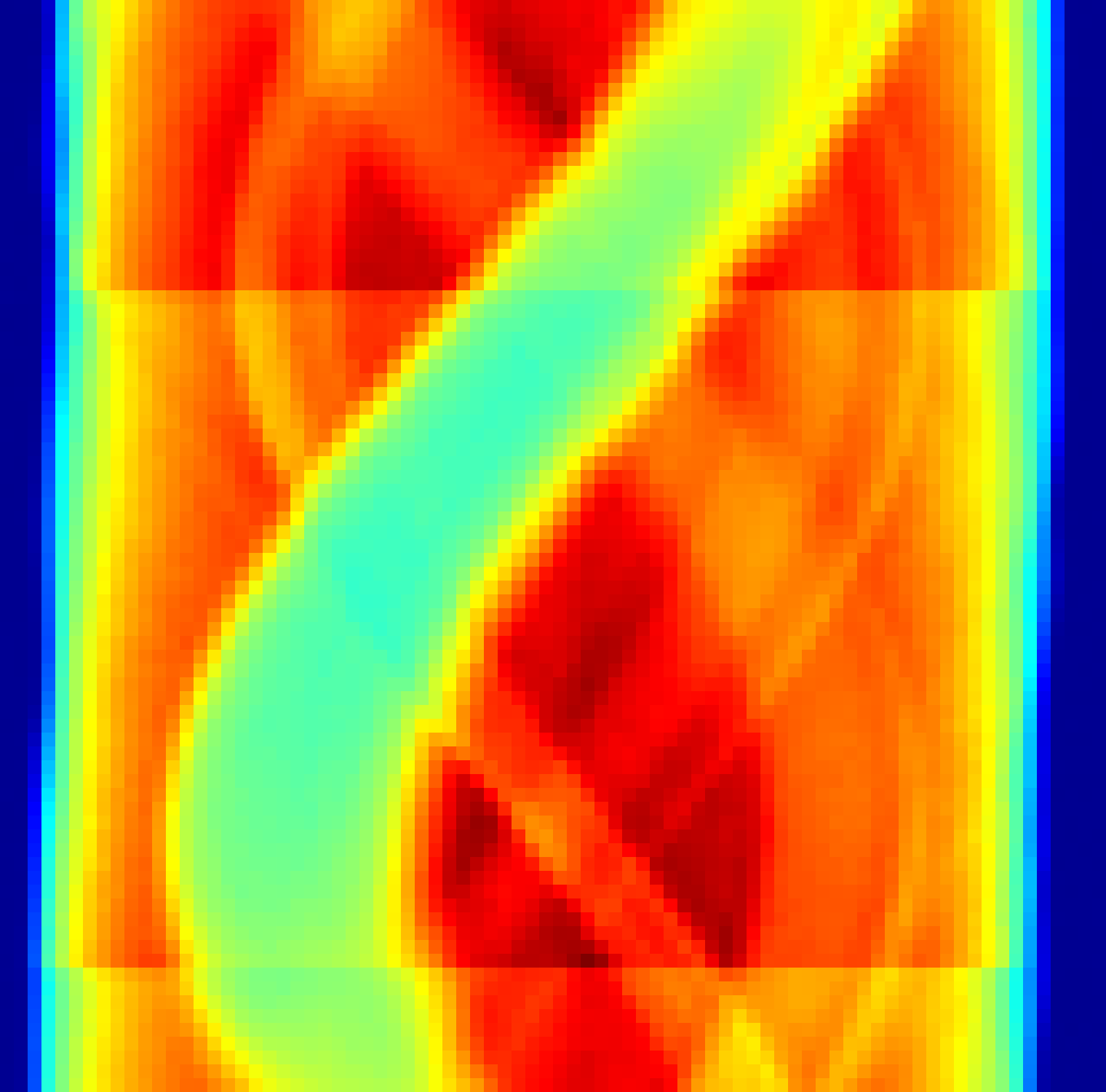}%
\caption[A sinogram of the EDD data.]{A sinogram consisting of 79 rotation rows $\times$ 80
translation columns for the EDD data. The displayed parameter is the total intensity of the
individual patterns. The observed shift is due to mechanical instability of the
translation-rotation stage.} \label{EDDSino}
\end{figure}

\begin{itemize}

\item
Wax: No standard diffraction pattern was available to compare with the experimental data and
identify the Bragg reflections. Therefore, the identification of this phase is based on the spatial
distribution in the tomographic images.

\item
Aluminium: Peaks 1, 2, and 3 of this phase are approximately in the same position as peaks 2, 3 and
4 of wax. In fact the aluminium peaks at these three positions have negative amplitude as though
they were absorption peaks. This negative-peak feature was observed for other peaks with other
phases. The following peaks of aluminium [FM3-M] are identified \cite{TougaitN2004}: ($E$ keV,
$hkl$, $d$ \AA) = (37.5, 200, 2.0280), (68.9, 111, 2.3417), (74.8, 200, 2.0280) and (80.0, 200,
2.0280). Reflections of aluminium at about $E=$ 27.3 and 42.0 keV were also detected but no data
could be retrieved from ICSD.

\item
PEEK: No standard diffraction pattern was available to compare with the experimental data and
identify the Bragg reflections. Therefore, the identification of this phase is based on the spatial
distribution in the tomographic images.

\item
Bone ash: The following peaks of calcium phosphate Ca$_3$(PO$_4$)$_2$-[R3CH] (which is the main
component of bone ash) are identified \cite{YashimaSKH2003}: ($E$ keV, $hkl$, $d$ \AA) = (46.5,
211, 3.4016), (48.2, 119, 3.2507), (51.0, 125, 3.1071), (52.3, 300, 3.0124) and (60.9, 220,
2.6088).

\item
Iron oxide (hematite Fe$_2$O$_3$): The following peaks of hematite [R3-CH] are identified
\cite{PailheWGD2008}: ($E$ keV, $hkl$, $d$ \AA) = (59.2, 104, 2.6996), (63.7, 110, 2.5175), (73.0,
113, 2.2066), (88.2, 024, 1.8412) and (96.1, 211, 1.6364).

\item
Glass: An amorphous scattering signal was observed over several energy positions with no obvious
dominant peak. Examples of this signal (which associates signals from other phases) are given in
the last row of Figure \ref{TomoEddAll}.

\end{itemize}

\clearpage
\subsection{Comparison and General Conclusions} \label{AddEDDCompare}

In this section, we compare the ADD and EDD imaging data and try to reach general conclusions about
these techniques.

\begin{figure} [!h]
\centering
\includegraphics
[scale=1] {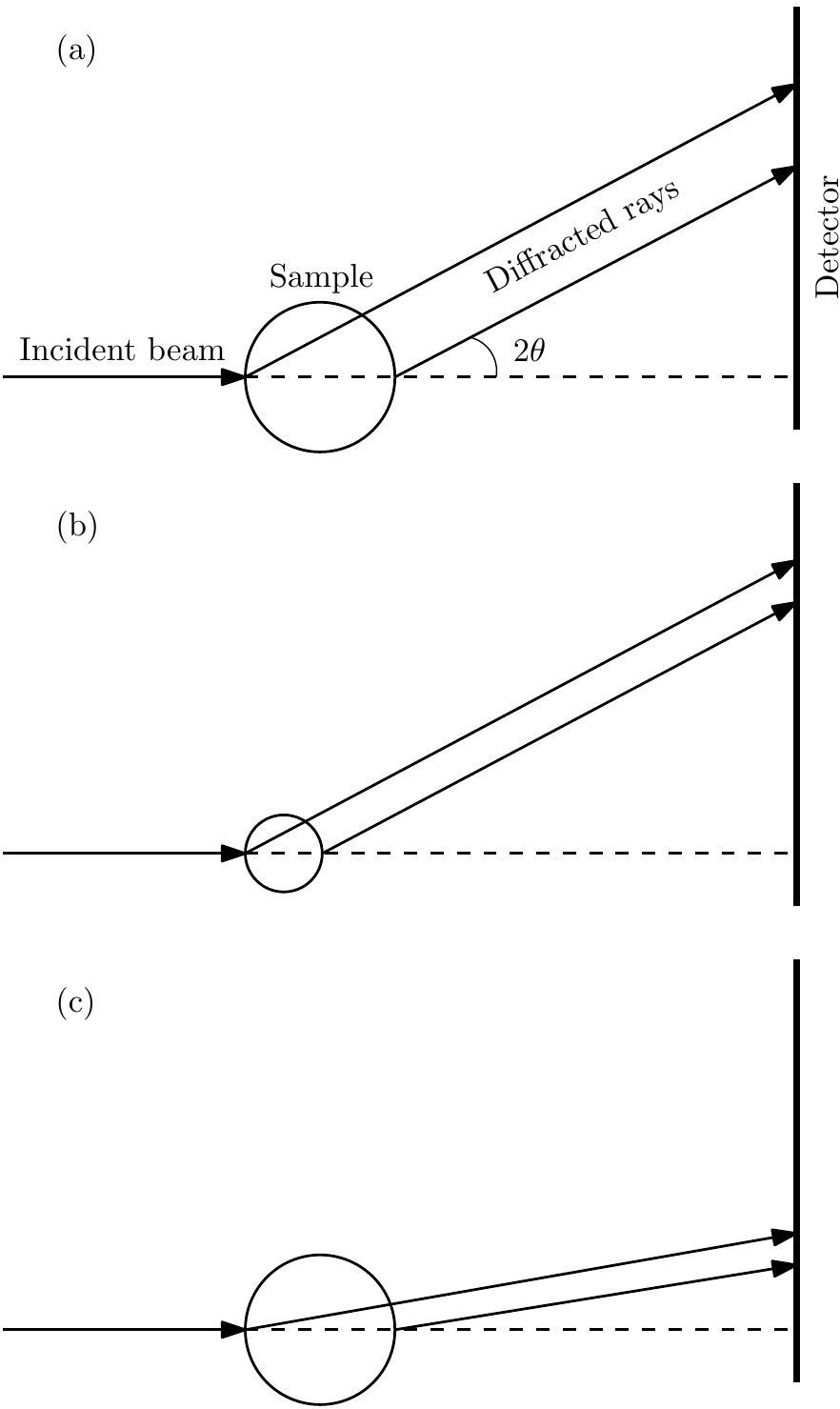}%
\caption[Effect of sample size and $2\SA$ magnitude on the spread of signal]{A schematic diagram
illustrating the effect of sample size and scattering angle on the signal spread which is
manifested in peak broadening and splitting. The large signal spread observed in (a), due to the
large sample size and large scattering angle, is diminished in (b) by reducing the size of the
sample and in (c) by reducing the scattering angle.} \label{SizeAngleEffect}
\end{figure}

\begin{itemize}

\item
Negative peaks were observed in both ADD and EDD data. Its most prominent occurrence is with
aluminium and wax where strong negative peaks of aluminium appear in the same positions as the
normal peaks of wax and vice versa (e.g. at \sad\ $\simeq$ 1.33 and 3.22 in ADD data, and at
$E$(keV) $\simeq$ 37.5 for EDD data). This feature was also observed between wax and components
other than aluminium. Some of the results and images presented in this section were obtained by
fitting these negative peaks like normal peaks then taking the absolute value of the Gaussian area.

\item
Peak broadening and splitting were observed in both ADD and EDD data but they were more prominent
in the ADD data. These effects, which are manifested in the raw diffraction data and hence in the
reconstructed data, can be explained by the large sample size and large scattering angle, as
illustrated graphically in Figure \ref{SizeAngleEffect}. Although the size of the sample is the
same for the ADD and EDD data, the (fixed) scattering angle of the EDD data is small (=4.5$^\circ$)
and hence the spread of the detected signal is less, resulting in reduced peak broadening and
splitting. However, the use of a small scattering angle can result in another problem that is peak
overlap. It should be remarked that these problems can be corrected for by employing correction
algorithms at the data processing and analysis stage. However, this is not a trivial task and hence
it is left for future research as it is beyond the remit of this thesis.

\item
To avoid peak splitting effects and the occurrence of repetitive peaks, we have adopted a minimum
gap between neighbouring peaks for them to be considered as two different peaks.

\item
The general impression is that the ADD data are more complicated and have more peaks than the EDD
data. This is due partly to the broadening and splitting effects which are more severe in the ADD
data, as already indicated. It may also be attributed in part to the fact that the diffracting
lozenge size and shape in an EDD scan is constant whereas the sampling volume is variable in an ADD
scan.

\item
The quality of the ADD and EDD data varies from one component to another although EDD results are
generally clearer and simpler. Due to the variation in radiation sources, detectors and other
equipment, it is difficult to make a generalised comparison of the two techniques based on the
results of just these experiments.

\item
A prominent feature of ADD data is the appearance of streaks in many images (e.g. Figure
\ref{TomoAl} \sad\ = 6.96 and Figure \ref{TomoIron} \sad\ = 5.74). These streaks change orientation
depending on the channels chosen to represent the peak. This feature may be explained by the fact
that the sampling volume, and the nature and length of the path of reflected signal in the ADD
technique, are variable. Streaks may also be attributed to peak distortion due to peak overlap,
peak splitting, background noise and curve-fitting strategy (peak range, position and
restrictions).

\item
An important problem in diffraction-based imaging, which requires correction at processing and
analysis stage, is the absorption of reflected signals inside the sample. This problem is
particularly serious when the sample is large due to the long signal paths. This reduces
signal-to-noise ratio and degrades the quality of the diffraction patterns. Also, the difference in
the length and nature of the signal paths, which may be more severe in the ADD technique due to
unparallel paths, results in uneven absorption and hence further distortion in the patterns. Uneven
absorption is more grave when the sample is large with complex structure and morphology. In such
cases, correction for absorption may become impossible to introduce. Because absorption is
energy-dependent, the EDD technique may suffer from selective absorption which can be another
source of pattern distortion. Some solutions to the problems associated with absorption have been
given by Lazzari \etal\ \cite{LazzariJSB2009, LazzariJB2010}.

\end{itemize}

\section{Prospects for Diffraction-Based Imaging} \label{Prospects}

The examples given in this thesis, and the related cited work, are clear demonstration of the
capability and potential of diffraction-based imaging; both in angle dispersive and energy
dispersive modes. As data acquisition and detector technologies are rapidly improving with the wide
availability of bright light sources, the future of these imaging techniques is very promising. The
rapid improvement in computational technology and easy access to massive computational resources is
another factor that can contribute to the development and widespread use of diffraction-based
imaging. In particular, EDD imaging can also exploit other signals, such as fluorescence and
scattering bands, and hence enjoys yet further potential. Multi-pixel detector technology, such as
HEXITEC multi-TEDDI detectors, will be a contributing factor to realising a huge increase in the
speed of data acquisition, and thereby an improved spatial and temporal resolution since this
technology will enable the acquisition of images with a finer pixel size and in shorter time
intervals. An obvious beneficiary of these technologies will be dynamic and time-dependent studies
and tomographic imaging. The main difficulties to overcome include data processing and analysis
capabilities to match the massive accumulation of data associated with these new imaging
techniques. An investment in software development will be a top priority for future progress. The
resultant data processing should consider pixel size and shape corrections which are required to
account for the finite size and distorted shape of the sampling elements.

\section{Manchester ERD HEXITEC Detector Data Sample} \label{Manch}

\begin{figure} [!h]
\centering
\includegraphics
[scale=1.5] {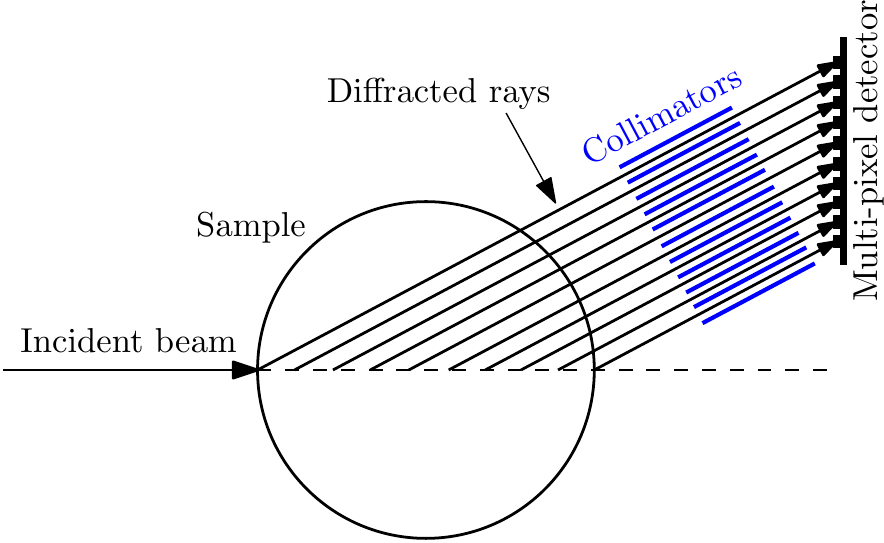}%
\caption[Schematic illustration of multi-pixel detector]{A schematic diagram illustrating the
working principle of a multi-pixel detector for which several parts of the sample are being
simultaneously sampled.} \label{MultiPixel}
\end{figure}

The TEDDI method in its current form does not lend itself well to the fast recording of tomographic
diffraction data. The method has several attractive attributes, e.g. the recording of fluorescence
signals, suitability for studying larger objects, and so on. However, there is a real prospect that
the length of data collection can be significantly reduced by parallelising signal acquisition
through the use of pixelated energy sensitive detectors, array collimators, and large parallel
beams. The \HEXITECl\ (\HEXITECs) project \cite{HEXITEC} is concerned with making this happen.
Figure \ref{MultiPixel} illustrates the basic principle, showing a 1D arrangement of collimators
and pixelated detector to collect signals from several sampling volumes simultaneously. The current
functioning HEXITEC system has a $20\times20$ 2D arrangement of pixelated detector and collimators.
This gives rise to 400 sampling volumes, so in principle one can collect data 400 times more
rapidly during a tomography experiment. The current objective for the HEXITEC project is to
construct an $80\times80$ operational device; this device will accelerate data acquisition by a
huge factor. It should be remarked that EasyDD has played a crucial role in the development of
these multi-pixel detectors as it has been used by the developers as a tool to map and visualise
the spectral responses of each pixel and perform data summarisation and analysis. At the time of
writing this thesis, the $20\times20$ HEXITEC detector has only recently become available and has
yet to be used operationally in diffraction mode. However, preliminary tests have been conducted in
a more primitive radiography mode. In the following we present a first test of the use of EasyDD
for processing data produced by this multi-pixelated detector; this involved binning the raw data
into histograms, followed by appropriate visualisation and analysis.

\begin{figure} [!h]
\centering %
\subfigure []%
{
\begin{minipage}[b]{0.45\textwidth} %
\centering \includegraphics[width=2.8in, trim = 0 0 0 0] {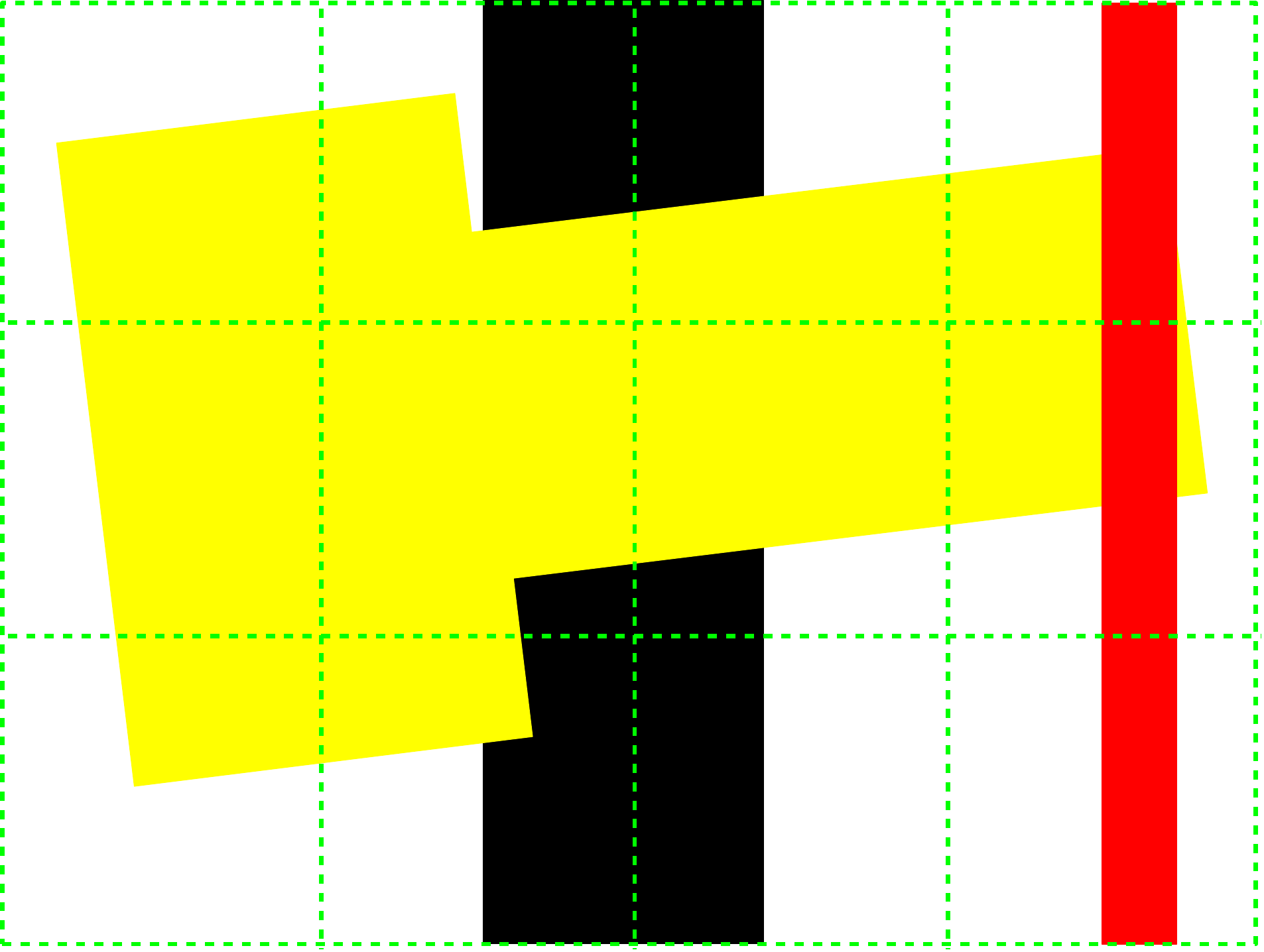}
\end{minipage}}
\hspace{0.5cm} %
\subfigure []%
{
\begin{minipage}[b]{0.45\textwidth} %
\centering \includegraphics[width=2.8in, trim = 0 0 0 0] {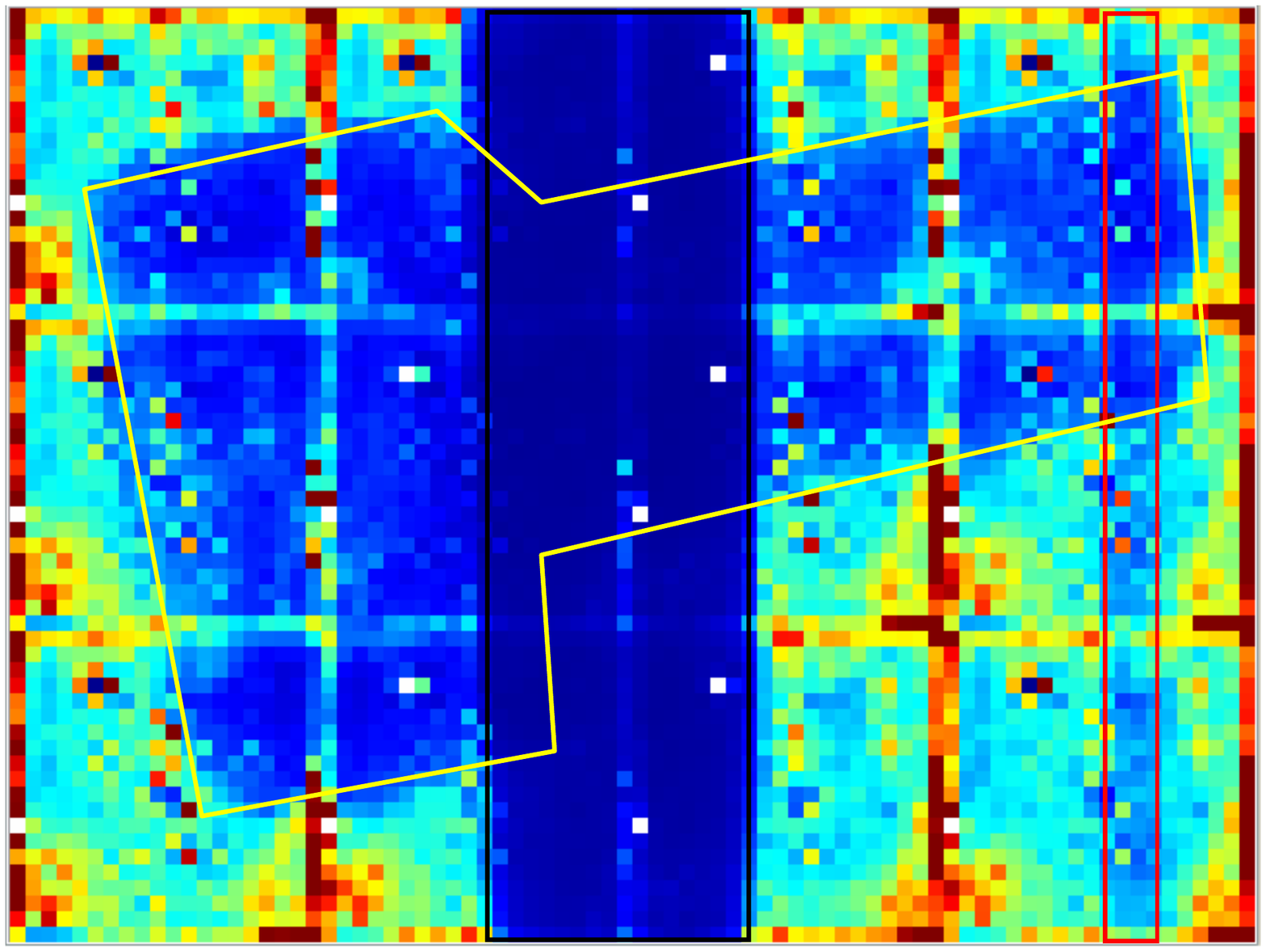}
\end{minipage}} \vspace{-0.3cm}
\caption[Manchester HEXITEC data]%
{A schematic illustration (a) and a total intensity image (b) of the phantom object, obtained using
the Manchester ERD HEXITEC detector. The phantom object consists of a cylinder of wax into which a
bolt (yellow), a lead sheet (black) and a copper wire (red) are inserted. \label{ManData}}
\end{figure}

This preliminary experiment was carried out at the Henry Moseley X-ray Imaging Facility
\cite{Moseley} in the Materials Science Centre at the University of Manchester\footnote{The data
and basic description of this experiment was provided by Conny Hansson of the University of
Manchester.}. The sample is a specially designed phantom object which consists of a cylinder of wax
in which three metal objects are inserted: an M3 bolt, a sheet of lead, and a copper wire. A
schematic diagram of the phantom object can be seen in Figure \ref{ManData} (a). A very simple
experimental setup was used in this experiment: the sample was placed in the path of the X-ray beam
and the detector was positioned directly behind the sample. The detector used was a prototype
$20\times20$ ERD HEXITEC multi-pixel detector produced by the Detector Development Group at
Rutherford Appleton Laboratory (Paul Seller \etal) for the University of Manchester. The entire
sample was imaged by moving the detector position in an array of 3 rows times 4 columns as
indicated in Figure \ref{ManData} (a). To identify the change in spectral response due to the
passing of X-rays through the sample, two array sets (i.e. each with 3$\times$4) were collected,
one with the sample in between the source and detector and one without the sample in between. The
collected raw data (i.e. with and without sample) consist of about 130 gigabytes. Using EasyDD, the
data were binned in histograms and hence reduced in size to about 35 megabytes. The collected
radiographic signals, which consist mainly of absorption patterns, for the second case (i.e.
presence of sample) are presented in Figure \ref{ManData} (b) where the colour-coded total
intensity of the individual pixels is mapped. Although the results of this experiment appear to be
generally of a low quality (due mainly to the data collection technique, since the detector was
used in a radiographic mode) they nevertheless demonstrate feasibility and great promise for future
diffraction imaging.

%

%% file: Conc.tex
%
%
\chapter{Conclusions} \label{Conclusions}

The EasyDD project is a first step in the right direction for the future development of
computational tools to deal with the growing demand on data processing and analysis capabilities.
Thanks to recent developments in the technology of radiation sources and data acquisition systems,
this approach is an important endeavour in developing software that can cope with the massive and
ever-increasing size of data collections that are generated in modern multi-tasking scientific
experiments. EasyDD has already proved to be a crucial component within a number of key studies,
e.g. \cite{LazzariJSB2009} and \cite{EspinosaOJBJe2009}. Without the efficiency and speed of
EasyDD, and without its effective strategies, such as the multi-batch processing approach, some of
these studies may not have happened.

In the course of this PhD project, several experimental data collections on alumina-supported metal
catalysts have been processed and analysed by the author using EasyDD. These include large data
collections on molybdenum oxide, nickel nitrate calcination, nickel nitrate reduction, and nickel
chloride based preparations. In the process, several chemical and crystallographic phases have been
identified and mapped in terms of spatial location and time/temperature, thereby leading to
important scientific implications.  Amongst the many systems discussed in this thesis, the nickel
chloride based system emerged as an outstanding example of what can be achieved in following its
evolution (precursor, intermediates and final active phase) in terms of time, temperature,
crystallite size and spatial distribution.  Further, the high level of detail extracted enabled us
to elucidate the overall evolving chemistry while also revealing new information on the physical
state of the catalyst (crystallite size) and providing evidence and suggesting a mechanism for the
inhomogeneity of crystallite size. This particular study stands out as a vivid example with which
to demonstrate the capabilities and potential of these X-ray imaging and analysis techniques.

During this overall study a novel approach for processing and analysing massive data collections
has emerged in which phase distribution pattern diagrams, combined with stack plots and standard
diffraction patterns from powder diffraction databases, play a key role in summarising and
presenting huge data sets in manageable form as an essential step to identifying phases and
tracking the evolution of complex physical and chemical systems. As such, these developmental
studies take diffraction-imaging capabilities way beyond  those of previous landmark studies (e.g.
Bleuet \etal\ \cite{BleuetWDSHW2008} and Espinosa-Alonso \etal\ \cite{EspinosaOJBJe2009})
particularly for \insitu\ and dynamic phase transformation studies. It is anticipated that these
developments should have a significant future impact, at the scientific, technological and
industrial level, within several fields of research such as catalysis, in operando studies, phase
transformations, dynamic stress imaging of construction/biological materials etc.  However, phase
identification strategies for dynamic transformation studies of chemical and crystallographic
systems still require further development as these systems are usually very complex and involve
many phases.

%% file: Bibl.tex
\renewcommand{\refname}{}

%

%% file: Samp.tex
%
%
\chapter[Samples of EasyDD Code]{} \label{SampleCode}

\vspace{-1cm}

\begin{spacing}{2}
{\LARGE \bf Samples of EasyDD Code}
\end{spacing}

\vspace{0.5cm}

\noindent In this appendix, we provide two snippets of EasyDD source code for two algorithms:
\bacpro\ and channels alignment.

\section{Back Projection Snippet} \label{SampleCodeBP}

{\singlespace {\scriptsize
\begin{verbatim}
void MainWindow::backProjectTab()
{
    if(tabWidget->tabText(tabWidget->currentIndex()).contains("BP"))
    {
        QMessageBox::about(this, tr("Message"), "Back projection has been performed on this tab.
                                                 Use 'Back Projection Shift'.");
        return;
    }

    createBackProDialog();

    if(!bPOKFlag)
        return;

    if(allTabs.size()==0)
    {
        QMessageBox::about(this, tr("Message"), "There is no tab.");
        return;
    }

    if(bPStartAngle == bPStopAngle)
    {
        QMessageBox::about(this, tr("Message"), "Start and stop angles must be different.");
        return;
    }

    int tabIndex = this->tabWidget->currentIndex();
    int rotSide, traSide;

    if(bPRotTraFlag)
    {
        rotSide = allTabs[tabIndex].size();
        traSide  = allTabs[tabIndex][0]->Spectra.size();
    }
    else
    {
        traSide  = allTabs[tabIndex].size();
        rotSide = allTabs[tabIndex][0]->Spectra.size();
    }

    if(rotSide > traSide)
    {
        QMessageBox::about(this, tr("Message"), "The algorithm currently work only when rotation
                                                 side <= translation side.");
        return;
    }

    double bPStepAngle = (bPStopAngle-bPStartAngle)/(rotSide-1);
    QVector <double> anglesVec;

    if(bPReadAnglesFileFlag)
    {
        bPAnglesFileName = bPAnglesFileName.trimmed();
        QFile anglesFile(bPAnglesFileName);
        anglesFile.open(QIODevice::ReadOnly);
        QTextStream In(&anglesFile);

        if(!anglesFile.exists())
        {
            QMessageBox::about(this, tr("Message"), "The angles file `" + bPAnglesFileName +
                                                    "' does not exist.");
            return;
        }

        QString ANG;

        while (!In.atEnd())
        {
            In >> ANG;
            ANG.simplified();
            if(ANG.isEmpty())
                continue;

            anglesVec.push_back(ANG.toDouble()*PI/180.0);
        }
    }
    else
    {
        for(int a=0; a<rotSide; ++a)
            anglesVec.push_back((bPStartAngle+a*bPStepAngle)*PI/180.0);
    }

    if(anglesVec.size() != rotSide)
    {
        QMessageBox::about(this, tr("Message"), "Number of angles not equal to side size.");
        return;
    }

    int nn2 = 2 * pow(2, ceil(log(1.0*traSide)/log(2.0)));
    Vec_IO_DP dataVec(nn2);
    QVector <double> Filt;

    if(bPFFTFlag == 1)
    {
        if(bPFilterTypeNum == 0)
        {
            QMessageBox::about(this, tr("Message"), "Filter should be chosen to proceed.");
            return;
        }
        else if(bPFilterTypeNum == 1)
        {
            for(int uu=0; uu<traSide; ++uu)
                Filt.push_back(fabs(sin(-PI+uu*(2*PI/traSide))));
        }
        else if(bPFilterTypeNum == 2)
        {
            for(int uu=0; uu<traSide; ++uu)
                Filt.push_back(fabs((-PI+uu*(2*PI/traSide))));
        }
    }

    allTables[tabIndex]->bPChanFlag = bPChannelsFlag;

    if(bPChannelsFlag == 1 || bPChannelsFlag == 2)
    {
        QVector < QVector <double> > filtPR;

        filtPR.resize(traSide);

        if(bPChannelsFlag == 1)
        {
            if(bPRotTraFlag)
            {
                for(int aa=0; aa<rotSide; ++aa)
                    for(int bb=0; bb<traSide; ++bb)
                        filtPR[aa].push_back(allTabs[tabIndex][aa]->Spectra[bb].Intensity);
            }
            else
            {
                for(int aa=0; aa<traSide; ++aa)
                    for(int bb=0; bb<rotSide; ++bb)
                        filtPR[bb].push_back(allTabs[tabIndex][aa]->Spectra[bb].Intensity);
            }
        }
        else if(bPChannelsFlag == 2)
        {
            if(bPStartChannel > bPEndChannel)
            {
                QMessageBox::about(this, tr("Message"), "Wrong channels range.");
                return;
            }

            if(bPEndChannel >= minMaxNumOfChannels[tabIndex].first)
            {
                QMessageBox::about(this, tr("Message"), "Wrong channels range.\nCommon channel
                    index range is 0-" + QString::number(minMaxNumOfChannels[tabIndex].first-1)
                    + ".");
                return;
            }

            if(bPRotTraFlag)
            {
                for(int aa=0; aa<rotSide; ++aa)
                    for(int bb=0; bb<traSide; ++bb)
                    {
                        double sum = 0.0;

                        for(int cc=bPStartChannel; cc<=bPEndChannel; ++cc)
                            sum += allTabs[tabIndex][aa]->Spectra[bb].dataPoints[cc].y();

                        filtPR[aa].push_back(sum);
                    }
            }
            else
            {
                for(int aa=0; aa<traSide; ++aa)
                    for(int bb=0; bb<rotSide; ++bb)
                    {
                        double sum = 0.0;

                        for(int cc=bPStartChannel; cc<=bPEndChannel; ++cc)
                            sum += allTabs[tabIndex][aa]->Spectra[bb].dataPoints[cc].y();

                        filtPR[bb].push_back(sum);
                    }
            }
        }

        if(bPFFTFlag == 1)
        {
            if(bPFftDftFlag)
            {
                for(int hh=0; hh<filtPR.size(); ++hh)
                {
                    int gg;
                    long int KK = 0;
                    for(gg=0; gg<filtPR[hh].size(); ++gg)
                    {
                        dataVec[KK]   = filtPR[hh][gg];
                        dataVec[KK+1] = 0.0;
                        KK += 2;
                    }
                    for(int tt=KK; tt<dataVec.size(); ++tt)
                        dataVec[tt] = 0.0;

                    fourierFFT(dataVec, 1);

                    KK = 0;
                    for(gg=0; gg<filtPR[hh].size(); ++gg)
                    {
                        dataVec[KK]   = dataVec[KK] * Filt[gg];
                        dataVec[KK+1] = dataVec[KK+1] * Filt[gg];
                        KK += 2;
                    }

                    fourierFFT(dataVec, -1);

                    KK = 0;
                    for(gg=0; gg<filtPR[hh].size(); ++gg)
                    {
                        filtPR[hh][gg] = dataVec[KK]/(nn2/2.0);
                        KK += 2;
                    }
                }
            }

            else
            {
                for(int hh=0; hh<filtPR.size(); ++hh)
                {
                    Vec_DP x1(filtPR[hh].size()), y1(filtPR[hh].size());
                    int gg;
                    for(gg=0; gg<filtPR[hh].size(); ++gg)
                    {
                        x1[gg]   = filtPR[hh][gg];
                        y1[gg]   = 0.0;
                    }

                    myFFT(1, filtPR[hh].size(), x1, y1);

                    for(gg=0; gg<filtPR[hh].size(); ++gg)
                    {
                        x1[gg]   = x1[gg] * Filt[gg];
                        y1[gg]   = y1[gg] * Filt[gg];
                    }

                    myFFT(-1, filtPR[hh].size(), x1, y1);

                    for(gg=0; gg<filtPR[hh].size(); ++gg)
                    {
                        filtPR[hh][gg] = x1[gg];
                    }
                }
            }
        }

        double midIndex = (1.0*traSide+1.0)/2.0;

        QVector < QVector <double> > BPI, xpr, ypr, filtIndex, BPIa;
        BPI.resize(traSide);
        xpr.resize(traSide);
        ypr.resize(traSide);
        filtIndex.resize(traSide);
        BPIa.resize(traSide);

        for(int b=0; b<BPI.size(); ++b)
            for(int c=0; c<BPI.size(); ++c)
            {
                BPI[b].push_back(0.0);
                xpr[b].push_back(1.0*c-midIndex+1.0);
                ypr[b].push_back(1.0*b-midIndex+1.0);
                filtIndex[b].push_back(0.0);
                BPIa[b].push_back(0.0);
            }

        QVector < int > tempI;
        QVector < QPair <int, int> > pairI;

        QVector < QVector < QPair <int, int> > >spota;
        QVector < QVector < int > >nfi;

        for(unsigned int mm = 0 ; mm < anglesVec.size() ; mm++)
        {
            QVector < QVector <int> > F;
            spota.push_back (pairI);
            nfi.push_back (tempI);

            for(unsigned int i = 0 ; i < traSide ; i++)
            {
                F.push_back(tempI);

                for(unsigned int j = 0 ; j < traSide ; j ++)
                {
                    double filtidx = floor(0.5+(midIndex+xpr[i][j]*sin(anglesVec[mm])-ypr[i][j]
                                           *cos(anglesVec[mm])));
                    F[i].push_back(filtidx);
                }
            }

            for(int k = 0  ; k < traSide; k ++)
            {
                for(unsigned int l = 0 ; l < traSide ; l ++)
                {
                    if (F[l][k] > 0 && F[l][k] <= traSide)
                    {
                        QPair <int,int> tem;
                        tem.first = l;
                        tem.second = k;

                        spota[mm].push_back(tem);
                        nfi[mm].push_back(F[l][k]-1);
                    }
                }
            }
        }

        for(unsigned int ii = 0  ; ii < rotSide ; ii++)
        {
            for(unsigned int jj = 0; jj <  spota[ii].size(); ++jj)
            {
                int r = spota[ii][jj].first;
                int c = spota[ii][jj].second;
                int ll = nfi[ii][jj];
                BPI[r][c] += filtPR[ii][ll];
            }
        }

        bPColorFlag[tabIndex] = 1;
        bPTabFlag[tabIndex] = 0;
        tempIntensityStore[tabIndex].resize(0);
        tempIntensityStore[tabIndex].resize(allTabs[tabIndex].size());

        for(int ss=0; ss<allTabs[tabIndex].size(); ++ss)
            for(int tt=0; tt<allTabs[tabIndex][ss]->Spectra.size(); ++tt)
                tempIntensityStore[tabIndex][ss].push_back(allTabs[tabIndex][ss]->
                                                            Spectra[tt].Intensity);

        if(traSide != rotSide)
            tabResize(tabIndex);

        for(int JJ=0; JJ<allTabs[tabIndex].size(); ++JJ)
            for(int KK=0; KK<allTabs[tabIndex][JJ]->Spectra.size(); ++KK)
            {
                if(allTabs[tabIndex][JJ]->Spectra[KK].dataPoints.size() == 0)
                {
                    allTabs[tabIndex][JJ]->Spectra[KK].dataPoints.resize
                                                        (minMaxNumOfChannels[tabIndex].first);
                    for(int MM=0; MM<minMaxNumOfChannels[tabIndex].first; ++MM)
                    {
                        allTabs[tabIndex][JJ]->Spectra[KK].dataPoints[MM].setX(0.0);
                        allTabs[tabIndex][JJ]->Spectra[KK].dataPoints[MM].setY(0.0);
                    }
                }
            }

        for(int aa=0; aa<BPI.size(); ++aa)
            for(int bb=0; bb<BPI[aa].size(); ++bb)
                allTabs[tabIndex][aa]->Spectra[bb].Intensity = BPI[aa][bb];

        QString newname = "BP" + QString::number(tabIndex+1);
        this->tabWidget->setTabText(tabIndex, newname);

        this->update();
    }

    else
    {
        if(bPComChanMesFlag)
        {
            bPComChanMesFlag = 0;
            QMessageBox::about(this, tr("Message"), "Back projection will be carried out for
                            common channels only, i.e. 0-"
                            + QString::number(minMaxNumOfChannels[tabIndex].first-1) + ".");
        }

        QVector < QVector < QVector <double> > > allVec;
        double midIndex = (1.0*traSide+1.0)/2.0;

        QVector < QVector <double> > xpr, ypr, filtIndex, BPIa;
        xpr.resize(traSide);
        ypr.resize(traSide);
        filtIndex.resize(traSide);
        BPIa.resize(traSide);

        for(int b=0; b<traSide; ++b)
            for(int c=0; c<traSide; ++c)
            {
                xpr[b].push_back(1.0*c-midIndex+1.0);
                ypr[b].push_back(1.0*b-midIndex+1.0);
                filtIndex[b].push_back(0.0);
                BPIa[b].push_back(0.0);
            }

        QVector < int > tempI;
        QVector < QPair <int, int> > pairI;

        QVector < QVector < QPair <int, int> > >spota;
        QVector < QVector < int > >nfi;

        for(unsigned int mm = 0 ; mm < anglesVec.size() ; mm++)
        {
            QVector < QVector <int> > F;
            spota.push_back (pairI);
            nfi.push_back (tempI);

            for(unsigned int i = 0 ; i < traSide ; i++)
            {
                F.push_back(tempI);
                for(unsigned int j = 0 ; j < traSide ; j ++)
                {
                    double filtidx = floor(0.5+(midIndex+xpr[i][j]*sin(anglesVec[mm])-ypr[i][j]
                                           *cos(anglesVec[mm])));
                    F[i].push_back(filtidx);
                }
            }

            for(int k = 0  ; k < traSide; k ++)
            {
                for(unsigned int l = 0 ; l < traSide ; l ++)
                {
                    if (F[l][k] > 0 && F[l][k] <= traSide)
                    {
                        QPair <int,int> tem;
                        tem.first = l;
                        tem.second = k;
                        spota[mm].push_back(tem);
                        nfi[mm].push_back(F[l][k]-1);
                    }
                }
            }
        }

        QProgressDialog progress(this);
        progress.setLabelText(tr("Performing back projection for each individual channel in
                                                    current tab.").arg("Tab Back Projection"));
        progress.setValue(0);
        progress.setRange(0, minMaxNumOfChannels[tabIndex].first-1);
        progress.show();

        unsigned long int count = 0;

        for(int oo=0; oo<minMaxNumOfChannels[tabIndex].first; ++oo)
        {
            progress.setValue(count);
            qApp->processEvents();
            if(progress.wasCanceled())
                return;
            count++;

            QVector < QVector <double> > BPI;
            BPI.resize(traSide);
            for(int dd=0; dd<traSide; ++dd)
                for(int zz=0; zz<traSide; ++zz)
                    BPI[dd].push_back(0.0);

            QVector < QVector <double> > filtPR;

            filtPR.resize(rotSide);

            if(bPRotTraFlag)
            {
                for(int aa=0; aa<rotSide; ++aa)
                    for(int bb=0; bb<traSide; ++bb)
                        filtPR[aa].push_back(allTabs[tabIndex][aa]->
                                        Spectra[bb].dataPoints[oo].y());
            }
            else
            {
                for(int aa=0; aa<traSide; ++aa)
                    for(int bb=0; bb<rotSide; ++bb)
                        filtPR[bb].push_back(allTabs[tabIndex][aa]->
                                        Spectra[bb].dataPoints[oo].y());
            }

            if(bPFFTFlag == 1)
            {
                if(bPFftDftFlag)
                {
                    for(int hh=0; hh<filtPR.size(); ++hh)
                    {
                        int gg;
                        long int KK = 0;
                        for(gg=0; gg<filtPR[hh].size(); ++gg)
                        {
                            dataVec[KK]   = filtPR[hh][gg];
                            dataVec[KK+1] = 0.0;
                            KK += 2;
                        }
                        for(int tt=KK; tt<dataVec.size(); ++tt)
                            dataVec[tt] = 0.0;

                        fourierFFT(dataVec, 1);

                        KK = 0;
                        for(gg=0; gg<filtPR[hh].size(); ++gg)
                        {
                            dataVec[KK]   = dataVec[KK] * Filt[gg];
                            dataVec[KK+1] = dataVec[KK+1] * Filt[gg];
                            KK += 2;
                        }

                        fourierFFT(dataVec, -1);

                        KK = 0;
                        for(gg=0; gg<filtPR[hh].size(); ++gg)
                        {
                            filtPR[hh][gg] = dataVec[KK]/(nn2/2.0);
                            KK += 2;
                        }
                    }
                }

                else
                {
                    for(int hh=0; hh<filtPR.size(); ++hh)
                    {
                        Vec_DP x1(filtPR[hh].size()), y1(filtPR[hh].size());
                        int gg;
                        for(gg=0; gg<filtPR[hh].size(); ++gg)
                        {
                            x1[gg]   = filtPR[hh][gg];
                            y1[gg]   = 0.0;
                        }

                        myFFT(1, filtPR[hh].size(), x1, y1);

                        for(gg=0; gg<filtPR[hh].size(); ++gg)
                        {
                            x1[gg]   = x1[gg] * Filt[gg];
                            y1[gg]   = y1[gg] * Filt[gg];
                        }

                        myFFT(-1, filtPR[hh].size(), x1, y1);

                        for(gg=0; gg<filtPR[hh].size(); ++gg)
                        {
                            filtPR[hh][gg] = x1[gg];
                        }
                    }
                }
            }

            for(unsigned int ii = 0  ; ii < rotSide ; ii++)
            {
                for(unsigned int jj = 0; jj <  spota[ii].size(); ++jj)
                {
                    int r = spota[ii][jj].first;
                    int c = spota[ii][jj].second;
                    int ll = nfi[ii][jj];
                    BPI[r][c] += filtPR[ii][ll];
                }
            }

            for(int x=0; x<BPI.size(); ++x)
            {
                for(int y=0; y<BPI[x].size(); ++y)
                {
                    BPI[x][y] = BPI[x][y]/(anglesVec.size()*1.0);
                }
            }

            allVec.push_back(BPI);
        }

        if(traSide != rotSide)
            tabResize(tabIndex);

        for(int ww=0; ww<traSide; ++ww)
            for(int xx=0; xx<traSide; ++xx)
            {
                allTabs[tabIndex][ww]->Spectra[xx].bPDataPoints = allTabs[tabIndex][ww]->
                                                                    Spectra[xx].dataPoints;
                allTabs[tabIndex][ww]->Spectra[xx].bPDataPoints.resize(allVec.size());

                for(int dd=0; dd<allVec.size(); ++dd)
                {
                    allTabs[tabIndex][ww]->Spectra[xx].bPDataPoints[dd].setX(dd);
                    allTabs[tabIndex][ww]->Spectra[xx].bPDataPoints[dd].setY(allVec[dd][ww][xx]);
                }
            }

        QString newname = "BP" + QString::number(tabIndex+1);
        this->tabWidget->setTabText(tabIndex, newname);
        backProjectTabShift();
    }
}
\end{verbatim}
}}

\section{Channels Alignment Snippet} \label{SampleCodeCA}

{\singlespace {\scriptsize
\begin{verbatim}
void MainWindow::alignChann()
{
    if(allTabs.size()==0)
    {
        QMessageBox::about(this, tr("Message"), "There is no tab.");
        return;
    }

    int inVoDi, a, b, c, d, e;
    int tabIdx = tabWidget->currentIndex();
    QVector < int > inVoDiVec;

    createAlignChanDialog();
    if(!alignOKDialogFlag)
        return;

    int parIdx = parIdxAlign;
    parIdx--;

    if(infFacAlign < 1 || infFacAlign > 50 || disNumAlign < 0
       || disNumAlign > allTabs[tabIdx][0]->Spectra.size()*infFacAlign/2 || parIdx < 0)
    {
        QMessageBox::about(this, tr("Message"), "Inappropriate input data. Operation aborted!");
        return;
    }

    for(a=0; a<allTabs[tabIdx].size(); ++a)
        for(b=0; b<allTabs[tabIdx][a]->Spectra.size(); ++b)
            if(!allTabs[tabIdx][a]->Spectra[b].isRefined
               || allTabs[tabIdx][a]->Spectra[b].refinedPars.size() < parIdxAlign)
            {
                QMessageBox::about(this, tr("Message"), "Not all voxels are refined OR parameter
                                                            index is wrong. Operation aborted!");
                return;
            }

    QProgressDialog progress(this);
    progress.setLabelText(tr("Aligning Channels").arg("Aligning"));
    progress.setValue(0);
    progress.show();

    clock_t startProgram(clock());

    QVector < QVector < double > > parVec;
    parVec.resize(allTabs[tabIdx].size());
    double temCO;

    for(a=0; a<allTabs[tabIdx].size(); ++a)
        for(b=0; b<allTabs[tabIdx][a]->Spectra.size(); ++b)
        {
            temCO = qMin(cutOffAlign, allTabs[tabIdx][a]->Spectra[b].refinedPars[parIdx]);
            parVec[a].push_back(temCO);
        }

    int newNumCols = 1+infFacAlign*(allTabs[tabIdx][0]->Spectra.size()-1);

    QVector < QVector < double > > chanInt;
    chanInt.resize(parVec.size());
    for(a=0; a<chanInt.size(); ++a)
        chanInt[a].resize(newNumCols);

    for(a=0; a<parVec.size(); ++a)
    {
        for(b=0; b<parVec[a].size()-1; ++b)
        {
            chanInt[a][b*infFacAlign] = parVec[a][b];
            for(c=(b*infFacAlign)+1; c<((b+1)*infFacAlign); ++c)
            {
                chanInt[a][c] = chanInt[a][b*infFacAlign] + (parVec[a][b+1] - parVec[a][b])
                                * (c-(b*infFacAlign)) / infFacAlign;
            }
        }
        chanInt[a].last() = parVec[a].last();
    }

    double center = chanInt[0].size() /2.0;
    double cenMas = 0;
    double sum = 0;

    for(b=0; b<chanInt[0].size(); ++b)
    {
        cenMas += b*chanInt[0][b];
        sum += chanInt[0][b];
    }

    cenMas /= sum;
    inVoDi = (int)(cenMas - center);

    inVoDiVec.push_back(-inVoDi);

    if(inVoDi > 0)
    {
        for(b=0; b<inVoDi; ++b)
        {
            double val = chanInt[0][0];
            chanInt[0].remove(0);
            chanInt[0].push_back(val);
        }
    }
    else
    {
        for(b=0; b<abs(inVoDi); ++b)
        {
            double val = chanInt[0].last();
            chanInt[0].remove(chanInt[0].size()-1);
            chanInt[0].prepend(val);
        }
    }

    for(a=1; a<chanInt.size(); ++a)
    {
        QVector < QPair < double, int> > LSVec;
        QVector < double > chanIntC;
        chanIntC.resize(chanInt[a].size());

        for(b=0; b<chanIntC.size(); ++b)
            chanIntC[b] = chanInt[a][b];

        for(b=0; b<disNumAlign; ++b)
        {
            double val = chanIntC[0];
            chanIntC.remove(0);
            chanIntC.push_back(val);
        }

        for(b=-disNumAlign; b<=disNumAlign; ++b)
        {
            double lsSum = 0;

            for(c=0; c<chanIntC.size(); ++c)
            {
                lsSum += (chanInt[0][c] - chanIntC[c]) * (chanInt[0][c] - chanIntC[c]);
            }

            QPair <double, int> lsIdx;
            lsIdx.first = lsSum;
            lsIdx.second = b;
            LSVec.push_back(lsIdx);

            double val = chanIntC.last();
            chanIntC.remove(chanIntC.size()-1);
            chanIntC.prepend(val);
        }

        qSort(LSVec.begin(), LSVec.end());
        inVoDi = LSVec[0].second;

        inVoDiVec.push_back(inVoDi);

        if(inVoDi < 0)
        {
            for(b=0; b<abs(inVoDi); ++b)
            {
                double val = chanInt[a][0];
                chanInt[a].remove(0);
                chanInt[a].push_back(val);
            }
        }
        else
        {
            for(b=0; b<inVoDi; ++b)
            {
                double val = chanInt[a].last();
                chanInt[a].remove(chanInt[a].size()-1);
                chanInt[a].prepend(val);
            }
        }
    }

    quint32 minNumChan = 9999999;
    long int maxNumChan = -9999999;

    for(b=0; b<allTabs[tabIdx].size(); ++b)
    {
        for(c=0; c<allTabs[tabIdx][b]->Spectra.size(); ++c)
        {
            minNumChan = (minNumChan < allTabs[tabIdx][b]->Spectra[c].dataPoints.size()) ?
                                minNumChan : allTabs[tabIdx][b]->Spectra[c].dataPoints.size();
            maxNumChan = (maxNumChan > allTabs[tabIdx][b]->Spectra[c].dataPoints.size()) ?
                                maxNumChan : allTabs[tabIdx][b]->Spectra[c].dataPoints.size();
        }
    }

    progress.setRange(0, minNumChan);
    quint16 count = 0;

    for(b=0; b<minNumChan; ++b)
    {
        progress.setValue(count);
        qApp->processEvents();
        if(progress.wasCanceled())
            return;
        count++;

        for(c=0; c<allTabs[tabIdx].size(); ++c)
        {
            for(d=0; d<allTabs[tabIdx][c]->Spectra.size()-1; ++d)
            {
                chanInt[c][d*infFacAlign] = allTabs[tabIdx][c]->Spectra[d].dataPoints[b].y();
                for(e=(d*infFacAlign)+1; e<((d+1)*infFacAlign); ++e)
                {
                    chanInt[c][e] = chanInt[c][d*infFacAlign]
                                    + (allTabs[tabIdx][c]->Spectra[d+1].dataPoints[b].y()
                                    - allTabs[tabIdx][c]->Spectra[d].dataPoints[b].y())
                                    * (e-(d*infFacAlign)) / infFacAlign;
                }
            }
            chanInt[c].last() = allTabs[tabIdx][c]->Spectra.last().dataPoints[b].y();
        }

        for(c=0; c<inVoDiVec.size(); ++c)
        {
            if(inVoDiVec[c] < 0)
            {
                for(d=0; d<abs(inVoDiVec[c]); ++d)
                {
                    double val = chanInt[c][0];
                    chanInt[c].remove(0);
                    chanInt[c].push_back(val);
                }
            }
            else
            {
                for(d=0; d<inVoDiVec[c]; ++d)
                {
                    double val = chanInt[c].last();
                    chanInt[c].remove(chanInt[c].size()-1);
                    chanInt[c].prepend(val);
                }
            }
        }

        for(c=0; c<allTabs[tabIdx].size(); ++c)
        {
            for(d=0; d<(allTabs[tabIdx][c]->Spectra.size()-1); ++d)
            {
                allTabs[tabIdx][c]->Spectra[d].dataPoints[b].setY(0);

                for(e=0; e<infFacAlign; ++e)
                    allTabs[tabIdx][c]->Spectra[d].dataPoints[b].ry() +=
                                                                    chanInt[c][(d*infFacAlign)+e];
            }
            allTabs[tabIdx][c]->Spectra.last().dataPoints[b].setY(chanInt[c].last());
        }
    }

    progress.setValue(minNumChan);

    double elapsedProgram = (double)(clock() - startProgram) / CLOCKS_PER_SEC;
    QMessageBox::about(this, tr("Message"), "Total time (sec): " +QString::number(elapsedProgram));
}
\end{verbatim}
}}

%